\documentclass[10pt, journal]{IEEEtran}
\usepackage{cite}
\usepackage{amsmath,amssymb,amsfonts}
\usepackage{algorithmic}
\usepackage{graphicx}
\usepackage{textcomp}
\usepackage{xcolor}
\usepackage{amssymb}
\usepackage{indentfirst}
\usepackage{amsmath}
\usepackage{bm}
\usepackage{booktabs}
\usepackage{subfigure}
\usepackage{epsfig}
\usepackage{setspace}
\usepackage{algorithm}
\usepackage{changepage}
\usepackage{amsthm} 
\usepackage{soul,color}
\usepackage{multirow}
\usepackage{hyperref}
\newtheorem{theorem}{Theorem}
\newtheorem{lemma}{Lemma}
\newtheorem{proposition}{Proposition}
\newtheorem{corollary}{Corollary}
\theoremstyle{definition}
\newtheorem{remark}{Remark}

\IEEEaftertitletext{\vspace{-2\baselineskip}}
\ifCLASSINFOpdf
\else
\fi
\allowdisplaybreaks

\begin{document}
\title{A Revisit to Ordered Statistics Decoding:\\ Distance Distribution and Decoding Rules}

\author{Chentao~Yue,~\IEEEmembership{Student Member,~IEEE,}
        Mahyar~Shirvanimoghaddam,~\IEEEmembership{Senior Member,~IEEE,}
        Branka~Vucetic,~\IEEEmembership{Life Fellow,~IEEE,}
        and~Yonghui~Li,~\IEEEmembership{Fellow,~IEEE}% <-this % stops a space
\thanks{Chentao Yue, Mahyar Shirvanimoghaddam, Branka Vucetic, and Yonghui Li are with the School of Electrical and Information Engineering, the University of Sydney, NSW 2006, Australia (email:\{chentao.yue, mahyar.shm, branka.vucetic, yonghui.li\}@sydney.edu.au)

This paper was presented in part at IEEE International Symposium on Information Theory (ISIT), Paris, France, July 2019. This work was supported by the Australian Research Council through the Discovery Projects under Grants DP180100606 and DP190101988.

This paper has been accepted for publication by IEEE. DOI (identifier) 10.1109/TIT.2021.3078575. Copyright (c) 2017 IEEE. Personal use of this material is permitted.  However, permission to use this material for any other purposes must be obtained from the IEEE by sending a request to pubs-permissions@ieee.org.
}
}

\markboth{}%
{}



\maketitle

\begin{abstract}
This paper revisits the ordered statistics decoding (OSD). It provides a comprehensive analysis of the OSD algorithm by characterizing the statistical properties, evolution and the distribution of the Hamming distance and weighted Hamming distance from codeword estimates to the received sequence in the reprocessing stages of the OSD algorithm. We prove that the Hamming distance and weighted Hamming distance distributions can be characterized as mixture models capturing the decoding error probability and code weight enumerator. Simulation and numerical results show that our proposed statistical approaches can accurately describe the distance distributions. Based on these distributions and with the aim to reduce the decoding complexity, several techniques, including stopping rules and discarding rules, are proposed, and their decoding error performance and complexity are accordingly analyzed. Simulation results for decoding various eBCH codes demonstrate that the proposed techniques can significantly reduce the decoding complexity with a negligible loss in the decoding error performance.
\end{abstract}


% \fi
\IEEEpeerreviewmaketitle

\section{Introduction}

\IEEEPARstart{S}{ince} 1948, when Shannon introduced the notion of channel capacity \cite{Shannon}, researchers have been looking for powerful channel codes that can approach this limit. Low density parity check (LDPC) and Turbo codes have been shown to perform very close to the Shannon's limit at large block lengths and have been widely applied in the 3rd and 4th generations of mobile standards \cite{lin2004ECC}. The Polar code proposed by Arikan in 2008 \cite{polar2009} has attracted much attention in the last decade and has been chosen as the standard coding scheme for the fifth generation (5G) enhanced mobile broadband (eMBB) control channels and the physical broadcast channel. Polar codes take advantage of a simple successive cancellation decoder, which is optimal for asymptotically large code block lengths \cite{Mahyar2019ShortCode}. 
    
Short code design and the related decoding algorithms have rekindled a great deal of interest among industry and academia recently \cite{liva2016codeSurvey,2016Performancecomparison}. This interest was triggered by the stringent requirements of the new ultra-reliable and low-latency communications (URLLC) service for mission critical IoT (Internet of Things) services, including the hundred-of-microsecond time-to-transmit latency, block error rates of $10^{-5}$, and the bit-level granularity of the codeword size and code rate. These requirements mandate the use of short block-length codes; therefore, conventionally moderate/long codes may not be suitable \cite{Mahyar2019ShortCode}.
    
Several candidate channel codes such as LDPC, Polar, tail-biting convolutional code (TB-CC), and Turbo codes, have been considered for URLLC data channels \cite{Mahyar2019ShortCode}. While some of these codes perform closely to the Shannon's limit at asymptotically long block lengths, they usually suffer from performance degradation if the code length is short, e.g., Turbo codes with iterative decoding in short and moderate block lengths show a gap of more than 1 dB to the finite-length performance benchmark \cite{lin2004ECC}, where the benchmark is referred to as the error probability bound developed in \cite{PPV2010l} for finite block lengths. TB-CC can eliminate the rate loss of conventional convolutional codes due to the zero tail termination, but its decoding process is more complex than that of conventional codes \cite{Mahyar2019ShortCode}. Although LDPC codes have already been selected for eMBB data channels in 5G, recent investigations showed that there exist error floors for LDPC codes constructed using the base graph at high signal-to-noise ratios \cite{liva2016codeSurvey,Mahyar2019ShortCode} at moderate and short block lengths; hardly satisfying ultra-reliability requirements. Polar codes outperform LDPC codes with no error floor at short block lengths, but for short codes, it still falls short of the finite block length capacity bound \cite{Mahyar2019ShortCode}, i.e, the maximal channel coding
rate achievable at a given block length and error probability \cite{PPV2010l}.
    
Short Bose-Chaudhuri-Hocquenghem (BCH) codes have gained the interest of the research community recently \cite{NewOSD-5GNR, liva2016codeSurvey, Mahyar2019ShortCode,Chentao2019Hamming,Chentao2019SDD}, as they closely approach the finite length 
bound. As a class of powerful cyclic codes that are constructed using polynomials over finite fields \cite{lin2004ECC}, BCH codes have large minimum distances, but its maximum likelihood decoding is highly complex, introducing a significant delay at the receiver.
    
The ordered statistics decoding (OSD) was proposed in 1995, as an approximation of the maximum likelihood (ML) decoder for linear block codes \cite{Fossorier1995OSD} to reduce the decoding complexity. For a linear block code $\mathcal{C}(n,k)$, with minimum distance $d_{\mathrm{H}}$, it has been proven that an OSD with the order of $m = \lceil d_{\mathrm{H}}/4-1\rceil$ is asymptotically optimum approaching the same performance as the ML decoding \cite{Fossorier1995OSD}. However, the decoding complexity of an order-$m$ OSD can be as high as $O(k^m)$\cite{Fossorier1995OSD}. To meet the latency demands of the URLLC, OSD is being considered as a suitable decoding method for short block length BCH codes \cite{2012Segment,dhakal2016error,NewOSD-5GNR, Chentao2019SDD}. However, to make the OSD suitable for practical URLLC applications, the complexity issue needs to be addressed.

In OSD, the bit-wise log-likelihood ratios (LLRs) of the received symbols are sorted in descending order, and the order of the received symbols and the columns of the generator matrix are permuted accordingly. Gaussian elimination over the permuted generator matrix is performed to transform it to a systematic form. Then, the first $k$ positions, referred to as the most reliable basis (MRB), will be XORed with a set of the test error patterns (TEP) with the Hamming weight up to a certain degree, where the maximum Hamming weight of TEPs is referred to as the decoding order. Then the vectors obtained by XORing the MRB are re-encoded using the permuted generator matrix to generate candidate codeword estimates. This is referred to as the reprocessing and will continue until all the TEPs with the Hamming weights up to the decoding order are processed. Finally, the codeword estimate with the minimum distance from the received signal is selected as the decoding output.

Most of the previous work has focused on improving OSD and some significant progress has been achieved. Some published papers considered the information outside of the MRB positions to either improve the error performance or reduce complexity \cite{Fossorier2007OSDbias,wu2007preprocessing_and_diversification,Wu2007OSDMRB,2012Segment,jin2006probabilisticConditions,NewOSD-5GNR}. The approach of decoding using different biased LLR values was proposed in \cite{Fossorier2007OSDbias} to refine the error performance of low-order OSD algorithms. This decoding approach performs reprocessing for several iterations with different biases over LLR within MRB positions and achieves a better decoding error performance than the original low-order OSD. However, extra decoding complexity is introduced through the iterative process. Skipping and stopping rules were introduced in \cite{wu2007preprocessing_and_diversification} and \cite{Wu2007OSDMRB} to prevent unpromising candidates, which are unlikely to be the correct output. The decoder in \cite{wu2007preprocessing_and_diversification} utilizes two preprocessing rules and a multibasis scheme to achieve the same error rate performance as an order-$(w+2)$ OSD, but with the complexity of an order-$w$ OSD. This algorithm decomposes a TEP by a sub-TEP and an unit vector, and much additional complexity is introduced in processing sub-TEPs. Authors in \cite{Wu2007OSDMRB} proposed a skipping rule based on the likelihood of the current candidate, which significantly reduces the complexity. An order statistics based list decoding proposed in \cite{2012Segment} cuts the MRB to several partitions and performs independent OSD over each of them to reduce the complexity, but it overlooks the candidates generated across partitions and suffers a considerable error performance degradation. A fast OSD algorithm which combines the discarding rules from \cite{Wu2007OSDMRB} and the stopping criterion from \cite{jin2006probabilisticConditions} was proposed in \cite{NewOSD-5GNR}, which can reduce the complexity from $O(k^m)$ to $O(k^{m-2})$ at high signal-to-noise ratios (SNRs). The latest improvement of OSD is the Segmentation-Discarding Decoding (SDD) proposed in \cite{Chentao2019SDD}, where a segmentation technique is used to reduce the frequency of checking the stopping criterion and a group of candidates can be discarded with one condition check satisfied. Some papers also utilized the information outside MRB to obtain further refinement \cite{FossorierBoxandMatch,Fossorier2002IISR}. The Box-and-Match algorithm (BMA) approach can significantly reduce the decoding complexity by using the ``match'' procedure \cite{FossorierBoxandMatch}, which defines a control band (CB) and identifies each TEP based on CB, and the searching and matching of candidates are implemented by memory spaces called ``boxes''. However, BMA introduces a considerable amount of extra computations in the ``match'' procedure and it is not convenient to implement. The iterative information set reduction (IISR) technique was proposed in \cite{Fossorier2002IISR} to reduce the complexity of OSD. IISR applies a permutation over the positions around the boundary of MRB and generates a new MRB after each reprocessing. This technique can reduce the complexity with a slight degradation of the error performance and has the potential to be combined with other techniques mentioned above.
	    
Many of the above approaches utilize the distance from the codeword estimates to the received symbols, either Hamming or weighted Hamming distance, to design their techniques. For example, there is a distance-based optimal condition designed in the BMA \cite{FossorierBoxandMatch}, where the reprocessing rule is designed based on the distance between sub-TEPs and received symbols in \cite{wu2007preprocessing_and_diversification}, and skipping and stopping rules introduced in \cite{wu2007preprocessing_and_diversification} and \cite{Wu2007OSDMRB} are also designed based on the distance, etc. Despite the improvements in decoding complexity, these algorithms lack a rigorous error performance and complexity analysis. Till now, it is still unclear how the Hamming distance or the weighted Hamming distance evolves during the reprocessing stage of the OSD algorithm. Although some attempts were made to analyze the error performance of the OSD algorithm and its alternatives \cite{Fossorier1995OSD,Fossorier2002ErrorAnalysis,fossorier1996first,dhakal2016error}, the Hamming distance and weighted Hamming distance were left unattended. If the evolution of the Hamming distance and weighted Hamming distance in the reprocessing stage are known, more insights of how those decoding approaches improve the decoding performance could be obtained. Furthermore, those decoding conditions can be designed in an optimal manner and their performance and complexity can be analyzed more carefully.
	
In this paper, we revisit the OSD algorithm and investigate the statistical distribution of both Hamming distance and weighted Hamming distance between codeword estimates and the received sequence in the reprocessing stage of OSD. With the knowledge of the distance distribution, several decoding techniques are proposed and their complexity and error performance are analyzed. The main contributions of this work are summarized below. 
\begin{itemize}
\item We derive the distribution of the Hamming distance in the 0-reprocessing of OSD and extend the result to any order $i$-reprocessing by considering the ordered discrete statistics. We verify that the distribution of the Hamming distance can be described by a mixed model of two random variables related to the number of channel errors and the code weight enumerator, respectively, and the weight of the mixture is determined by the channel condition in terms of signal-to-noise ratio (SNR). Simulation and numerical results show that the proposed statistical approach can describe the distribution of Hamming distance of any order reprocessing accurately. In addition, the normal approximation of the Hamming distance distribution is derived.

\item We derive the distribution of the weighted Hamming distance in the 0-reprocessing of OSD and extend the result to any order $i$-reprocessing by considering the ordered continuous statistics. It is shown that the weighted Hamming distribution is also a mixture of two different distributions, determined by the error probability of the ordered sequence and the code weight enumerator, respectively. The exact expression of the weighted Hamming distribution is difficult to calculate numerically due to a large number of integrals, thus a normal approximation of the weighted Hamming distance distribution is introduced. Numerical and simulation results verify the tightness of the approximation.
	    
\item Based on the distance distributions, we propose several decoding techniques. Based on the Hamming distance, a hard individual stopping rule (HISR), a hard group stopping rule (HGSR), and a hard discarding rule (HDR) are proposed and analyzed. It can be indicated that in OSD, the Hamming distance can also be a good metric of the decoding quality. Simulation results show that with the proposed hard rules, the decoding complexity can be reduced with a slight degradation in the error performance. Based on the weighted Hamming distance distribution, soft decoding techniques, namely the soft individual stopping rule (SISR), the soft group stopping rule (SGSR), and the soft discarding rule (SDR) are proposed and analyzed. Compared with hard rules, these soft rules are more accurate to identify promising candidates and determine when to terminate the decoding with some additional complexity. For different performance-complexity trade-off requirements, the above decoding techniques (hard rules and soft rules) can be implemented with a suitable parameter selection.

\item We further show that when the code has a binomial-like weight spectrum, the proposed techniques can be implemented with linear or quadratic complexities in terms of the message length. Accordingly, the overall asymptotic complexity of OSD employing the proposed techniques is analyzed. Simulations show that the proposed techniques outperform the state of the art in terms of the TEP-reduction capability and the run-time of decoding a single codeword.
\end{itemize}
	
The rest of this paper is organized as follows. Section \ref{Sec::Preliminaries} describes the preliminaries of OSD. In Section \ref{sec::OrderStat}, statistical approaches are introduced for analyzing ordered sequences in OSD. The Hamming distance and weighted Hamming distance distributions are introduced and analyzed in Sections \ref{sec::HDdis} and \ref{sec::WHD}, respectively. Then, the hard and soft decoding techniques are proposed and analyzed in Section \ref{sec::HDdistech} and \ref{Sec::SoftTech}, respectively. Section \ref{sec::Discussion} discusses the practical implementation and complexities of the proposed techniques. Finally, Section \ref{Sec::Conclusion} concludes the paper.
    
\emph{Notation}: In this paper, we use an uppercase letter, e.g., $X$, to represent a random variable and $[X]_u^v$ to denote a sequence of random variables, i.e., $[X]_u^v = [X_u,X_{u+1},\ldots,X_v]$. Lowercase letters like $x$ are used to indicate the values of scalar variables or the sample of random variables, e.g., $x$ is a sample of random variable $X$. The mean and variance of a random variable $X$ is denoted by $\mathbb{E}[X]$ and $\sigma_{X}^2$, respectively. The probability density function ({$\mathrm{pdf}$}) and cumulative distribution function ($\mathrm{cdf}$) of a continuous random variable $X$ are denoted by $f_{X}(x)$ and $F_{X}(x)$, respectively, and the probability mass function ($\mathrm{pmf}$) of a discrete random variable $Y$ is denoted by $p_{Y}(y) \triangleq \mathrm{Pr}(Y=y)$, where $\mathrm{Pr}(\cdot)$ is the probability of an event. Unless otherwise specified, we use $f_{X}(x|Z=z)$ to denote the conditional $\mathrm{pdf}$ of a continuous random variable $X$ conditioning on the event $\{Z=z\}$, and accordingly the conditional means and variances of $X$ are denoted by $\mathbb{E}[X|Z\!=\!z]$ and $\sigma_{X|Z\!=\!z}^2$, respectively. Similarly, the conditional $\mathrm{pmf}$ of a discrete variable $Y$ are represented as $p_{Y}(y|Z\!=\!z)$. We use a bold letter, e.g., $\mathbf A$, to represent a matrix, and a lowercase bold letter, e.g., $\mathbf{a}$, to denote a row vector. We also use $[a]_u^v$ to denote a row vector containing element $a_{\ell}$ for $u\le \ell\le v$, i.e., $[a]_u^v = [a_u,a_{u+1},\ldots,a_v]$. We use superscript $^\mathrm{T}$ to denote the transposition of a matrix or vector, e.g., $\mathbf A^\mathrm{T}$ and $\mathbf a^\mathrm{T}$, respectively. Furthermore, we use a calligraphic uppercase letter to denote a probability distribution, e.g., binomial distribution $\mathcal{B}(n,p)$ and normal distribution $\mathcal{N}(\mu,\sigma^2)$, or a set, e.g., $\mathcal{A}$. In particular, $\mathbb{N}$ denotes the set of all natural numbers.

\section{Preliminaries} \label{Sec::Preliminaries}
We consider a binary linear block code $\mathcal{C}(n,k)$ with binary phase shift keying (BPSK) modulation over an additive white Gaussian Noise (AWGN) channel, where $k$ and $n$ denote the information block and codeword length, respectively. Let $ \mathbf{b} = [b]_{1}^k$ and $\mathbf{c} = [c]_{1}^n$ denote the information sequence and codeword, respectively. Given the generator matrix $\mathbf{G}$ of code $\mathcal{C}(n,k)$, the encoding operation can be described as $\mathbf{c} = \mathbf{b}\cdot\mathbf{G}$. At the channel output, the received signal (also referred to as the noisy signal) is given by $\mathbf{r} = \mathbf{s} + \mathbf{w}$, where $\mathbf{s} = [s]_{1}^n$ denotes the sequence of modulated symbols with $s_{u} = (-1)^{c_{u}}\in \{\pm 1\}$, $1\leq u \leq n$, and $\mathbf{w}= [w]_{1}^n$ is the AWGN vector with zero mean and variance $N_0/2$, for $N_0$ being the single side-band power spectrum density. The signal-to-noise ratio (SNR) is then given by $\gamma=2/N_0$.

At the receiver, the bit-wise hard decision vector $\mathbf{y}= [y]_{1}^n$ can be obtained according to the following rule:
	\begin{equation} \label{equ::Prelim::HD rule}
    	y_{u}=
    	\begin{cases}
    		1,& \text{for} \ r_{u}<0, 1\leq u \leq n\\
    		0,& \text{for} \ r_{u}\geq 0, 1\leq u \leq n
    	\end{cases}
	\end{equation}
	where $y_{u}$ is the hard-decision estimation of codeword bit $c_{u}$.

    In general, if the codewords in $\mathcal{C}(n,k)$ have equal transmission probability, the log-likelihood-ratio (LLR) of the $u$-th symbol of the received signal can be calculated as $l_{u} \triangleq \ln \frac{\mathrm{Pr}(c_{u}=1|r_{u})}{\mathrm{Pr}(c_{u}=0|r_{u})}$, which can be further simplified to $l_{u} = 4r_{u}/N_{0}$ if BPSK symbols are transmitted. We consider the scaled magnitude of LLR as the reliability corresponding to bitwise decision, defined by $\alpha_{u} = |r_{u}|$, where $|\cdot|$ is the absolute operation. Utilizing the bit reliability, the soft-decision decoding can be effectively conducted using the OSD algorithm \cite{Fossorier1995OSD}. In OSD, a permutation $\pi_1$ is performed to sort the received signal $\mathbf{r}$ and the corresponding columns of the generator matrix in descending order of their reliabilities. The sorted received symbols and the sorted hard-decision vector are denoted by $\mathbf{r}^{(1)}=\pi_{1}(\mathbf{r})$ and $\mathbf{y}^{(1)}=\pi_{1}(\mathbf{y})$, respectively, and the corresponding reliability vector and permuted generator matrix are denoted by $\bm{\alpha}^{(1)}=\pi_{1}(\bm{\alpha})$ and $\mathbf{G}^{(1)} = \pi_1(\mathbf{G})$, respectively.

    Next, the systematic form matrix $\mathbf{\widetilde G} = [\mathbf{I}_k \  \mathbf{\widetilde{P}}]$ is obtained by performing Gaussian elimination on $\mathbf{G}^{(1)}$, where $\mathbf{I}_k$ is a $k$-dimensional identity matrix and $\mathbf{\widetilde{P}}$ is the parity sub-matrix. An additional permutation $\pi_{2}$ may be performed during Gaussian elimination to ensure that the first $k$ columns are linearly independent. The permutation $\pi_{2}$ will inevitably disrupt the descending order property of $\bm{\alpha}^{(1)}$ to some extent; nevertheless, it has been shown that the disruption is minor\cite{Fossorier1995OSD}. Accordingly, the received symbols, the hard-decision vector, the reliability vector, and the generator matrix are sorted to $\widetilde{\mathbf{r}} = \pi_{2}(\pi_{1}(\mathbf{r}))$, $\widetilde{\mathbf{y}} = \pi_{2}(\pi_{1}(\mathbf{y}))$, $\widetilde{\bm{\alpha}} = \pi_{2}(\pi_{1}(\bm{\alpha}))$, and $\mathbf{\widetilde G} = \pi_{2}(\pi_{1}(\mathbf{G}))$, respectively.

    After the Gaussian elimination and permutations, the first $k$ index positions of $\widetilde{\mathbf{y}}$ are associated with the MRB \cite{Fossorier1995OSD}, which is denoted by $\widetilde{\mathbf{y}}_{\mathrm{B}} =[\widetilde y]_1^k$, and the rest of positions are associated with the redundancy part. A test error pattern $\mathbf{e} = [e]_1^k$ is added to $\widetilde{\mathbf{y}}_{\mathrm{B}}$ to obtain one codeword estimate by re-encoding as follows.
	\begin{equation}
		\widetilde{\mathbf{c}}_{\mathbf{e}} = \left(\widetilde{\mathbf{y}}_{\mathrm{B}}\oplus \mathbf{e}\right)\mathbf{\widetilde G} = \left[\widetilde{\mathbf{y}}_{\mathrm{B}}\oplus \mathbf{e} \  ~\left(\widetilde{\mathbf{y}}_{\mathrm{B}}\oplus \mathbf{e}\right)\mathbf{\widetilde{P}}\right] ,
	\end{equation}
	where $\widetilde{\mathbf{c}}_{\mathbf{e}} = [\widetilde{c}_{\mathbf{e}}]_1^n$ is the ordered codeword estimate with respect to TEP $\mathbf{e}$. 

	In OSD, TEPs are checked in increasing order of their Hamming weights; that is, in the $i$-reprocessing, all TEPs of Hamming weight $i$ will be generated and re-encoded. The maximum Hamming weight of TEPs is limited to $m$, which is referred to as the decoding order of OSD. Thus, for an order-$m$ decoding, maximum $\sum_{i=0}^{m}\binom{k}{i}$ TEPs will be re-encoded to find the best codeword estimate. For BPSK modulation, finding the best ordered codeword estimate $\widetilde{\mathbf{c}}_{opt}$ is equivalent to minimizing the weighted Hamming distance (WHD) between $\widetilde{\mathbf{c}}_{\mathbf{e}}$ and $\widetilde{\mathbf{y}}$, which is defined as \cite{valembois2002comparison}
	\begin{equation} \label{equ::Prelim::WHD_define}
		 d^{(\mathrm{W})}(\widetilde{\mathbf{c}}_{\mathbf{e}},\widetilde{\mathbf{y}}) \triangleq \sum_{\substack{1 \leq u \leq n \\ \widetilde{c}_{\mathbf{e},u}\neq \widetilde{y}_{u}}} \widetilde{\alpha}_{u}.
	\end{equation}
    Here, we also define the Hamming distance between $\widetilde{\mathbf{c}}_{\mathbf{e}}$ and $\widetilde{\mathbf{y}}$ as
	\begin{equation} \label{equ::Prelim::HDis_define}
	    d^{(\mathrm{H})}(\widetilde{\mathbf{c}}_{\mathbf{e}},\widetilde{\mathbf{y}}) \triangleq || \widetilde{\mathbf{c}}_{\mathbf{e}}\oplus\widetilde{\mathbf{y}} ||,
	\end{equation}
    where $||\cdot||$ is the $\ell_1$-norm. For simplicity of notations, we denote the WHD and Hamming distance between $\widetilde{\mathbf{c}}_{\mathbf{e}}$ and $\widetilde{\mathbf{y}}$ by $d^{(\mathrm{W})}_{\mathbf{e}} = d^{(\mathrm{W})}(\widetilde{\mathbf{c}}_{\mathbf{e}},\widetilde{\mathbf{y}})$ and $d_{\mathbf{e}}^{(\mathrm{H})} = d^{(\mathrm{H})}(\widetilde{\mathbf{c}}_{\mathbf{e}},\widetilde{\mathbf{y}})$, respectively. Furthermore, we alternatively use $w(\mathbf{e})$ to denote the Hamming weight of a binary vector $\mathbf{e}$, e.g., $w(\mathbf{e}) = ||\mathbf{e}||$.  Finally, the estimate $\hat{\mathbf{c}}_{opt}$ corresponding to the initial received sequence $\mathbf{r}$, is obtained by performing inverse permutations over $\widetilde{\mathbf{c}}_{opt}$, i.e.
	$\hat{\mathbf{c}}_{opt} = \pi_{1}^{-1}(\pi_{2}^{-1}(\widetilde{\mathbf{c}}_{opt}))$.

\section{Ordered Statistics in OSD} \label{sec::OrderStat}

\subsection{Distributions of received Signals}
For the simplicity of analysis and without loss of generality, we assume an all-zero codeword from $\mathcal{C}(n,k)$ is transmitted. Thus, the $u$-th symbol of the AWGN channel output $\mathbf{r}$ is given by $r_u = 1 + w_u$, $1\leq u\leq n$. Channel output $\mathbf{r}$ is observed by the receiver and the bit-wise reliability is then calculated as $\alpha_u = |1+w_u|$, $1\leq u \leq n$.  Let us consider the $u$-th reliability as a random variable denoted by $A_u$, then the sequence of random variables representing the reliabilities is denoted by $[A]_1^n$. Accordingly, after the permutations, the random variables of ordered reliabilities $\widetilde{\bm{\alpha}} = [\widetilde\alpha]_1^n$ are denoted by $[\widetilde{A}]_1^n$. Similarly, let $[R]_1^n$ and $[\widetilde R]_1^n$ denote sequences of random variables representing the received symbols before and after permutations, respectively. Note that $[A]_1^n$ and $[R]_1^n$ are two sequences of independent and identically distributed (i.i.d.) random variables. Thus, the $\mathrm{pdf}$ of $R_u$, $1 \leq u \leq {n}$, is given by
        \begin{equation} \label{equ::OrderStat::pdfofR}
            f_R(r) = \frac{1}{\sqrt{\pi N_0}}e^{-\frac{(r-1)^2}{N_0}} ,
        \end{equation}	
        and the $\mathrm{pdf}$ of $A_u$, $1 \leq u \leq {n}$, is given by
        \begin{equation} 
            f_{A}(\alpha)=
            \begin{cases}
                0,                           & \text{if} \ \alpha<0,\\
                \frac{e^{-\frac{(\alpha+1)^2}{N_0}}}{\sqrt{\pi N_0}} + \frac{e^{-\frac{(\alpha-1)^2}{N_0}}}{\sqrt{\pi N_0}},        & \text{if} \ \alpha\geq 0.
            \end{cases}
        \end{equation}
        Given the $Q$-function defined by $Q(x) \!=\! \frac{1}{\sqrt{2\pi }} \!\int _{x}^{\infty} \exp(\!-\frac{u^2}{2})du$, the $\mathrm{cdf}$ of $A_u$ can be derived as 
        \begin{equation} 
        	F_{A}(\alpha)=
        	\begin{cases}
        		0,                           & \text{if} \ \alpha<0,\\
        		1 -  Q(\frac{\alpha+1}{\sqrt{N_0/2}}) -  Q(\frac{\alpha-1}{\sqrt{N_0/2}}),       & \text{if} \ \alpha\geq 0.
        	\end{cases}
        \end{equation}
        
        By omitting the second permutation in Gaussian elimination, the $\mathrm{pdf}$ of the $u$-th order reliability $\widetilde{A}_u$ can be derived as \cite{papoulis2002probability}
        \begin{equation}  \label{equ::OrderStat::pdfAu}
        \begin{split}
           	f_{\widetilde{A}_u}(\widetilde{\alpha}_u) =& \frac{n!}{(u-1)!(n-u)!}\\
           	\cdot & (1-F_{A}(\widetilde{\alpha}_u))^{u-1}  F_{A}(\widetilde{\alpha}_u)^{n-u}  f_{A}(\widetilde{\alpha}_u) .         
        \end{split}
        \end{equation}
        For simplicity, the permutation $\pi_2$ is omitted in the subsequent analysis in this paper, since the influence of $\pi_2$ in OSD is minor\footnote{The second permutation $\pi_2$ occurs only when the first $k$ columns of $\pi_1(\mathbf{G})$ are not linearly independent. As shown in \cite[Eq. (59)]{Fossorier1995OSD}, the probability that permutation $\pi_2$ is occurring is very small. Also, even if $\pi_2$ occurs, the number of operations of $\pi_2$ is much less than the number of operations of $\pi_1$ \cite{Fossorier1995OSD}. Therefore, we omit $\pi_2$ in the following analysis for simplicity.}. Similar to (\ref{equ::OrderStat::pdfAu}), the joint $\mathrm{pdf}$ of $\widetilde{A}_u$ and $\widetilde{A}_v$, $1\leq u<v \leq n$, can be derived as follows.
        \begin{equation} 
        \label{equ::OrderStat::jointpdfAij}
        	\begin{split}		
        		f_{\widetilde{A}_u,\widetilde{A}_v}(\widetilde{\alpha}_u,\widetilde \alpha_v) =& \frac{n!}{(u-1)!(v-u-1)!(n-v)!} \\
        		\cdot & (1\!-\! F_{A}(\widetilde{\alpha}_u)) ^{u\!-\!1} \left(F_{A}(\widetilde{\alpha}_u) \!-\! F_{A}(\widetilde \alpha_v)\right)^{v\!-\!u\!-\!1}  \\
        		 \cdot & F_{A}(\widetilde \alpha_v)^{n-v} f_{A}(\alpha_u) f_{A}(\widetilde \alpha_v)  \mathbf{1}_{[0,\widetilde{\alpha}_u]}(\widetilde \alpha_v) ,
        	\end{split}
        \end{equation}	
        where $\mathbf{1}_{\mathcal{X}}(x) = 1$ if $x \in \mathcal{X}$ and $\mathbf{1}_{\mathcal{X}}(x) = 0$, otherwise. For the sequence of ordered received signals $[\widetilde R]_1^n$, the $\mathrm{pdf}$ of $\widetilde{R}_i$ and the joint $\mathrm{pdf}$ of $\widetilde{R}_i$ and $\widetilde{R}_j$, $0 \leq u<v \leq n$,  are respectively given by
        \begin{equation}  \label{equ::OrderStat::pdfRi}
        \begin{split}
            f_{\widetilde R_u}(\widetilde r_u) =& \frac{n!}{(u-1)!(n-u)!} \\
            \cdot &(1-F_{A}(|\widetilde r_u|))^{u-1}   F_{A}(|\widetilde r_u|)^{n-u}  f_R(\widetilde r_u),        
        \end{split}
        \end{equation}
        and
        \begin{equation} 
        \label{equ::OrderStat::jointpdfRij}
        	\begin{split}		
        		f_{\widetilde R_u,\widetilde R_v}(\widetilde r_u,\widetilde r_v) =& \frac{n!}{(u-1)!(v-u-1)!(n-v)!} \\
        		 \cdot& (1\!-\!F_{A}(|\widetilde r_u|)) ^{u\!-\!1} (F_{A}(|\widetilde r_u|) \!-\! F_{A}(|\widetilde r_v|))^{v\!-\!u\!-\!1}  \\
        		 \cdot&  F_{A}(|\widetilde r_v|)^{n-v} f_{R}(\widetilde r_u) f_{R}(\widetilde r_v)  \mathbf{1}_{[0,|\widetilde r_u|]}(|\widetilde r_v|) .
        	\end{split}
        \end{equation}	
        
    \begin{table*} [t]
    \centering
    \begin{minipage}{\textwidth}
        \begin{equation}  \label{equ::OrderStat::Eab}
        	p_{E_a^b}(j) = 
        		\begin{cases}
        			\displaystyle\int_{0}^{\infty}\!\!\!\int_{0}^{\infty}\!\! \binom{b-a+1}{j} p(x,y)^j (1 - p(x,y))^{b\!-\!a\!+\!1\!-\!j} f_{\widetilde{A}_{a-1},\widetilde{A}_{b+1} }(x,y)dydx , \  & \textup{for $a > 1$  and  $b < n$} \\
    	            \displaystyle\int_{0}^{\infty}\binom{n-a+1}{j} p(x,0)^j (1 - p(x,0))^{n-a+1-j} f_{\widetilde{A}_{a-1}}(x)dx , \  &  \textup{for $a > 1$  and  $b = n$}\\        			
    	            \displaystyle\int_{0}^{\infty}\binom{b}{j} p(\infty,y)^j (1 - p(\infty,y))^{b-j} f_{\widetilde{A}_{b+1} }(y)dy , \ &  \textup{for $a = 1$  and  $b < n$} \\
    	            \displaystyle \binom{n}{j} \left(1 - Q\left(\frac{-2}{\sqrt{2N_0}}\right)\right)^j Q\left(\frac{-2}{\sqrt{2N_0}}\right)^{n-j} , \  & \textup{for $a = 1$  and  $b = n$}\\
        		\end{cases}
    	\end{equation}		
    \medskip
    \hrule
    \end{minipage}
    \end{table*}
    
    In OSD, the ordered received sequence is divided into MRB and redundancy parts as defined in Section \ref{Sec::Preliminaries}. Then, the reprocessing re-encodes the MRB bits with TEPs to generate entire codeword estimates with redundancy bits. Thus, it is necessary to find the number of errors within these two parts (i.e., MRB and the redundancy part) separately, since they will affect the distance between codeword estimates and the received sequence in different ways, which will be further investigated in the subsequent sections. First of all, the statistics of the number of errors in the ordered hard-decision vector $\widetilde{\mathbf{y}}$ is summarized in the following Lemma.
    
    \begin{lemma}  \label{lem::OrderStat::Eab}
    Let random variable $E_a^b$ denote the number of errors in the positions from $a$ to $b$, $1\leq a < b \leq n$ over the ordered hard-decision vector $\widetilde{\mathbf{y}}$. The probability mass function $p_{E_a^b}(j)$ of $E_a^b$, for $0\!\leq\! j\! \leq\! b-a+1$, is given by (\ref{equ::OrderStat::Eab}) on the top of this page, where $f_{\widetilde{A}_a}(x)$ and $f_{\widetilde{A}_a,\widetilde{A}_b}(x,y)$ are given by (\ref{equ::OrderStat::pdfAu}) and (\ref{equ::OrderStat::jointpdfAij}), respectively, and $p(x,y)$ is given by
        	\begin{equation}  \label{equ::OrderStat::Pxy::quote}
        		p(x,y) = \frac{Q(\frac{-2x-2}{\sqrt{2N_0}}) -  Q(\frac{-2y-2}{\sqrt{2N_0}}) }{Q(\frac{-2x-2}{\sqrt{2N_0}}) -  Q(\frac{-2y-2}{\sqrt{2N_0}})  + Q(\frac{2y-2}{\sqrt{2N_0}}) -  Q(\frac{2x-2}{\sqrt{2N_0}}) } .
        	\end{equation}	
    	\end{lemma}  
        \begin{IEEEproof}
            
            Let us first consider the case when $a > 1$ and $b < n$, and other cases can be easily extended. Asssume the $(a-1)$-th and $(b+1)$-th ordered reliabilities are given by $\widetilde{A}_{a-1}  = x$ and $\widetilde{A}_{b+1}  =y$, respectively. Then, it can be obtained that the ordered received symbols $[\widetilde{R}_a,\widetilde{R}_{a+1},\ldots,\widetilde{R}_b] = [\widetilde{R}]_a^b$ satisfy
            \begin{equation}
                x\geq|\widetilde{R}_a| \geq |\widetilde{R}_{a+1}| \geq \ldots \geq |\widetilde{R}_{b-1}| \geq |\widetilde{R}_{b}| \geq y.
            \end{equation}
            
            Because $[\widetilde{R}]_1^n$ is obtained by permuting $[R]_1^n$, these $b-a+1$ ordered random variables $[\widetilde{R}]_a^b$ uniquely correspond to $b-a+1$ unsorted random variables $[R_{\ell_a},R_{\ell_{a+1}}\ldots,R_{\ell_b}] = [R_\ell]_a^b$. In other words, for an $\widetilde{R}_{u}$, $a\leq u \leq b$, there exists an $R_{\ell_u}$, $1\leq \ell_u \leq n$, that satisfies $\widetilde{R}_{u} = R_{\ell_u}$.

            From the correspondence, there are $b-a+1$ unsorted reliabilities $[R_\ell]_a^b \in [R]_1^n$ satisfying $x \geq |R_{\ell_u}| \geq y$, where $1\leq \ell_u \leq n$ and $a<u<b$. Because $[R]_1^n$ are i.i.d. random variables, for an arbitrary $R_{\ell_u}\in [R_\ell]_a^b$, the probability that $R_{\ell_u}$ results in an incorrect bit in $[y_\ell]_a^b$ conditioning on $\widetilde{A}_{a-1}  = x$ and $\widetilde{A}_{b+1}  =y$ is given by 
            \begin{equation}
                p(x,y) = \frac{\mathrm{Pr}(-x \leq R_{\ell_u} \leq -y)}{\mathrm{Pr}(-x \leq R_{\ell_u} \leq -y) + \mathrm{Pr}(y \leq R_{\ell_u} \leq x) }.
            \end{equation}
            
            It can be seen that $\mathrm{Pr}(-x \leq R_{\ell_u} \leq -y) = Q(\frac{-2x-2}{\sqrt{2N_0}}) - Q(\frac{-2y-2}{\sqrt{2N_0}})$ and $\mathrm{Pr}(y \leq R_{\ell_u} \leq x) = Q(\frac{2y-2}{\sqrt{2N_0}}) - Q(\frac{2x-2}{\sqrt{2N_0}})$, which are respectively given by the areas of the shadowed parts on the left and right sides of the zero point in Fig. \ref{Fig::III::Lemma1pxy}. Thus, by comparing the areas of two shadowed parts, the probability $p(x,y)$ can be derived as
        	\begin{equation} \label{equ::OrderStat::Pxy}
        		\begin{split}
        			p(x,y)  & = \mathrm{Pr}(R_{\ell_u} < 0 \ | \ { x \geq |R_{\ell_u}| \geq y})\\
        			&= \frac{ Q(\frac{-2x-2}{\sqrt{2N_0}}) - Q(\frac{-2y-2}{\sqrt{2N_0}})}{ Q(\frac{-2x-2}{\sqrt{2N_0}}) - Q(\frac{-2y-2}{\sqrt{2N_0}}) + Q(\frac{2y-2}{\sqrt{2N_0}}) - Q(\frac{2x-2}{\sqrt{2N_0}})} .
        		\end{split}
        	\end{equation}
        	
            \begin{figure}
            	\begin{center}
            		\includegraphics[scale=0.35] {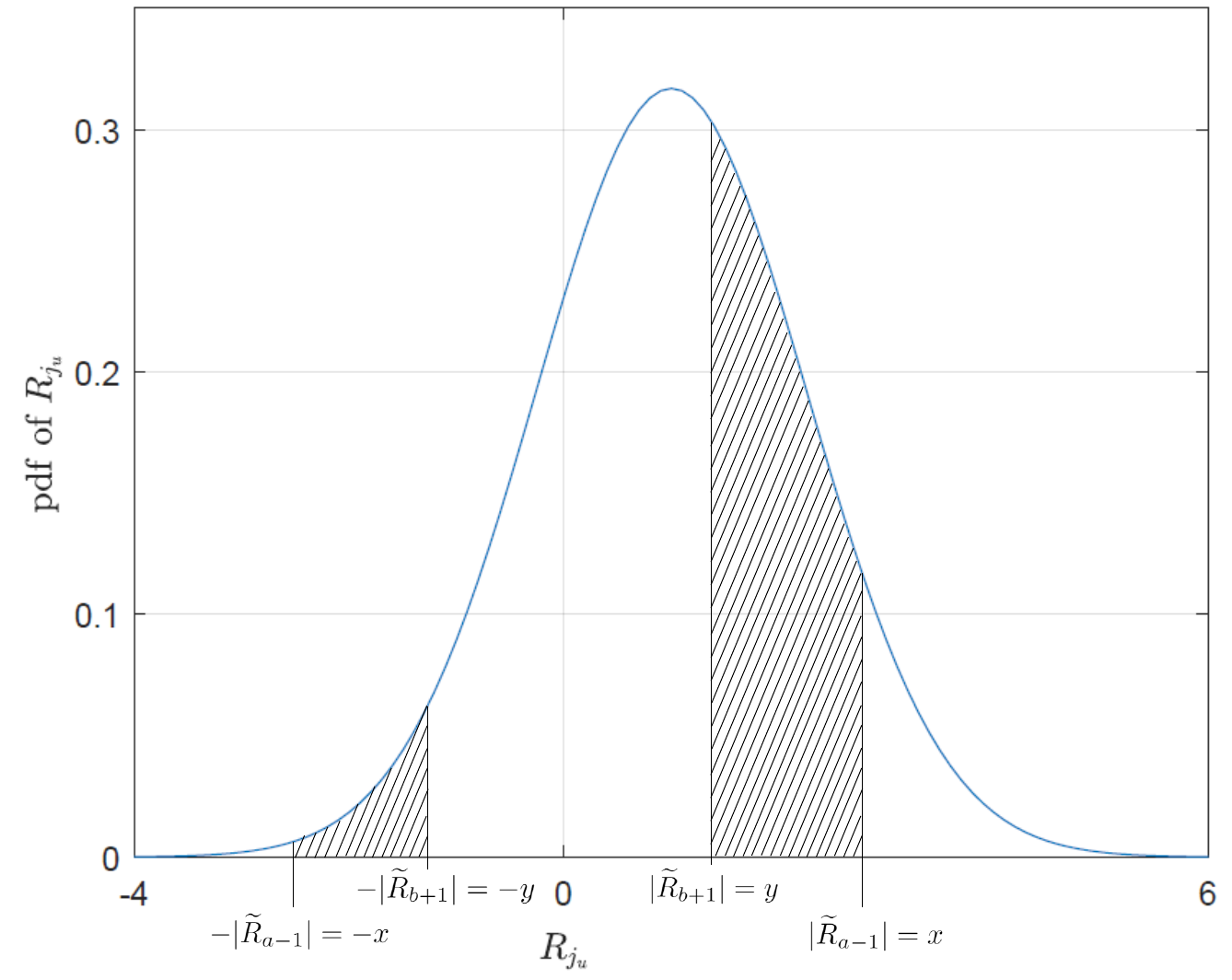}
            		\caption{Demonstration of obtaining $p(x,y)$ in (\ref{equ::OrderStat::Pxy}).}
            		\label{Fig::III::Lemma1pxy}
            	\end{center}
            \end{figure}
        	
            Therefore, conditioning on $\widetilde{A}_{a-1}  = x$ and $\widetilde{A}_{b+1}  =y$, the probability that $[R_{l}]_a^b$ results in exact $j$ errors in $[y_\ell]_a^b$ is given by 
            \begin{equation}  \label{equ::OrderStat::Eab::cond}
                p_{E_a^b}(j|x,y) = \binom{b-a+1}{j}p(x,y)^j(1 - p(x,y))^{b-a+1-j}. 
            \end{equation} 
            It can be noticed that (\ref{equ::OrderStat::Eab::cond}) depends on $x$ and $y$, i.e., the values of $\widetilde{A}_{a-1}$ and $\widetilde{A}_{b+1}$, respectively. By integrating (\ref{equ::OrderStat::Eab::cond}) over x and y with $f_{\widetilde{A}_{a-1},\widetilde{A}_{b+1}}(x,y)$,  we can easily obtain $p_{E_a^b}(j)$ for the case \{$a > 1$ and $b < n$\}.
            
            For the case when $a > 1$ and $b = n$, we can simply assume that $\widetilde{A}_{a-1}  = x$. Then, it can be obtained that the ordered received symbols $[\widetilde{R}_a,\widetilde{R}_{a+1},\ldots,\widetilde{R}_n]$ satisfy
            \begin{equation}
                x\geq|\widetilde{R}_a| \geq |\widetilde{R}_{a+1}| \geq \ldots \geq |\widetilde{R}_{n}| \geq 0.
            \end{equation}
            Using the relationship between ordered and unsorted random variables, there are $n-a+1$ unsorted random variables $R_{\ell_u}$, $a\leq u \leq n$, satisfying $ x \geq |R_{\ell_u}| \geq 0$. For each $R_{\ell_u}$, the probability that it results in an incorrect bit in $[y_\ell]_a^n$ is given by $p(x,0)$. Finally, by integrating $\binom{n-a+1}{j}p(x,0)^j(1 - p(x,0))^{n-a+1-j}$ over $x$, the case \{$a > 1$, $b = n$\} is obtained. 
            
            Similarly, the case \{$a = 1$, $b < n$\} of (\ref{equ::OrderStat::Eab}) can be obtained by assuming $\widetilde{A}_{b+1}   = y$, and considering there are $b$ unsorted random variables $[R_\ell]_1^b$ satisfying  $ \infty \geq |R_{\ell_u}| \geq y$ and having average error probability $p(\infty,y)$. Then, the case \{$a = 1$, $b < n$\} of (\ref{equ::OrderStat::Eab}) can be derived by integrating $\binom{b}{j}p(\infty,y)^j(1 - p(\infty,y))^{b-j}$ over $\widetilde{A}_{b+1} = y$ with the $\mathrm{pdf}$ $f_{\widetilde{A}_{b+1}}(y)$. 
            
            If $a = 1$ and $b = n$, the event \{there are $j$\ errors in $\widetilde{\mathbf{y}}$\} is equivalent to \{there are $j$\ errors in $\mathbf{y}$\}, since $\widetilde{\mathbf{y}}$ is obtained by permuting $\mathbf{y}$. Thus, $p_{E_a^b}(j) = p_{E_1^n}(j)$ can be simply obtained by $p_{E_a^b}(j) = \binom{n}{j} \left(1 - Q(\frac{-2}{\sqrt{2N_0}})\right)^j Q(\frac{-2}{\sqrt{2N_0}})^{n-j}$. On the other hand, it can also be obtained by considering that there are $n$ unsorted random variables having error probability $p(\infty,0)$, because $p(\infty,0) = 1 - Q(\frac{-2}{\sqrt{2N_0}})$.
            
        \end{IEEEproof}

    Please note that the case $\{a = 1,b < n\}$ of Lemma \ref{lem::OrderStat::Eab} was also investigated in the previous work \cite[Eq. (16)]{dhakal2016error}.
        
\begin{figure} [t]
	    	\vspace{-0.6em}
            \centering
            \vspace{-1ex}
            \subfigure[linear scale]
            {
                \includegraphics[scale = 0.6]{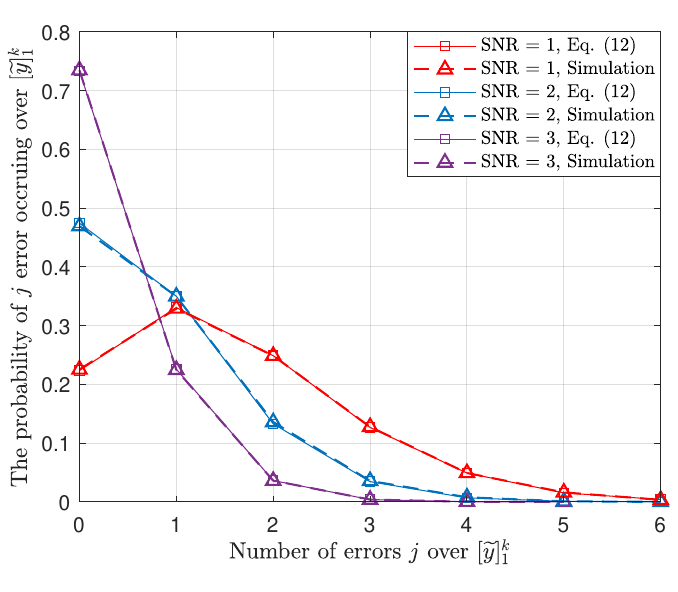}
                \label{Fig::OrderStat::E1k::lin}
            }
            \subfigure[logarithmic scale]
            {
                \includegraphics[scale = 0.6]{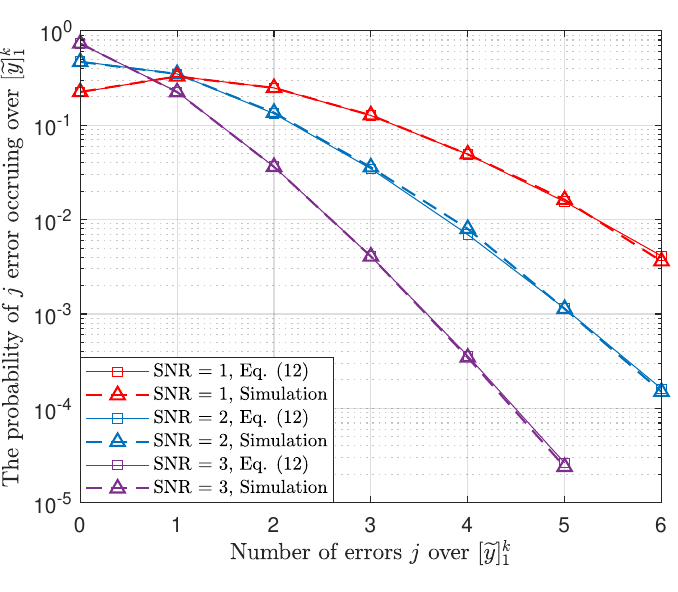}
                \label{Fig::OrderStat::E1k::log}
            }

        	\vspace{-0.4em}
            \caption{The probability of $j$ errors occurring over $[1,k]$ positions of $\widetilde{\mathbf{y}}$ in decoding the eBCH (128, 64, 22) code at different SNRs.}
            \label{Fig::OrderStat::E1k}
        	\vspace{-0.5em}
        \end{figure}

    We show the $\mathrm{pmf}$ of $E_1^k$ for a $(128,64,22)$ eBCH code at different SNRs in Fig. \ref{Fig::OrderStat::E1k}. As can be seen, Lemma \ref{lem::OrderStat::Eab} can precisely describe the $\mathrm{pmf}$ of the number of errors over the ordered hard-decision vector $\widetilde{\mathbf{y}}$. Moreover, it can be observed from the distribution of $E_{1}^{k}$ that the probability of having more than $\min\{ \lceil d_{\mathrm{H}}/4-1 \rceil,k\}$ errors is relatively low at high SNRs, which is consistent with the results in \cite{Fossorier1995OSD}, where $d_{\mathrm{H}}$ is the minimum Hamming distance of $\mathcal{C}(n,k)$. For the demonstrated $(128,64,22)$ eBCH code, the OSD decoding with order $\min\{\lceil d_{\mathrm{H}}/4-1 \rceil,k\} = 5$ is nearly maximum-likelihood\cite{Fossorier1995OSD}.
        
\subsection{Properties of Ordered Reliabilities and Approximations} \label{sec::OrderStat::App}
    Motivated by \cite{fossorier1996first}, we give an approximation of the ordered reliabilities in OSD using the central limit theorem, which can be utilized to simplify the WHD distributions in the following sections. We also show that the event $\{E_1^{k} = j\}$ tends to be independent of the event \{the $\ell$-th ($\ell>k$) position of $\widetilde{\mathbf{y}}$ is in error\} when SNR is high. Furthermore, despite the independence shown in the high SNR regime, for the strict dependency between ordered reliabilities $\widetilde{A}_u$ and $\widetilde{A}_v$, $1\leq u<v \leq n$, we prove that the covariance $\mathrm{cov}(\widetilde{A}_u,\widetilde{A}_v)$ is non-negative.
    
    For the ordered reliability random variables $[\widetilde{A}]_1^n$, the distribution of $\widetilde{A}_u$, $1\leq u \leq n$, can be approximated by a normal distribution $\mathcal{N}(\mathbb{E}[\widetilde{A}_u],\sigma_{\widetilde{A}_u}^2)$ with the $\mathrm{pdf}$ given by
        \begin{equation} \label{equ::OrderStat::Aapp}
            f_{\widetilde{A}_u} (\widetilde{\alpha}_u) \approx \frac{1}{\sqrt{2\pi \sigma_{\widetilde{A}_u}^2}}\exp\left(-\frac{(\widetilde{\alpha}_u-\mathbb{E}[\widetilde{A}_u])^2}{2\sigma_{\widetilde{A}_u}^2}\right),
        \end{equation}
        where
        \begin{equation} \label{equ::OrderStat::Amean}
              \mathbb{E}[\widetilde{A}_u] = F_{A}^{-1}(1-\frac{u}{n})
        \end{equation}
        and
        \begin{equation} \label{equ::OrderStat::Avar}
        \begin{split}
             \sigma_{\widetilde{A}_u}^2 &= \pi N_0\frac{(n-u)u}{n^3} \\
             &\cdot\left(\exp\left(\!-\frac{(\mathbb{E}[\widetilde{A}_u]+1)^2}{N_0}\!\right) + \exp\left(\!-\frac{(\mathbb{E}[\widetilde{A}_u]-1)^2}{N_0}\right)\!\right)^{-2}\!\!.   
        \end{split}
        \end{equation} 
        Details of the approximation can be found in Appendix \ref{App::Aapp}. Similarly, the joint distribution of $\widetilde{A}_u$ and $\widetilde{A}_v$, $0\leq u<v\leq n$, can be approximated to a bivariate normal distribution with the following joint $\mathrm{pdf}$
        \begin{equation}\label{equ::OrderStat::jointAapp}
        \begin{split}
             f_{\widetilde{A}_u,\widetilde{A}_v}& (\widetilde{\alpha}_u,\widetilde{\alpha}_v) \approx \frac{1}{2 \pi \sigma_{\widetilde{A}_u} \sigma_{\widetilde{A}_v | \widetilde{A}_u=\widetilde{\alpha}_u}}\\
             &\cdot\exp\left(\!-\frac{(\widetilde{\alpha}_u \!-\!\mathbb{E}[\widetilde{A}_u])^2}{2\sigma_{\widetilde{A}_u}^2}\!-\!\frac{(\widetilde{\alpha}_v \!-\! \mathbb{E}[\widetilde{A}_v | \widetilde{A}_u\!=\!\widetilde{\alpha}_u])^2}{2\sigma_{\widetilde{A}_v | \widetilde{A}_u=\widetilde{\alpha}_u}^2}\right),      
        \end{split}
        \end{equation}
        where
        \begin{equation} \label{equ::OrderStat::jointAmean}
            \mathbb{E}[\widetilde{A}_v | {\widetilde{A}_u=\widetilde{\alpha}_u}] =  \gamma_{\widetilde{\alpha}_u}^{-1}\left(\frac{v-u}{{n}-u}\right),
        \end{equation}
        and
        \begin{equation} \label{equ::OrderStat::jointAvar}
            \begin{split}
                &\sigma_{\widetilde{A}_v  | \widetilde{A}_u=\widetilde{\alpha}_u}^2 = \pi N_0 \frac{(n-v)(v-u)}{(n-u)^3}\\
                &\cdot \left(\!\frac{\exp\left(\frac{-(\mathbb{E}[\widetilde{A}_v | {\widetilde{A}_u\!=\!\widetilde{\alpha}_u}]-1)^2}{N_0}\right)\!+\!\exp\left(\frac{-(\mathbb{E}[\widetilde{A}_v | {\widetilde{A}_u\!=\!\widetilde{\alpha}_u}]+1)^2}{N_0}\right)}{F_A(\widetilde{\alpha}_u)}\right) ^{\!-\!2}\!\!.
            \end{split}
        \end{equation}
            In (\ref{equ::OrderStat::jointAmean}), $\gamma_{\widetilde{\alpha}_u}\left(t\right)$ is defined as follows
        \begin{equation}
            \gamma_{\widetilde{\alpha}_u}\left(t\right) = \frac{F_{ A}(\widetilde{\alpha}_u) -  F_{ A}(t)}{F_{ A}(\widetilde{\alpha}_u)}.
        \end{equation}
         Details of this approximation are summarized in Appendix \ref{App::jointAapp}. Note that although (\ref{equ::OrderStat::Aapp}) and (\ref{equ::OrderStat::jointAapp}) provide approximations of the distributions regarding ordered reliabilities $\widetilde{A}_u$ and $\widetilde{A}_v$, the means and variances given by (\ref{equ::OrderStat::Amean}), (\ref{equ::OrderStat::Avar}), (\ref{equ::OrderStat::jointAmean}), and (\ref{equ::OrderStat::jointAvar}) are determined with a rigorous derivation without approximations, as shown in Appendix \ref{App::Aapp} and \ref{App::jointAapp}.
         
\begin{figure}
        	\begin{center}
        		\includegraphics[scale=0.65] {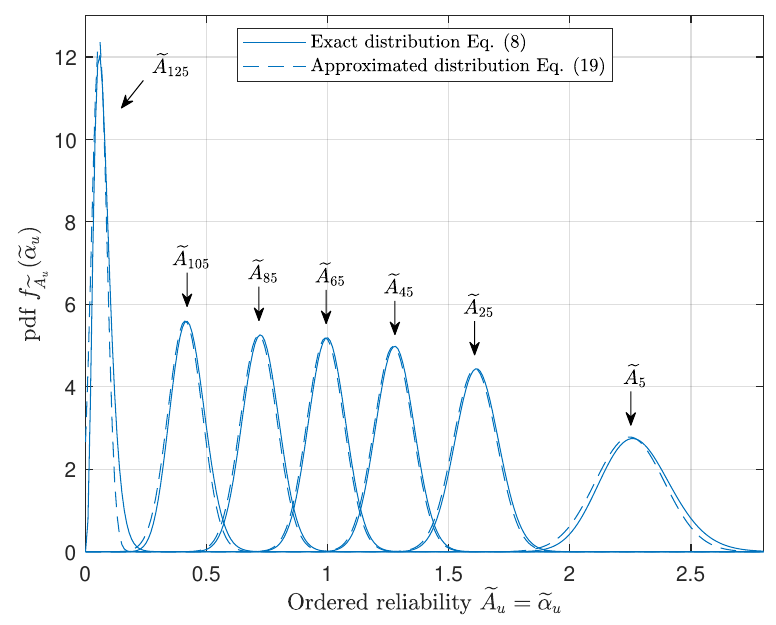}
        		\caption{The approximation of the distribution of the $u^{th}$ ordered reliability in decoding a $(128,64,22)$ eBCH code when SNR = 3 dB. }
        		\label{Fig::III::Aapp}
        	\end{center}
        \end{figure}

        We show the distributions of ordered reliabilities in the decoding of a $(128,64,22)$ eBCH code in Fig. \ref{Fig::III::Aapp}. As can be seen, the normal distribution $\mathcal{N}(\mathbb{E}[\widetilde{A}_u],\sigma_{\widetilde{A}_u}^2)$ with the mean and variance given by (\ref{equ::OrderStat::Amean}) and (\ref{equ::OrderStat::Avar}), respectively, provides a good approximation to (\ref{equ::OrderStat::pdfAu}) for a wide range of $u$. Particularly, the approximation of the distribution of the $u$-th reliability $\widetilde{A}_u$ is tight when $u$ is not close to 1 or $n$. Specifically, when $u = n/2$ (by assuming $n$ is even, similar analysis can be drawn for $u = \lfloor n/2 \rfloor$ if $n$ is odd), it can be seen that $\widetilde{A}_{\frac{n}{2}}$ is the median of the $n$ samples $[\alpha_1,\alpha_2,\ldots,\alpha_{n}]$ of random variable $A$. Thus, when $n$ is large, $\widetilde{A}_{\frac{n}{2}}$ is asymptotically normal with mean $m_{A}$ and variance $\frac{1}{4nf_{A}(m_{A})^2}$\cite{rider1960variance}, where $m_{A}$ is the median of the distribution of $A$, defined as a real number satisfying
        \begin{equation} \label{equ::OrderStat::MedianA}
            \int_{-\infty}^{m_{A}} f_{A}(x) dx \geq \frac{1}{2} \ \text{and} \ \int_{m_{A}}^{\infty} f_{A}(x) dx \geq \frac{1}{2}.
        \end{equation}
        Because $f_{A}(x)$ is a continuous $\mathrm{pdf}$, it can be directly obtained that $m_{A} = F_{A}^{-1}(\frac{1}{2})$ from (\ref{equ::OrderStat::MedianA}), that is, $m_{A}$ is also given by (\ref{equ::OrderStat::Amean}) when $u = n/2$. Then, substituting $u = n/2$ and $m_{A} = F_{A}^{-1}(\frac{1}{2}) = \mathbb{E}[\widetilde{A}_{\frac{n}{2}}]$ into (\ref{equ::OrderStat::Avar}), it can be obtained that 
        \begin{equation}
        \begin{split}
            \sigma_{\widetilde{A}_{\frac{n}{2}}}^2 \!&=\! \frac{\pi N_0}{4n} \!\left(\!\exp\left(-\frac{(m_{A}\!+\!1)^2}{N_0}\right) \!+\! \exp\left(-\frac{(m_{A}\!-\!1)^2}{N_0}\right) \! \right)^{-2} \\
            &= \frac{1}{4nf_{A}(m_{A})^2}.            
        \end{split}
        \end{equation}
        Therefore, it can be concluded that (\ref{equ::OrderStat::Aapp}) with mean (\ref{equ::OrderStat::Amean}) and variance (\ref{equ::OrderStat::Avar}) provides a tight approximation for $\widetilde{A}_{\frac{n}{2}}$, which is consistent with the results given in \cite{rider1960variance}.
        
        Next, we give more results regarding the distributions of the ordered reliabilities. Based on the mean of $\widetilde{A}_v$ conditioning on ${\widetilde{A}_u=\widetilde{\alpha}_u}$, i.e., $\mathbb{E}[\widetilde{A}_v | {\widetilde{A}_u=\widetilde{\alpha}_u}]$ given by (\ref{equ::OrderStat::jointAmean}), we observe that 
        \begin{equation}
            \frac{F_{A}(\mathbb{E}[\widetilde{A}_v |{\widetilde{A}_u = \widetilde{\alpha}_u}])}{F_{A}(\widetilde{\alpha}_u)} = \frac{n-v}{n-u}.
        \end{equation}
        In the asymptotic scenario, where the SNR goes to infinity, we have
        \begin{equation} \label{equ::OrderStat::meanIndepence}
        \begin{split}
            \lim_{N_0 \to 0} F_{A} & (\mathbb{E}[\widetilde{A}_v |  {\widetilde{A}_u = \widetilde{\alpha}_u}])  \overset{(a)}{=} \frac{n-v}{n-u}{F_{A}(\mathbb{E}[\widetilde{A}_u])}\\
            &= \left(\frac{n-v}{n-u}\right) \left(\frac{n-u}{n}\right) = F_{A}(\mathbb{E}[\widetilde{A}_v]) ,            
        \end{split}
        \end{equation}
        where the step (a) follows from that $\widetilde{A}_{u}$ concentrates on the mean when $N_0 \to 0$. Eq. (\ref{equ::OrderStat::meanIndepence}) implies that $\mathbb{E}[\widetilde{A}_v |\widetilde{A}_u = \widetilde{\alpha}_u]$ tends toward $\mathbb{E}[\widetilde{A}_v]$ when the SNR is high enough. Similarly for the variance, we obtain
         \begin{equation} \label{equ::varIndepence}
                \frac{\sigma_{\widetilde{A}_v | {\widetilde{A}_u \!=\! \widetilde{\alpha}_u}}^2 }{\sigma_{\widetilde{A}_v}^2} \!=\! \frac{(n\!-\!v)(v\!-\!u)}{(n-u)^3} \cdot  \frac{(n\!-\!u)^2}{n^2} \cdot  \frac{n^3}{(n\!-\!v)v} = \frac{(v\!-\!u)n}{(n\!-\!u)v},
         \end{equation}
         which implies that $\sigma_{\widetilde{A}_v | \widetilde{A}_u = \widetilde{\alpha}_u}^2 \approx  \sigma_{\widetilde{A}_v}^2$ when $u \ll v$. Combining (\ref{equ::OrderStat::meanIndepence}) and (\ref{equ::varIndepence}), we can conclude that at high SNRs and when $u \ll v$, ordered reliabilities $\widetilde{A}_u$ and $\widetilde{A}_v$ tend to be independent of each other, i.e., $f_{\widetilde{A}_u,\widetilde{A}_v}(\widetilde{\alpha}_u,\widetilde{\alpha}_v)\approx f_{\widetilde{A}_u}(\widetilde{\alpha}_u) f_{\widetilde{A}_v}(\widetilde{\alpha}_v)$.
         
        Based on Lemma \ref{lem::OrderStat::Eab} and the distribution of ordered reliabilities, $\mathrm{Pr}(E_1^{k} = j )$ and the probability that the $\ell$-th position of $\widetilde{\mathbf{y}}$ is in error, denoted by $\mathrm{Pe}(\ell)$, are respectively given by 
         \begin{equation}
         \begin{split}
             	\mathrm{Pr}(E_1^{k} \!=\! j )  &= p_{E_1^{k}}(j)\\
             	&= \int_{0}^{\infty}\binom{k}{j} p(\infty,y)^j (1 - p(\infty,y))^{k-j} f_{\widetilde{A}_{k}}(y)dy , 
         \end{split}
         \end{equation}
         and
         \begin{equation} \label{equ::OrderStat::bitPe}
             	\mathrm{Pe}(\ell) = \int_{0}^{\infty} \frac{f_R(-x)}{f_R(x)+f_R(-x)} f_{\widetilde{A}_{\ell}}(x)dx .
         \end{equation}
         At high SNRs and when $\ell \gg k$, we further obtain that
         \begin{align} \label{equ::OrderStat::EPeIndenIndepence}
                & \mathrm{Pr}(E_1^{k} \!=\! j ) \mathrm{Pe}(\ell) \notag\\
                &\qquad=  \int_{0}^{\infty}\binom{k}{j}p(\infty,y)^j (1 - p(\infty,y))^{k-j} f_{\widetilde{A}_{k}}(y)dy \notag\\
                & \qquad\qquad\qquad \cdot \int_{0}^{\infty}  \frac{f_R(-x)}{f_R(x)+f_R(-x)} f_{\widetilde{A}_{\ell}}(x)dx \notag\\
                & \qquad \approx \int_{0}^{\infty}\int_{0}^{\infty}\binom{k}{j}p(\infty,y)^j (1 - p(\infty,y))^{k-j}\\
                & \qquad\qquad\qquad \cdot\left(\frac{f_R(-x)}{f_R(x)+f_R(-x)}\right) f_{\widetilde{A}_k,\widetilde{A}_{\ell}}(x,y)dxdy \notag\\
                &\qquad= \mathrm{Pr}(\{E_1^{k} = j\} \cap \{\text{the } \ell \text{-th bit of }\widetilde{\mathbf{y}}\text{ is in error}\}). \notag
         \end{align}
        Eq. (\ref{equ::OrderStat::EPeIndenIndepence}) holds because $f_{\widetilde{A}_{\ell},\widetilde{A}_{k}}(\widetilde{\alpha}_{\ell},\widetilde{\alpha}_{k})\approx f_{\widetilde{A}_{\ell}}(\widetilde{\alpha}_{\ell}) f_{\widetilde{A}_{k}}(\widetilde{\alpha}_{k})$. From (\ref{equ::OrderStat::EPeIndenIndepence}) we can see that the event $\{E_1^{k} = j\}$ tends to be independent of the event $\{\text{the } \ell \text{-th bit of }\widetilde{\mathbf{y}}\text{ is in error}\}$ when $\ell \gg k$ and at high SNRs. This conclusion is in fact consistent with the conclusion presented in \cite{fossorier1996first} that despite $\widetilde{R}_u$ and $\widetilde{R}_v$, $1\leq u < v \leq n  $, are statistically dependent, their respective error probabilities tend to be independent, for $ n$ large enough and $ n\gg u$.
        
        In the following Lemma, we show that despite $\widetilde{A}_u$ and $\widetilde{A}_v$ tend to be independent when SNR is high and $u\ll v$, their covariance $\mathrm{cov}(\widetilde{A}_u,\widetilde{A}_v)$ is non-negative for any $u$ and $v$, $1 \leq u<v\leq n$.
        \begin{lemma} \label{Lem::positiveCov}
            For any $u$ and $v$, $1\leq u <v \leq n$, the covariance of reliabilities $\widetilde{A}_{u}$ and $\widetilde{A}_{v}$ satisfies $\mathrm{cov}(\widetilde{A}_u,\widetilde{A}_v) \geq 0$.
        \end{lemma}
        \begin{IEEEproof}
        For the reliabilities before and after ordering, we have $\sum_{u=1}^{n}\widetilde{A}_u = \sum_{u=1}^{n} A_u$ and $\sum_{u=1}^{n}\widetilde{A}_u^2 = \sum_{u=1}^{n} A_u^2$ and by taking expectation on both sides, we obtain the following inequality
            \begin{equation}
                \mathbb{E}[\widetilde{A}_u^2]+\mathbb{E}[\widetilde{A}_v^2] \leq \sum_{u=1}^{n}\mathbb{E}[\widetilde{A}_u^2] =  \sum_{u=1}^{n}\mathbb{E}[A_u^2] = n\mathbb{E}[A^2] < \infty,
            \end{equation}
            where the last inequality is due to the fact that the second moment of normal distribution exists and is finite. 
            %, the second moment $\mathbb{E}[A^2]$ of $A_u$ satisfies
            % \begin{equation}
            %     \begin{split}
            %          \mathbb{E}[A^2]& = \int_{0}^{\infty} x^2 f_{A}(x)dx \\
            %          &= \int_{0}^{\infty} x^2 \phi(\frac{x+1}{\sqrt{N_0/2}}) dx + \int_{0}^{\infty} x^2 \phi(\frac{x-1}{\sqrt{N_0/2}}) dx < \infty
            %     \end{split}
            % \end{equation}
            % where $\phi(x)$ is the $\mathrm{pdf}$ of the standard normal distribution. 
            Then, following the argument in  \cite[Theorem 2.1]{bickel1967SomeContributions} for the ordered statistics, the covariance of the $u$-th variable and $v$-th variable is non-negative if the sum of corresponding second moments is finite. This completes the proof.
        \end{IEEEproof}
 
\section{The Hamming Distance in OSD}  \label{sec::HDdis}
\subsection{0-Reprocessing Case}
Let us first consider the Hamming distance $d_0^{(\mathrm{H})} = d^{(\mathrm{H})}(\widetilde{\mathbf{c}}_0,\widetilde{\mathbf{y}})$ in the 0-reprocessing where no TEP is added to MRB positions before re-encoding, i.e., $\widetilde{\mathbf{c}}_0 = \widetilde{\mathbf{y}}_{\mathrm{B}}\widetilde{\mathbf{G}}$. To find the distribution of 0-reprocessing Hamming distance, we now regard it as a random variable denoted by $D_{0}^{(\mathrm{H})}$, and accordingly $d_{0}^{(\mathrm{H})}$ is the sample of $D_{0}^{(\mathrm{H})}$.
                
 Let us re-write $\widetilde{\mathbf{y}}$ and $\widetilde{\mathbf{c}}_0$ as $\widetilde{\mathbf{y}} = [\widetilde{\mathbf{y}}_{\mathrm{B}} \ \  \widetilde{\mathbf{y}}_\mathrm{P}] $ and $\widetilde{\mathbf{c}}_0 = [\widetilde{\mathbf{c}}_{0,\mathrm{B}} \ \  \widetilde{\mathbf{c}}_{0,\mathrm{P}}]$, respectively, where subscript $\mathrm{B}$ and $\mathrm{P}$ denote the first $k$ positions and the remaining positions of a length-$n$ vector, respectively. Also, let us define $\widetilde{\mathbf{c}} = \pi_2(\pi_1(\mathbf{c})) =  [\widetilde{\mathbf{c}}_{\mathrm{B}} \ \  \widetilde{\mathbf{c}}_{\mathrm{P}}] $ representing the transmitted codeword after permutations, which is unknown to the decoder but useful in the analysis later. Accordingly, we define $\widetilde{\mathbf{e}} = [\widetilde{\mathbf{e}}_{\mathrm{B}} \ \ \widetilde{\mathbf{e}}_{\mathrm{P}}]$ as the permuted hard-decision error, i.e., $\widetilde{\mathbf{e}} = \widetilde{\mathbf{c}} \oplus  \widetilde{\mathbf{y}}$. For an arbitrary permuted codeword $\widetilde{\mathbf{c}}' = [\widetilde{\mathbf{c}}'_{\mathrm{B}} \ \ \widetilde{\mathbf{c}}'_{\mathrm{P}}]$ from $\mathcal{C}(n,k)$, where $\widetilde{\mathbf{c}}'$ is generated by an information vector $\mathbf{b}'$ with Hamming weight $w(\mathbf{b}') = q$ and the permuted generator matrix $\widetilde{\mathbf{G}}$, i.e., $\widetilde{\mathbf{c}}' = \mathbf{b}'\widetilde{\mathbf{G}}$, we further define $p_{\mathbf{c}_{\mathrm{P}}}(u,q)$ as the probability of $w(\widetilde{\mathbf{c}}'_{\mathrm{P}}) = u$ when $w(\mathbf{b}') = q$ i.e., $p_{\mathbf{c}_{\mathrm{P}}}(u,q) = \mathrm{Pr}(w(\widetilde{\mathbf{c}}'_{\mathrm{P}})\! =\! u | w(\mathbf{b}') \!=\! q)$. It can be seen that $p_{\mathbf{c}_{\mathrm{P}}}(u,q)$ is characterized by the structure of the generator matrix $\mathbf{G}$ of $\mathcal{C}(n,k)$, which is independent of the channel conditions.
        
        In the 0-reprocessing, the Hamming distance $D_0^{(\mathrm{H})}$ is affected by both the number of errors in $\widetilde{\mathbf{y}}_\mathrm{P}$ and also the Hamming weights of the parity part $\widetilde{\mathbf{c}}_{\mathrm{P}}'$ of permuted codewords $\widetilde{\mathbf{c}}'$ from $\mathcal{C}(n,k)$ simultaneously, which is explained in the following Lemma.

    	\begin{lemma} \label{lem::HDdis::0phase}
    		After the 0-reprocessing of decoding a linear block code $\mathcal{C}(n,k)$, the Hamming distance $D_0^{(\mathrm{H})}$ between $\widetilde{\mathbf{y}}$ and $\widetilde{\mathbf{c}}_0$ is given by
    		\begin{equation} \label{equ::HDdis::0phase::cases}
    			D_{0}^{(\mathrm{H})}= 
    			\begin{cases}
    			    E_{k+1}^n,    & \mathrm{w.p.}~p_{E_1^k}(0) ,  \\
    			    W_{\mathbf{c}_{\mathrm{P}}},  &  \mathrm{w.p.}~ 1 - p_{E_1^k}(0),
    			\end{cases}
    		\end{equation}
            where $E_{k+1}^n$ is the random variable defined by (\ref{equ::OrderStat::Eab}) in Lemma \ref{lem::OrderStat::Eab} and $p_{E_1^k}(0)$ is given by
    		\begin{equation} \label{equ::HDdis::0phase::weightPe}
    			p_{E_1^k}(0) = \int_{0}^{\infty} (1 - p(\infty,y))^{k} f_{\widetilde A_{k+1}}(y)dy.
    		\end{equation} 
    		$W_{\mathbf{c}_{\mathrm{P}}}$ is a discrete random variable whose $\mathrm{pmf}$ is given by
    		\begin{equation}   \label{equ::HDdis::0phase::pdfWcp}
    		\begin{split}
     		    p_{W_{\mathbf{c}_{\mathrm{P}}}}(j) &= \sum_{u = 0}^{n-k}\sum_{v = 0}^{n-k}\frac{\binom{u}{\delta}\binom{n-k-u}{v-\delta}}{\binom{n-k}{v}}\cdot p_{d}(u) \cdot p_{E_{k+1}^{n}}(v)
     		    \\ &\cdot \mathbf{1}_{\mathbb{N}\bigcap [0,\min(u,v)]}(\delta),
    		\end{split}
    		\end{equation}
    		where $\delta = (u+v-j)/2$, and
    		\begin{equation} \label{equ::HDdis::0phase::pd}
    		    p_{d}(u) =\frac{1}{1 - p_{E_1^k}(0)} \sum_{q=1}^{k}p_{E_1^k}(q)p_{\mathbf{c}_{\mathrm{P}}}(u,q),
    		\end{equation}
    		$p_{\mathbf{c}_{\mathrm{P}}}(u,q) $ is defined as the probability of $w(\widetilde{\mathbf{c}}_{\mathrm{P}}') = u$ for an arbitrary permuted codeword $\mathbf{c}'$ from $\mathcal{C}(n,k)$, and here the codeword $\widetilde{\mathbf{c}}'$ is generated by an information vector with Hamming weight $q$.
    	\end{lemma}
    	\begin{IEEEproof}
            The hard-decision results can be represented by
            \begin{equation}
    			\widetilde{\mathbf y} = [\widetilde{\mathbf y}_{\mathrm{B}}\ \ \widetilde{\mathbf y}_{\mathrm{P}}] = [\widetilde{\mathbf{c}}_{\mathrm{B}}\oplus \widetilde{\mathbf{e}}_{\mathrm{B}} \ \ \widetilde{\mathbf{c}}_{\mathrm{P}}\oplus \widetilde{\mathbf{e}}_{\mathrm{P}}],
    		\end{equation}
    		where $\widetilde{\mathbf{e}}_{\mathrm{B}} $ and $\widetilde{\mathbf{e}}_{\mathrm{P}}$ are respectively the errors over MRB and the parity part introduced by the hard-decision decoding. If $\widetilde{\mathbf{e}}_{\mathrm{B}} = \mathbf{0}$, the 0-reprocessing result is given by $\widetilde{\mathbf{c}}_0 = [\widetilde{\mathbf{c}}_{\mathrm{B}} \oplus \mathbf 0]\widetilde{\mathbf G} = [\widetilde{\mathbf{c}}_{\mathrm{B}} \  \ \widetilde{\mathbf{c}}_{\mathrm{P}}]$. Therefore, the Hamming distance is obtained as
    		\begin{equation}
  	            D_{0}^{(\mathrm{H})} = \lVert \widetilde{\mathbf{y}} \oplus \mathbf {\widetilde c}_0  \rVert   = \lVert \widetilde{\mathbf{c}}_{\mathrm{P}}\oplus \widetilde{\mathbf{c}}_{\mathrm{P}}\oplus \widetilde{\mathbf{e}}_{\mathrm{P}} \rVert = E_{k+1}^{n}   .
    		\end{equation}
            The probability of event $\{\widetilde{\mathbf{e}}_{\mathrm{B}} = \mathbf 0\}$ is simply given by $p_{E_1^k}(0)$ according to Lemma 1. 
            
            If there are errors in $\widetilde{\mathbf y}_\mathrm{B}$, i.e., $\widetilde{\mathbf{e}}_{\mathrm{B}} \neq \mathbf 0$, the 0-reprocessing result is given by $\widetilde{\mathbf{c}}_0 = [\widetilde{\mathbf{c}}_{\mathrm{B}} \oplus  \widetilde{\mathbf{e}}_{\mathrm{B}}]\widetilde{\mathbf G} = [\widetilde{\mathbf{c}}_{0,\mathrm{B}} \ \  \widetilde{\mathbf{c}}_{0,\mathrm{P}}]$. Thus, $D_{0}^{(\mathrm{H})}$ is obtained as 
    		\begin{equation} \label{equ::HDdis::0phase::eb!=0}
       		    D_{0}^{(\mathrm{H})} =  \lVert \widetilde{\mathbf{y}} \oplus \mathbf {\widetilde c}_0  \rVert = \lVert \widetilde{\mathbf{c}}_{0,\mathrm{P}} \oplus \widetilde{\mathbf{c}}_{\mathrm{P}}\oplus \widetilde{\mathbf{e}}_{\mathrm{P}} \rVert .     
    		\end{equation}
    		
    		    Let $\widetilde{\mathbf{d}}_{0} = [\widetilde{\mathbf{d}}_{0,\mathrm{B}} \ \ \widetilde{\mathbf{d}}_{0,\mathrm{P}}] = [\widetilde{d}_{0}]_{1}^{n}$, where $\widetilde{\mathbf{d}}_{0,\mathrm{B}} = [\widetilde{d}_{0}]_1^k$ is an all-zero vector and $\widetilde{\mathbf{d}}_{0,\mathrm{P}} = \widetilde{\mathbf{c}}_{0,\mathrm{P}}\! \oplus\! \widetilde{\mathbf{c}}_{\mathrm{P}}\!\oplus\! \widetilde{\mathbf{e}}_{\mathrm{P}}$. Because $\mathcal{C}(n,k)$ is a linear block codes, $\widetilde{\mathbf{c}}_{0,\mathrm{P}}' = \widetilde{\mathbf{c}}_{0,\mathrm{P}} \oplus \widetilde{\mathbf{c}}_{\mathrm{P}} = [\widetilde{c}_{0}']_{k+1}^{n}$ is also the parity part of a codeword of $\mathcal{C}(n,k)$. In fact, it can be also observed that $\widetilde{\mathbf{c}}_{0}' = \widetilde{\mathbf{e}}_{\mathrm{B}}\widetilde{\mathbf{G}} = [\widetilde{\mathbf{e}}_{\mathrm{B}} \ \ \widetilde{\mathbf{c}}_{0,\mathrm{P}}' ]$. Let us define a random variable $W_{\mathbf{c}_{\mathrm{P}}}$ representing the Hamming weight of $\widetilde{\mathbf{d}}_{0,\mathrm{P}} = \widetilde{\mathbf{c}}_{0,\mathrm{P}}' \oplus \widetilde{\mathbf{e}}_{\mathrm{P}}$. When $\widetilde{\mathbf{e}}_{\mathrm{B}} \neq \mathbf 0$, it can be seen that $D_{0}^{(\mathrm{H})} = W_{\mathbf{c}_{\mathrm{P}}}$. 
    		    
    		    Therefore, because $\widetilde{\mathbf{d}}_{0,\mathrm{P}} = \widetilde{\mathbf{c}}_{0,\mathrm{P}}' \oplus \widetilde{\mathbf{e}}_{\mathrm{P}}$, the $\mathrm{pmf}$ of $W_{\mathbf{c}_{\mathrm{P}}}$ is determined by both $\widetilde{\mathbf{c}}_{0,\mathrm{P}}'$ and $\widetilde{\mathbf{e}}_{\mathrm{P}}$. By observing that $\widetilde{\mathbf{c}}_{0}' = \widetilde{\mathbf{e}}_{\mathrm{B}}\widetilde{\mathbf{G}}$ and that each column of $\mathbf{G}$ has an equal probability to be permuted to other columns of $\widetilde{\mathbf{G}}$ when receiving a new signal from the channel, the probability $\mathrm{Pr}(w(\widetilde{\mathbf{c}}_{0,\mathrm{P}}') = u)$ can be given by $p_{\mathbf{c}_{\mathrm{P}}}(u,w(\widetilde{\mathbf{e}}_{\mathrm{B}}))$, i.e., the probability that the Hamming weight of the parity part of a codeword is given by $u$, where the codeword is generated by an information vector with Hamming weight $w(\widetilde{\mathbf{e}}_{\mathrm{B}})$. Furthermore, because $\widetilde{\mathbf{e}}_{\mathrm{B}}$ is in fact the errors in MRB introduced by the hard decision, the $\mathrm{pmf}$ of $w(\widetilde{\mathbf{e}}_{\mathrm{B}})$ is simply given by (\ref{equ::OrderStat::Eab}) introduced in Lemma \ref{lem::OrderStat::Eab}. Finally, let $p_{d}(u)$ denote the $\mathrm{pmf}$ of $w(\widetilde{\mathbf{c}}_{0,\mathrm{P}}')$, $p_{d}(u)$ can be derived using the law of total probability, i.e.,
    		    \begin{equation}
    		        p_{d}(u) = \frac{1}{1 - p_{E_1^k}(0)} \sum_{q=1}^{k}p_{E_1^k}(q)p_{\mathbf{c}_{\mathrm{P}}}(u,q).
    		    \end{equation}
    		    Hereby, we obtain (\ref{equ::HDdis::0phase::pd}). 
    		    
    		    Next, recall that $\widetilde{\mathbf{d}}_{0,\mathrm{P}} = \widetilde{\mathbf{c}}_{0,\mathrm{P}}' \oplus \widetilde{\mathbf{e}}_{\mathrm{P}}$. To obtain the $\mathrm{pmf}$ of $W_{\mathbf{c}_{\mathrm{P}}}$, i.e., the Hamming weight of $\widetilde{\mathbf{d}}_{0,\mathrm{P}} = \widetilde{\mathbf{c}}_{0,\mathrm{P}}' \oplus \widetilde{\mathbf{e}}_{\mathrm{P}}$, let us first define the probability of $w(\widetilde{\mathbf{d}}_{0,\mathrm{P}}) = j$ conditioning on $w(\widetilde{\mathbf{c}}_{0,\mathrm{P}}') = u$ and $w(\widetilde{\mathbf{e}}_{\mathrm{P}}) = v$, simply denoted by $p_{W_{\mathbf{c}_{\mathrm{P}}}}(j|u,v)$. Since each column of $\mathbf{G}$ has an equal probability to be permuted to other columns of $\widetilde{\mathbf{G}}$ when receiving a new signal from the channel, each bit in $\widetilde{\mathbf{c}}_{0,\mathrm{P}}'$ has an equal probability to be nonzero. Furthermore, recalling the arguments in Lemma \ref{lem::OrderStat::Eab}, conditioning on $\widetilde{A}_{k-1} = x$, each bit in $\widetilde{\mathbf{e}}_{\mathrm{P}}'$ has an equal probability $p(x,0)$ to be nonzero. Thus, $p_{W_{\mathbf{c}_{\mathrm{P}}}}(j|u,v)$ is given by
    		    \begin{equation}
    		        p_{W_{\mathbf{c}_{\mathrm{P}}}}(j|u,v) = \frac{\binom{u}{\delta}\binom{n-k-u}{v-\delta}}{\binom{n-k}{v}}\cdot \mathbf{1}_{\mathbb{N}\bigcap [0,\min(u,v)]}(\delta), 
    		    \end{equation}
    		    where $\delta = \frac{u+v-j}{2}$ represents the number of nonzero bits that are unflipped from $\widetilde{\mathbf{c}}_{0,\mathrm{P}}'$ to $\widetilde{\mathbf{e}}_{\mathrm{P}}$. Finally, by using the law of total probability for all possible values of $w(\widetilde{\mathbf{c}}_{0,\mathrm{P}}') = u$ and $w(\widetilde{\mathbf{e}}_{\mathrm{P}}) = v$, and $\widetilde{A}_{k-1} = x$ we can finally obtain $p_{W_{\mathbf{c}_{\mathrm{P}}}}(j)$ as
    		    \begin{align}
    		        p_{W_{\mathbf{c}_{\mathrm{P}}}}\!(j) \!=\! & \int_{0}^{\infty} \sum_{u = 0}^{n-k}\sum_{v = 0}^{n-k}\frac{\binom{u}{\delta}\binom{n-k-u}{v - \delta}}{\binom{n-k}{v}}\cdot \mathbf{1}_{\mathbb{N}\bigcap [0,\min(u,v)]}(\delta) \notag\\
    		        \cdot& p_{d}(u) \binom{n\!-\!k}{v} p(x,\!0)^{v} (1\!-\! p(x,\!0))^{n\!-\!k\!-\!v} f_{\widetilde{A}_{k\!-\!1}}\!(x) dx \notag\\
    		        \overset{(a)}{=} & \sum_{u = 0}^{n-k}\sum_{v = 0}^{n-k}\frac{\binom{u}{\delta}\binom{n-k-u}{v-\delta}}{\binom{n-k}{v}} p_{d}(u)  p_{E_{k+1}^{n}}(v)\\
    		        \cdot & \mathbf{1}_{\mathbb{N}\bigcap [0,\min(u,v)]}(\delta),\notag
    		    \end{align}
    		    where step (a) follows from that $p_{E_{k+1}^{n}}(v) = \int_{0}^{\infty}\binom{n-k}{v}\cdot\\ p(x,0)^{v} (1- p(x,0))^{n-k-v} f_{\widetilde{A}_{k-1}}(x) dx$, as introduced in Lemma \ref{lem::OrderStat::Eab}.
    		    Recall that the probability of event $\{\widetilde{\mathbf{e}}_{\mathrm{B}} \neq \mathbf 0\}$ can be derived as $ 1- p_{E_1^{k}}(0)$ according to Lemma \ref{lem::OrderStat::Eab}, and $D_{0}^{(\mathrm{H})} = W_{\mathbf{c}_{\mathrm{P}}}$ when $\widetilde{\mathbf{e}}_{\mathrm{B}} \neq \mathbf 0$, then Lemma \ref{lem::HDdis::0phase} is proved.
    	\end{IEEEproof}
    	
    	From (\ref{equ::HDdis::0phase::weightPe}), we can see that the probability $p_{E_1^k}(0)$ is a functions of $k$, $n$, the and noise power $N_0$. If $k$ and $n$ are fixed, $p_{E_1^k}(0)$ is a monotonically increasing function of SNR. This implies that the channel condition determines the weight of the composition of the Hamming distance.	Combining Lemma \ref{lem::OrderStat::Eab} and Lemma \ref{lem::HDdis::0phase}, the distribution of $D_{0}^{(\mathrm{H})}$ is summarized in the following Theorem. 
    	\begin{theorem} \label{the::HDdis::0phase}
    		Given a linear block code $\mathcal C(n,k)$, the $\mathrm{pmf}$ of the Hamming distance between $\widetilde{\mathbf y}$ and $\widetilde{\mathbf{c}}_0$, $D_{0}^{(\mathrm{H})}$, is given by
    		\begin{equation} \label{equ::HDdis::0phase}
    			p_{D_{0}^{(\mathrm{H})}}(j)= p_{E_1^k}(0)p_{E_{k+1}^{n}}(j) + \left(1 - p_{E_1^k}(0)\right)p_{W_{\mathbf{c}_{\mathrm{P}}}}(j) ,
    		\end{equation}
    		where $p_{E_1^k}(0)$ is given by (\ref{equ::HDdis::0phase::weightPe}), and $p_{E_{k+1}^{n}}(j)$ and $p_{W_{\mathbf{c}_{\mathrm{P}}}}(j)  $ are the $\mathrm{pmf}$s of random variables $E_{k+1}^{n}$ and $W_{\mathbf{c}_{\mathrm{P}}}$ given by (\ref{equ::OrderStat::Eab}) and (\ref{equ::HDdis::0phase::pdfWcp}), respectively.
    	\end{theorem}
    	\begin{IEEEproof}
    		The $\mathrm{pmf}$ of $D_{0}^{(\mathrm{H})}$ can be derived in the form of conditional probability as
    		\begin{equation} 
    		\begin{split}
    		    p_{D_{0}^{(\mathrm{H})}}(j)& = \mathrm{Pr}(\widetilde{\mathbf{e}}_{\mathrm{B}} = \mathbf{0}) p_{D_{0}^{(\mathrm{H})}}(j|\widetilde{\mathbf{e}}_{\mathrm{B}} = \mathbf{0}) \\
    		    &+ \mathrm{Pr}(\widetilde{\mathbf{e}}_{\mathrm{B}} \neq \mathbf{0}) p_{D_{0}^{(\mathrm{H})}}(j|\widetilde{\mathbf{e}}_{\mathrm{B}} \neq \mathbf{0}).    
    		\end{split}
    		\end{equation}
    		
    		From the Lemma \ref{lem::HDdis::0phase}, we can see that $\mathrm{Pr}(\widetilde{\mathbf{e}}_{\mathrm{B}} = \mathbf{0})$ and $\mathrm{Pr}(\widetilde{\mathbf{e}}_{\mathrm{B}} \neq \mathbf{0})$ are given by $p_{E_1^k}(0)$ and $1 - p_{E_1^k}(0)$, respectively, and the conditional $\mathrm{pmf}$ $p_{D_{0}^{(\mathrm{H})}}(j|\widetilde{\mathbf{e}}_{\mathrm{B}} = \mathbf{0})$ and $p_{D_{0}^{(\mathrm{H})}}(j|\widetilde{\mathbf{e}}_{\mathrm{B}} \neq \mathbf{0})$ are given by $p_{E_{k+1}^{n}}(j)$ and $p_{W_{\mathbf{c}_{\mathrm{P}}}}(j)$, respectively. Therefore, the $\mathrm{pmf}$ of $D_{0}^{(\mathrm{H})}$ can be obtained as (\ref{equ::HDdis::0phase}).
    	\end{IEEEproof}
    	
    	 {\color{black}It is important to note that in (\ref{equ::HDdis::0phase}), $p_{E_{k+1}^{n}}(j)$ is given by (\ref{equ::OrderStat::Eab}) in Lemma \ref{lem::OrderStat::Eab} when $a=k+1$ and $b=n$, and $p_{W_{\mathbf{c}_{\mathrm{P}}}}(j)$ is affected  by $p_{\mathbf{c}_{\mathrm{P}}}(j,q)$. Here $p_{\mathbf{c}_{\mathrm{P}}}(j,q)$ is defined as the probability that the parity-part Hamming weight of an arbitrary codeword from $\mathcal{C}(n,k)$ is given by $j$, where the permuted codeword is generated by an information vector with Hamming weight $q$. As can be seen, $p_{\mathbf{c}_{\mathrm{P}}}(j,q)$ is determined by the code structure and weight enumerator. One can find $p_{\mathbf{c}_{\mathrm{P}}}(j,q)$ if the codebook of $\mathcal{C}(n,k)$ is known or via computer search. It is beyond the scope of this paper to theoretically determine $p_{\mathbf{c}_{\mathrm{P}}}(j,q)$ for a specific code; nevertheless, in Section \ref{sec::HDdis::Numerical}, we will show examples of $p_{D_{0}^{(\mathrm{H})}}(j)$ for some well-known codes.}
	
\subsection{$i$-Reprocessing Case}
        In this section, we extend the analysis provided for the Hamming distance in 0-reprocessing in Theorem \ref{the::HDdis::0phase} to any order-$i$ reprocessing, $0 < i \leq m$, where $m$ is the predetermined maximum reprocessing order of the OSD algorithm. Let us define a random variable $D_i^{(\mathrm{H})}$ representing the minimum Hamming distance between codeword estimates and $\widetilde{\mathbf{y}}$ after the first $i$ reprocessings of an order-$m$ OSD have been performed, and $d_i^{(\mathrm{H})}$ is the sample of $D_i^{(\mathrm{H})}$. For the simplicity of expression, for integers $u,v$ and $w$ satisfying $0\leq u < v\leq w$, we introduce a new notation as follows 
        \begin{equation} \label{equ::HDdis::defineBiSum}
            b_{u:v}^w = \sum_{j=u}^v \binom{w}{j}.
        \end{equation}
     
        In an order-$m$ OSD, the decoder first performs the 0-reprocessing and then performs the following stages of reprocessing with the increasing order $i$, $1 \leq i \leq m$. As defined, $D_i^{(\mathrm{H})}$ is the minimum of the Hamming weights between $\sum_{j=0}^{i}\binom{k}{j}$ codeword estimates and $\widetilde{\mathbf{y}}$. {\color{black}To characterize the distribution of $D_i^{(\mathrm{H})}$, we make an important assumption that the Hamming weights of any two codeword estimates generated in OSD are independent, and elaborate on the rationality and limits of this assumption in Remark \ref{rem::HDdis::iphase}. Under this assumption, we summarize the distribution of $D_i^{(\mathrm{H})}$ as follows, started from Lemma \ref{lem::HDdis::iphase::eB&eWeight} and concluded by Theorem \ref{the::HDdis::iphase}.}
        {\color{black}
        \begin{lemma} \label{lem::HDdis::iphase::eB&eWeight}
            In an order-$m$ OSD, assume that the number of errors over MRB introduced by the hard decision, denoted by $w(\widetilde{\mathbf{e}}_{\mathrm{B}})$, satisfies $w(\widetilde{\mathbf{e}}_{\mathrm{B}})>i$. Then, for an arbitrary TEP $\mathbf{e}$ satisfying $w(\mathbf{e})\leq i$ ($0\leq i \leq m$), the Hamming weight of $\mathbf{e}\oplus\widetilde{\mathbf{e}}_{\mathrm{B}}$, denoted by a random variable $W_{\mathbf{e},\widetilde{\mathbf{e}}_{\mathrm{B}}}$, has the conditional $\mathrm{pmf}$ given by
            \begin{equation} \label{equ::HDdis::iphase::eB&eWeight}
            \begin{split}
                 & p_{W_{\mathbf{e},\widetilde{\mathbf{e}}_{\mathrm{B}}}}(j|w(\widetilde{\mathbf{e}}_{\mathrm{B}})>i) \\
                 & \qquad= \sum_{u = i+1}^{k}\sum_{v = 0}^{i} \frac{\binom{u}{\delta}\binom{k-u}{v-\delta}}{\binom{k}{v}} \cdot\frac{p_{E_{1}^{k}}(u)}{1-\sum_{q = 0}^{i}p_{E_{1}^{k}}(q)} \cdot \frac{\binom{k}{v}}{b_{0:i}^{k}}\\
                 &\qquad \cdot \mathbf{1}_{\mathbb{N}\bigcap [0,\min(u,v)]}(\delta), 
            \end{split}
            \end{equation}
             where $\delta = \frac{u+v-j}{2}$ and $p_{E_{1}^{k}}(u)$ is given by (\ref{equ::OrderStat::Eab}).
        \end{lemma}
        \begin{IEEEproof}
            As introduced in Lemma \ref{lem::OrderStat::Eab}, the probability $\mathrm{Pr}(w(\widetilde{\mathbf{e}}_{\mathrm{B}}) = u | w(\widetilde{\mathbf{e}}_{\mathrm{B}}) > i)$ is given by
            \begin{equation}
                \mathrm{Pr}(w(\widetilde{\mathbf{e}}_{\mathrm{B}}) = u | w(\widetilde{\mathbf{e}}_{\mathrm{B}}) > i) = \frac{p_{E_{1}^{k}}(u)}{1-\sum_{q = 0}^{i}p_{E_{1}^{k}}(q)}.
            \end{equation}
            Furthermore, the probability $\mathrm{Pr}(w(\mathbf{e})=v)$ for selecting an arbitrary TEP with the maximal Hamming weight $i$ is given by
            \begin{equation}
                \mathrm{Pr}(w(\mathbf{e}) = v) = \frac{\binom{k}{v}}{b_{0:i}^{k}}.
            \end{equation}
            Similar to (\ref{equ::HDdis::0phase::pdfWcp}), summing up the conditional probabilities $\mathrm{Pr}(w(\widetilde{\mathbf{e}}_{\mathrm{B}}   \oplus \mathbf{e})\!=\!j \ |\ w(\widetilde{\mathbf{e}}_{\mathrm{B}}) \!=\! u, w(\widetilde{\mathbf{e}}_{\mathrm{B}}) \!>\! i, w(\mathbf{e}) \!=\! v)$ with coefficients $\mathrm{Pr}(w(\widetilde{\mathbf{e}}_{\mathrm{B}}) \!=\! u | w(\widetilde{\mathbf{e}}_{\mathrm{B}}) \!>\! i) \mathrm{Pr}(w(\mathbf{e}) = v)$, Eq. (\ref{equ::HDdis::iphase::eB&eWeight}) can be finally obtained.
        \end{IEEEproof}
        Based on Lemma \ref{lem::HDdis::iphase::eB&eWeight}, we can directly show that for an integer $u$, $0\leq u\leq k$, the conditional $\mathrm{pmf}$ $p_{W_{\mathbf{e},\widetilde{\mathbf{e}}_{\mathrm{B}}}}(j|w(\widetilde{\mathbf{e}}_{\mathrm{B}})= u)$ is given by
        \begin{equation}
            p_{W_{\mathbf{e},\widetilde{\mathbf{e}}_{\mathrm{B}}}}(j|w(\widetilde{\mathbf{e}}_{\mathrm{B}})\!=\! u) = \sum_{v = 0}^{i} \frac{\binom{u}{\delta}\binom{k\!-\!u}{v\!-\!\delta}}{\binom{k}{v}} \cdot \frac{\binom{k}{v}}{b_{0:i}^{k}}\cdot\mathbf{1}_{\mathbb{N}\bigcap [0,\min(u,v)]}(\delta).
        \end{equation}
        where $\delta = \frac{u+v-j}{2}$.
        
        Then, let a random variable $W_{\mathbf{e},\mathbf{c}_{\mathrm{P}}}$ denote the Hamming weight of $\widetilde{\mathbf{c}}_{\mathbf{e},\mathrm{P}}' \oplus \widetilde{\mathbf{e}}_{\mathrm{P}}$ for an arbitrary TEP $\mathbf{e}$ processed in the first $i$ reprocessings of OSD, where $\widetilde{\mathbf{c}}_{\mathbf{e},\mathrm{P}}'$ is the parity part of $\widetilde{\mathbf{c}}_{\mathbf{e}}' = [\mathbf{e}\oplus\widetilde{\mathbf{e}}_{\mathrm{B}}]\widetilde{\mathbf{G}}$. We obtain the conditional $\mathrm{pmf}$ of $W_{\mathbf{e},\mathbf{c}_{\mathrm{P}}}$ when $w(\widetilde{\mathbf{e}}_{\mathrm{B}})= u$ and $w(\widetilde{\mathbf{e}}_{\mathrm{P}})= v$ in the following lemma. 
        \begin{lemma} \label{lem::HDdis::iphase::Wecp}
            When the number of errors over $\widetilde{\mathbf{y}}_{\mathrm{B}}$ is given by $w(\widetilde{\mathbf{e}}_{\mathrm{B}})= u$ and the number of errors over $\widetilde{\mathbf{y}}_{\mathrm{P}}$ is given by $w(\widetilde{\mathbf{e}}_{\mathrm{P}})= v$, for an arbitrary TEP $\mathbf{e}$ in an order-$m$ OSD, the Hamming weight of $\widetilde{\mathbf{c}}_{\mathbf{e},\mathrm{P}}' \oplus \widetilde{\mathbf{e}}_{\mathrm{P}}$, denoted by the random variable $W_{\mathbf{e},\mathbf{c}_{\mathrm{P}}}$, has the conditional $\mathrm{pmf}$ $ p_{W_{\mathbf{e},\mathbf{c}_{\mathrm{P}}}}(j| u,v)$ given by
        \begin{equation} \label{equ::HDdis::iphase::Wecp}
        \begin{split}
            p_{W_{\mathbf{e},\mathbf{c}_{\mathrm{P}}}}(j| u,v) &= \sum_{\ell = 0}^{n-k} \frac{\binom{v}{\delta}\binom{n-k-v}{\ell-\delta}}{\binom{n-k}{\ell}} \sum_{q = 0}^{k}p_{W_{\mathbf{e},\widetilde{\mathbf{e}}_{\mathrm{B}}}}(q|w(\widetilde{\mathbf{e}}_{\mathrm{B}})= u) \\
            &\cdot p_{\mathbf{c}_{\mathrm{P}}}(\ell, q)\cdot \mathbf{1}_{\mathbb{N}\bigcap [0,\min(\ell,v)]}(\delta),
        \end{split}
        \end{equation}
        where $\delta = \frac{\ell+v-j}{2}$.
        \end{lemma}
        \begin{IEEEproof}
            Based on Lemma \ref{lem::HDdis::iphase::eB&eWeight}, the probability $\mathrm{Pr}(w(\widetilde{\mathbf{c}}_{\mathbf{e},\mathrm{P}}') = \ell| w(\widetilde{\mathbf{e}}_{\mathrm{B}})= u)$ is given by
            \begin{equation}
                \begin{split}
                    &\mathrm{Pr}(w(\widetilde{\mathbf{c}}_{\mathbf{e},\mathrm{P}}') \!=\! \ell|w(\widetilde{\mathbf{e}}_{\mathrm{B}}) \!=\! u) \\
                    & \qquad = \sum_{q = 0}^{k}p_{W_{\mathbf{e},\widetilde{\mathbf{e}}_{\mathrm{B}}}}(q|w(\widetilde{\mathbf{e}}_{\mathrm{B}})= u) p_{\mathbf{c}_{\mathrm{P}}}(\ell, q)    
                \end{split}
            \end{equation}
            Then, similar to (\ref{equ::HDdis::0phase::pdfWcp}), summing up the conditional probabilities $\mathrm{Pr}(w(\widetilde{\mathbf{c}}_{\mathbf{e},\mathrm{P}}' \oplus \widetilde{\mathbf{e}}_{\mathrm{P}})=j \ |\ w(\widetilde{\mathbf{c}}_{\mathbf{e},\mathrm{P}}') = \ell, w(\widetilde{\mathbf{e}}_{\mathrm{P}})= v)$ with coefficients $ \mathrm{Pr}(w(\widetilde{\mathbf{c}}_{\mathbf{e},\mathrm{P}}') = \ell| w(\widetilde{\mathbf{e}}_{\mathrm{B}})= u)$, (\ref{equ::HDdis::iphase::Wecp}) can be obtained.
        \end{IEEEproof}
        
        For the simplicity of notation, we denote $p_{W_{\mathbf{e},\mathbf{c}_{\mathrm{P}}}}\!(j|w(\widetilde{\mathbf{e}}_{\mathrm{B}})\!>\\ i,w(\widetilde{\mathbf{e}}_{\mathrm{P}})= v) $ as $ p_{W_{\mathbf{e},\mathbf{c}_{\mathrm{P}}}}(j| i^{(>)},v)$. Following Lemma \ref{lem::HDdis::iphase::eB&eWeight} and Lemma \ref{lem::HDdis::iphase::Wecp}, $p_{W_{\mathbf{e},\mathbf{c}_{\mathrm{P}}}}(j| i^{(>)},v)$ is given by 
        \begin{equation}
        \begin{split}
              &p_{W_{\mathbf{e},\mathbf{c}_{\mathrm{P}}}}(j| i^{(>)},v)\\
              &= \sum_{\ell = 0}^{n-k} \frac{\binom{v}{\delta}\binom{n-k-v}{\ell-\delta}}{\binom{n-k}{\ell}} \sum_{q = 0}^{k}p_{W_{\mathbf{e},\widetilde{\mathbf{e}}_{\mathrm{B}}}}(q|w(\widetilde{\mathbf{e}}_{\mathrm{B}}) > i) p_{\mathbf{c}_{\mathrm{P}}}(\ell, q)\\
              &\cdot \mathbf{1}_{\mathbb{N}\bigcap [0,\min(\ell,v)]}(\delta),
        \end{split}
        \end{equation}
        where $\delta = \frac{\ell+v-j}{2}$.
        
        Based on the results and notations introduced in Lemma \ref{lem::HDdis::iphase::eB&eWeight} and Lemma \ref{lem::HDdis::iphase::Wecp}, the distribution of the minimum Hamming distance $D_i^{(\mathrm{H})}$ after the $i$-reprocessing of an order-$m$ OSD is then given in the following Theorem.
        
        \begin{theorem} \label{the::HDdis::iphase}
        	Given a linear block code $\mathcal{C}(n,k)$, the $\mathrm{pmf}$ of the minimum Hamming distance $D_i^{(\mathrm{H})}$ after the $i$-reprocessing of an order-$m$ OSD decoding is given by
        	\begin{equation}  \label{equ::HDdis::iphase}
        	    \begin{split}
                	p_{\! D_i^{(\mathrm{H})}}(j)\!= & \sum\limits_{u=0}^{i}p_{E_1^{k}}(u)\sum\limits_{v=0}^{n-k}  p_{E_{k+1}^{n}}(v) p_{EW}(j|u,v) \\
                	+ & \!\left(\! 1\!-\! \sum_{u=0}^{i}p_{\!E_1^{k}}(u)\!\right)\!\sum\limits_{v=0}^{n-k}  p_{\!E_{k\!+\!1}^{n}}\!(v) p_{\!\widetilde W_{\mathbf{c}_{\mathrm{P}}}}(j\!-\!i,\!b_{0:i}^{k} | i^{(>)}\!\!,v) 	
        	    \end{split}
        	\end{equation}    	
    	where $p_{EW}(j|u,v)$ is given by
    	\begin{equation} \label{equ::HDdis::iphase::EW}
    	    p_{\!EW}(j|u,\!v) \!=\! 
    	    \begin{cases}
    	         \sum\limits_{\ell=u\!+\!v}^{n\!-\!k}p_{\widetilde W_{\mathbf{c}_{\mathrm{P}}}}(\ell,b_{1,i}^{k}| u,v), &\ \text{for} \ \ j = u\!+\! v, \\
    	         p_{\widetilde W_{\mathbf{c}_{\mathrm{P}}}}(j,b_{1,i}^{k}| u,v), &\ \text{for} \  \ 1\!\leq \! j \!<\! u \!+\! v, \\ 
    	         0 , &\ \text{otherwise}.
    	    \end{cases}
    	\end{equation}
    	$p_{\widetilde W_{\mathbf{c}_{\mathrm{P}}}}(j,b| u,v)$ is given by
        \begin{equation} \label{equ::HDdis::iphase::Wcp}
            p_{\widetilde W_{\mathbf{c}_{\mathrm{P}}}}(j,b| u,v) = b \int_{F_{W_{\mathbf{e},\mathbf{c}_{\mathrm{P}}}}(j|u,v)-p_{W_{\mathbf{e},\mathbf{c}_{\mathrm{P}}}}(j|u,v)}^{F_{W_{\mathbf{e},\mathbf{c}_{\mathrm{P}}}}(j|u,v)} (1-\ell)^{b-1} d\ell ,
        \end{equation}
        and $F_{W_{\mathbf{e},\mathbf{c}_{\mathrm{P}}}}(j|u,v)$ and $p_{W_{\mathbf{e},\mathbf{c}_{\mathrm{P}}}}(j|u,v)$ are the conditional $\mathrm{cdf}$ and $\mathrm{cdf}$ of random variable $W_{\mathbf{e},\mathbf{c}_{\mathrm{P}}}$ introduced in Lemma \ref{lem::HDdis::iphase::Wecp}, respectively.
        \end{theorem}  
        
        \begin{IEEEproof}
            The proof is provided in Appendix \ref{app::proof::HDdis::iphase}.
        \end{IEEEproof}
        
        \begin{remark} \label{rem::HDdis::iphase}
            Theorem \ref{the::HDdis::iphase} is developed based on the assumption that the Hamming weights of any two codeword estimates generated in OSD are independent. In other words, the Hamming weights of any linear combination of the rows of $\widetilde{\mathbf{G}}$ are independent. This assumption is reasonable when the Hamming weight of each row of $\widetilde{\mathbf{G}}$ is not much lower than $n-k$. However, when the Hamming weight of each row of $\widetilde{\mathbf{G}}$ is much lower than $n-k$, dependencies will possibly occur between the Hamming weights of two codewords who share the rows of $\widetilde{\mathbf{G}}$ as the basis, especially for codeword estimates generated by TEPs with low Hamming weights. In this case, (\ref{equ::HDdis::iphase}) will show discrepancies with the actual distributions of $D_{i}^{(\mathrm{H})}$, and (\ref{equ::HDdis::iphase::Wcp}) needs to be modified for considering discrete ordered statistics with correlations between variables. Therefore, Theorem \ref{the::HDdis::iphase} may not be compatible with the codes with small minimum distance $d_{\mathrm{H}}$ or with sparse generator matrix $\mathbf{G}$, because the rows of the generator matrix of these codes tend to have lower Hamming weights.
        \end{remark}
        }

\subsection{Approximations and Numerical Examples} \label{sec::HDdis::Numerical}
        In this section, we simplify and approximate the Hamming distance distributions given in Theorem \ref{the::HDdis::0phase} and \ref{the::HDdis::iphase} when the weight spectrum of $\mathcal{C}(n,k)$ can be well approximated by the binomial distribution. Then, we verify Theorem \ref{the::HDdis::0phase} and \ref{the::HDdis::iphase} by comparing simulation results and numerical results for Polar and eBCH codes.
	
        Recalling the $\mathrm{pmf}$ of $0$-reprocessing Hamming distance $D_{0}^{(\mathrm{H})}$ given by (\ref{equ::HDdis::0phase}), random variables $E_{1}^{k}$ and $W_{\mathbf{c}_{\mathrm{P}}}$ need to be approximated separately. {\color{black}Starting from $E_{1}^{k}$, we first define a binomial random variable $X_{u} \sim \mathcal B(n-k,p(u\Delta x,0))$, where $u$ is a non-negative integer, $\Delta x$ is the infinitesimal of $x$ and $p(x,0)$ is given by (\ref{equ::OrderStat::Pxy}). $X_u$ in fact represents the number of errors resulted by $(n-k)$ unsorted received symbols $[R]_1^{n-k}$ satisfying $0 \leq [|R|]_1^{n-k} \leq u\Delta x$. Since $X_{u}$ is binomial, the mean and variance of $X_{u}$ can be found as follows}
    	\begin{equation} 
          \mathbb{E}[X_{u}] = (n-k) p(u\Delta x,0)
        \end{equation}
        and
    	\begin{equation} 
          \sigma_{X_{u}}^2 = (n - k)p(u\Delta x,0)(1-p(u\Delta x,0)),
        \end{equation}    
        respectively. When $(n - k)$ is large, $X_{u}$ can be naturally approximated by the normal distribution with the following $\mathrm{pdf}$
        \begin{equation} 
            f_{X_{u}}(y) = \frac{1}{\sqrt{2\pi \sigma_{X_{u}}^2}}\exp\left(-\frac{(y-\mathbb{E}[X_{u}])^2}{2\sigma_{X_{u}}^2}\right).
        \end{equation}
          {\color{black}According to the case of $\{a\geq 1, b= n \}$ of (\ref{equ::OrderStat::Eab}), consider converting the integral operation into a summation of infinitesimal quantities, then the $\mathrm{pmf}$ of random variable $E_{k+1}^{n}$ given by (\ref{equ::OrderStat::Eab}) can be represented by the linear combination of $ f_{X_{u}}(y)$ for $u = 0,1,\ldots,\infty$ with weights $f_{\widetilde{A}_{k}}(u\Delta x)\Delta x$, i.e., }
        \begin{equation}
          p_{E_{k+1}^{n}}(j) = \sum_{u=0}^{\infty}  f_{\widetilde{A}_{k}}(u\Delta x) \Delta x f_{X_{u}}(j).
        \end{equation}
        Therefore, we regard $p_{E_{k+1}^{n}}(j)$ as the infinite mixture model of Gaussian distributions. Accordingly, the mean is given by 
    	\begin{equation} 
            \begin{split}
                \mathbb{E}[E_{k+1}^{n}] &= \sum_{u=0}^{\infty}(n-k)p(u\Delta x,0)  f_{\widetilde{A}_{k+1}}(u\Delta x) \Delta x  \\
                &= \int_{0}^{\infty}(n-k)p(x, 0)  f_{\widetilde{A}_{k}}( x) d x ,
            \end{split}
        \end{equation}
    	and the variance is given by
    	\begin{equation}
            \begin{split}
                \sigma_{E_{k+1}^{n}}^2 = & \int_{0}^{\infty}(n-k)(2p(x, 0)-p(x, 0)^2)  f_{\widetilde{A}_{k}}( x , y) d x 
        		 \\
        		 &- \left(\int_{0}^{\infty}(n-k)p(x, 0)  f_{\widetilde{A}_{k}}( x) d x \right)^2 .
            \end{split}
        \end{equation}
    	Furthermore, based on the argument of infinite Gaussian mixture model and observing that $E_{k+1}^{n}$ is unimodal, we approximate the distribution of $E_{k+1}^{n}$ by a normal distribution $\mathcal{N}(\mathbb{E}[E_{k+1}^{n}],\sigma_{E_{k+1}^{n}}^2)$, the $\mathrm{pdf}$ of which is given by
    	\begin{equation} \label{equ::HDdis::EknApp}
            f_{E_{k+1}^{n}}(x) = \frac{1}{\sqrt{2\pi \sigma_{E_{k+1}^{n}}^2}}\exp\left(-\frac{(x-\mathbb{E}[E_{k+1}^{n}]^2}{2\sigma_{E_{k+1}^{n}}^2}\right).
        \end{equation}
         We will show later via numerical examples that the approximation (\ref{equ::HDdis::EknApp}) could be accurate.
         Note that (\ref{equ::HDdis::EknApp}) can be further tightened by truncating the function and restricting the support to $x\geq 0$. However, because the value of $\int_{-\infty}^{0} f_{E_{k+1}^{n}}(x)$ is negligible and for the simplicity of expression, we keep (\ref{equ::HDdis::EknApp}) in its current form.
        
        For the random variable $W_{\mathbf{c}_{\mathrm{P}}}$ whose $\mathrm{pmf}$ is given by (\ref{equ::HDdis::0phase::pdfWcp}), obtaining an approximation is difficult. Hence, we consider simplifying and approximating $W_{\mathbf{c}_{\mathrm{P}}}$ only when the weight spectrum of $\mathcal{C}(n,k)$ can be tightly approximated by the binomial distribution \footnote{There are many kinds of codes whose weight distribution can be approximated by a binomial distribution\cite{macwilliams1977codingtheory}, e.g., BCH codes etc.}. {\color{black}Assume $\mathcal{C}(n,k)$ is a linear block code with the minimum weight $d_{\mathrm{H}}$ and weight distribution $\{|\mathcal{A}_0|,|\mathcal{A}_1|,\ldots,|\mathcal{A}_{n}|\}$, where $\mathcal{A}_u$ is the 
       set of codewords with the Hamming weight $u$, and $|\mathcal{A}_u|$ is the cardinality of $\mathcal{A}_u$. Then, the probability that a codeword has weight $u$ can be represented by the truncated binomial distribution, i.e.
    	\begin{equation}  \label{equ::HDdis::BinSpectrum}
    		\frac{|\mathcal{A}_u |}{2^k} \approx \frac{1}{\psi2^{n}}\binom{n}{u} \ \  \text{for}\ \ u = 0 \text{ or } u \geq d_{\mathrm{H}},
    	\end{equation}
    	where $\psi = 1 - \sum_{u = 1}^{d_{\mathrm{H}} - 1} \binom{n}{u}2^{-n}$ is the normalization coefficient such that $\sum_{u=d_{\mathrm{H}}}^n\mathcal{A}_u=2^k$. For such a code $\mathcal{C}(n,k)$ whose weight spectrum is well approximated by (\ref{equ::HDdis::BinSpectrum}), we can obtain that when $\sum_{u = 1}^{d_{\mathrm{H}} - 1} \binom{n}{u}2^{-n}$ is negligible (i.e., when $n \gg d_{\mathrm{H}}$ and $\psi \approx 1$). Thus, $p_{\mathbf{c}_{\mathrm{P}}}(u,q)$ in (\ref{equ::HDdis::0phase::pd}) can be approximated to 
		\begin{equation} \label{equ::HDdis::0phase::pcpApp}
			p_{\mathbf{c}_{\mathrm{P}}}(u,q) \approx \frac{1}{2^{n-k}}\binom{n-k}{u},
		\end{equation}
		and it is approximately independent of $q$. In this case, $p_{d}(u)$ given by (\ref{equ::HDdis::0phase::pd}) can be approximated as
		\begin{equation} \label{equ::HDdis::0phase::pdApp}
			p_{d}(u) \approx \frac{1}{2^{n-k}}\binom{n-k}{u}.
		\end{equation}
        Then, substituting (\ref{equ::HDdis::0phase::pdApp}) into (\ref{equ::HDdis::0phase::pdfWcp}), the $\mathrm{pmf}$ $p_{W_{\mathbf{c}_{\mathrm{P}}}}$ can be approximated  as
        \begin{equation} \label{equ::HDdis::0phase::WcpApp}
        \begin{split}
            p_{W_{\mathbf{c}_{\mathrm{P}}}}(j) \!\! \overset{(a)}{\approx} \! & \int_{0}^{\infty}\binom{n-k}{j} \left(\frac{1}{2}p(x,0)+\frac{1}{2}(1-p(x,0))\right)^{j}\\
            &\cdot\left(\!1 \!-\! \frac{1}{2}p(x,0)\!-\!\frac{1}{2}(1\!-\!p(x,0))\right)^{n\!-\!k\!-\!j}\!\!\!f_{\widetilde{A}_{k-1}}(x) dx \\
            \overset{(b)}{=} & \binom{n-k}{j} \left(\frac{1}{2}\right)^{j}\left(1 - \frac{1}{2}\right)^{n-k-j} = p_{d}(j),
        \end{split}
        \end{equation}
        where step (a) takes $p_{E_{k+1}^{n}}(j) = \int_{0}^{\infty}\binom{n-k}{j} p(x,0)^{j} (1- p(x,0))^{n-k-j} f_{\widetilde{A}_{k-1}}(x) dx$ and substitutes $p_{d}(u)$ with $p_{d}(2\delta-v+j)$, and step (b) follows from that $\frac{1}{2}p(x,0)-\frac{1}{2}(1-p(x,0)) = \frac{1}{2}$.
        Therefore, when $\mathcal{C}(n,k)$ has the weight spectrum described by (\ref{equ::HDdis::BinSpectrum}), $p_{W_{\mathbf{c}_{\mathrm{P}}}}(j)$ can be approximated by a normal random variable $\mathcal{N} (\frac{1}{2}(n-k),\frac{1}{4}(n-k))$ with the $\mathrm{pdf}$
        \begin{equation} \label{equ::HDdis::0phase::WcpNormalApp}
            f_{W_{\mathbf{c}_{\mathrm{P}}}}(x) = \frac{1}{\sqrt{\frac{1}{2}\pi(n-k)}}\exp\left(-\frac{(x-\frac{1}{2}(n-k))^2}{\frac{1}{2}(n-k)}\right).
        \end{equation}
       }
        Finally, when $\mathcal{C}(n,k)$ has the weight spectrum described by (\ref{equ::HDdis::BinSpectrum}), the $\mathrm{pmf}$ of the Hamming distance in 0-reprocessing, i.e., $p_{D_0^{(\mathrm{H})}}(x)$, introduced in Theorem \ref{the::HDdis::0phase} can be approximated by $f_{D_0^{(\mathrm{H})}}(x)$, which is the $\mathrm{pdf}$ of a mixture of two normal distributions given by
        \begin{equation} \label{equ::HDdis::0phase::App}
            f_{D_0^{(\mathrm{H})}}(x) = p_{E_1^k}(0)  f_{E_{k+1}^{n}}(x) + (1 - p_{E_1^k}(0)) f_{W_{\mathbf{c}_{\mathrm{P}}}}(x),
        \end{equation}
        where $f_{E_{k+1}^{n}}(x)$ and $f_{W_{\mathbf{c}_{\mathrm{P}}}}(x)$ are respectively given by (\ref{equ::HDdis::EknApp}) and (\ref{equ::HDdis::0phase::WcpNormalApp}).
        
        When $\mathcal{C}(n,k)$ has the weight spectrum described by (\ref{equ::HDdis::BinSpectrum}), the distribution of the Hamming distance after $i$-reprocessing introduced in Theorem \ref{the::HDdis::iphase} can also have a continuous approximation based on the results of 0-reprocessing and continuous ordered statistics.  {\color{black} Similar to obtaining (\ref{equ::HDdis::0phase::WcpApp}), the $\mathrm{pmf}$ $p_{W_{\mathbf{e},\mathbf{c}_{\mathrm{P}}}}(j| u,v)$ given by (\ref{equ::HDdis::iphase::Wecp}) can also be approximated to 
        \begin{equation}
            p_{W_{\mathbf{e},\mathbf{c}_{\mathrm{P}}}}(j| u,v) \approx  \frac{1}{2^{n-k}}\binom{n-k}{u},
        \end{equation}
        which is independent of $u$ and $v$, and can be further approximated by a normal random variable $\mathcal{N} (\frac{1}{2}(n-k),\frac{1}{4}(n-k))$ with the $\mathrm{pdf}$ $f_{W_{\mathbf{e},\mathbf{c}_{\mathrm{P}}}}(x) = f_{W_{\mathbf{c}_{\mathrm{P}}}}(x)$.} Replacing $p_{W_{\mathbf{e},\mathbf{c}_{\mathrm{P}}}}(j| u,v)$ and $p_{E_{k+1}^{n}}(j)$ with $f_{W_{\mathbf{c}_{\mathrm{P}}}}(x)$ and $f_{E_{k+1}^{n}}(j)$ respectively in (\ref{equ::HDdis::iphase}), and converting discrete ordered statistics to continuous ordered statistics in (\ref{equ::HDdis::iphase::Wcp}), the $\mathrm{pmf}$ of $D_i^{(\mathrm{H})}$ given by (\ref{equ::HDdis::iphase}) can be approximated by
    	\begin{align}   \label{equ::HDdis::iphase::App}
        	f_{D_i^{(\mathrm{H})}}(x)=& \sum\limits_{u=0}^{i} p_{E_1^{k}}(u)  \left(f_{E_{k+1}^{n}}(x-u)\int_x^\infty f_{\widetilde W_{\mathbf{c}_{\mathrm{P}}}}(v,b_{1:i}^{k})dv \right. \notag\\
        	& \left.+ f_{\widetilde W_{\mathbf{c}_{\mathrm{P}}}}(x-u,b_{1:i}^{k})\int_x^{\infty} f_{E_{k+1}^{n}}(v)dv \right)\notag \\
        	&+ \left(1-\sum_{u=0}^{i}p_{E_1^{k}}(u) \right)   f_{\widetilde W_{\mathbf{c}_{\mathrm{P}}}}(x-i,b_{0:i}^{k}), 		    
    	\end{align}    	
        where
    	\begin{equation}
    		f_{\widetilde W_{\mathbf{c}_{\mathrm{P}}}}(x,b)=b \cdot f_{W_{\mathbf{c}_{\mathrm{P}}}}(x) \left(1 - \int_{-\infty}^{x}f_{W_{\mathbf{c}_{\mathrm{P}}}}(v)dv \right)^{b-1}.
    	\end{equation}

\begin{figure}
    		\begin{center}
    			\includegraphics[scale=0.55] {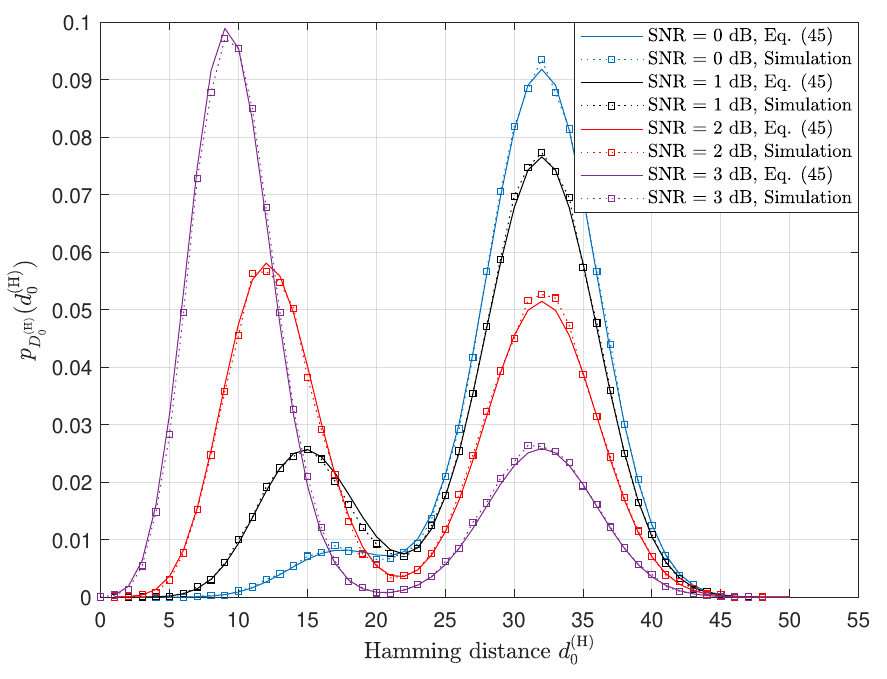}
    			\caption{The distributions of $D_{0}^{(\mathrm{H})}$ in decoding $(128,64,22)$ eBCH code at different SNRs.}
    			\label{Fig::IV::BCH128-HD-0Phase}
    		\end{center}
    	\end{figure}

    	 {\color{black} We take the decoding of eBCH codes and Polar codes as examples to verify the accuracy of Hamming distance distributions (\ref{equ::HDdis::0phase}) and (\ref{equ::HDdis::iphase}).} We first show the distribution of $D_{0}^{(\mathrm{H})}$ in decoding $(128,64,22)$ eBCH code in Fig. \ref{Fig::IV::BCH128-HD-0Phase}. As the SNR increases, it can be seen that the distribution will concentrate towards left (i.e., $D_{0}^{(\mathrm{H})}$ becomes smaller), which indicates that the decoding error decreases as well.
    	
\begin{figure}
    		\begin{center}
    			\includegraphics[scale=0.55] {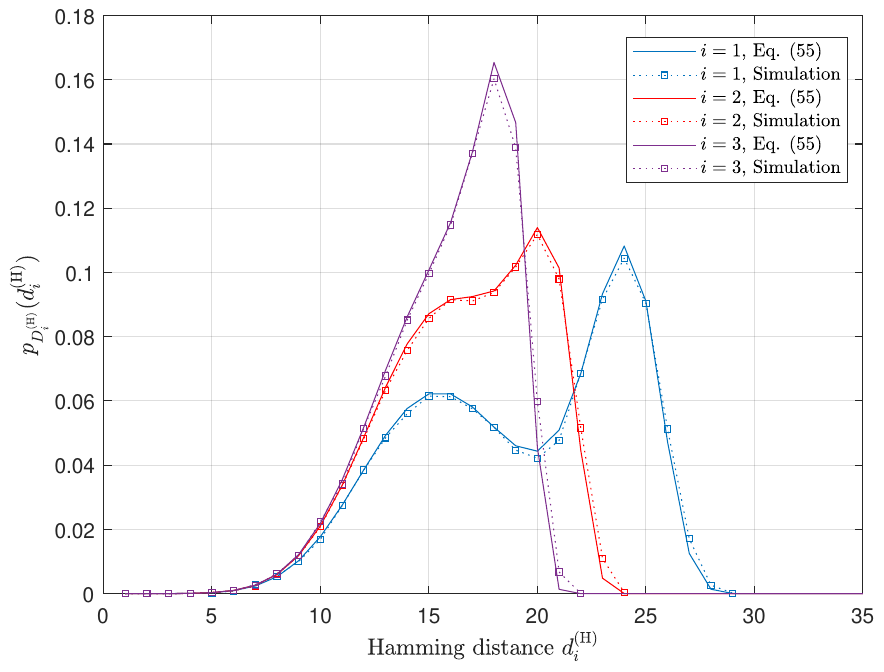}
    			\caption{The distributions of $D_{i}^{(\mathrm{H})}$ in decoding $(128,64,22)$ eBCH code, SNR = 1 dB. }
    			\label{Fig::IV::BCH128-HD-iPhase}
    		\end{center}
    	\end{figure}

    	We also show the distribution of $D_{i}^{(\mathrm{H})}$, $i=1,2,3$, in decoding $(128,64,22)$ eBCH code in Fig. \ref{Fig::IV::BCH128-HD-iPhase}. From (\ref{equ::HDdis::iphase}), we can see that the distribution of $D_{i}^{\mathrm{H}}$ is also a mixture of two random distributions, and the weight of mixture is given by $\sum_{u=0}^{i} p_{E_1^k}(u)$ and $1 - \sum_{u=0}^{i} p_{E_1^k}(u)$, respectively. It is known that an order-$i$ OSD can correct maximum $i$ errors in the MRB positions, therefore the decoding performance is determined by the probability that the number of errors in MRB is less than $i$ \cite{dhakal2016error}, which is given by $\sum_{u=0}^{i} p_{E_1^k}(u)$. From the simulation results in Fig. \ref{Fig::IV::BCH128-HD-iPhase}, it can be seen that the weight of the first term of (\ref{equ::HDdis::iphase}) increases as the decoding order increases, which implies that the decoding performance is improved with higher reprocessing order.
    	
\begin{figure}
    		\begin{center}
    			\includegraphics[scale=0.55] {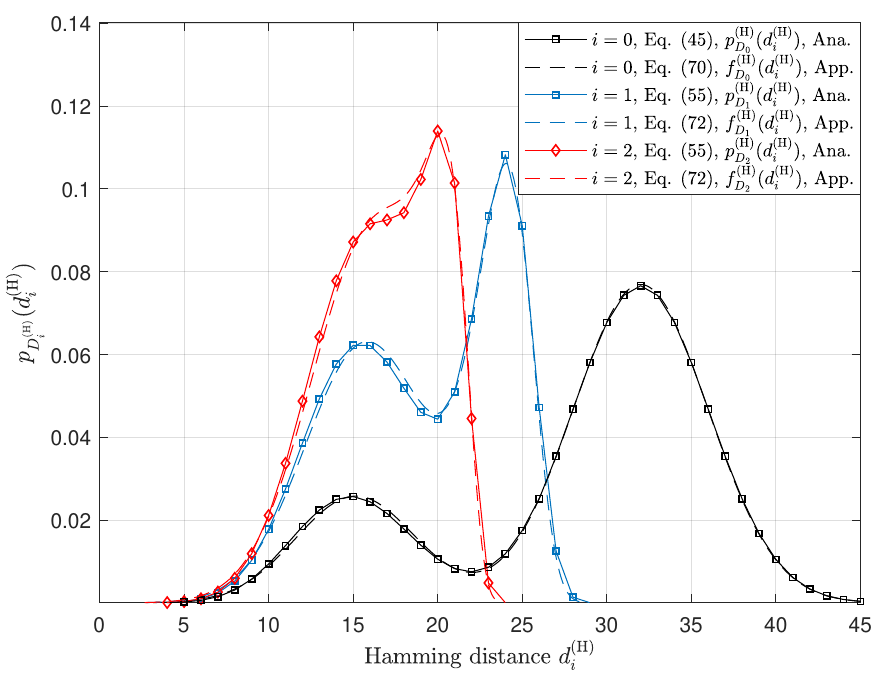}
    			\caption{The Normal approximations of the distributions of $D_{i}^{(\mathrm{H})}$ in decoding $(128,64,22)$ eBCH code, SNR = 1 dB, $i=0,1,2$.}
    			\label{Fig::IV::BCH128-HD-App}
    		\end{center}
    	\end{figure}

        Because the weight spectrum of $(128,64,22)$ eBCH code can be well approximated by the binomial distribution, we verify the accuracy of the approximations obtained in (\ref{equ::HDdis::0phase::App}) and (\ref{equ::HDdis::iphase::App}) for the distributions of $D_{0}^{(\mathrm{H})}$ and $D_{i}^{(\mathrm{H})}$ in decoding $(128,64,22)$ eBCH code in Fig. \ref{Fig::IV::BCH128-HD-App}. It can be seen that the normal approximation of Hamming distance distribution is tight, especially for low order reprocessings.
    	
\begin{figure}
    		\begin{center}
    			\includegraphics[scale=0.55] {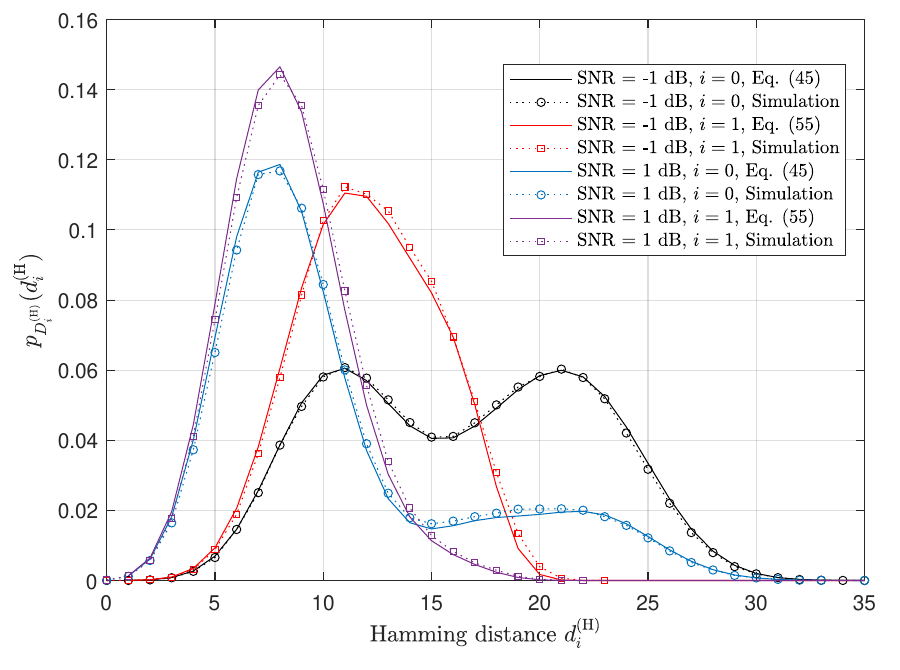}
    			\caption{{\color{black}The distributions of $D_{i}^{(\mathrm{H})}$ distribution in decoding $(64,21,16)$ Polar code, $i=0,1$.}}
    			\label{Fig::IV::Polar64-HD}
    		\end{center}
    	\end{figure}

    	For the case that the binomial distribution cannot approximate the weight spectrum of the code, we take the $(64,21,16)$ Polar code as an example to verify Theorem \ref{the::HDdis::0phase} and Theorem \ref{the::HDdis::iphase}. As depicted in Fig \ref{Fig::IV::Polar64-HD}, the $\mathrm{pmfs}$ given by (\ref{equ::HDdis::0phase}) and (\ref{equ::HDdis::iphase}) can accurately describe the distributions of $D_{0}^{(\mathrm{H})}$ and $D_{i}^{(\mathrm{H})}$, respectively. Note that in the numerical computation, we determine $p_{\mathbf{c}_{\mathrm{P}}}(\ell,q)$ in (\ref{equ::HDdis::iphase::Wecp}) by computer search. One can further determine $p_{\mathbf{c}_{\mathrm{P}}}(\ell,q)$ theoretically based on the code structure to enable an accurate calculation of (\ref{equ::HDdis::iphase::Wecp}).

\section{The Weighted Hamming Distance in OSD} \label{sec::WHD}

In this section, we characterize the distribution of the WHD in the OSD algorithm. Compared to the Hamming distance, WHD plays a more critical role in the OSD decoding since it is usually applied as the metric in finding the best codeword estimate. Given the distribution of WHD, we can acquire more information about a codeword candidate generated by the re-encoding and benefit the decoder design. 
    
The accurate characterization of the WHD distribution involves the linear combination of a large number of dependent and non-identical random variables. In what follows, we first introduce the exact expression of WHD distribution in 0-reprocessing, and then give a  normal approximation using the approximation we derived in Section \ref{sec::OrderStat::App}. The results of 0-reprocessing will be further extended to the general $i$-reprocessing OSD case.
    
\subsection{WHD distribution in the 0-reprocessing}
Let $\widetilde{\mathbf{c}}_{0}$ denote the codeword estimate after the 0-reprocessing. The WHD between $\widetilde{\mathbf{c}}_{0}$ and $\widetilde{\mathbf{y}}$ is defined as
        \begin{equation} \label{equ::WHD::0phase::define}
             d_{0}^{(\mathrm{W})} = d^{(\mathrm{W})}(\widetilde{\mathbf{c}}_0,\widetilde{\mathbf{y}})  \triangleq \sum_{\substack{1\leq u<\leq n \\ \widetilde{c}_{0,u}\neq  \widetilde{y}_{u}}} \widetilde{\alpha}_{u}.
        \end{equation}
Let $D_{0}^{(\mathrm{W})}$ denote the random variable of 0-reprocessing WHD, and $d_{0}^{(\mathrm{W})}$ is the sample of $D_{0}^{(\mathrm{W})}$. Consider a vector $\mathbf{t}_h^{\mathrm{P}} = [t^{\mathrm{P}}]_1^h$ with length $h$, $0\leq h \leq (n-k)$, representing a set of position indices satisfying $(k+1)\leq t_1^{\mathrm{P}}<t_2^{\mathrm{P}} <\ldots < t_h^{\mathrm{P}}\leq n$. Assume that $\mathcal{T}_h^{\mathrm{P}} = \left\{\mathbf{t}_h^{\mathrm{P}}\right\}$ is the set of all the vectors $\mathbf{t}_h^{\mathrm{P}}$ with length $h$, thus the cardinality of $\mathcal{T}_h^{\mathrm{P}}$ is $\binom{n-k}{h}$. Let $\mathbf{z}_{\mathbf{t}_h^{\mathrm{P}}}$ denote a length-$(n-k)$ binary vector which has nonzero elements only in the positions indexed by $\mathbf{t}_h^{\mathrm{P}}- k$. Let us also define a new random variable $\widetilde{A}_{\mathbf{t}_h^{\mathrm{P}}}$ representing the sum of reliabilities corresponding to the position indices $\mathbf{t}_h^{\mathrm{P}}$, i.e., $\widetilde{A}_{\mathbf{t}_h^{\mathrm{P}}} = \sum\limits_{u=1}^{h} \widetilde{A}_{t_u^{\mathrm{P}}}$, and the $\mathrm{pdf}$ of $\widetilde { A}_{\mathbf{t}_h^{\mathrm{P}}}$ is denoted by $f_{\widetilde{A}_{\mathbf{t}_h^{\mathrm{P}}}}(x)$.

    Assuming that the probability $p_{\mathbf{c}_{\mathrm{P}}}(u,q)$ with respect to $\mathcal{C}(n,k)$ is known, we characterize the distribution of 0-reprocessing WHD in Lemma \ref{lem::WHD::0phase::Pc} and Theorem \ref{the::WHD::0phase} as follows.

        \begin{lemma} \label{lem::WHD::0phase::Pc}
            Given a linear block code $\mathcal{C}(n, k)$ and its respective $p_{\mathbf{c}_{\mathrm{P}}}(u,q)$, consider the probability $\mathrm{Pr}(\widetilde{\mathbf{c}}_{0,\mathrm{P}}' \oplus \widetilde{\mathbf{e}}_{\mathrm{P}} = \mathbf{z}_{\mathbf{t}_h^{\mathrm{P}}}|\widetilde{\mathbf{e}}_{\mathrm{B}}\neq \mathbf{0})$, denoted by $\mathrm{Pc}(\mathbf{t}_h^{\mathrm{P}})$, where $\widetilde{\mathbf{c}}_{0,\mathrm{P}}'$ is the parity part of $\widetilde{\mathbf{c}}_{0}' = \widetilde{\mathbf{e}}_{\mathrm{B}}\widetilde{\mathbf{G}}$ and $\widetilde{\mathbf{e}}_{\mathrm{B}}\neq \mathbf{0}$. Then, $\mathrm{Pc}(\mathbf{t}_h^{\mathrm{P}})$ is given by
            \begin{equation} \label{equ::WHD::0phase::Pct}
                \mathrm{Pc}(\mathbf{t}_h^{\mathrm{P}}) = \!\!\sum_{\mathbf{x}\in \{0,1\}^{n-k}}\!\! \mathrm{Pr}(\widetilde{\mathbf{c}}_{0}' = \mathbf{z}_{\mathbf{t}_h^{\mathrm{P}}}\oplus\mathbf{x}|\widetilde{\mathbf{e}}_{\mathrm{B}}\neq \mathbf{0})\mathrm{Pr}(\widetilde{\mathbf{e}}_{\mathrm{P}} = \mathbf{x}),
            \end{equation}
            where $\mathbf{x}\ = [x]_1^{n-k}$ is a length-$(n-k)$ binary vector, and $\mathrm{Pr}(\widetilde{\mathbf{c}}_{0}' = \mathbf{z}_{\mathbf{t}_h^{\mathrm{P}}}\oplus\mathbf{x}|\widetilde{\mathbf{e}}_{\mathrm{B}}\neq \mathbf{0})$ and $\mathrm{Pr}(\widetilde{\mathbf{e}}_{\mathrm{P}} = \mathbf{x})$ are respectively given by
            \begin{equation}  \label{equ::WHD::0phase::Pct::cx}
                \mathrm{Pr}(\widetilde{\mathbf{c}}_{0}' \!=\! \mathbf{z}_{\mathbf{t}_h^{\mathrm{P}}}\oplus\mathbf{x}|\widetilde{\mathbf{e}}_{\mathrm{B}}\neq \mathbf{0}) \!=\! \sum_{q = 1}^{k} \frac{p_{E_1^k}(q) p_{\mathbf{c}_{\mathrm{P}}}(w(\mathbf{z}_{\mathbf{t}_h^{\mathrm{P}}}\oplus\mathbf{x}),q)}{(1\!-\!p_{E_1^k}(0))\binom{n-k}{w(\mathbf{z}_{\mathbf{t}_h^{\mathrm{P}}}\oplus\mathbf{x})}},
            \end{equation}
            and 
            \begin{equation}  \label{equ::WHD::0phase::Pct::ex}
            \begin{split}
                \mathrm{Pr}(\widetilde{\mathbf{e}}_{\mathrm{P}} \!=\! \mathbf{x}) &=  \underbrace{\int_{0}^{\infty}\!\! \cdots }_{n-k-w(\mathbf{x})}\underbrace{\int_{-\infty}^{0}\!\! \cdots }_{w(\mathbf{x})} \left(\frac{n!}{k!}F_{A}(x_{k+1})^{k}\prod_{v=k}^{n} f_{R}(x_v) \right.\\
                &\cdot \left.\prod_{v=k+1}^{n} \mathbf{1}_{[0,|x_{v-1}|]}(|x_v|) \right) \prod_{\substack{k < v \leq n\\ z_v=0}} dx_v \prod_{\substack{k < v \leq n\\z_v \neq 0}} dx_v   .
            \end{split}
            \end{equation}
        \end{lemma}
        \begin{IEEEproof}
              For a specific vector $\mathbf{z}_{\mathbf{t}_h^{\mathrm{P}}}$, there exist $2^{n-k}$ possible pairs of $\widetilde{\mathbf{c}}_{0,\mathrm{P}}'$ and $\widetilde{\mathbf{e}}_{\mathrm{P}}$ that satisfy $\widetilde{\mathbf{c}}_{0,\mathrm{P}}' \oplus \widetilde{\mathbf{e}}_{\mathrm{P}} = \mathbf{z}_{\mathbf{t}_h^{\mathrm{P}}}$. To see this, we assume that there exists an arbitrary length-$(n-k)$ binary vector $\mathbf{x}$, then it can be noticed that $\mathbf{z}_{\mathbf{t}_h^{\mathrm{P}}} = \mathbf{x} \oplus \mathbf{x} \oplus \mathbf{z}_{\mathbf{t}_h^{\mathrm{P}}}$. Therefore, (\ref{equ::WHD::0phase::Pct}) can be obtained by considering the probability $\mathrm{Pr}(\widetilde{\mathbf{c}}_{0}' = \mathbf{x} \oplus \mathbf{z}_{\mathbf{t}_h^{\mathrm{P}}}|\widetilde{\mathbf{e}}_{\mathrm{B}}\neq \mathbf{0})\mathrm{Pr}(\widetilde{\mathbf{e}}_{\mathrm{P}} = \mathbf{x})$.
              
              When symbols with random noises are being received and the generator matrix is permuted accordingly, each column of the generator matrix has an equal probability of being permuted to any other columns. Thus, if $w(\widetilde{\mathbf{c}}_{0,\mathrm{P}}') = w(\mathbf{x} \oplus \mathbf{z}_{\mathbf{t}_h^{\mathrm{P}}})$, it can be seen that 
              \begin{equation}
                  \mathrm{Pr}\left(\widetilde{\mathbf{c}}_{0,\mathrm{P}}'\! = \! \mathbf{x} \oplus \mathbf{z}_{\mathbf{t}_h^{\mathrm{P}}} | w(\widetilde{\mathbf{c}}_{0,\mathrm{P}}') \!=\! w(\mathbf{x} \oplus \mathbf{z}_{\mathbf{t}_h^{\mathrm{P}}})\right)  = \frac{1}{\binom{n-k}{w(\mathbf{x} \oplus \mathbf{z}_{\mathbf{t}_h^{\mathrm{P}}})}}.
              \end{equation}
              Then, by observing that $\mathrm{Pr}(w(\widetilde{\mathbf{c}}_{0,\mathrm{P}}') = w(\mathbf{x} \oplus \mathbf{z}_{\mathbf{t}_h^{\mathrm{P}}})|\widetilde{\mathbf{e}}_{\mathrm{B}}\neq \mathbf{0}) = \frac{1}{1-p_{E_1^k}(0)}\sum\limits_{q = 1}^{k} p_{E_1^k}(q) p_{\mathbf{c}_{\mathrm{P}}}(w(\mathbf{x} \oplus \mathbf{z}_{\mathbf{t}_h^{\mathrm{P}}}),q)$, finally $\mathrm{Pr}(\widetilde{\mathbf{c}}_{0}' = \mathbf{x} \oplus \mathbf{z}_{\mathbf{t}_h^{\mathrm{P}}}|\widetilde{\mathbf{e}}_{\mathrm{B}}\neq \mathbf{0})$ can be determined as (\ref{equ::WHD::0phase::Pct::cx}).
              
              The probability $\mathrm{Pr}(\widetilde{\mathbf{e}}_{\mathrm{P}} = \mathbf{x})$ can be determined by considering the joint error probability of parity bits of $\widetilde{\mathbf{y}}$, which can be obtained by the joint distribution of ordered received symbols $[\widetilde{R}]_{k+1}^{n}$. According to the ordered statistics theory \cite{balakrishnan2014order}, the joint $\mathrm{pdf}$ of $[\widetilde{R}]_{k+1}^{n}$, denoted by $f_{[\widetilde{R}]_{k+1}^{n}}(x_{k+1},\ldots,x_{n})$, can be derived as
              \begin{equation}
              \begin{split}
                    &f_{[\widetilde{R}]_{k+1}^{n}}(x_{k+1},\ldots,x_{n}) \\
                    &= \frac{n!}{k!}F_{A}(x_{k+1})^{k}\cdot\prod_{v=k}^{n} f_{R}(x_v) \prod_{v=k+1}^{n} \mathbf{1}_{[0,|x_{v-1}|]}(|x_v|).   
              \end{split}
              \end{equation}
              Therefore, $\mathrm{Pr}(\widetilde{\mathbf{e}}_{\mathrm{P}} = \mathbf{x}) $ can be finally determined as 
              \begin{equation}
              \begin{split}
                    &\mathrm{Pr}(\widetilde{\mathbf{e}}_{\mathrm{P}} = \mathbf{x}) \\ & = \!\!\! \underbrace{\int_{0}^{\infty}\!\! \cdots }_{n\!-\!k\!-\!w(\mathbf{x})}\underbrace{\int_{-\infty}^{0} \!\!\cdots }_{w(\mathbf{x})}f_{[\widetilde{R}]_{k+1}^{n}}\!(x_{k+1},\ldots,x_{n}) \!\! \prod_{\substack{k < v \leq n\\ z_v=0}}\!\! dx_v \!\!\prod_{\substack{k < v \leq n\\z_v \neq 0}}\!\! dx_v .    
              \end{split}
              \end{equation}
              
              Finally, summing up the probability $\mathrm{Pr}(\widetilde{\mathbf{c}}_{0}' = \mathbf{x} \oplus \mathbf{z}_{\mathbf{t}_h^{\mathrm{P}}})\cdot\\\mathrm{Pr}(\widetilde{\mathbf{e}}_{\mathrm{P}} = \mathbf{x})$ for $2^{n-k}$ different $\mathbf{x}$, (\ref{equ::WHD::0phase::Pct::ex}) is obtained.
              
        \end{IEEEproof}
        
        \begin{theorem} \label{the::WHD::0phase}
            Given a linear block code $\mathcal{C}(n, k)$ and its respective $p_{\mathbf{c}_{\mathrm{P}}}(u,q)$, the $\mathrm{pdf}$ of the weighted Hamming distance $D_0^{(\mathrm{W})}$ between $\widetilde{\mathbf{y}}$ and $\widetilde{\mathbf{c}}_{0}$ after the 0-reprocessing is given by
        	\begin{equation}  \label{equ::WHD::0phase}
        	\begin{split}
             	f_{D_0^{(\mathrm{W})}}(x) &= \sum_{h=0}^{n-k}\sum_{ \mathbf{t}_h^{\mathrm{P}} \in \mathcal{T}_h^{\mathrm{P}}} \mathrm{Pe}( \mathbf{t}_h^{\mathrm{P}}) f_{\widetilde{A}_{ \mathbf{t}_h^{\mathrm{P}}}}(x) \\
             	&+ \sum_{h=0}^{n-k}\sum_{ \mathbf{t}_h^{\mathrm{P}} \in \mathcal{T}_h^{\mathrm{P}}} (1-p_{E_1^k}(0)) \mathrm{Pc}( \mathbf{t}_h^{\mathrm{P}}) f_{\widetilde{A}_{ \mathbf{t}_h^{\mathrm{P}}}}(x),   
        	\end{split}
        	\end{equation}    	
        	 where $p_{E_1^k}(0)$ is given by (\ref{equ::HDdis::0phase::weightPe}), $f_{\widetilde{A}_{\mathbf{t}_h^{\mathrm{P}}}}(x)$ is the $\mathrm{pdf}$ of the sum of reliabilities corresponding to the position indices $\mathbf{t}_h^{\mathrm{P}}$, i.e., $\widetilde{A}_{\mathbf{t}_h^{\mathrm{P}}} = \sum_{u=1}^{h} \widetilde{A}_{t_u^{\mathrm{P}}}$,  $\mathrm{Pe}(\mathbf{t}_h^{\mathrm{P}})$ is given by
            \begin{equation}
            \begin{split}
                  \mathrm{Pe}(\mathbf{t}_h^{\mathrm{P}})&= \underbrace{\int_{0}^{\infty}\!\! \cdots }_{n\!-\!h}\underbrace{\int_{-\infty}^{0}\!\! \cdots }_{h} \left(n!\! \prod_{v=1}^{n}\! f_{R}(x_v)\! \prod_{v=2}^{n}\! \mathbf{1}_{[0,|x_{v-1}|]}(|x_v|) \right)  \\
                  &\cdot\prod_{\substack{0 < v \leq n\\v\in \mathbf{t}_h^{\mathrm{P}}}} dx_v \prod_{\substack{0 < v \leq n\\v\notin \mathbf{t}_h^{\mathrm{P}}}} dx_v ,            
            \end{split}
            \end{equation} 
             {\color{black} and $\mathrm{Pc}(\mathbf{t}_h^{\mathrm{P}})$ is given by (\ref{equ::WHD::0phase::Pct}).}
        \end{theorem}  
        \begin{IEEEproof}
            The proof is provided in Appendix \ref{app::proof::WHD::0phase}.
        \end{IEEEproof}
        
    \subsection{WHD distribution in the $i$-Reprocessing}   \label{sec::WHD-iphase}
        In this part, we introduce the distribution of the recorded minimum WHD after the $i$-reprocessing ($0\leq i\leq m$) in the order-$m$ OSD, i.e., the minimum WHD among the $0,1,\cdots,i$ reprocessings. We define the random variable $D_i^{(\mathrm{W})}$ representing this minimum WHD, and random variable $D_{\mathbf{e}}^{(\mathrm{W})}$ representing the WHD between $\widetilde{\mathbf{c}}_{\mathbf{e}}$ and $\widetilde{\mathbf{y}}$. Accordingly, $d_{i}^{(\mathrm{W})}$ and $d_{\mathbf{e}}^{(\mathrm{W})}$ are the samples of $D_i^{(\mathrm{W})}$ and $D_{\mathbf{e}}^{(\mathrm{W})}$, respectively.
    	
    	Consider a vector $\mathbf{t}_{\ell}^{\mathrm{B}} = [t^{\mathrm{B}}]_1^{\ell}$, $0\leq \ell \leq i$, representing a set of position indices within the MRB part which satisfy $1\leq t_1^{\mathrm{B}} < t_2^{\mathrm{B}} <\ldots < t_{\ell}^{\mathrm{B}} \leq k$. Assume that $\mathcal{T}_{\ell}^{\mathrm{B}} = \left\{\mathbf{t}_{\ell}^{\mathrm{B}}\right\}$ is the set of all vectors $\mathbf{t}_{\ell}^{\mathrm{B}}$ with length $\ell$, thus the cardinality of $\mathcal{T}_{\ell}^{\mathrm{B}}$ is given by $\binom{k}{\ell}$. Let us consider a new indices vector $\mathbf{t}_{\ell}^{h}$ defined as $\mathbf{t}_{\ell}^{h} = [\mathbf{t}_{\ell}^{\mathrm{B}} \  \ \mathbf{t}_h^{\mathrm{P}}]$ with length $\ell+h$, and let the random variable $\widetilde{A}_{\mathbf{t}_{\ell}^{h}}$ denote the sum of reliabilities corresponding to the position indices $\mathbf{t}_{\ell}^{h}$, i.e., $\widetilde{A}_{\mathbf{t}_{\ell}^{h}} = \sum_{u=1}^{\ell}\widetilde{A}_{\mathbf{t}_u^{\mathrm{B}}} +  \sum_{u=1}^{h} \widetilde{A}_{\mathbf{t}_u^{\mathrm{P}}}$, with the $\mathrm{pdf}$ $f_{\widetilde{A}_{\mathbf{t}_{\ell}^{h}}}(x)$. Furthermore, let $\mathbf{z}_{t_{\ell}^{\mathrm{B}}}$ denote a length-$k$ binary vector whose nonzero elements are indexed by $t_{\ell}^{\mathrm{B}}$. Thus, $\mathbf{z}_{\mathbf{t}_{\ell}^{h}} = [\mathbf{z}_{t_{\ell}^{\mathrm{B}}} \ \ \mathbf{z}_{t_{h}^{\mathrm{P}}}]$ is a length-$n$ binary vector with nonzero elements indexed by $\mathbf{t}_{\ell}^{h}$. Next, we investigate the distribution of $D_i^{(\mathrm{W})}$, started with Lemma \ref{lem::WHD::iphase::eB=e} and concluded in Theorem \ref{the::WHD::iphase}.
        
        First, we give the $\mathrm{pdf}$ of $D_{\mathbf{e}}^{(\mathrm{W})}$ on the condition that some TEP $\mathbf{e}$ eliminates the error pattern $\widetilde{\mathbf{e}}_{\mathrm{B}}$ over $\widetilde{\mathbf{y}}_{\mathrm{B}}$, which is summarized in the following Lemma.
    	\begin{lemma} \label{lem::WHD::iphase::eB=e}
        	Given a linear block code $\mathcal{C}(n,k)$, if the errors $\widetilde{\mathbf{e}}_{\mathrm{B}}$ over $\widetilde{\mathbf{y}}_{\mathrm{B}}$ are eliminated by a TEP $\mathbf{e}$ after the $i$-reprocessing ($0\leq i\leq m$) of an order-$m$ OSD, the $\mathrm{pdf}$ of the weighted Hamming distance between $\widetilde{\mathbf{c}}_{\mathbf{e}}$ and $\widetilde{\mathbf{y}}$, $D_{\mathbf{e}}^{(\mathrm{W})}$, is given by 
	        \begin{equation} \label{equ::WHD::iphase::eb=e}
                f_{D_{\mathbf{e}}^{(\mathrm{W})}}(x|\widetilde{\mathbf{e}}_{\mathrm{B}} \!=\! \mathbf{e}) = \frac{1}{\sum\limits_{v=0}^i \! p_{E_1^{k}}(\!v\!) }\sum_{\ell=0}^{i}\sum_{h=0}^{n-k} \!\sum_{\substack{\mathbf{t}_{\ell}^{h}\\\mathbf{t}_{\ell}^{\mathrm{B}} \in \mathcal{T}_{\ell}^{\mathrm{B}} \\ \mathbf{t}_{h}^{\mathrm{P}} \in \mathcal{T}_h^{\mathrm{P}} }}\! \mathrm{Pe}( \mathbf{t}_{\ell}^{h})  f_{\widetilde{A}_{ \mathbf{t}_{\ell}^{h}}}(x),
            \end{equation}
            where $\mathrm{Pe}( \mathbf{t}_{\ell}^{h})$ is given by 
            \begin{equation} \label{equ::WHD::iphase::eb=e::Pet}
            \begin{split}
                \mathrm{Pe}( \mathbf{t}_{\ell}^{h}) \!&=\!\! \underbrace{\int_{0}^{\infty} \!\!\cdots}_{n\!-\!h\!-\!\ell}\underbrace{\int_{-\infty}^{0} \!\!\cdots}_{h\!+\!\ell} \left(n! \prod_{v=1}^{n} f_{R}(x_v) \prod_{v=2}^{n} \mathbf{1}_{[0,|x_{v-1}|]}(|x_v|) \right) \\
                &\cdot\prod_{\substack{1 \leq v \leq {n}\\v\in \mathbf{t}_{\ell}^{h}}} dx_v \prod_{\substack{1 \leq v \leq {n}\\v\notin \mathbf{t}_{\ell}^{h}}} dx_v ,   
            \end{split}
            \end{equation} 
            and $f_{\widetilde{A}_{\mathbf{t}_{\ell}^{h}}}(x)$ is the $\mathrm{pdf}$ of $\widetilde{A}_{\mathbf{t}_{\ell}^{h}}= \sum\limits_{u=1}^{\ell}\widetilde{A}_{\mathbf{t}_u^{\mathrm{B}}} +  \sum\limits_{u=1}^{h} \widetilde{A}_{\mathbf{t}_u^{\mathrm{P}}}$.
        \end{lemma}  
        \begin{IEEEproof}
            The proof is provided in Appendix \ref{app::proof::WHD::iphase::eB=e}. 
        \end{IEEEproof}
        
    From Lemma \ref{lem::WHD::iphase::eB=e} and its proof, we can see that if errors in MRB positions are eliminated by a TEP, the WHD is determined by the errors in MRB part and the parity part. In contrast, if the errors are not eliminated by a TEP, both the error over $\widetilde{\mathbf{y}}$ and the code weight enumerator affect the WHD. We summarize this conclusion in the following Lemma.
      {\color{black}   
    \begin{lemma} \label{lem::WHD::iphase::eB!=e}
        Given a linear block code $\mathcal{C}(n, k)$ with the probability $p_{\mathbf{c}_{\mathrm{P}}}(u,q)$, if the errors over the MRB $\widetilde{\mathbf{y}}_{\mathrm{B}}$ are \textbf{not} eliminated by any TEPs in the first $i$ ($0\leq i \leq m$) reprocessings of an order-$m$ OSD, for a random TEP $\mathbf{e}$, the weighted Hamming distance between $\widetilde{\mathbf{c}}_{\mathbf{e}}$ and $\widetilde{\mathbf{y}}$ is given by 
	        \begin{equation} \label{equ::WHD::iphase::eB!=e}
                f_{D_{\mathbf{e}}^{(\mathrm{W})}}(x|\widetilde{\mathbf{e}}_{\mathrm{B}}\neq\mathbf{e}) = \sum_{\ell=0}^{i}\sum_{h=0}^{n-k} \sum_{\substack{\mathbf{t}_{\ell}^{h}\\\mathbf{t}_{\ell}^{\mathrm{B}} \in \mathcal{T}_{\ell}^{\mathrm{B}} \\ \mathbf{t}_{h}^{\mathrm{P}} \in \mathcal{T}_h^{\mathrm{P}} }} \mathrm{Pc}(\mathbf{t}_{\ell}^{h})  f_{\widetilde{A}_{\mathbf{t}_{\ell}^{h}}}(x),
            \end{equation}
            where $\mathrm{Pc}(\mathbf{t}_{\ell}^{h})$ is given by 
            \begin{equation} \label{equ::WHD::iphase::eB!=e::Pc}
                  \mathrm{Pc}(\mathbf{t}_{\ell}^{h}) = \frac{1}{b_{0:i}^{k}}\cdot\sum_{\mathbf{x}\in \{0,1\}^{n-k}}\mathrm{Pr}(\widetilde{\mathbf{c}}_{\mathbf{e},\mathrm{P}}' =  \mathbf{z}_{\mathbf{t}_h^{\mathrm{P}}}\oplus\mathbf{x})\mathrm{Pr}(\widetilde{\mathbf{e}}_{\mathrm{P}} = \mathbf{x}),
            \end{equation} 
            where $ \mathbf{x}$ is a length-$(n-k)$ binary vector. The probability $\mathrm{Pr}(\widetilde{\mathbf{c}}_{\mathbf{e}}' =  \mathbf{z}_{\mathbf{t}_h^{\mathrm{P}}}\oplus\mathbf{x})$ is given by
            \begin{equation} \label{equ::WHD::iphase::eB!=e::zx}
            \begin{split}
                &\mathrm{Pr}(\widetilde{\mathbf{c}}_{\mathbf{e},\mathrm{P}}' =  \mathbf{z}_{\mathbf{t}_h^{\mathrm{P}}}\oplus\mathbf{x}) \\
                &\qquad = \frac{1}{\binom{n-k}{w( \mathbf{z}_{\mathbf{t}_h^{\mathrm{P}}}\oplus\mathbf{x})}}\sum_{q = 1}^{k} p_{W_{\mathbf{e},\widetilde{\mathbf{e}}_{\mathrm{B}}}}(q|\mathbf{e} \!=\! \mathbf{z}_{t_{\ell}^{\mathrm{B}}}) p_{\mathbf{c}_{\mathrm{P}}}(w( \mathbf{z}_{\mathbf{t}_h^{\mathrm{P}}}\!\oplus\!\mathbf{x}),q).
            \end{split}
            \end{equation}
            $p_{W_{\mathbf{e},\widetilde{\mathbf{e}}_{\mathrm{B}}}}(q|\mathbf{e} = \mathbf{z}_{t_{\ell}^{\mathrm{B}}})$ is the conditional $\mathrm{pmf}$ of $W_{\mathbf{e},\widetilde{\mathbf{e}}_{\mathrm{B}}}$ given by
            \begin{equation}
                p_{W_{\mathbf{e},\widetilde{\mathbf{e}}_{\mathrm{B}}}}(q|\mathbf{e} = \mathbf{z}_{t_{\ell}^{\mathrm{B}}}) = \sum_{\substack{\mathbf{x} \in \{0,1\}^{k}\\ w(\mathbf{z}_{t_{\ell}^{\mathrm{B}}}\oplus\mathbf{x})=q}}\mathrm{Pr}(\widetilde{\mathbf{e}}_{\mathrm{B}} = \mathbf{x}),
            \end{equation}
            where $\mathbf{x} = [x]_1^k$ is a length-$k$ binary vector satisfying $w(\mathbf{z}_{t_{\ell}^{\mathrm{B}}}\oplus\mathbf{x})=q$, and
            \begin{equation} \label{equ::WHD::iphase::eB!=e::Pc::eB=eksi}
            \begin{split}
                \mathrm{Pr}(\widetilde{\mathbf{e}}_{\mathrm{B}} \!=\! \mathbf{x}) &\!=\!\!\! \underbrace{\int_{0}^{\infty}\!\!\! \cdots }_{k \!-\! w(\mathbf{x})}\underbrace{\int_{-\infty}^{0}\! \!\!\cdots }_{w(\mathbf{x})} \!\left(\!\!n! \!\prod_{v=1}^{n}\! f_{R}(x_v)\! \prod_{v=2}^{n} \!\mathbf{1}_{[0,|x_{v-1}|]}(|x_v|)\! \right) \\
                &\cdot\prod_{\substack{0 < v \leq k\\  x_{v} \neq 0 }} dx_v \prod_{\substack{0 < v \leq k\\ x_{v} =0}} dx_v .
            \end{split}
            \end{equation}
            Furthermore, the probability $\mathrm{Pr}(\widetilde{\mathbf{e}}_{\mathrm{P}} = \mathbf{x})$ is given by (\ref{equ::WHD::0phase::Pct::ex}), and
            $f_{\widetilde{A}_{\mathbf{t}_{\ell}^{h}}}(x)$ is the $\mathrm{pdf}$ of $\widetilde{A}_{\mathbf{t}_{\ell}^{h}}= \sum\limits_{v=1}^{\ell}\widetilde{A}_{t_v^{\mathrm{B}}} +  \sum\limits_{v=1}^{h} \widetilde{A}_{t_v^{\mathrm{P}}}$.

        \end{lemma} 
        \begin{IEEEproof}
            The proof is provided in Appendix \ref{app::proof::WHD::iphase::eB!=e}.
        \end{IEEEproof}
        It is worth noting that $q \neq 0$ in (\ref{equ::WHD::iphase::eB!=e::zx}), therefore $\mathbf{e} \neq \widetilde{\mathbf{e}}_{\mathrm{B}}$, i.e., the errors over the MRB are not eliminated by any TEPs.
        
        We can directly extend the result in Lemma \ref{lem::WHD::iphase::eB!=e} to find the conditional $\mathrm{pdf}$ of the $D_{\mathbf{e}}^{(\mathrm{W})}$ conditioning on $\{w(\widetilde{\mathbf{e}}_{\mathrm{B}}) \!\neq\! \mathbf{e},w(\widetilde{\mathbf{e}}_{\mathrm{B}}) \!\leq \! i\}$ as 
        \begin{equation} \label{equ::WHD::iphase::eB!=e::eB<=i}
        \begin{split}
            f_{D_{\mathbf{e}}^{(\mathrm{W})}}&(x|\widetilde{\mathbf{e}}_{\mathrm{B}}\!\neq\!\mathbf{e},w(\widetilde{\mathbf{e}}_{\mathrm{B}})\! \leq\! i) \\
            &= \sum_{\ell=0}^{i}\sum_{h=0}^{n-k} \sum_{\substack{\mathbf{t}_{\ell}^{h}\\\mathbf{t}_{\ell}^{\mathrm{B}} \in \mathcal{T}_{\ell}^{\mathrm{B}} \\ \mathbf{t}_{h}^{\mathrm{P}} \in \mathcal{T}_h^{\mathrm{P}} }} \mathrm{Pc}(\mathbf{t}_{\ell}^{h}|w(\widetilde{\mathbf{e}}_{\mathrm{B}})\! \leq\! i)  f_{\widetilde{A}_{\mathbf{t}_{\ell}^{h}}}(x),    
        \end{split}
        \end{equation}
        where the conditional probability $\mathrm{Pc}(\mathbf{t}_{\ell}^{h}|w(\widetilde{\mathbf{e}}_{\mathrm{B}})\! \leq\! i)$ is obtained similar to (\ref{equ::WHD::iphase::eB!=e::Pc}), but with $p_{W_{\mathbf{e},\widetilde{\mathbf{e}}_{\mathrm{B}}}}(q|\mathbf{e} = \mathbf{z}_{t_{\ell}^{\mathrm{B}}})$ replaced by $p_{W_{\mathbf{e},\widetilde{\mathbf{e}}_{\mathrm{B}}}}(q|\mathbf{e} \!=\! \mathbf{z}_{t_{\ell}^{\mathrm{B}}},w(\widetilde{\mathbf{e}}_{\mathrm{B}})\! \leq\! i)$ given by
        \begin{equation}
        \begin{split}
               p_{W_{\mathbf{e},\widetilde{\mathbf{e}}_{\mathrm{B}}}}&(q|\mathbf{e} \!=\! \mathbf{z}_{t_{\ell}^{\mathrm{B}}},w(\widetilde{\mathbf{e}}_{\mathrm{B}})\! \leq\! i) \\
               &= \sum_{\substack{\mathbf{x} \in \{0,1\}^{k}\\ w(\mathbf{z}_{t_{\ell}^{\mathrm{B}}}\oplus\mathbf{x})=q\\ w(\mathbf{x})\leq i}}\mathrm{Pr}(\widetilde{\mathbf{e}}_{\mathrm{B}} = \mathbf{x})\left(\sum_{u=0}^{i}p_{E_1^k}(u)\right)^{-1}, 
        \end{split}
        \end{equation}
        Similar to (\ref{equ::WHD::iphase::eB!=e::eB<=i}), we can also obtain $f_{D_{\mathbf{e}}^{(\mathrm{W})}}(x|\widetilde{\mathbf{e}}_{\mathrm{B}}\!\neq\!\mathbf{e},w(\widetilde{\mathbf{e}}_{\mathrm{B}})\! >\! i)$ as 
        \begin{equation} \label{equ::WHD::iphase::eB!=e::eB>i}
        \begin{split}
            f_{D_{\mathbf{e}}^{(\mathrm{W})}}&(x|\widetilde{\mathbf{e}}_{\mathrm{B}}\!\neq\!\mathbf{e},w(\widetilde{\mathbf{e}}_{\mathrm{B}})\! >\! i) \\
            &= \sum_{\ell=0}^{i}\sum_{h=0}^{n-k} \sum_{\substack{\mathbf{t}_{\ell}^{h}\\\mathbf{t}_{\ell}^{\mathrm{B}} \in \mathcal{T}_{\ell}^{\mathrm{B}} \\ \mathbf{t}_{h}^{\mathrm{P}} \in \mathcal{T}_h^{\mathrm{P}} }} \mathrm{Pc}(\mathbf{t}_{\ell}^{h}|w(\widetilde{\mathbf{e}}_{\mathrm{B}})\! >\! i)  f_{\widetilde{A}_{\mathbf{t}_{\ell}^{h}}}(x),
        \end{split}
        \end{equation}
        by considering 
        \begin{equation}
        \begin{split}
            p_{W_{\mathbf{e},\widetilde{\mathbf{e}}_{\mathrm{B}}}}&(q|\mathbf{e} \!=\! \mathbf{z}_{t_{\ell}^{\mathrm{B}}},w(\widetilde{\mathbf{e}}_{\mathrm{B}})\! >\! i) \\
            &= \sum_{\substack{\mathbf{x} \in \{0,1\}^{k}\\ w(\mathbf{z}_{t_{\ell}^{\mathrm{B}}}\oplus\mathbf{x})=q\\ w(\mathbf{x})> i}}\mathrm{Pr}(\widetilde{\mathbf{e}}_{\mathrm{B}} = \mathbf{x})\left(1 - \sum_{u=0}^{i}p_{E_1^k}(u)\right)^{-1},  
        \end{split}
        \end{equation}
        For the sake of brevity, we omit the proofs of (\ref{equ::WHD::iphase::eB!=e::eB<=i}) and (\ref{equ::WHD::iphase::eB!=e::eB>i}) because their proofs are similar to that of Lemma \ref{lem::WHD::iphase::eB!=e}.
        }
        
        Lemma \ref{lem::WHD::iphase::eB=e} and Lemma \ref{lem::WHD::iphase::eB!=e} give the $\mathrm{pdf}$ of the WHD after the $i$-reprocessing in an order-$m$ OSD under two different conditions. However, it is worthy of noting that in Lemma \ref{lem::WHD::iphase::eB=e} and Lemma \ref{lem::WHD::iphase::eB!=e}, even though we assume that the errors are eliminated by one TEP $\mathbf{e}$, the specific pattern of $\mathbf{e}$ is unknown and is not included in the assumption. It is reasonable because the decoder cannot know which TEP can exactly eliminate the error, but only output the decoding result by comparing the distances. Combining Lemma \ref{lem::WHD::iphase::eB=e} and Lemma \ref{lem::WHD::iphase::eB!=e} and considering ordered statistics over a sequence of random variable $D_{\mathbf{e}}^{(\mathrm{W})}$, we next characterize the distribution of the minimum WHD $D_{i}^{(\mathrm{W})}$ after the $i$-reprocessing of an order-$m$ OSD.

        On the conditions that 1) the errors in MRB are \textbf{not} eliminated by any test error patterns and 2) $w(\mathbf{e}_{\mathrm{B}}) \leq i$, in the first $i$ ($0\leq i \leq m$) reprocessings of an order-$m$ OSD, we first consider the correlations between two random variables $D_{\mathbf{e}}^{(\mathrm{W})}$ and $D_{\hat{\mathbf{e}}}^{(\mathrm{W})}$, where $\mathbf{e}$ and $\hat{\mathbf{e}}$ are two arbitrary TEPs that are checked in decoding one received signal sequence, satisfying $\mathbf{e} \neq \widetilde{\mathbf{e}}_{\mathrm{B}}$, $\hat{\mathbf{e}} \neq \widetilde{\mathbf{e}}_{\mathrm{B}}$, and $\mathbf{e} \neq \hat{\mathbf{e}}$. Thus, pdfs of $D_{\mathbf{e}}^{(\mathrm{W})}$ and $D_{\hat{\mathbf{e}}}^{(\mathrm{W})}$ are both given by the mixture model described by (\ref{equ::WHD::iphase::eB!=e::eB<=i}) with the $\mathrm{pdf}$ $f_{D_{\mathbf{e}}^{(\mathrm{W})}}(x|\widetilde{\mathbf{e}}_{\mathrm{B}}\!\neq\!\mathbf{e},w(\widetilde{\mathbf{e}}_{\mathrm{B}})\! \leq\! i)$. However, $D_{\mathbf{e}}^{(\mathrm{W})}$ and $D_{\hat{\mathbf{e}}}^{(\mathrm{W})}$ are not independent random variables, because $D_{\mathbf{e}}^{(\mathrm{W})}$ and $D_{\hat{\mathbf{e}}}^{(\mathrm{W})}$ are both linear combinations of $[\widetilde{A}]_1^{n}$ which are dependent variables. For $[\widetilde{A}]_1^n$, we define the mean matrix $\widetilde{\mathbf{E}}_{n\times n}$ as 

        \begin{equation} \label{lem::WHD::iphase::MeanMx}
            \widetilde{\mathbf{E}}_{n\times n}= 
            \begin{bmatrix}
                \mathbb{E}[\widetilde{A}_1]^2 & \mathbb{E}[\widetilde{A}_1]\mathbb{E}[\widetilde{A}_2] & \cdots & \mathbb{E}[\widetilde{A}_1]\mathbb{E}[\widetilde{A}_{n}] \\
                \mathbb{E}[\widetilde{A}_2]\mathbb{E}[\widetilde{A}_1] & \mathbb{E}[\widetilde{A}_2]^2 & \cdots & \mathbb{E}[\widetilde{A}_2]\mathbb{E}[\widetilde{A}_{n}]  \\
                \vdots & \vdots & \ddots & \vdots \\
                \mathbb{E}[\widetilde{A}_{n}]\mathbb{E}[\widetilde{A}_1] & \mathbb{E}[\widetilde{A}_{n}]\mathbb{E}[\widetilde{A}_2] & \cdots & \mathbb{E}[\widetilde{A}_{n}]^2
            \end{bmatrix} ,
        \end{equation}
        and the covariance matrix $\widetilde{\mathbf\Sigma}_{{n}\times {n}}$ as
        \begin{equation} \label{lem::WHD::iphase::VarMx}
            \widetilde{\mathbf\Sigma} _{{n}\times {n}}\! = \!
            \begin{bmatrix}
            	\mathrm{cov}(\widetilde{A}_1,\widetilde{A}_1) & \mathrm{cov}(\widetilde{A}_1,\widetilde{A}_2) & \cdots & \mathrm{cov}(\widetilde{A}_1,\widetilde{A}_{n})\\ 
            	\mathrm{cov}(\widetilde{A}_2,\widetilde{A}_1) & \mathrm{cov}(\widetilde{A}_2,\widetilde{A}_2) & \cdots & \mathrm{cov}(\widetilde{A}_2,\widetilde{A}_{n})\\ 
            	\vdots & \vdots & \ddots  &  \vdots \\ 
            	\mathrm{cov}(\widetilde{A}_{n},\widetilde{A}_1) & \mathrm{cov}(\widetilde{A}_{n},\widetilde{A}_2) & \cdots & \mathrm{cov}(\widetilde{A}_{n},\widetilde{A}_{n})
            \end{bmatrix}.
        \end{equation}
        
        Consider two different position indices vectors $\mathbf{t}_{\ell}^{h} = [\mathbf{t}_{\ell}^{\mathrm{B}} \ \ \mathbf{t}_{h}^{\mathrm{P}}]$ and $\hat{\mathbf{t}}_{\hat{\ell}}^{\hat{h}} = [\hat{\mathbf{t}}_{\hat{\ell}}^{\mathrm{B}} \ \ \hat{\mathbf{t}}_{\hat{h}}^{\mathrm{P}}]$. For their corresponding random variables $\widetilde{A}_{\mathbf{t}_{\ell}^{h}}$, and $\widetilde{A}_{\hat{\mathbf{t}}_{\hat{\ell}}^{\hat{h}}}$ representing the sum of reliabilities of positions in $\mathbf{t}_{\ell}^{h}$ and $\hat{\mathbf{t}}_{\hat{\ell}}^{\hat{h}}$, respectively, the covariance of $\widetilde{A}_{\mathbf{t}_{\ell}^{h}}$ and $\widetilde{A}_{\hat{\mathbf{t}}_{\hat{\ell}}^{\hat{h}}}$ is given by
        \begin{equation}
        \begin{split}
              \mathrm{cov}\left(\widetilde{A}_{\mathbf{t}_{\ell}^{h}},\widetilde{A}_{\hat{\mathbf{t}}_{\hat{\ell}}^{\hat{h}}}\right) &= \sum_{u=1}^\ell\sum_{v=1}^{\hat{\ell}} \widetilde{\mathbf\Sigma}_{{ t}_{u}^{\mathrm{B}},{\hat{t}}_{v}^{\mathrm{B}}} + \sum_{u=1}^h\sum_{v=1}^{\hat{h}} \widetilde{\mathbf\Sigma}_{ t_u^{\mathrm{P}}, {\hat{t}}_v^{\mathrm{P}}} \\
              &+ \sum_{u=1}^\ell\sum_{v=1}^{\hat{h}} \widetilde{\mathbf\Sigma}_{ t_u^{\mathrm{B}}, {\hat{t}}_v^{\mathrm{P}}} + 
            \sum_{u=1}^{\hat{\ell}}\sum_{v=1}^{h} \widetilde{\mathbf\Sigma}_{ {\hat{t}}_u^{\mathrm{B}}, t_v^{\mathrm{P}}} .      
        \end{split}
        \end{equation}

         \begin{table*} [t]
        \centering
        \begin{minipage}{\textwidth}
            \begin{equation} \label{equ::WHD::iphase::eB!=e::eB<=i::Cov}
                    \begin{split}
                        \mathrm{cov}\left(D_{\mathbf{e}}^{(\mathrm{W})},D_{\hat{\mathbf{e}}}^{(\mathrm{W})}\right) =\sum_{\ell=0}^i\sum_{h=0}^{n-k}\sum_{\substack{\mathbf{t}_{\ell}^{h}\\\mathbf{t}_{\ell}^{\mathrm{B}} \in \mathcal{T}_l^{\mathrm{B}} \\ \mathbf{t}_{h}^{\mathrm{P}} \in \mathcal{T}_h^{\mathrm{P}} }}\sum_{\hat{\ell}=0}^{i}\sum_{\hat{h}=0}^{n-k}\sum_{\substack{\hat{\mathbf{t}}_{\hat{\ell}}^{\hat{h}}\\\hat{\mathbf{t}}_{\hat{\ell}}^{\mathrm{B}} \in \mathcal{T}_{\hat{\ell}}^{\mathrm{B}} \\ \hat{\mathbf{t}}_{\hat{h}}^{\mathrm{P}} \in \mathcal{T}_{\hat{h}}^{\mathrm{P}} }} \mathrm{Pc}(\mathbf{t}_{\ell}^{h}|w(\widetilde{\mathbf{e}}_{\mathrm{B}})\! \leq\! i) \ \mathrm{Pc}(\hat{\mathbf{t}}_{\hat{\ell}}^{\hat{h}}|w(\widetilde{\mathbf{e}}_{\mathrm{B}})\! \leq\! i) \ \mathrm{cov}(\widetilde{A}_{\mathbf{z}_{\ell,h}},\widetilde{A}_{\hat{\mathbf{x}}_{\hat{\ell},\hat{h}}}),
                    \end{split}
            \end{equation}	
        \medskip
        \hrule
        \end{minipage}
        \end{table*}
        
         However, $D_{\mathbf{e}}^{(\mathrm{W})}$ and $D_{\hat{\mathbf{e}}}^{(\mathrm{W})}$ are linear combinations of the same samples $[\widetilde{\alpha}]_1^{n}$ because $\mathbf{e}$ and $\hat{\mathbf{e}}$ are two different TEPs used in decoding one received signal sequence. Thus, the covariance of $D_{\mathbf{e}}^{(\mathrm{W})}$ and $D_{\hat{\mathbf{e}}}^{(\mathrm{W})}$ cannot be simply obtained by combining $\mathrm{cov}(\widetilde{A}_{\mathbf{t}_{\ell}^{h}},\widetilde{A}_{\hat{\mathbf{t}}_{\hat{\ell}}^{\hat{h}}})$ for all possible $\mathbf{t}_{\ell}^{h}$ and $\hat{\mathbf{t}}_{\hat{\ell}}^{\hat{h}}$. For example, if $\widetilde{\mathbf{d}}_{\mathbf{e}} = [1,1,0]$ and $\widetilde{\mathbf{d}}_{\hat{\mathbf{e}}} = [1,0,1]$ for $n = 3$, i.e., $D_{\mathbf{e}}^{(\mathrm{W})} = \widetilde{\alpha}_1 +  \widetilde{\alpha}_2$ and $D_{\mathbf{e}}^{(\mathrm{W})} = \widetilde{\alpha}_1 +  \widetilde{\alpha}_3$, we can observe that the covariance of $D_{\mathbf{e}}^{(\mathrm{W})}$ and $D_{\hat{\mathbf{e}}}^{(\mathrm{W})}$ will only be determined by $\mathrm{cov}(\widetilde{A}_2,\widetilde{A}_3)$, and $\widetilde{\alpha}_1$ will be considered as a constant which will not affect the correlations. Accordingly, we can find the covariance of $D_{\mathbf{e}}^{(\mathrm{W})}$ and $D_{\hat{\mathbf{e}}}^{(\mathrm{W})}$ as (\ref{equ::WHD::iphase::eB!=e::eB<=i::Cov}) on the top of the next page.

        where $\mathbf{z}_{\ell,h}$ is the position indices of the nonzero elements of $\mathbf{z}_{\mathbf{t}_{\ell}^{h}}\odot [\mathbf{z}_{\mathbf{t}_{\ell}^{h}} \oplus \mathbf{z}_{\hat{\mathbf{t}}_{\hat{\ell}}^{\hat{h}}}]$, and $\hat{\mathbf{x}}_{\hat{\ell},\hat{h}}$ is the position indices of the nonzero elements of $\mathbf{z}_{\hat{\mathbf{t}}_{\hat{\ell}}^{\hat{h}}}\odot [\mathbf{z}_{\mathbf{t}_{\ell}^{h}} \oplus \mathbf{z}_{\hat{\mathbf{t}}_{\hat{\ell}}^{\hat{h}}}]$, where $\odot$ is the Hadamard product of vectors. It can be seen that $\mathbf{z}_{\ell,h}$ in fact represents the positions indexed by $\mathbf{z}_{\mathbf{t}_{\ell}^{h}}$ but not by $\mathbf{z}_{\hat{\mathbf{t}}_{\hat{\ell}}^{\hat{h}}}$. Then, because $D_{\mathbf{e}}^{(\mathrm{W})}$ and $D_{\hat{\mathbf{e}}}^{(\mathrm{W})}$ follow the same distribution, they have the same mean $\mathbb{E}[D_{\mathbf{e}}^{(\mathrm{W})}|\widetilde{\mathbf{e}}_{\mathrm{B}} \neq \mathbf{e}]$ and variance $\sigma^2_{D_{\mathbf{e}}^{(\mathrm{W})}|\widetilde{\mathbf{e}}_{\mathrm{B}} \neq \mathbf{e}}$, which can be simply obtained as
        \begin{equation} \label{equ::WHD::iphase::eB!=e::eB<=i::Mean}
        \begin{split}
                \mathbb{E}[D_{\mathbf{e}}^{(\mathrm{W})}|&\widetilde{\mathbf{e}}_{\mathrm{B}} \!\neq\! \mathbf{e},w(\widetilde{\mathbf{e}}_{\mathrm{B}}) \!\leq\! i] \\
                &= \int_{0}^{\infty} x f_{D_{\mathbf{e}}^{(\mathrm{W})}}(x|\widetilde{\mathbf{e}}_{\mathrm{B}}\!\neq\!\mathbf{e},w(\widetilde{\mathbf{e}}_{\mathrm{B}})\! \leq\! i) dx 
        \end{split}
        \end{equation}
        and
        \begin{equation} \label{equ::WHD::iphase::eB!=e::eB<=i::Var}
        \begin{split}
              &\sigma^2_{D_{\mathbf{e}}^{(\mathrm{W})}|\widetilde{\mathbf{e}}_{\mathrm{B}} \neq \mathbf{e},w(\widetilde{\mathbf{e}}_{\mathrm{B}}) \leq i} \\
              &= \!\int_{0}^{\infty} x^2 \!\! f_{D_{\mathbf{e}}^{(\mathrm{W})}}(x|\widetilde{\mathbf{e}}_{\mathrm{B}}\!\neq\!\mathbf{e},w(\widetilde{\mathbf{e}}_{\mathrm{B}})\! \leq\! i) dx - \mathbb{E}[D_{\mathbf{e}}^{(\mathrm{W})}|\widetilde{\mathbf{e}}_{\mathrm{B}} \neq \mathbf{e}] ^2,  
        \end{split}
        \end{equation}
        respectively, where $f_{D_{\mathbf{e}}^{(\mathrm{W})}}(x|\widetilde{\mathbf{e}}_{\mathrm{B}}\!\neq\!\mathbf{e},w(\widetilde{\mathbf{e}}_{\mathrm{B}})\! \leq\! i)$ is the $\mathrm{pdf}$ given by (\ref{equ::WHD::iphase::eB!=e::eB<=i}).
        Therefore, on the conditions that $\{\mathbf{e} \neq \widetilde{\mathbf{e}}_{\mathrm{B}},\hat{\mathbf{e}} \neq \widetilde{\mathbf{e}}_{\mathrm{B}},\mathbf{e} \neq \hat{\mathbf{e}}\}$, we derive the correlation coefficient $\rho_{1}$ between $D_{\mathbf{e}}^{(\mathrm{W})}$ and $D_{\hat{\mathbf{e}}}^{(\mathrm{W})}$ as
        \begin{equation} \label{equ::WHD::iphase::eB!=e::eB<=i::Rho}
            \rho_{1} = \frac{ \mathrm{cov}\left(D_{\mathbf{e}}^{(\mathrm{W})},D_{\hat{\mathbf{e}}}^{(\mathrm{W})}\right)}{\sigma^2_{D_{\mathbf{e}}^{(\mathrm{W})}|\widetilde{\mathbf{e}}_{\mathrm{B}} \neq \mathbf{e},w(\widetilde{\mathbf{e}}_{\mathrm{B}}) \leq i}}.
        \end{equation}
        
        On the conditions that 1) the errors in MRB are \textbf{not} eliminated by any test error patterns and 2) $w(\mathbf{e}_{\mathrm{B}}) > i$, we can also obtain $\mathbb{E}[D_{\mathbf{e}}^{(\mathrm{W})}|\widetilde{\mathbf{e}}_{\mathrm{B}} \!\neq\! \mathbf{e},w(\widetilde{\mathbf{e}}_{\mathrm{B}})\! > \! i] $ and $\sigma^2_{D_{\mathbf{e}}^{(\mathrm{W})}|\widetilde{\mathbf{e}}_{\mathrm{B}} \neq \mathbf{e},w(\widetilde{\mathbf{e}}_{\mathrm{B}}) > i}$ similar to (\ref{equ::WHD::iphase::eB!=e::eB<=i::Mean}) and (\ref{equ::WHD::iphase::eB!=e::eB<=i::Var}), respectively.  {\color{black} Furthermore, we use $\rho_{2}$ to denote the correlation coefficient between $D_{\mathbf{e}}^{(\mathrm{W})}$ and $D_{\hat{\mathbf{e}}}^{(\mathrm{W})}$ conditioning on $w(\mathbf{e}_{\mathrm{B}}) > i$, which can be obtained similar to (\ref{equ::WHD::iphase::eB!=e::eB<=i::Rho}) by replacing $\mathrm{Pc}(\mathbf{t}_{\ell}^{h}|w(\widetilde{\mathbf{e}}_{\mathrm{B}})\! \leq\! i) $ and $ \mathrm{Pc}(\hat{\mathbf{t}}_{\hat{\ell}}^{\hat{h}}|w(\widetilde{\mathbf{e}}_{\mathrm{B}})\! \leq\! i)$ with $\mathrm{Pc}(\mathbf{t}_{\ell}^{h}|w(\widetilde{\mathbf{e}}_{\mathrm{B}})\! >\! i)$ and $\mathrm{Pc}(\hat{\mathbf{t}}_{\hat{\ell}}^{\hat{h}}|w(\widetilde{\mathbf{e}}_{\mathrm{B}})\! >\! i)$, respectively.}

        With the help of the correlation coefficients $\rho_1$ and $\rho_2$ and combining Lemma \ref{lem::WHD::iphase::eB=e} and Lemma \ref{lem::WHD::iphase::eB!=e}, we can have the insight that the distribution of the minimum WHD in an order-$m$ OSD can be derived by considering the ordered statistics over dependent random variables of WHDs. However, for the $\mathrm{pdf}$ of ordered dependent random variable with an arbitrary distribution, only the recurrence relations can be found and the explicit expressions are unsolvable \cite{balakrishnan2014order}. Therefore, we here seek the distribution of the minimum WHD under a stronger assumption that the distribution of $D_{\mathbf{e}}^{(\mathrm{W})}$ is normal, where the dependent ordering of arbitrary statistics can be simplified to ordered statistics of exchangeable normal variables. This assumption follows from that the WHDs are linear combinations of the ordered reliabilities, and the distribution will tend to normal if the code length $n$ is large. Under this assumption, we summarize the $\mathrm{pdf}$ of the minimum WHD $D_i^{(\mathrm{W})}$ after the $i$-reprocessing of an order-$m$ OSD, denoted by  $f_{D_i^{(\mathrm{W})}}(x)$, in the following Theorem.
         {\color{black}
        \begin{theorem} \label{the::WHD::iphase}
        	Given a linear block code $\mathcal{C}(n,k)$, the $\mathrm{pdf}$ of the minimum weighted Hamming distance $D_i^{(\mathrm{W})}$ between $\widetilde{\mathbf{y}}$ and $\widetilde{\mathbf{c}}_{opt}$ after the $i$-reprocessing ($0\leq i\leq m$) of an order-$m$ OSD decoding is given by
            	\begin{equation} \label{equ::WHD::iphase}
            	    \begin{split}
            	        f_{D_i^{(\mathrm{W})}}(x) &=  \!\sum_{v=0}^{i}p_{E_1^{k}}(v)\! \\
            	        &\cdot \left(f_{D_{\mathbf{e}}^{(\mathrm{W})}}(x|\widetilde{\mathbf{e}}_{\mathrm{B}}\! = \!\mathbf{e})\!\!\int_{x}^{\infty}\!\!\!f_{\widetilde{D}_i^{(\mathrm{W})}}\left(u, b_{1:i}^{k}|w(\widetilde{\mathbf{e}}_{\mathrm{B}}) \!\leq\! i\right) du \right.\\
            	        &+ \!\left. f_{\widetilde{D}_i^{(\mathrm{W})}}\left(u, b_{1:i}^{k}|w(\widetilde{\mathbf{e}}_{\mathrm{B}}) \!\leq\! i\right) \!\!\int_{x}^{\infty}\!\!\!f_{D_{\mathbf{e}}^{(\mathrm{W})}}(u|\widetilde{\mathbf{e}}_{\mathrm{B}}\! =\! \mathbf{e})du \! \right) \\
                        & +  \left(1 - \sum_{v=0}^{i}p_{E_1^{k}}(v)\right) f_{\widetilde{D}_i^{(\mathrm{W})}}\left(u, b_{0:i}^{k}|w(\widetilde{\mathbf{e}}_{\mathrm{B}}) > i\right) ,
            	    \end{split}
            	\end{equation}
            	where $f_{\widetilde{D}_i^{(\mathrm{W})}}(x,b|w(\widetilde{\mathbf{e}}_{\mathrm{B}}) \leq i)$ and $ f_{\widetilde{D}_i^{(\mathrm{W})}}(x,b|w(\widetilde{\mathbf{e}}_{\mathrm{B}}) > i)$ are given by (\ref{equ::WHD::iphase::DependOrder1}) and (\ref{equ::WHD::iphase::DependOrder2}) on the top of the next page, respectively, and
    	         \begin{table*} [t]
                    \centering
                    \begin{minipage}{\textwidth}
                    \begin{equation} \label{equ::WHD::iphase::DependOrder1}
            		\begin{split}
    	                f_{\widetilde{D}_i^{(\mathrm{W})}}(x,b|w(\widetilde{\mathbf{e}}_{\mathrm{B}}) \leq i) &= \int_{-\infty}^{\infty} \left(\sqrt{1-\rho_{1}} \ \sigma_{D_{\mathbf{e}}^{(\mathrm{W})}|\widetilde{\mathbf{e}}_{\mathrm{B}} \neq \mathbf{e},w(\widetilde{\mathbf{e}}_{\mathrm{B}}) \leq i}\right)^{-1} \\
    	                &\cdot f_{\phi} \left(\frac{(x - \mathbb{E}[D_{\mathbf{e}}^{(\mathrm{W})}|\widetilde{\mathbf{e}}_{\mathrm{B}} \neq \mathbf{e},w(\widetilde{\mathbf{e}}_{\mathrm{B}}) \leq i])/\sigma_{D_{\mathbf{e}}^{(\mathrm{W})}|\widetilde{\mathbf{e}}_{\mathrm{B}} \neq \mathbf{e},w(\widetilde{\mathbf{e}}_{\mathrm{B}}) \leq i}   + \sqrt\rho_1 z}{\sqrt{1- \rho_1}},b\right) \phi (z) \ dz, 	    
            		\end{split}
    	            \end{equation}
    	            
              		\begin{equation} \label{equ::WHD::iphase::DependOrder2}
                		\begin{split}
        	                f_{\widetilde{D}_i^{(\mathrm{W})}}(x,b|w(\widetilde{\mathbf{e}}_{\mathrm{B}}) > i) &= \int_{-\infty}^{\infty} \left(\sqrt{1-\rho_{2}} \ \sigma_{D_{\mathbf{e}}^{(\mathrm{W})}|\widetilde{\mathbf{e}}_{\mathrm{B}} \!\neq\! \mathbf{e},w(\widetilde{\mathbf{e}}_{\mathrm{B}})\! >\! i}\right)^{-1} \\
        	                &\cdot f_{\phi} \left(\frac{(x - \mathbb{E}[D_{\mathbf{e}}^{(\mathrm{W})}|\widetilde{\mathbf{e}}_{\mathrm{B}} \!\neq\! \mathbf{e},w(\widetilde{\mathbf{e}}_{\mathrm{B}})\! >\! i])/\sigma_{D_{\mathbf{e}}^{(\mathrm{W})}|\widetilde{\mathbf{e}}_{\mathrm{B}} \!\neq\! \mathbf{e},w(\widetilde{\mathbf{e}}_{\mathrm{B}})\! >\! i}   + \sqrt\rho_2 z}{\sqrt{1- \rho_2}},b\right) \phi (z) \ dz, 	    
                		\end{split}
    	            \end{equation}	
                    \medskip
                    \hrule
                    \end{minipage}
                    \end{table*}
            	\begin{equation}
	            	f_{\phi}(x,b) = b \ \phi (x)\left(1 - \int_{-\infty}^{x} \phi (u)du\right)^{b-1},
            	\end{equation}	
	             $\phi(x)$ is the $\mathrm{pdf}$ of the standard normal distribution and $f_{D_{\mathbf{e}}^{(\mathrm{W})}}(x|\widetilde{\mathbf{e}}_{\mathrm{B}} \!=\! \mathbf{e})$ is given by (\ref{equ::WHD::iphase::eb=e}).
        \end{theorem}
        \begin{IEEEproof}
            The proof is provided in Appendix \ref{app::proof::WHD::iphase}.
        \end{IEEEproof}
        }
        
\subsection{Simplifications, Approximations, and Numerical Results} \label{subsec::appWHD}

    Theorem \ref{the::WHD::0phase} and Theorem \ref{the::WHD::iphase} investigate exact expressions of the $\mathrm{pdf}$s of the WHDs in the 0-reprocessing and after the $i$-reprocessing. However, calculating (\ref{equ::WHD::0phase}) and (\ref{equ::WHD::iphase}) is daunting as $f_{\widetilde{A}_{\mathbf{t}_h^{\mathrm{P}}}}(x)$ and $f_{\widetilde{A}_{\mathbf{t}_{\ell}^{h}}}(x)$ are the $\mathrm{pdf}$s of the summations of non-i.i.d. reliabilities and characterizing $\mathrm{Pe}(\mathbf{t}_h^{\mathrm{P}})$ and $\mathrm{Pc}(\mathbf{t}_{\ell}^{h})$ involves calculating a large number of integrals. 
        
     {\color{black}In this section, we consider simplifying and approximating (\ref{equ::WHD::0phase}) and (\ref{equ::WHD::iphase}) by assuming that the probability $p_{\mathbf{c}_{\mathrm{P}}}(u,q)$ of $\mathcal{C}(n,k)$ is known and has been determined from the codebook. First, we investigate the probability that a parity bit of a codeword estimate in OSD is non-zero, as summarized in the following Lemma.
    
    \begin{lemma} \label{lem::WHD::App::PcPbit}
        Let $p_{\mathbf{c}_{\mathrm{P}}}^{\mathrm{bit}}(\ell,q)$ denote the probability that the $\ell$-th bit ($k < \ell \leq n$) of $\widetilde{\mathbf{c}}' = \mathbf{b}'\widetilde{\mathbf{G}} = [\widetilde{c}']_1^{n}$ is nonzero when $w(\mathbf{b}') = q$, i.e., $p_{\mathbf{c}_{\mathrm{P}}}^{\mathrm{bit}}(\ell,q) = \mathrm{Pr}(\widetilde{c}'_{\ell} \neq 0  | w(\mathbf{b}') = q)$, then  $p_{\mathbf{c}_{\mathrm{P}}}^{\mathrm{bit}}(\ell,q)$ can be derived as
        \begin{equation} \label{equ::WHD::App::PcPbit1}
            p_{\mathbf{c}_{\mathrm{P}}}^{\mathrm{bit}}(\ell,q) = \sum_{u=0}^{n-k} \frac{u}{n-k}\cdot p_{\mathbf{c}_{\mathrm{P}}}(u,q).
        \end{equation}
        Furthermore, let $p_{\mathbf{c}_{\mathrm{P}}}^{\mathrm{bit}}(\ell,h,q)$ denote the joint probability that the $\ell$-th and $h$-th bit ($k < \ell < h \leq n$) of $\widetilde{\mathbf{c}}'$ is nonzero when $w(\mathbf{b}') = q$, and $p_{\mathbf{c}_{\mathrm{P}}}^{\mathrm{bit}}(\ell,h,q)$ is given by
        \begin{equation}  \label{equ::WHD::App::PcPbit2}
             p_{\mathbf{c}_{\mathrm{P}}}^{\mathrm{bit}}(\ell,h,q) = \sum_{u=0}^{n-k} \frac{u(u-1)}{(n-k)(n-k-1)}\cdot p_{\mathbf{c}_{\mathrm{P}}}(u,q) .
        \end{equation}
    \end{lemma}
    \begin{IEEEproof}
        Considering that the columns of $\mathbf{G}$ are randomly permuted to the columns of $\widetilde{\mathbf{G}}$ whenever new noisy symbols are received, when $w(\widetilde{\mathbf{c}}'_{\mathrm{P}}) = u$ with the probability $p_{\mathbf{c}_{\mathrm{P}}}(u,q) $, each bit $\widetilde{c}'_{\ell}$ of $\widetilde{\mathbf{c}}'$, $k<\ell\leq n$, has equal probability $\frac{u}{n-k}$ to be nonzero. Then, (\ref{equ::WHD::App::PcPbit1}) can be easily obtained, and (\ref{equ::WHD::App::PcPbit2}) can also be obtained similarly.
    \end{IEEEproof}
    Note that $p_{\mathbf{c}_{\mathrm{P}}}^{\mathrm{bit}}(\ell,q)$ and $p_{\mathbf{c}_{\mathrm{P}}}^{\mathrm{bit}}(\ell,h,q)$ are identical for all integers $\ell$ and $h$, $k < \ell<h\leq n$, because of the randomness of the permutation over $\mathbf{G}$. In other words, despite $\mathbf{G}$ is permuted according to the received signals, an arbitrary column of $\mathbf{G}$ has the same probability to be permuted to each column of $\widetilde{\mathbf{G}}$. Next, based on $p_{\mathbf{c}_{\mathrm{P}}}^{\mathrm{bit}}(\ell,q)$ and $p_{\mathbf{c}_{\mathrm{P}}}^{\mathrm{bit}}(\ell,h,q)$, we simplify and approximate the distributions given by (\ref{equ::WHD::0phase}) and (\ref{equ::WHD::iphase}), respectively.}
    
    \subsubsection{Simplification and Approximation of $D_{0}^{(\mathrm{W})}$}
    In what follows, first an approximation of $f_{\widetilde{A}_{\mathbf{t}_h^{\mathrm{P}}}}(x)$ based on the normal approximation of ordered reliabilities (previously derived in Section \ref{sec::OrderStat::App}) will be introduced, then the probability that the different bits between $\widetilde{\mathbf{c}}_{0}$ and $\widetilde{\mathbf{y}}$ are nonzero will be characterized, and finally (\ref{equ::WHD::0phase}) is simplified for practical computations. In addition, some numerical examples for decoding BCH and Polar codes using order-0 OSD will be illustrated.
        
    Recall that the random variable $\widetilde{A}_u$ of the $u$-th ordered reliability can be approximated by a normal random variable with the distribution $\mathcal{N}(\mathbb{E}[\widetilde{A}_i],\sigma_{\widetilde{A}_i}^2)$, thus $\widetilde{A}_{\mathbf{t}_h^{\mathrm{P}}} = \sum_{u=1}^{h} \widetilde{A}_{{t}_u^{\mathrm{P}}}$ can also be regarded as a normal random variable. Using the mean and covariance matrices introduced in (\ref{lem::WHD::iphase::MeanMx}) and (\ref{lem::WHD::iphase::VarMx}), respectively, the mean and variance of $\widetilde{A}_{\mathbf{t}_h^{\mathrm{P}}}$ are given by 
        \begin{equation}
            \mathbb{E}[\widetilde{A}_{\mathbf{t}_h^{\mathrm{P}}}] = \sum_{u = 1}^{h}\sqrt{\widetilde{\mathbf{E}}_{t_u^{\mathrm{P}},  t_u^{\mathrm{P}}}}
        \end{equation}
        and
        \begin{equation}
        	\sigma^2_{\widetilde{A}_{\mathbf{t}_h^{\mathrm{P}}}} = \sum_{u=1}^{h} \sum_{v=1}^{h} \widetilde{\mathbf\Sigma}_{ t_u^{\mathrm{P}},t_v^{\mathrm{P}}}.
        \end{equation}
        Therefore, $f_{\widetilde { A}_{\mathbf{t}_h^{\mathrm{P}}}}(x)$ can be approximated by a normal distribution $\mathcal{N}(\mathbb{E}[\widetilde{A}_{\mathbf{t}_h^{\mathrm{P}}}] , \sigma^2_{\widetilde{A}_{\mathbf{t}_h^{\mathrm{P}}}} )$ with the $\mathrm{pdf}$ given by
    	\begin{equation}
        	f_{\widetilde{A}_{\mathbf{t}_h^{\mathrm{P}}}}(x) = \frac{1}{\sqrt{2\pi \sigma^2_{\widetilde { A}_{\mathbf{t}_h^{\mathrm{P}}}}}} \exp\left(-\frac{(x - \mathbb{E}[\widetilde { A}_{\mathbf{t}_h^{\mathrm{P}}}])^2}{2\sigma^2_{\widetilde { A}_{\mathbf{t}_h^{\mathrm{P}}}}}\right).
        \end{equation}
        
         Then, let us consider the probability that the $\ell$-th ($k<\ell \leq n$) bit of $\widetilde{\mathbf{d}}_{0} = [\widetilde{d}_{0}]_1^{n} = \widetilde{\mathbf{c}}_{0}\oplus \widetilde{\mathbf{y}}$ is nonzero, i.e., $\mathrm{Pr}(\widetilde{d}_{0,\ell}\neq 0)$. As discussed in Lemma \ref{lem::HDdis::0phase}, when $\widetilde{\mathbf{e}}_{\mathrm{B}} = \mathbf{0}$, $\widetilde{\mathbf{d}}_{0,\mathrm{P}}$ equals to $\widetilde{\mathbf{e}}_{\mathrm{P}}$ and $\mathrm{Pr}(\widetilde{d}_{0,\ell}\neq 0)$ can be simply characterized the error probability of the $\ell$-th bit of $\widetilde{\mathbf{y}}$. Whereas, when $\widetilde{\mathbf{e}}_{\mathrm{B}} \neq \mathbf{0}$, $\widetilde{\mathbf{d}}_{0,\mathrm{P}}$ is given by $\widetilde{\mathbf{d}}_{0,\mathrm{P}} = \widetilde{\mathbf{c}}_{0,\mathrm{P}}'\oplus \widetilde{\mathbf{e}}_{\mathrm{P}}$, where $\widetilde{\mathbf{c}}_{0}' = \widetilde{\mathbf{e}}_{\mathrm{B}} \widetilde{\mathbf{G}} = [\widetilde{\mathbf{e}}_{\mathrm{B}} \ \ \widetilde{\mathbf{c}}_{0,\mathrm{P}}']$. Therefore, $\mathrm{Pr}(\widetilde{d}_{0,\ell}\neq 0)$ is obtained as 
        \begin{equation} \label{equ::WHD::App::d0bit1}
            \begin{split}
                \mathrm{Pr}&(\widetilde{d}_{0,\ell}\neq 0) \\
                &= \mathrm{Pr}(\widetilde{c}_{0,\ell}'\neq 0)\mathrm{Pr}(\widetilde{e}_{\ell}= 0) + \mathrm{Pr}(\widetilde{c}_{0,\ell}'= 0)\mathrm{Pr}(\widetilde{e}_{\ell}\neq 0)  \\
                &\overset{(a)}{=}  \frac{1}{1- p_{E_1^k}(0)}\sum_{q=1}^{k}p_{E_1^k}(q)\left(p_{\mathbf{c}_{\mathrm{P}}}^{\mathrm{bit}}(\ell,q)(1-\mathrm{Pe}(\ell)) \right.\\
                &+ \left.(1 - p_{\mathbf{c}_{\mathrm{P}}}^{\mathrm{bit}}(\ell,q))\mathrm{Pe}(\ell) \right), 
            \end{split}
        \end{equation}
        where $p_{\mathbf{c}_{\mathrm{P}}}^{\mathrm{bit}}(\ell,q)$ is given by (\ref{equ::WHD::App::PcPbit2}) and step (a) takes $\mathrm{Pe}(\ell) = \mathrm{Pr}(\widetilde{e}_{\ell}\neq 0) $. When $\widetilde{\mathbf{e}}_{\mathrm{B}} \neq \mathbf{0}$, the joint nonzero probabilities of the $\ell$-th and the $h$-th ($k<\ell <h \leq n$) bits of $\widetilde{\mathbf{d}}_{0}$, i.e., $\mathrm{Pr}(\widetilde{d}_{0,\ell}\neq 0, \widetilde{d}_{0,h}\neq 0)$, is given by
        \begin{equation} \label{equ::WHD::App::d0bit2}
        \begin{split}
            \mathrm{Pr}\{\widetilde{d}_{0,\ell} \neq 0,& \widetilde{d}_{h} \neq 0\} \\
            =&  \mathrm{Pr}\{\widetilde{c}_{0,\ell}' \neq 0, \widetilde{c}_{0,h}' \neq 0\}\mathrm{Pr}\{\widetilde{e}_{\ell} = 0, \widetilde{e}_{h} = 0\} \\
            +&\mathrm{Pr}\{\widetilde{c}_{0,\ell}' \neq 0, \widetilde{c}_{0,h}' = 0\}\mathrm{Pr}\{\widetilde{e}_{\ell} = 0, \widetilde{e}_{h} \neq 0\} 
            \\
            +& \mathrm{Pr}\{\widetilde{c}_{0,\ell}' = 0, \widetilde{c}_{0,h}' \neq 0\}\mathrm{Pr}\{\widetilde{e}_{\ell} \neq 0, \widetilde{e}_{h} = 0\} \\
            +&\mathrm{Pr}\{\widetilde{c}_{0,\ell}' = 0, \widetilde{c}_{0,h}' = 0\}\mathrm{Pr}\{\widetilde{e}_{\ell} \neq 0, \widetilde{e}_{h} \neq 0\}.
        \end{split}
        \end{equation}
        In (\ref{equ::WHD::App::d0bit2}), $\mathrm{Pr}\{\widetilde{c}_{0,\ell}' \neq 0, \widetilde{c}_{0,h}' \neq 0\}\mathrm{Pr}\{\widetilde{e}_{\ell} = 0, \widetilde{e}_{h} = 0\}$ is determined as
        \begin{equation} \label{equ::WHD::App::d0bit2::oneTerm}
        \begin{split}
            &\mathrm{Pr}\{\widetilde{c}_{0,\ell}' \neq 0, \widetilde{c}_{0,h}' \neq 0\}\mathrm{Pr}\{\widetilde{e}_{\ell} = 0, \widetilde{e}_{h} = 0\} \\
            &\!=\! \sum_{q=1}^{k} \frac{p_{E_1^k}(q)p_{\mathbf{c}_{\mathrm{P}}}^{\mathrm{bit}}(\ell,h,q)}{1 \!-\! p_{E_1^k}(0)} \!\! \int_{0}^{\infty}\!\!\!\int_{0}^{\infty}\!\!\! f_{\widetilde R_{\ell},\widetilde R_{h}}(\widetilde r_{\ell},\widetilde r_{h}) d\widetilde r_{\ell} \ d\widetilde r_{h}, 
        \end{split}
        \end{equation}
        where $f_{\widetilde R_{\ell},\widetilde R_{h}}(\widetilde r_{\ell},\widetilde r_{h})$ is the joint $\mathrm{pdf}$ of two ordered received symbols, which is given by (\ref{equ::OrderStat::jointpdfRij}). Other terms of (\ref{equ::WHD::App::d0bit2}) can be determined similar to (\ref{equ::WHD::App::d0bit2::oneTerm}).
   
        Next, similar to the Hamming distance distribution in 0-reprocessing, we approximate (\ref{equ::WHD::0phase}) by considering the large-number Gaussian mixture model. Let $f_
       {D_{0}^{(\mathrm{w})}}(x|w(\widetilde{\mathbf{e}}_{\mathrm{B}})\!=\!0)$ denote the first mixture component in (\ref{equ::proof::WHD::0phase::cond}), i.e.,
        \begin{equation} \label{equ::WHD::0phase::eB=0}
        \begin{split}
             & f_{D_{0}^{(\mathrm{w})}}(x|w(\widetilde{\mathbf{e}}_{\mathrm{B}})\!=\!0) \\
             &= \sum_{h=0}^{n-k}\sum_{\substack{\mathbf{t}_{h}^{\mathrm{P}} \\ \mathbf{t}_h^{\mathrm{P}} \in \mathcal{T}_h^{\mathrm{P}}}}   \mathrm{Pr}\left(\widetilde{\mathbf{d}}_{0,\mathrm{P}} = \mathbf{z}_{\mathbf{t}_h^{\mathrm{P}}} |  w(\widetilde{\mathbf{e}}_{\mathrm{B}})=0\right) f_{\widetilde{A}_{\mathbf{t}_h^{\mathrm{P}}}}(x).     
        \end{split}
        \end{equation}
        $f_{D_{0}^{(\mathrm{w})}}(x|w(\widetilde{\mathbf{e}}_{\mathrm{B}})\!=\!0)$ is also the $\mathrm{pdf}$ of $D_{0}^{(\mathrm{w})}$ conditioning on $\{w(\widetilde{\mathbf{e}}_{\mathrm{B}})=0\}$.
        Also, let $\mathbf{t}_h^{\mathrm{P}(u)}$ denote the vector $\mathbf{t}_h^{\mathrm{P}}$ that contains the element ``$u$'' and $\mathbf{t}_h^{\mathrm{P}(u,v)}$ denote the vector $\mathbf{t}_h^{\mathrm{P}}$ that contains both ``$u$'' and ``$v$'', i.e., $\mathbf{t}_h^{\mathrm{P}(u)} = \{\mathbf{t}_h^{\mathrm{P}} | \ \exists \ \ell,1 \!\leq\! \ell \leq h, \ t_{\ell}^{\mathrm{P}} \!=\! u\}$ and $\mathbf{t}_h^{\mathrm{P}(u,v)} = \{\mathbf{t}_h^{\mathrm{P}} | \ \exists \ \ell \text{ and } j, 1 \!\leq\! \ell \!<\! j \!\leq\! h,\ t_{\ell}^{\mathrm{P}} \!=\! u , \ t_j^{\mathrm{P}} \!=\! v  \}$. {\color{black}Then, the mean of the first mixture component $f_
       {D_{0}^{(\mathrm{w})}}(x|w(\widetilde{\mathbf{e}}_{\mathrm{B}})\!=\!0)$ can be derived and approximated as
        \begin{align} \label{equ::WHD::0phase::eB=0::Mean}
                &\mathbb{E}[D_{0}^{(\mathrm{w})}|w(\widetilde{\mathbf{e}}_{\mathrm{B}})\!=\!0] \notag\\
                & = \sum_{u=k+1}^{n} \sum_{h=0}^{n-k}\!\sum_{\substack{\mathbf{t}_{h}^{\mathrm{P}(u)} \\ \mathbf{t}_h^{\mathrm{P}(u)} \in \mathcal{T}_h^{\mathrm{P}}}}\!\!\mathrm{Pr}\left(\widetilde{\mathbf{d}}_{0,\mathrm{P}} = \mathbf{z}_{\mathbf{t}_h^{\mathrm{P}(u)}} |  w(\widetilde{\mathbf{e}}_{\mathrm{B}})\!=\!0 \right) \sqrt{\widetilde{\mathbf{E}}_{u,u}} \notag\\
                & = \sum_{u=k+1}^{n} \mathrm{Pe}(u| E_1^{k} = 0)\sqrt{\widetilde{\mathbf{E}}_{u,u}} \notag\\
                & \overset{(a)}{\approx} \sum_{u=k+1}^{n} \mathrm{Pe}(u)\sqrt{\widetilde{\mathbf{E}}_{u,u}} \ ,
        \end{align}
        where $\mathrm{Pe}(u)$ is the bit-wise error probability given by (\ref{equ::OrderStat::bitPe}) and step (a) follows the independence between $\mathrm{Pe}(u)$ and $E_{1}^{k}$, as introduced in (\ref{equ::OrderStat::EPeIndenIndepence}).} Similarly, the variance of mixture component $f_
       {D_{0}^{(\mathrm{w})}}(x|w(\widetilde{\mathbf{e}}_{\mathrm{B}})\!=\!0)$ can be derived and approximated as 
        \begin{align}  \label{equ::WHD::0phase::eB=0::Var}
        		\sigma_{D_{0}^{(\mathrm{w})}|w(\widetilde{\mathbf{e}}_{\mathrm{B}})=0}^{2} 
    			= & \sum_{u=k+1}^{{n}} \sum_{v=k+1}^{{n}} \mathrm{Pe}(u,v|E_1^{k} = 0) \left[\widetilde{\mathbf{E}} + \widetilde{\mathbf\Sigma}\right]_{u,v} \notag\\
    			& - \left(\mathbb{E}[D_{0}^{(\mathrm{w})}|w(\widetilde{\mathbf{e}}_{\mathrm{B}})\!=\!0]\right)^2 \notag\\
    			\approx & \sum_{u=k+1}^{{n}} \sum_{v=k+1}^{{n}} \mathrm{Pe}(u,v) \left[\widetilde{\mathbf{E}} + \widetilde{\mathbf\Sigma}\right]_{u,v} \\
    			& - \left(\mathbb{E}[D_{0}^{(\mathrm{w})}|w(\widetilde{\mathbf{E}}_{\mathrm{B}})\!=\!0]\right)^2, \notag
        \end{align}
        where $\mathrm{Pe}(u,v)$ is the joint probability that the $u$-th and $v$-th positions of $\widetilde{\mathbf{y}}$ are both in error. When $u=v$, $\mathrm{Pe}(u,v)$ is simply given by $\mathrm{Pe}(u)$. Otherwise, $\mathrm{Pe}(u,v)$ is given by  $\int_{-\infty}^0 \int_{-\infty}^0 f_{\widetilde R_u,\widetilde R_v}(x,y) dxdy$.
        
        Next, let $f_{D_{0}^{(\mathrm{w})}}(x|w(\widetilde{\mathbf{e}}_{\mathrm{B}})\!\neq\!0)$ denote the second mixture component in (\ref{equ::proof::WHD::0phase::cond}), i.e., 
        \begin{equation}
        \begin{split}
              f_{D_{0}^{(\mathrm{w})}}&(x|w(\widetilde{\mathbf{e}}_{\mathrm{B}})\!\neq\!0)\\
              &= \sum_{h=0}^{n-k}\sum_{\substack{\mathbf{t}_{h}^{\mathrm{P}} \\ \mathbf{t}_h^{\mathrm{P}} \in \mathcal{T}_h^{\mathrm{P}}}}   \mathrm{Pr}\left(\widetilde{\mathbf{d}}_{0,\mathrm{P}} = \mathbf{z}_{\mathbf{t}_h^{\mathrm{P}}} |  w(\widetilde{\mathbf{e}}_{\mathrm{B}})\neq 0\right) f_{\widetilde{A}_{\mathbf{t}_h^{\mathrm{P}}}}(x).   
        \end{split}
        \end{equation}
        {\color{black}$f_{D_{0}^{(\mathrm{w})}}(x|w(\widetilde{\mathbf{e}}_{\mathrm{B}})\!\neq\!0)$ is also the $\mathrm{pdf}$ of $D_{0}^{(\mathrm{w})}$ conditioning on $\{w(\widetilde{\mathbf{e}}_{\mathrm{B}})\neq 0\}$. For simplicity, we denote $\mathrm{Pr}(\widetilde{d}_{0,\ell}\neq 0)$ and $\mathrm{Pr}\{\widetilde{d}_{0,\ell} \neq 0, \widetilde{d}_{h} \neq 0\}$ obtained in (\ref{equ::WHD::App::d0bit1}) and (\ref{equ::WHD::App::d0bit2}) as $\mathrm{Pc}_{0}(\ell)$ and $\mathrm{Pc}_{0}(\ell,h)$, respectively.  Using the similar approach of obtaining (\ref{equ::WHD::0phase::eB=0::Mean}) and (\ref{equ::WHD::0phase::eB=0::Var}) and considering $\mathrm{Pr}\left(\widetilde{\mathbf{d}}_{0,\mathrm{P}} = \mathbf{z}_{\mathbf{t}_h^{\mathrm{P}}} |  w(\widetilde{\mathbf{e}}_{\mathrm{B}})\neq 0\right) = \mathrm{Pc}(\mathbf{t}_h^{\mathrm{P}})$, the mean and variance of $f_{D_{0}^{(\mathrm{w})}}(x|w(\widetilde{\mathbf{e}}_{\mathrm{B}})\!\neq\!0)$ can be derived as
        \begin{equation} \label{equ::WHD::App::0phase::Mean}
        	\begin{split}
        		 \mathbb{E}[D_{0}^{(\mathrm{w})}|w(\widetilde{\mathbf{e}}_{\mathrm{B}})\!\neq \!0] = \sum_{u =k+1}^{n}\mathrm{Pc}_{0}(u) \sqrt{\widetilde{\mathbf{E}}_{u,u}}.
        	\end{split}
        \end{equation}
        and 
        \begin{equation}
        	\begin{split}
        		\sigma_{D_{0}^{(\mathrm{w})}|w(\widetilde{\mathbf{e}}_{\mathrm{B}})\neq 0}^{2} &= \sum_{u=k+1}^{n} \sum_{v=k+1}^{n}\mathrm{Pc}_{0}(u,v) \left[\widetilde{\mathbf{E}} + \widetilde{\mathbf\Sigma}\right]_{u,v}  \\
        		&- \left(\mathbb{E}[D_{0}^{(\mathrm{w})}|w(\widetilde{\mathbf{e}}_{\mathrm{B}})\!\neq \!0]\right)^2 \ ,
        	\end{split}
        \end{equation}
        respectively, where $\mathrm{Pc}_{0}(u) = \mathrm{Pr}(\widetilde{d}_{0,u}\neq 0)$ is given by (\ref{equ::WHD::App::d0bit1}) and $\mathrm{Pc}_{0}(u,v) = \mathrm{Pr}(\widetilde{d}_{0,u}\neq 0,\widetilde{d}_{0,v}\neq 0)$ is given by (\ref{equ::WHD::App::d0bit2}) for $u\neq v$. In particular, $\mathrm{Pc}_{0}(u,v) = \mathrm{Pc}_{0}(u) $ when $u=v$. }
        
        Because $D_{0}^{(\mathrm{w})}$ can be regarded as a linear combination of a number of random variables $[\widetilde{A}]_1^{n}$ when $n$ is large, we approximate the $\mathrm{pdf}$ of $D_{0}^{(\mathrm{w})}$ by a combination of two normal distributions, whose $\mathrm{pdf}$ is given by 
        \begin{equation}  \label{equ::WHD::App::0phase}
            \begin{split}
               & f_{D_0^{(\mathrm{W})}}(x) \\
                &=  \!p_{\!E_1^k}(0) f_{\!D_{0}^{(\mathrm{w})}}\!(x|w(\widetilde{\mathbf{e}}_{\mathrm{B}})\!\!=\!\!0)\!+\! (1 \!-\! p_{E_1^k}(0)) f_{\!D_{0}^{(\mathrm{w})}}\!(x|w(\widetilde{\mathbf{e}}_{\mathrm{B}})\!\!\neq\!\!0) \\
               & \approx  \frac{p_{E_1^k}(0)}{\sqrt{2\pi 	\sigma_{D_{0}^{(\mathrm{w})}|w(\widetilde{\mathbf{e}}_{\mathrm{B}})\!= \!0}^{2}}}\exp\left(-\frac{(x-\mathbb{E}[D_{0}^{(\mathrm{w})}|w(\widetilde{\mathbf{e}}_{\mathrm{B}})\! = \!0])^2}{2	\sigma_{D_{0}^{(\mathrm{w})}|w(\widetilde{\mathbf{e}}_{\mathrm{B}})\! = \!0}^{2}}\right) \\
               & +  \frac{1 - p_{E_1^k}(0)}{\sqrt{2\pi 	\sigma_{D_{0}^{(\mathrm{w})}|w(\widetilde{\mathbf{e}}_{\mathrm{B}})\!\neq \!0}^{2}}}\exp\left(-\frac{(x-\mathbb{E}[D_{0}^{(\mathrm{w})}|w(\widetilde{\mathbf{e}}_{\mathrm{B}})\!\neq \!0])^2}{2 	\sigma_{D_{0}^{(\mathrm{w})}|w(\widetilde{\mathbf{e}}_{\mathrm{B}})\!\neq \!0}^{2}}\right) .
            \end{split}
        \end{equation}
        
\begin{figure}
    		\begin{center}
    			\includegraphics[scale=0.55] {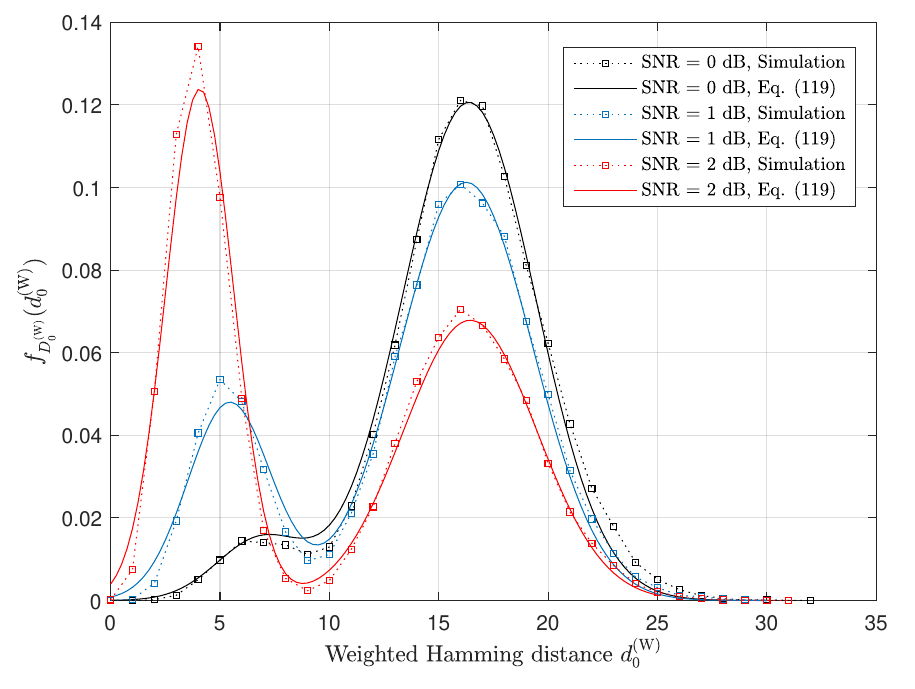}
    			\caption{The distribution of $D_{0}^{(\mathrm{W})}$ in decoding $(128,64,22)$ eBCH code.}
    			\label{Fig::IV::BCH128-WHD-0phase}
    		\end{center}
    	\end{figure}

\begin{figure}
    		\begin{center}
    			\includegraphics[scale=0.55] {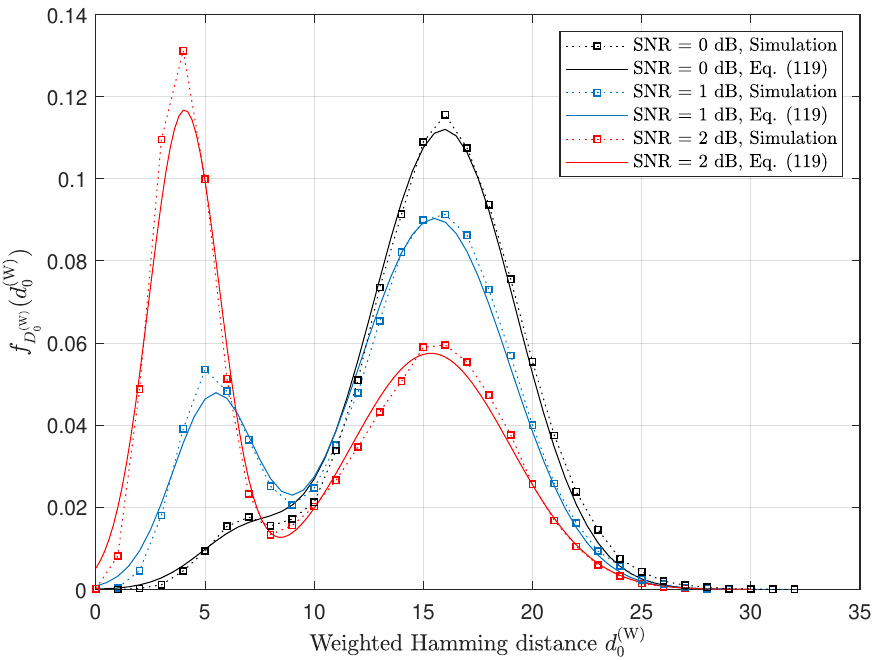}
    			\caption{{\color{black}The distribution of $D_{0}^{(\mathrm{W})}$ in decoding $(128,64,8)$ Polar code.}}
    			\label{Fig::IV::Polar128-WHD-0phase}
    		\end{center}
    	\end{figure}

        To verify (\ref{equ::WHD::App::0phase}), we show the distributions of $D_{0}^{(\mathrm{W})}$ for decoding the $(128,64,22)$ eBCH code and {\color{black}$(128,64,8)$ Polar code} in Fig. \ref{Fig::IV::BCH128-WHD-0phase} and Fig. \ref{Fig::IV::Polar128-WHD-0phase}, respectively, at different SNRs. It can be seen that (\ref{equ::WHD::App::0phase}) provides a promising approximation of the 0-reprocessing WHD distribution. Similar to the distribution of 0-reprocessing Hamming distance, the $\mathrm{pdf}$ of $D_{0}^{(\mathrm{H})}$ is also a mixture of two distributions. The weight of the left and right parts can be a reflection of the channel condition and decoding error performance since the weights of $f_{D_{0}^{(\mathrm{w})}}(x|w(\widetilde{\mathbf{e}}_{\mathrm{B}})\!=\!0)$ and $f_{D_{0}^{(\mathrm{w})}}(x|w(\widetilde{\mathbf{e}}_{\mathrm{B}})\!\neq\!0)$ in (\ref{equ::WHD::App::0phase}) are controlled by $p_{E_1^k}(0)$. It can be seen that the distribution concentrates towards the left when the channel SNR increases, indicating that the decoding error performance improves. {\color{black}From Fig. \ref{Fig::IV::BCH128-WHD-0phase} and Fig. \ref{Fig::IV::Polar128-WHD-0phase}, we can also observe that the discrepancies between the approximation (\ref{equ::WHD::App::0phase}) and the simulation results mainly exist on the left side of the curves, dominated by $f_{D_{0}^{(\mathrm{w})}}(x|w(\widetilde{\mathbf{e}}_{\mathrm{B}})\!=\!0)$. This is because 1) the support of $D_{0}^{(\mathrm{W})}$ is $[0,\infty)$ but (\ref{equ::WHD::App::0phase}) is obtained with complete normal distributions, and 2) $f_{D_{0}^{(\mathrm{w})}}(x|w(\widetilde{\mathbf{e}}_{\mathrm{B}})\!=\!0)$ is obtained by approximated mean and variance (e.g., step (a) of (\ref{equ::WHD::App::0phase::Mean})).}
        
        \subsubsection{Simplification and Approximation of $D_{i}^{(\mathrm{W})}$}
         In what follows, we first investigate the probability that the different bits between $\widetilde{\mathbf{c}}_{\mathbf{e}}$ and $\widetilde{\mathbf{y}}$ are nonzero, followed by simplifying and approximating the means, variances, and covariance introduced in Section \ref{sec::WHD-iphase}. Finally, we study the normal approximation of $D_i^{(\mathrm{W})}$ after the $i$-reprocessing of an order-$m$ OSD.
         
         {\color{black}As investigated in Theorem \ref{the::HDdis::iphase} and Theorem \ref{the::WHD::iphase}, when $\widetilde{\mathbf{e}}_{\mathrm{B}}\neq \mathbf{e}$, the difference pattern between $\widetilde{\mathbf{c}}_{\mathbf{e}}$ and $\widetilde{\mathbf{y}}$ can be given by $\widetilde{\mathbf{d}}_{\mathbf{e}} = \widetilde{\mathbf{c}}_{\mathbf{e}} \oplus \widetilde{\mathbf{y}}= [\mathbf{e} \ \ \widetilde{\mathbf{c}}_{\mathbf{e},\mathrm{P}}' \oplus \widetilde{\mathbf{e}}_{\mathrm{P}}]$, where $\widetilde{\mathbf{c}}_{\mathbf{e},\mathrm{P}}$ is the parity part of $\widetilde{\mathbf{c}}_{\mathbf{e}}' = [\widetilde{\mathbf{e}}_{\mathrm{B}}\oplus \mathbf{e}]\widetilde{\mathbf{G}}$. Next, we characterize the probability that $\ell$-th bit $\widetilde{d}_{\mathbf{e},\ell}$ of $\widetilde{\mathbf{d}}_{\mathbf{e}}$ is nonzero, conditioning on $\{w(\widetilde{\mathbf{e}}_{\mathrm{B}})\leq i\}$ and $\{w(\widetilde{\mathbf{e}}_{\mathrm{B}})> i\}$, respectively. Similar to (\ref{equ::WHD::App::d0bit1}), the probability $\mathrm{Pr}(\widetilde{d}_{\mathbf{e},\ell} \neq 0 | w(\widetilde{\mathbf{e}}_{\mathrm{B}})\!\leq\! i)$ is given by
        \begin{equation} \label{equ::WHD::App::debit1::eB<=i}
            \begin{split}
                &\mathrm{Pr}(\widetilde{d}_{\mathbf{e},\ell} \neq 0 | w(\widetilde{\mathbf{e}}_{\mathrm{B}})\leq i) \\
                &= \mathrm{Pr}(\widetilde{c}_{\mathbf{e},\ell}'\neq 0)\mathrm{Pr}(\widetilde{e}_{\ell}= 0) + \mathrm{Pr}(\widetilde{c}_{\mathbf{e},\ell}'= 0)\mathrm{Pr}(\widetilde{e}_{\ell}\neq 0)  \\
                &=  \sum_{q=1}^{k}p_{W_{\mathbf{e},\widetilde{\mathbf{e}}_{\mathrm{B}}}}(q|w(\widetilde{\mathbf{e}}_{\mathrm{B}})\leq i) \left(p_{\mathbf{c}_{\mathrm{P}}}^{\mathrm{bit}}(\ell,q)(1-\mathrm{Pe}(\ell)) \right. \\
                & \quad + \left. (1 - p_{\mathbf{c}_{\mathrm{P}}}^{\mathrm{bit}}(\ell,q))\mathrm{Pe}(\ell) \right), 
            \end{split}
        \end{equation}
        where $p_{W_{\mathbf{e},\widetilde{\mathbf{e}}_{\mathrm{B}}}}(q|w(\widetilde{\mathbf{e}}_{\mathrm{B}})\leq i) $ is the conditional $\mathrm{pmf}$ of the random variable $W_{\mathbf{e},\widetilde{\mathbf{e}}_{\mathrm{B}}}$ introduced in Lemma \ref{lem::HDdis::iphase::eB&eWeight}. Following Lemma \ref{lem::HDdis::iphase::eB&eWeight}, $p_{W_{\mathbf{e},\widetilde{\mathbf{e}}_{\mathrm{B}}}}(q|w(\widetilde{\mathbf{e}}_{\mathrm{B}})\leq i)$ is given by
        \begin{equation} \label{equ::WHD::App::eB&eWeight::eB<=i}
        \begin{split}
            p_{W_{\mathbf{e},\widetilde{\mathbf{e}}_{\mathrm{B}}}}\!(q|w(\widetilde{\mathbf{e}}_{\mathrm{B}})\!\leq\! i) &\!=\! \sum_{u = 0}^{i}\sum_{v = 0}^{i} \frac{\binom{u}{\delta}\binom{k-u}{v-\delta}}{\binom{k}{v}} \cdot\frac{p_{E_{1}^{k}}(u)}{\sum_{q = 0}^{i}p_{E_{1}^{k}}(q)} \\
            &\cdot \frac{\binom{k}{v}}{b_{0:i}^{k}} \cdot\mathbf{1}_{\mathbb{N}\bigcap[0,\min(u,v)]}(\delta), 
        \end{split}
        \end{equation}
        where $\delta= \frac{u+v-q}{2}$. The probability $\mathrm{Pr}(\widetilde{d}_{\mathbf{e},\ell} \neq 0 | w(\widetilde{\mathbf{e}}_{\mathrm{B}}) > i)$ is also given by (\ref{equ::WHD::App::debit1::eB<=i}) with replacing $p_{W_{\mathbf{e},\widetilde{\mathbf{e}}_{\mathrm{B}}}}(q|w(\widetilde{\mathbf{e}}_{\mathrm{B}})\leq i)$ with $p_{W_{\mathbf{e},\widetilde{\mathbf{e}}_{\mathrm{B}}}}(q|w(\widetilde{\mathbf{e}}_{\mathrm{B}}) \!>\! i)$, which is given by (\ref{equ::HDdis::iphase::eB&eWeight}). For simplicity, let us denote $\mathrm{Pr}(\widetilde{d}_{\mathbf{e},\ell} \neq 0 | w(\widetilde{\mathbf{e}}_{\mathrm{B}})\leq i)$ and $\mathrm{Pr}(\widetilde{d}_{\mathbf{e},\ell} \neq 0 | w(\widetilde{\mathbf{e}}_{\mathrm{B}})> i)$ by $\mathrm{Pc}_{\mathbf{e}}(\ell|i^{(\leq)})$ and $\mathrm{Pc}_{\mathbf{e}}(\ell|i^{(>)})$, respectively. Also, for probabilities $\mathrm{Pr}(\widetilde{d}_{\mathbf{e},\ell} =0 | w(\widetilde{\mathbf{e}}_{\mathrm{B}})\leq i)$ and $\mathrm{Pr}(\widetilde{d}_{\mathbf{e},\ell}= 0 | w(\widetilde{\mathbf{e}}_{\mathrm{B}})> i)$, we simply denote them by $\mathrm{Pc}_{\mathbf{e}}(\bar{\ell}|i^{(\leq)})$ and $\mathrm{Pc}_{\mathbf{e}}(\bar{\ell}|i^{(>)})$.
        
        The joint probability $\mathrm{Pr}(\widetilde{d}_{\mathbf{e},\ell} \neq 0, \widetilde{d}_{\mathbf{e},h} \neq 0 | w(\widetilde{\mathbf{e}}_{\mathrm{B}})\leq i)$ can be determined similar to (\ref{equ::WHD::App::d0bit2}), i.e., 
        \begin{equation} \label{equ::WHD::App::debit2::eB<=i}
            \begin{split}
                &\mathrm{Pr}(\widetilde{d}_{\mathbf{e},\ell} \neq 0, \widetilde{d}_{\mathbf{e},h} \neq 0 | w(\widetilde{\mathbf{e}}_{\mathrm{B}})\!\leq\! i) \\
                &= \mathrm{Pr}\{\widetilde{c}_{\mathbf{e},\ell}' \neq 0, \widetilde{c}_{\mathbf{e},h}' \neq 0| w(\widetilde{\mathbf{e}}_{\mathrm{B}})\!\leq\! i\}\mathrm{Pr}\{\widetilde{e}_{\ell} = 0, \widetilde{e}_{h} = 0\} \\
                &+\mathrm{Pr}\{\widetilde{c}_{\mathbf{e},\ell}' \neq 0, \widetilde{c}_{\mathbf{e},h}' = 0| w(\widetilde{\mathbf{e}}_{\mathrm{B}})\!\leq\! i\}\mathrm{Pr}\{\widetilde{e}_{\ell} = 0, \widetilde{e}_{h} \neq 0\} 
                \\
                &+ \mathrm{Pr}\{\widetilde{c}_{\mathbf{e},\ell}' = 0, \widetilde{c}_{\mathbf{e},h}' \neq 0| w(\widetilde{\mathbf{e}}_{\mathrm{B}})\!\leq\! i\}\mathrm{Pr}\{\widetilde{e}_{\ell} \neq 0, \widetilde{e}_{h} = 0\} \\
                &+\mathrm{Pr}\{\widetilde{c}_{\mathbf{e},\ell}' = 0, \widetilde{c}_{\mathbf{e},h}' = 0| w(\widetilde{\mathbf{e}}_{\mathrm{B}})\!\leq\! i\}\mathrm{Pr}\{\widetilde{e}_{\ell} \neq 0, \widetilde{e}_{h} \neq 0\}.
            \end{split}
        \end{equation}
        By considering $p_{W_{\mathbf{e},\widetilde{\mathbf{e}}_{\mathrm{B}}}}(q|w(\widetilde{\mathbf{e}}_{\mathrm{B}})\!\leq\! i)$ in (\ref{equ::WHD::App::eB&eWeight::eB<=i}) and $p_{\mathbf{c}_{\mathrm{P}}}^{\mathrm{bit}}(\ell,h,q)$ in (\ref{equ::WHD::App::PcPbit2}), (\ref{equ::WHD::App::debit2::eB<=i}) can be computed. We omit the expanded expression of (\ref{equ::WHD::App::debit2::eB<=i}) here for the sake of brevity. Furthermore, the probability $\mathrm{Pr}(\widetilde{d}_{\mathbf{e},\ell} \neq 0, \widetilde{d}_{\mathbf{e},h} \neq 0 | w(\widetilde{\mathbf{e}}_{\mathrm{B}})\!>\! i)$ can be obtained similar to (\ref{equ::WHD::App::debit2::eB<=i}), by replacing $p_{W_{\mathbf{e},\widetilde{\mathbf{e}}_{\mathrm{B}}}}(q|w(\widetilde{\mathbf{e}}_{\mathrm{B}})\leq i)$ with $p_{W_{\mathbf{e},\widetilde{\mathbf{e}}_{\mathrm{B}}}}(q|w(\widetilde{\mathbf{e}}_{\mathrm{B}})> i)$ given by (\ref{equ::HDdis::iphase::eB&eWeight}). For simplicity of notation, we denote $\mathrm{Pr}(\widetilde{d}_{\mathbf{e},\ell} \neq 0, \widetilde{d}_{\mathbf{e},h} \neq 0 | w(\widetilde{\mathbf{e}}_{\mathrm{B}})\!\leq\! i)$ and $\mathrm{Pr}(\widetilde{d}_{\mathbf{e},\ell} \neq 0, \widetilde{d}_{\mathbf{e},h} \neq 0 | w(\widetilde{\mathbf{e}}_{\mathrm{B}})\!>\! i)$ as $\mathrm{Pc}_{\mathbf{e}}(\ell,h|i^{(\leq)})$ and $\mathrm{Pc}_{\mathbf{e}}(\ell,h|i^{(>)})$, respectively. In addition, we use $\mathrm{Pc}_{\mathbf{e}}(\bar{\ell},h|i^{(\leq)})$ and $\mathrm{Pc}_{\mathbf{e}}(\ell,\bar{h}|i^{(>)})$ to denote $\mathrm{Pr}(\widetilde{d}_{\mathbf{e},\ell} = 0, \widetilde{d}_{\mathbf{e},h} \neq 0 | w(\widetilde{\mathbf{e}}_{\mathrm{B}})\!\leq\! i) $ and $\mathrm{Pr}(\widetilde{d}_{\mathbf{e},\ell} \neq 0, \widetilde{d}_{\mathbf{e},h} = 0 | w(\widetilde{\mathbf{e}}_{\mathrm{B}})\!>\! i) $, respectively.
        
        Based on the probabilities $\mathrm{Pc}_{\mathbf{e}}(\ell|i^{(\leq)})$, $\mathrm{Pc}_{\mathbf{e}}(\ell|i^{(>)})$ $\mathrm{Pc}_{\mathbf{e}}(\ell,h|i^{(\leq)})$ and $\mathrm{Pc}_{\mathbf{e}}(\ell,h|i^{(>)})$ introduced above, we next simplify and approximate the distribution of $D_{i}^{(\mathrm{W})}$. We first consider the WHD $D_{\mathbf{e}}^{(\mathrm{W})}$ conditioning on  $\widetilde{\mathbf{e}}_{\mathrm{B}} = \mathbf{e}$ introduced in Lemma \ref{lem::WHD::iphase::eB=e}.  According to (\ref{equ::OrderStat::EPeIndenIndepence}), the mean of $D_{\mathbf{e}}^{(\mathrm{W})}$ conditioning on $\widetilde{\mathbf{e}}_{\mathrm{B}} = \mathbf{e}$ can be approximated as
        \begin{equation}  \label{equ::WHD::App::eB=e::Mean::App}
            \begin{split}
                &\mathbb{E}[D_{\mathbf{e}}^{(\mathrm{W})}|\widetilde{\mathbf{e}}_{\mathrm{B}} \!=\! \mathbf{e}] \\
                & = \sum_{\ell=0}^{i}\sum_{h = 0}^{n-k} \sum_{\substack{\mathbf{t}_{\ell}^{h}\\\mathbf{t}_{\ell}^{\mathrm{B}} \in \mathcal{T}_l^{\mathrm{B}} \\ \mathbf{t}_{h}^{\mathrm{P}} \in \mathcal{T}_h^{\mathrm{P}} }} \mathrm{P}(\mathbf{t}_{\ell}^{h})   \left(\sum_{u=1}^\ell \sqrt{\widetilde{\mathbf{E}}_{t_u^{\mathrm{B}},t_u^{\mathrm{B}}}} + \sum_{u=1}^h \sqrt{\widetilde{\mathbf{E}}_{t_u^{\mathrm{P}},t_u^{\mathrm{P}}}} \right) \\
            &  = \sum_{u=1}^{n} \mathrm{Pe}(u | E_1^{k} \!\leq\! i) \sqrt{\widetilde{\mathbf{E}}_{u,u}} \\
            & \overset{(a)}{\approx}  \!\!\left(\!1 \!-\! \frac{p_{E_1^{k}}(i)}{\sum_{v=0}^{i} p_{E_1^{k}}(v)}\!\right)\!\!\sum_{u=1}^{k} \mathrm{Pe}(u) \!\sqrt{\widetilde{\mathbf{E}}_{u,u}} \!+\!\! \sum_{u=k\!+\!1}^{n}\!\! \mathrm{Pe}(u) \!\sqrt{\widetilde{\mathbf{E}}_{u,u}},
            \end{split}
        \end{equation}
        where step (a) follows from that $\mathrm{Pe}(u | E_1^{k} \!\leq\! i) \approx \frac{\mathrm{Pe}(u)\mathrm{Pr}(E_1^{k} \leq i-1)}{\mathrm{Pr}(E_1^{k} \leq i)}$ for $u$, $1\leq u \leq k$, and $\mathrm{Pe}(u | E_1^{k} \!\leq\! i) \approx \mathrm{Pe}(u )$ for $u$, $k+1\leq u \leq n$, according to (\ref{equ::OrderStat::EPeIndenIndepence}). Similarly, the variance of $D_{\mathbf{e}}^{(\mathrm{W})}$ is approximated as
        \begin{equation} \label{equ::WHD::App::eB=e::Var::App}
            \begin{split}
                & \sigma_{D_{\mathbf{e}}^{(\mathrm{W})}|\widetilde{\mathbf{e}}_{\mathrm{B}} = \mathbf{e}}^2
                \\
                &\approx  \left(1 - \frac{p_{E_1^{k}}(i)+p_{E_1^{k}}(i\!-\!1)}{\sum_{\ell=0}^{i} p_{E_1^{k}}(\ell)}\right) \sum_{u=1}^{k}\sum_{v=1}^{k} \mathrm{Pe}(u,v)[\widetilde{\mathbf{E}} + \widetilde{\mathbf\Sigma}]_{u,v} \\
                &+\sum_{u=k+1}^{n}\sum_{v=k+1}^{n} \mathrm{Pe}(u,v)[\widetilde{\mathbf{E}} + \widetilde{\mathbf\Sigma}]_{u,v} \\
                &+  2 \left(1 - \frac{p_{E_1^{k}}(i)}{\sum_{\ell=0}^{i} p_{E_1^{k}}(\ell)}\right) \sum_{u=1}^{k}\sum_{v=k+1}^{n}\mathrm{Pe}(u,v)[\widetilde{\mathbf{E}} + \widetilde{\mathbf\Sigma}]_{u,v}\\
                &-\left(\mathbb{E}[D_{\mathbf{e}}^{(\mathrm{W})}|\widetilde{\mathbf{e}}_{\mathrm{B}}\! =\! \mathbf{e}]\right)^2.
            \end{split}
        \end{equation}
        where $\mathrm{Pe}(u,v) = \mathrm{Pe}(u)$ for $u=v$. Then, because the $\mathrm{pdf}$ $f_{D_{\mathbf{e}}^{(\mathrm{W})}}(x|\widetilde{\mathbf{e}}_{\mathrm{B}} \!=\! \mathbf{e})$ given by (\ref{equ::WHD::iphase::eb=e}) is a large-number Gaussian mixture model, we formulate it as the $\mathrm{pdf}$ of a Gaussian distribution $\mathcal N(\mathbb{E}[D_{\mathbf{e}}^{(\mathrm{W})}|\widetilde{\mathbf{e}}_{\mathrm{B}}\! =\! \mathbf{e}], \sigma_{D_{\mathbf{e}}^{(\mathrm{W})}|\widetilde{\mathbf{e}}_{\mathrm{B}} = \mathbf{e}}^2)$ denoted by $f_{D_{\mathbf{e}}^{(\mathrm{W})}}^{\mathrm{app}}(x|\widetilde{\mathbf{e}}_{\mathrm{B}} \!=\! \mathbf{e})$ i.e.,
	    \begin{equation} \label{equ::WHD::App::De::eb=e::App}
	    \begin{split}
  	        &f_{D_{\mathbf{e}}^{(\mathrm{W})}}^{\mathrm{app}}(x|\widetilde{\mathbf{e}}_{\mathrm{B}} \!=\! \mathbf{e})\\
  	        &=\! 
	        \frac{1}{\sqrt{2\pi \sigma_{\!D_{\mathbf{e}}^{(\mathrm{W})}|\widetilde{\mathbf{e}}_{\mathrm{B}} = \mathbf{e}}^2}} \!\exp\!\left(\!\!-\frac{(x\!-\!\mathbb{E}[D_{\mathbf{e}}^{(\mathrm{W})}\!|\widetilde{\mathbf{e}}_{\mathrm{B}}\! =\! \mathbf{e}])^2}{2\sigma_{D_{\mathbf{e}}^{(\mathrm{W})}|\widetilde{\mathbf{e}}_{\mathrm{B}} = \mathbf{e}}^2}\!\!\right).  
	    \end{split}
	    \end{equation}
	    
	    Next, we simplify the mean and variance of $D_{\mathbf{e}}^{(\mathrm{W})}$ conditioning on  $\widetilde{\mathbf{e}}_{\mathrm{B}} \neq \mathbf{e}$ and $w(\widetilde{\mathbf{e}}_{\mathrm{B}})\leq i $, as introduced in Lemma \ref{lem::WHD::iphase::eB!=e}, as well as to characterize the related covariance. Considering the probability $\mathrm{Pr}(\widetilde{d}_{\mathbf{e},\ell} \neq 0 | w(\widetilde{\mathbf{e}}_{\mathrm{B}})\leq i)$, the conditional mean of $D_{\mathbf{e}}^{(\mathrm{W})}$, previously given by (\ref{equ::WHD::iphase::eB!=e::eB<=i::Mean}), can be simplified as
        \begin{equation} \label{equ::WHD::App::eB!=e::eB<=i::Mean::App}
        \begin{split}
            \mathbb{E}&[D_{\mathbf{e}}^{(\mathrm{W})}|\widetilde{\mathbf{e}}_{\mathrm{B}} \! \neq\! \mathbf{e},w(\widetilde{\mathbf{e}}_{\mathrm{B}})\!\leq\! i] \\
            &= \frac{b_{0:(i-1)}^{k-1}}{b_{0:i}^{k}}\sum_{u=1}^{k}\sqrt{\widetilde{\mathbf{E}}_{u,u}} + \sum_{u=k+1}^{n} \mathrm{Pc}_{\mathbf{e}}(u|i^{(\leq)})\sqrt{\widetilde{\mathbf{E}}_{u,u}},
        \end{split}
        \end{equation}
        where $\mathrm{Pc}_{\mathbf{e}}(u|i^{(\leq)}) = \mathrm{Pr}(\widetilde{d}_{\mathbf{e},u} \!=\! 0 | w(\widetilde{\mathbf{e}}_{\mathrm{B}})\!\leq\! i)$ is given by (\ref{equ::WHD::App::debit1::eB<=i}). Then, considering the joint probability $\mathrm{Pr}(\widetilde{d}_{\mathbf{e},\ell} \neq 0, \widetilde{d}_{\mathbf{e},h} \neq 0 | w(\widetilde{\mathbf{e}}_{\mathrm{B}})\leq i)$ and using the same approach of obtaining (\ref{equ::WHD::App::eB=e::Var::App}), the conditional variance of $D_{\mathbf{e}}^{(\mathrm{W})}$, previously given by (\ref{equ::WHD::iphase::eB!=e::eB<=i::Var}), can be simplified as
        \begin{align} \label{equ::WHD::App::eB!=e::eB<=i::Var::App}
                &\sigma^2_{D_{\mathbf{e}}^{(\mathrm{W})}|\widetilde{\mathbf{e}}_{\mathrm{B}} \neq \mathbf{e},w(\widetilde{\mathbf{e}}_{\mathrm{B}}) \leq i} \notag\\
                &=\sum_{u=1}^{k}\frac{b_{0:(i-1)}^{k-1}}{b_{0:i}^{k}}[\widetilde{\mathbf{E}}\!+\! \widetilde{\mathbf\Sigma}]_{u,u} + 2\sum_{u=1}^{k-1}\sum_{v=u+1}^{k}\frac{b_{0:(i-2)}^{k - 2}}{b_{0:i}^{k}}[\widetilde{\mathbf{E}}\!+\! \widetilde{\mathbf\Sigma}]_{u,v} \notag\\ 
                &+ 2\sum_{u=1}^{k}\sum_{v=k+1}^{n} \left(\frac{b_{0:(i-1)}^{k-1}}{b_{0:i}^{k}}\mathrm{Pc}_{\mathbf{e}}(v|i^{(\leq)})\right)[\widetilde{\mathbf{E}}\!+\! \widetilde{\mathbf\Sigma}]_{u,v}\notag\\
                &+ \sum_{u=k+1}^{n}\sum_{v=k+1}^{n}\mathrm{Pc}_{\mathbf{e}}(u,v|i^{(\leq)})[\widetilde{\mathbf{E}}\!+\! \widetilde{\mathbf\Sigma}]_{u,u}\notag\\
                &- \left(\mathbb{E}[D_{\mathbf{e}}^{(\mathrm{W})}|\widetilde{\mathbf{e}}_{\mathrm{B}} \! \neq\! \mathbf{e},w(\widetilde{\mathbf{e}}_{\mathrm{B}})\!\leq\! i]\right)^2,
        \end{align}
        where $\mathrm{Pc}_{\mathbf{e}}(u,v|i^{(\leq)}) = \mathrm{Pr}(\widetilde{d}_{\mathbf{e},\ell} \neq 0, \widetilde{d}_{\mathbf{e},h} \neq 0 | w(\widetilde{\mathbf{e}}_{\mathrm{B}})\!\leq\! i)$ is given by (\ref{equ::WHD::App::debit2::eB<=i}). In particular, $\mathrm{Pc}_{\mathbf{e}}(u,v|i^{(\leq)}) = \mathrm{Pc}_{\mathbf{e}}(u|i^{(\leq)})$ for $u=v$. On the conditions that $\widetilde{\mathbf{e}}_{\mathrm{B}} \neq \mathbf{e}$ and $w(\widetilde{\mathbf{e}}_{\mathrm{B}})\leq i $, we can also simplify the covariance given in (\ref{equ::WHD::iphase::eB!=e::eB<=i::Cov}) as 
         \begin{align} \label{equ::WHD::App::eB!=e::eB<=i::Cov::App}
                &\mathrm{cov}\left(D_{\mathbf{e}}^{(\mathrm{W})},D_{\hat{\mathbf{e}}}^{(\mathrm{W})}\right) \notag\\
                &=\sum_{u=1}^{k}\sum_{v=k\!+\!1}^{n}\!\!\left(\frac{b_{0:(i-1)}^{k-1} }{b_{0:i}^{k}}\cdot \frac{b_{0:i}^{k-1} }{b_{0:i}^{k}}\mathrm{Pc}_{\mathbf{e}}(\bar{u}|i^{(\leq)}) \mathrm{Pc}_{\mathbf{e}}(v|i^{(\leq)})\!\right)\!\widetilde{\mathbf\Sigma}_{u,v}\notag\\
                &+ 2\sum_{u=k+1}^{n-1}\sum_{v=u+1}^{n}\mathrm{Pc}_{\mathbf{e}}(\bar{u},v|i^{(\leq)})\ \mathrm{Pc}_{\mathbf{e}}(u,\bar{v}|i^{(\leq)})\ \widetilde{\mathbf\Sigma}_{u,v}\notag\\
                &+2\left(\frac{b_{0:(i-1)}^{k-2}}{b_{0:i}^{k}}\right)^2\sum_{u=1}^{k-1}\sum_{v=u+1}^{k}  \widetilde{\mathbf\Sigma}_{u,v} .
        \end{align}
        Utilizing (\ref{equ::WHD::App::eB!=e::eB<=i::Var::App}) and (\ref{equ::WHD::App::eB!=e::eB<=i::Cov::App}), the correlation efficiency $\rho_1$ given by (\ref{equ::WHD::iphase::eB!=e::eB<=i::Rho}) is numerically computed. Replacing probabilities $\mathrm{Pc}_{\mathbf{e}}(\cdot|i^{(\leq)})$ and $\mathrm{Pc}_{\mathbf{e}}(\cdot,\cdot|i^{(\leq)})$ with $\mathrm{Pc}_{\mathbf{e}}(\cdot|i^{(>)})$ and $\mathrm{Pc}_{\mathbf{e}}(\cdot,\cdot|i^{(>)})$ in (\ref{equ::WHD::App::eB!=e::eB<=i::Mean::App}), (\ref{equ::WHD::App::eB!=e::eB<=i::Var::App}), and (\ref{equ::WHD::App::eB!=e::eB<=i::Cov::App}), we can also obtain the mean $\mathbb{E}[D_{\mathbf{e}}^{(\mathrm{W})}|\widetilde{\mathbf{e}}_{\mathrm{B}} \! \neq\! \mathbf{e},w(\widetilde{\mathbf{e}}_{\mathrm{B}})\!>\! i]$, the variance $\sigma^2_{D_{\mathbf{e}}^{(\mathrm{W})}|\widetilde{\mathbf{e}}_{\mathrm{B}} \!\neq\! \mathbf{e},w(\widetilde{\mathbf{e}}_{\mathrm{B}})\! >\! i}$, and the covariance regarding $D_{\mathbf{e}}^{(\mathrm{W})}$ conditioning on $\{w(\widetilde{\mathbf{e}}_{\mathrm{B}}) \!>\! i,\widetilde{\mathbf{e}}_{\mathrm{B}} \!\neq\! \mathbf{e}\}$, and numerically calculate $\rho_2$. Finally, by substituting $f_{D_{\mathbf{e}}^{(\mathrm{W})}}^{\mathrm{app}}(x|\widetilde{\mathbf{e}}_{\mathrm{B}} \!=\! \mathbf{e})$ in (\ref{equ::WHD::App::De::eb=e::App}), the means and variances of $D_{\mathbf{e}}^{(\mathrm{W})}$, and $\rho_1$ and $\rho_2$ into (\ref{equ::WHD::iphase}), the distribution of the $D_{i}^{(\mathrm{W})}$ is finally approximated as 
    	\begin{align} \label{equ::WHD::App::iphase::App}
    	        f_{D_i^{(\mathrm{W})}}(x) &\approx  \sum_{v=0}^{i}p_{E_1^{k}}(v) \notag\\
    	        &\cdot \left(f_{D_{\mathbf{e}}^{(\mathrm{W})}}^{\mathrm{app}}(x|\widetilde{\mathbf{e}}_{\mathrm{B}}\! = \!\mathbf{e})\!\!\int_{x}^{\infty}\!\!\!f_{\widetilde{D}_i^{(\mathrm{W})}}\left(u, b_{1:i}^{k}|w(\widetilde{\mathbf{e}}_{\mathrm{B}}) \!\leq\! i\right) du \right. \notag\\
    	        &+\!\left. f_{\widetilde{D}_i^{(\mathrm{W})}}\left(u, b_{1:i}^{k}|w(\widetilde{\mathbf{e}}_{\mathrm{B}}) \!\leq\! i\right) \!\!\int_{x}^{\infty}\!\!\!f_{D_{\mathbf{e}}^{(\mathrm{W})}}^{\mathrm{app}}(u|\widetilde{\mathbf{e}}_{\mathrm{B}}\! =\! \mathbf{e})du \right) \notag\\
    	        &+  \left(1 - \sum_{v=0}^{i}p_{E_1^{k}}(v)\right) f_{\widetilde{D}_i^{(\mathrm{W})}}\left(u, b_{0:i}^{k}|w(\widetilde{\mathbf{e}}_{\mathrm{B}}) > i\right) ,
    	\end{align}
    	where $f_{\!\widetilde{D}_i^{(\mathrm{W})}}\!\left(u, b_{1:i}^{k}|w(\widetilde{\mathbf{e}}_{\mathrm{B}}) \!\leq\! i\right)$ and $f_{\!\widetilde{D}_i^{(\mathrm{W})}}\!\left(u, b_{0:i}^{k}|w(\widetilde{\mathbf{e}}_{\mathrm{B}}) \!>\! i\right)$ are respectively given by (\ref{equ::WHD::iphase::DependOrder1}) and (\ref{equ::WHD::iphase::DependOrder2}), and $f_{D_{\mathbf{e}}^{(\mathrm{W})}}^{\mathrm{app}}(x|\widetilde{\mathbf{e}}_{\mathrm{B}}\! =\! \mathbf{e})$ is given by (\ref{equ::WHD::App::De::eb=e::App}).}
	    
\begin{figure}
    		\begin{center}
    			\includegraphics[scale=0.55] {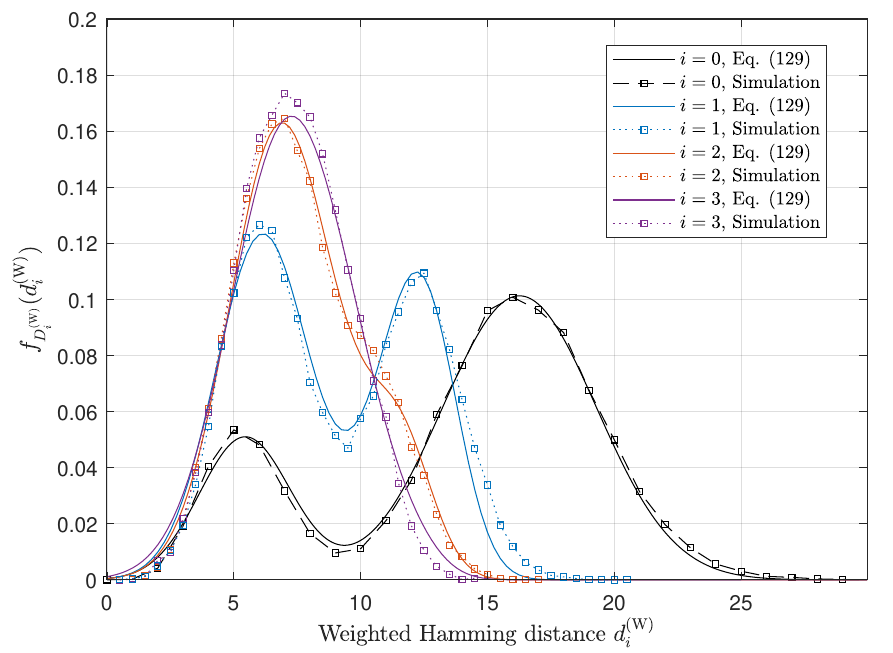}
    			\caption{The distribution of $D_{i}^{(\mathrm{W})}$ in decoding $(128,64,22)$ eBCH code when SNR = 1 dB.}
    			\label{Fig::IV::BCH128-WHD-iPhase}
    		\end{center}
    	\end{figure}

\begin{figure}
    		\begin{center}
    			\includegraphics[scale=0.55] {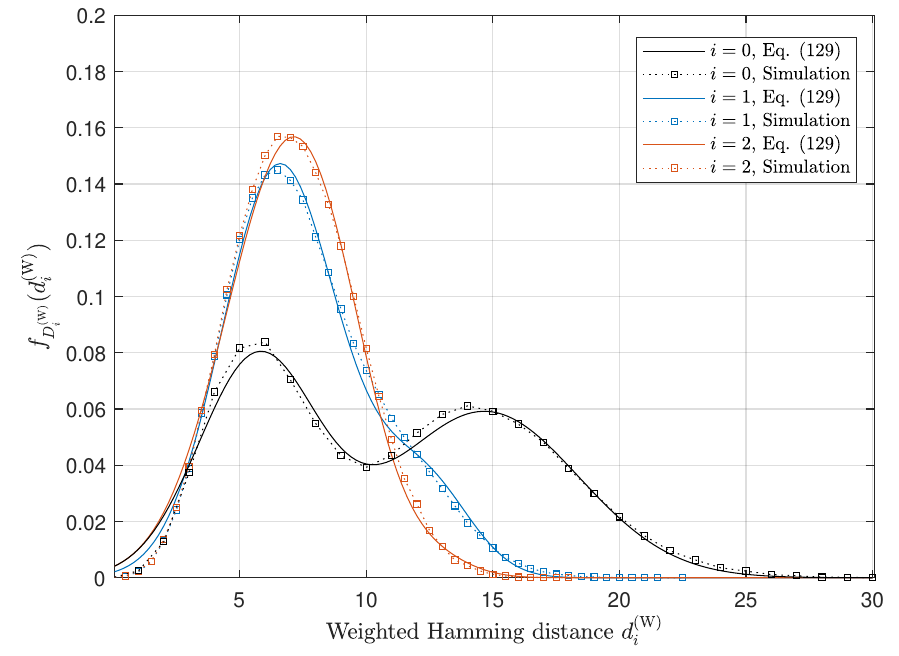}
    			\caption{{\color{black}The distribution of $D_{i}^{(\mathrm{W})}$ in decoding $(64,21,16)$ Polar code when SNR = 1 dB.}}
    			\label{Fig::IV::Polar64-WHD-iPhase}
    		\end{center}
    	\end{figure}

	    We enabled the numerical calculation of (\ref{equ::WHD::iphase}) by introducing the approximation (\ref{equ::WHD::App::iphase::App}). To verify (\ref{equ::WHD::App::iphase::App}), We compare the approximated distribution (\ref{equ::WHD::App::iphase::App}) of $D_{i}^{(\mathrm{W})}$ with the simulation results in decoding the (128,64,22) eBCH code and {\color{black}the (64,21,16) Polar code}, as depicted in Fig. \ref{Fig::IV::BCH128-WHD-iPhase} and Fig. \ref{Fig::IV::Polar64-WHD-iPhase}, respectively. As can be seen, (\ref{equ::WHD::App::iphase::App}) is a tight approximation of $f_{D_{i}^{(\mathrm{W})}}(x)$. Similar to the distribution of $D_{i}^{(\mathrm{H})}$, the $\mathrm{pdf}$ of $D_{i}^{(\mathrm{W})}$ also concentrates towards the left part when the reprocessing order increases. This is because the weight of the two combined components in $f_{D_{i}^{(\mathrm{W})}}(x)$ are given by $\sum_{v=0}^{i}p_{E_1^{k}}(v)$ and $1-\sum_{v=0}^{i}p_{E_1^{k}}(v)$, respectively. The extent to which the distribution concentrates towards the left reflects the improvement in the decoding performance, i.e., the more the distribution is concentrated to the left, the better the error performance. In addition, similar to the distribution of $D_{i}^{(\mathrm{H})}$, the distribution $D_{i}^{(\mathrm{W})}$ given by (\ref{equ::WHD::iphase}) or (\ref{equ::WHD::App::iphase::App}) is only compatible with codes with the minimum distance $d_{\mathrm{H}}$ not much lower than $n-k$, where the correlations between any two codeword estimates generated in OSD can be ignored. However, when $d_{\mathrm{H}}\ll n-k$ or the generator matrix is sparse, the result given by (\ref{equ::WHD::App::iphase::App}) will show discrepancies with the simulation results. 
	    
	    {\color{black}From Fig. \ref{Fig::IV::BCH128-WHD-iPhase} and Fig. \ref{Fig::IV::Polar64-WHD-iPhase}, we can also notice that although (\ref{equ::WHD::App::iphase::App}) provides a relatively tight approximation, there are still a few deviations between (\ref{equ::WHD::App::iphase::App}) and the simulation results. These deviations are mainly due to the reasons: 1) the approximation of ordered reliabilities enlarges the deviations of approximating $D_{i}^{(\mathrm{W})}$, as $D_{i}^{(\mathrm{W})}$ is composed of ordered reliabilities, 2) we approximately obtained the mean and variance of $D_{\mathbf{e}}^{(\mathrm{W})}$ for the simplicity of numerical calculations, e.g., step (a) of (\ref{equ::WHD::App::eB!=e::eB<=i::Mean::App}). Furthermore, the $\mathrm{pdf}$ in (\ref{equ::WHD::App::iphase::App}) is not truncated to consider only non-negative values of $D_{0}^{(\mathrm{W})}$ for the simplicity of expression. One can further improve the accuracy of (\ref{equ::WHD::App::iphase::App}) by considering the truncated distributions in the derivation.}
	    
\section{Hard-decision Decoding Techniques Based on the Hamming Distance Distribution} \label{sec::HDdistech}
For the OSD approach, the decoding complexity can be reduced by applying the discarding rule (DR) and stopping rule (SR). Given a TEP list, DRs are usually designed to identify and discard the unpromising TEPs, while SRs are typically designed to determine whether the best decoding result has been found and terminate the decoding process in advance. In this Section, we propose several SRs and DRs based on the derived Hamming distance distributions in Section \ref{sec::HDdis}. We mainly take BCH codes as examples to demonstrate the performance of the proposed conditions. The efficient decoding algorithms of BCH codes are of particular interest because they can hardly be decoded by using modern well-designed decoders (e.g., successive cancellation for Polar codes and belief propagation for LDPC). In Section \ref{sec::Discussion}, we will further show that the proposed techniques are especially effective for codes with binomial-like weight spectrum (e.g., the BCH code), in which case the SRs and the DRs can be efficiently implemented. 
    
\subsection{Hard Success Probability of Codeword Estimates}
Recalling the statistics of the Hamming distance $D_{0}^{(\mathrm{H})}$ proposed in Theorem \ref{the::HDdis::0phase}, the $\mathrm{pmf}$ of Hamming distance $D_0^{(H)}$ is a mixture of two random variables $E_{k+1}^{n}$ and $W_{\mathbf{c}_{\mathrm{P}}}$ which represent the number of errors in redundant positions and the Hamming weight of the redundant part of a codeword from $\mathcal{C}(n,k)$, respectively. Furthermore, from Lemma \ref{lem::HDdis::0phase}, it is clear that $E_{k+1}^{n}$ can represent the Hamming distance between $\widetilde{\mathbf{y}}$ and the $0$-reprocessing estimate $\widetilde{\mathbf{c}}_{0}$ if no errors occur in MRB positions and $W_{\mathbf{c}_{\mathrm{P}}}$ can represent the Hamming distance if there are some errors in the MRB positions.
                
It is known that 0-reprocessing of OSD can be regarded as the reprocessing of a special all-zero TEP $\mathbf 0$, where $\widetilde{\mathbf{y}}_{\mathrm{B}} \oplus \mathbf 0$ is re-encoded. Thus, Eq. (\ref{equ::HDdis::0phase}) in Theorem \ref{the::HDdis::0phase} is in fact the Hamming distance between $\hat{\mathbf{c}}_{\mathbf{e}}$ and $\mathbf y$ in the special case that $\mathbf{e} = \mathbf 0$. In order to obtain the SRs and DRs for an arbitrary TEP $\mathbf{e}$, we first introduce the following Corollary from Theorem \ref{the::HDdis::0phase}.
{\color{black}                
\begin{corollary} \label{cor::HDtech::CondDis::HDforTEPe}
    Given a linear block code $\mathcal{C}(n,k)$ and a specific TEP $\mathbf{e}$ satisfying $w(\mathbf{e})=v$, the $\mathrm{pmf}$ of the Hamming distance between $\widetilde{\mathbf{y}}$ and $\widetilde{\mathbf{c}}_{\mathbf{e}}$, i.e., $D_{\mathbf{e}}^{(\mathrm{H})}$, is given by
    		\begin{equation} \label{equ::HDtech::CondDis::HDforTEPe}
    		\begin{split}
    			p_{D_{\mathbf{e}}^{(\mathrm{H})}}(j)&=  \mathrm{Pe}(\mathbf{e}) p_{E_{k+1}^{n}}(j-v) \\
    			&+  (1\!-\!\mathrm{Pe}(\mathbf{e}))	p_{W_{\mathbf{e},\mathbf{c}_{\mathrm{P}}}}(j\!-\!v|w(\mathbf{e})\!=\!v),    
    		\end{split}
    		\end{equation}
    		for $j \geq w(\mathbf{e})$, where $\mathrm{Pe}(\mathbf{e})$ is given by
    		\begin{equation} \label{equ::HDtech::CondDis::Pe(e)}
    		\begin{split}
	    	 	    \mathrm{Pe}(\mathbf{e})  &= \underbrace{\int_{0}^{\infty} \cdots }_{k-w(\mathbf{e})}\underbrace{\int_{-\infty}^{0} \cdots }_{w(\mathbf{e})}\\
	    	 	    &\cdot\left(\frac{n!}{(n-k)!} F_{A}(|x_k|) \prod_{\ell=1}^{k} f_{R}(x_{\ell}) \prod_{\ell=2}^{k} \mathbf{1}_{[0,|x_{\ell-1}|]}(|x_{\ell}|)\right)\\
	    	 	    &\cdot\prod_{\substack{0 < \ell \leq k\\\mathbf{e}_{\ell} \neq 0 }} dx_{\ell} \prod_{\substack{0 < \ell \leq k\\\mathbf{e}_{\ell} = 0}} dx_{\ell} ,	    
    		\end{split}
    		\end{equation}
    		$p_{E_{k+1}^{n}}(j)$ is the $\mathrm{pmf}$ of random variable $E_{k+1}^{n}$ given by (\ref{equ::OrderStat::Eab}), and $p_{W_{\mathbf{e},\mathbf{c}_{\mathrm{P}}}}(j|w(\mathbf{e})=v)$ is the conditional $\mathrm{pmf}$ of random variable $W_{\mathbf{e},\mathbf{c}_{\mathrm{P}}}$ defined in Lemma \ref{lem::HDdis::iphase::Wecp}. The conditional $\mathrm{pmf}$ $p_{W_{\mathbf{e},\mathbf{c}_{\mathrm{P}}}}(j|w(\mathbf{e})=v)$ is given by
	        \begin{equation} \label{equ::HDtech::CondDis::Wecp::w(e)=v}
	        \begin{split}
                   &p_{W_{\mathbf{e},\mathbf{c}_{\mathrm{P}}}}(j| w(\mathbf{e})=v) \\
                   &= \sum_{\ell = 0}^{n-k}\sum_{u=0}^{n-k} \frac{\binom{u}{\delta}\binom{n-k-u}{\ell-\delta}}{\binom{n-k}{\ell}} \sum_{q = 0}^{k}\left(p_{W_{\mathbf{e},\widetilde{\mathbf{e}}_{\mathrm{B}}}}(q|w(\mathbf{e})=v) p_{\mathbf{c}_{\mathrm{P}}}(\ell, q)\right)\\
                   &\cdot p_{E_{k+1}^{n}}(u) \cdot \mathbf{1}_{\mathbb{N}\bigcap[0,\min(u,\ell)]}(\delta),         
	        \end{split}
            \end{equation}
    		where $\delta = \frac{\ell+u-j}{2}$, and $p_{W_{\mathbf{e},\widetilde{\mathbf{e}}_{\mathrm{B}}}}(q|w(\mathbf{e})=v)$ is the conditional $\mathrm{pmf}$ of random variable $W_{\mathbf{e},\widetilde{\mathbf{e}}_{\mathrm{B}}}$ introduced in Lemma \ref{lem::HDdis::iphase::eB&eWeight}, which is given by
    		\begin{equation}
    		    p_{W_{\mathbf{e},\widetilde{\mathbf{e}}_{\mathrm{B}}}}\!(q|w(\mathbf{e})\!\!=\!\!v) \!=\!\! \sum_{u = 0}^{k} \frac{\binom{u}{\delta'}\binom{k\!-\!u}{v\!-\!\delta'}}{\binom{k}{v}} p_{E_{1}^{k}}(u) \cdot \mathbf{1}_{\mathbb{N},[0,\min(u,v)]}(\delta'),
    		\end{equation}
    		for $\delta' = \frac{u+v-q}{2}$.
        \end{corollary}
\begin{IEEEproof}
Similar to (\ref{equ::HDdis::0phase}) in Theorem \ref{the::HDdis::0phase} with respect to the all-zero TEP $\mathbf 0$, the $\mathrm{pmf}$ of $D_{\mathbf{e}}^{(\mathrm{H})}$ with respect to a general TEP $\mathbf{e}$ can be derived by replacing $p_{E_1^k}(0)$ and $1 - p_{E_1^k}(0)$ by $\mathrm{Pe}(\mathbf{e})$ and $1-\mathrm{Pe}(\mathbf{e})$, respectively, where $\mathrm{Pe}(\mathbf{e})$ is the probability that only the nonzero positions of $\mathbf{e}$ are in error in $\widetilde{\mathbf{y}}_{\mathrm{B}}$, i.e., $\mathbf{e}$ can eliminate the errors in MRB. Furthermore, slightly different from $D_{0}^{(\mathrm{H})}$ given by (\ref{equ::HDdis::0phase::cases}), the Hamming distance $D_{\mathbf{e}}^{(\mathrm{H})}$ is given by $E_1^{k} + w(\mathbf{e})$ when $\widetilde{\mathbf{e}}_{\mathrm{B}} = \mathbf{e}$, because the Hamming distance contributed by MRB positions needs to be included. In contrast, when $\widetilde{\mathbf{e}}_{\mathrm{B}} \neq \mathbf{e}$, the difference pattern between $\widetilde{\mathbf{c}}_{\mathbf{e}}$ and $\widetilde{\mathbf{y}}$ is given by $\widetilde{\mathbf{d}}_{\mathbf{e}} = [\mathbf{e} \ \ (\widetilde{\mathbf{e}}_{\mathrm{B}}\oplus\mathbf{e})\widetilde{\mathbf{P}}]$ and $D_{\mathbf{e}}^{(\mathrm{H})}$ is given by $w(\mathbf{e}) + w(\widetilde{\mathbf{d}}_{\mathbf{e},\mathrm{P}})$. The Hamming weight $w(\widetilde{\mathbf{d}}_{\mathbf{e},\mathrm{P}})$ is described by the random variable $W_{\mathbf{e},\mathbf{c}_{\mathrm{P}}}$ introduced in Lemma \ref{lem::HDdis::iphase::Wecp}. The $\mathrm{pmf}$ of $W_{\mathbf{e},\mathbf{c}_{\mathrm{P}}}$ conditioning on $w(\mathbf{e})=v$, given by (\ref{equ::HDtech::CondDis::Wecp::w(e)=v}), can be easily obtained from (\ref{equ::HDdis::iphase::Wecp}).
\end{IEEEproof}
}

From Corollary \ref{cor::HDtech::CondDis::HDforTEPe}, we know that for the Hamming distance $D_{\mathbf{e}}^{\mathrm{H}}$ with respect to an arbitrary TEP $\mathbf{e}$, the $\mathrm{pmf}$ $p_{D_{\mathbf{e}}^{(\mathrm{H})}}(j)$ is also a mixture of two random variables $E_1^{k} + w(\mathbf{e})$ and $W_{\mathbf{e},\mathbf{c}_{\mathrm{P}}} + w(\mathbf{e})$, and the weight of the mixture is determined by probability $\mathrm{Pe}(\mathbf{e})$. In fact, $\mathrm{Pe}(\mathbf{e})$ is the probability that $\mathbf{e}$ could eliminate the MRB errors $\widetilde{\mathbf{e}}_{\mathrm{B}}$, and we refer to $\mathrm{Pe}(\mathbf{e})$ as the \textit{a priori correct probability} of the codeword estimate $\widetilde{\mathbf{c}}_{\mathbf{e}}$ with respect to $\mathbf{e}$. Nevertheless, based on (\ref{equ::HDtech::CondDis::HDforTEPe}) we can further find the probability that TEP $\mathbf{e}$ could eliminate the error pattern $\widetilde{\mathbf{e}}_{\mathrm{B}}$ when given the Hamming distance $d_{\mathbf{e}}^{(\mathrm{H})}$ (a sample of $D_{\mathbf{e}}^{(\mathrm{H})}$), which is referred to as the \textit{hard success probability} of $\widetilde{\mathbf{c}}_{\mathbf{e}}$. The hard success probability can be regarded as the \textit{a posterior} correct probability of  $\widetilde{\mathbf{c}}_{\mathbf{e}}$, given the value of $D_{\mathbf{e}}^{(\mathrm{H})}$. We characterize the hard success probability in the following Corollary.
        \begin{corollary} \label{cor::HDtech::CondDis::HDpsuc}
            Given a linear block code $\mathcal{C}(n,k)$ and TEP $\mathbf{e}$, if the Hamming distance between $\widetilde {\mathbf{c}}_{\mathbf{e}}$ and $\widetilde {\mathbf y}$ is calculated as $d_{\mathbf{e}}^{(\mathrm{H})}$,  the probability that the errors in MRB are eliminated by TEP $\mathbf{e}$ is given by
		    \begin{equation} \label{equ::HDtech::CondDis::HDpsuc}
    			\mathrm{P}_{\mathbf{e}}^{\mathrm{suc}}(d_{\mathbf{e}}^{(\mathrm{H})})= \mathrm{Pe}(\mathbf{e})\frac{p_{E_{k+1}^{n}}\left(d_{\mathbf{e}}^{(\mathrm{H})}-w(\mathbf{e})\right)}{p_{D_{\mathbf{e}}^{(\mathrm{H})}}\left(d_{\mathbf{e}}^{(\mathrm{H})}-w(\mathbf{e})\right)},
    	    \end{equation}
    	    where $p_{D_{\mathbf{e}}^{(\mathrm{H})}}(j)$ is the $\mathrm{pmf}$ given by (\ref{equ::HDtech::CondDis::HDforTEPe}).
        \end{corollary}
        \begin{IEEEproof}
            For the probability $\mathrm{P}_{\mathbf{e}}^{\mathrm{suc}}(d_{\mathbf{e}})$, we observe
            \begin{equation} \label{equ::HDtech::CondDis::HDpsuc_proof}
                \begin{split}
                    \mathrm{P}_{\mathbf{e}}^{\mathrm{suc}}(d_{\mathbf{e}}^{(\mathrm{H})}) \!=\! \frac{\mathrm{Pr}\!\left(\!D_{\mathbf{e}}^{(\mathrm{H})} \!\!=\! d_{\mathbf{e}}^{(\mathrm{H})}\!,\widetilde{\mathbf{e}}_{\mathrm{B}} \!=\! \mathbf{e}\!\right)}{\mathrm{Pr}\!\left(\!D_{\mathbf{e}}^{\!(\mathrm{H})} \!\!=\! d_{\mathbf{e}}^{(\mathrm{H})}\!,\widetilde{\mathbf{e}}_{\mathrm{B}} \!\!=\! \mathbf{e}\right)\!\! +\!\! \mathrm{Pr}\!\left(\!D_{\mathbf{e}}^{\!(\mathrm{H})} \!\!=\! d_{\mathbf{e}}^{(\mathrm{H})}\!,\widetilde{\mathbf{e}}_{\mathrm{B}} \!\!\neq\! \mathbf{e}\!\right)}\!,
                \end{split}
            \end{equation}
            where $\mathrm{Pr}\left(D_{\mathbf{e}}^{(\mathrm{H})} \!=\! d_{\mathbf{e}}^{(\mathrm{H})},\widetilde{\mathbf{e}}_{\mathrm{B}} \!=\! \mathbf{e}\right)$ is derived as $\mathrm {Pr}(\widetilde{\mathbf{e}}_{\mathrm{B}} \!=\! \mathbf{e}) \mathrm{Pr}(D_{\mathbf{e}}^{(\mathrm{H})} \!\!=\! d_{\mathbf{e}}^{(\mathrm{H})}|\widetilde{\mathbf{e}}_{\mathrm{B}} \!=\! \mathbf{e})$, and $\mathrm{Pr}\left(D_{\mathbf{e}}^{(\mathrm{H})} \!=\! d_{\mathbf{e}}^{(\mathrm{H})},\widetilde{\mathbf{e}}_{\mathrm{B}} \neq \mathbf{e}\right)$ is derived as $\mathrm{Pr}(\widetilde{\mathbf{e}}_{\mathrm{B}} \neq \mathbf{e}) \mathrm{Pr}(D_{\mathbf{e}}^{(\mathrm{H})} \!=\! d_{\mathbf{e}}^{(\mathrm{H})}|\widetilde{\mathbf{e}}_{\mathrm{B}} \neq \mathbf{e})$. From Corollary \ref{cor::HDtech::CondDis::HDforTEPe}, $\mathrm {Pr}(\widetilde{\mathbf{e}}_{\mathrm{B}}\! =\! \mathbf{e})$ is given by $\mathrm{Pe}(\mathbf{e})$, and $\mathrm{Pr}(D_{\mathbf{e}}^{(\mathrm{H})} \!=\! d_{\mathbf{e}}^{(\mathrm{H})}|\widetilde{\mathbf{e}}_{\mathrm{B}} \!=\! \mathbf{e})$ and $\mathrm{Pr}(D_{\mathbf{e}}^{(\mathrm{H})} \!=\! d_{\mathbf{e}}^{(\mathrm{H})}|\widetilde{\mathbf{e}}_{\mathrm{B}} \!\neq\! \mathbf{e})$ are in fact given by $p_{E_{k+1}^{n}}(d_{\mathbf{e}}^{(\mathrm{H})}\!-\!w(\mathbf{e}))$ and $p_{W_{\mathbf{e},\mathbf{c}_{\mathrm{P}}}}(d_{\mathbf{e}}^{(\mathrm{H})}\!-\!w(\mathbf{e}))$ in (\ref{equ::HDtech::CondDis::HDforTEPe}), respectively. Substituting $\mathrm{Pe}(\mathbf{e})$, $p_{E_{k+1}^{n}}(d_{\mathbf{e}}^{(\mathrm{H})}-w(\mathbf{e}))$ and $p_{W_{\mathbf{e},\mathbf{c}_{\mathrm{P}}}}(d_{\mathbf{e}}^{(\mathrm{H})}-w(\mathbf{e}))$ into (\ref{equ::HDtech::CondDis::HDpsuc_proof}), we obtain (\ref{equ::HDtech::CondDis::HDpsuc}).
        \end{IEEEproof}
                
\begin{figure}
    		\begin{center}
    			\includegraphics[scale=0.60] {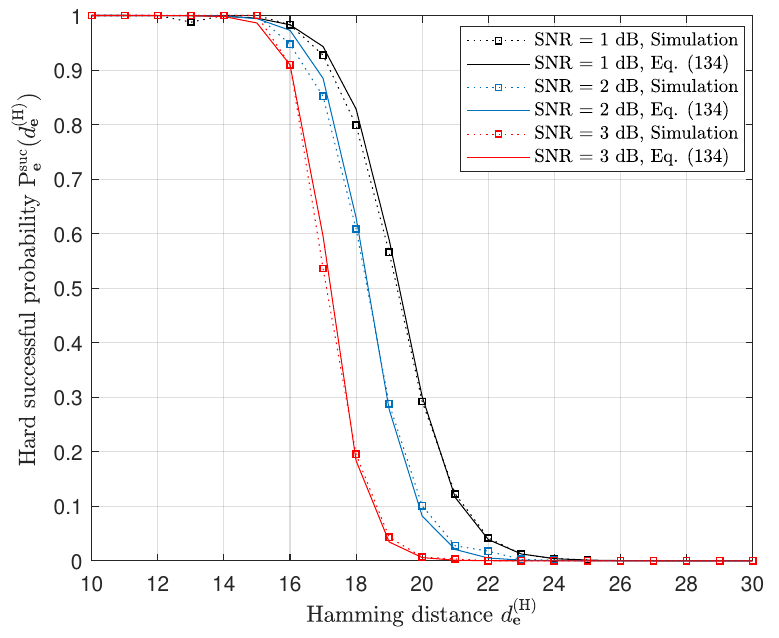}
    			\caption{{\color{black}$\mathrm{P}_{\mathbf{e}}^{\mathrm{suc}}(d_{\mathbf{e}}^{(\mathrm{H})})$ in decoding $(128,64,22)$ eBCH code at different SNR, for TEP $\mathbf{e} = [0,\ldots,0,1,1,0]$.}}
    			\label{Fig::VI::BCH128-HD-Psuc}
    		\end{center}
    	\end{figure}

         We show $\mathrm{P}_{\mathbf{e}}^{\mathrm{suc}}(d_{\mathbf{e}}^{(\mathrm{H})})$ as a function of $d_{\mathbf{e}}^{(\mathrm{H})}$ for TEP $\mathbf{e} = [0,\ldots,0,1,1,0]$ in decoding $(128,64,22)$ eBCH code in Fig. \ref{Fig::VI::BCH128-HD-Psuc}. As can be seen, when $d_{\mathbf{e}}^{(\mathrm{H})}$ decreases, the probability that errors in MRB are eliminated increases rapidly. In other words, the \textit{a posterior} correct probability of $\widetilde{\mathbf{c}}_{\mathbf{e}}$ increases as $d_{\mathbf{e}}^{(\mathrm{H})}$ decreases. It is of interest that although the WHD usually measures the likelihood of a codeword estimate to the hard-decision vector, the Hamming distance can also represent the likelihood. {\color{black}Because $\mathrm{Pe}(\mathbf{e})$ in (\ref{equ::HDtech::CondDis::HDpsuc}) involves large-number integrals, we adopted a numerical calculation with limited precision to keep the overall complexity affordable, which introduced the discrepancies between the simulation curves and the analytical curves shown in Fig. \ref{Fig::VI::BCH128-HD-Psuc}.} 
        
        Similarly, instead of calculating the success probability for each TEP, after the $i$-reprocessing $(0\leq i \leq m)$ of an order-$m$ OSD, we can obtain the minimum Hamming distance as $d_{i}^{(\mathrm{H})}$ and the locally best codeword estimate $\widetilde{\mathbf{c}}_{i}$. The \textit{a posterior} probability that the number of errors in MRB is less than or equal to $i$, i.e., $\mathrm{Pr}(w(\widetilde{\mathbf{e}}_{\mathrm{B}})\leq i | d_{i}^{(\mathrm{H})})$, can be evaluated. If $w(\widetilde{\mathbf{e}}_{\mathrm{B}})\leq i$, an order-$i$ OSD is capable of obtaining the correct decoding result. Thus, we refer to $\mathrm{Pr}(w(\widetilde{\mathbf{e}}_{\mathrm{B}})\leq i | d_{i}^{(\mathrm{H})})$ as the hard success probability $\mathrm{P}_{i}^{\mathrm{suc}}(d_{i}^{(\mathrm{H})})$ of $\widetilde{\mathbf{c}}_{i}$. This is summarized in the following Corollary.
        
        \begin{corollary} \label{cor::HDtech::CondDis::minHDpsuc}
            In an order-$m$ OSD of decoding a linear block code $\mathcal{C}(n,k)$, if the minimum Hamming distance between the codeword estimates and the hard-decision vector after $i$-reprocessing is given by $d_{i}^{(\mathrm{H})}$, the probability that the number of errors in MRB is less than or equal to $i$ is given by
            \begin{equation} \label{equ::HDtech::CondDis::minHDpsuc}
            \begin{split}
       			\mathrm{P}_{i}^{\mathrm{suc}}(d_{i}^{(\mathrm{H})}) &= 1-  \left(1 - \sum\limits_{u=0}^{i}p_{E_1^{k}}(u)\right)\\
       			&\cdot\frac{\sum\limits_{v=0}^{n-k}  p_{E_{k+1}^{n}}(v)  p_{\widetilde W_{\mathbf{c}_{\mathrm{P}}}}(d_{i}^{(\mathrm{H})}-i,b_{0:i}^{k} | i^{(>)},v) 	}{p_{D_i^{(\mathrm{H})}}(d_{i}^{(\mathrm{H})})}     
            \end{split}
		    \end{equation}
		    where $p_{D_{i}^{(\mathrm{H})}}(d)$ is given by (\ref{equ::HDdis::iphase}) and $p_{\widetilde W_{\mathbf{c}_{\mathrm{P}}}}(j-i,b_{0:i}^{k} | i^{(>)},v) $ is given by (\ref{equ::HDdis::iphase::Wcp}).
        \end{corollary}
        \begin{IEEEproof}
            Following the same steps as the proof of Corollary \ref{cor::HDtech::CondDis::HDpsuc} and using Theorem \ref{the::HDdis::iphase}, we can obtain (\ref{equ::HDtech::CondDis::minHDpsuc}).
        \end{IEEEproof}
      
\begin{figure}
    		\begin{center}
    			\includegraphics[scale=0.60] {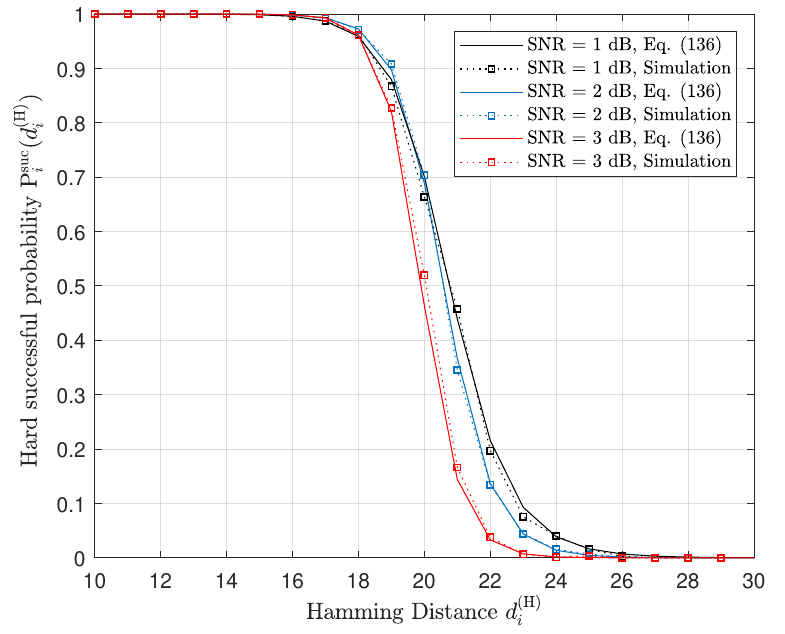}
    			\caption{{\color{black}$\mathrm{P}_{i}^{\mathrm{suc}}(d_{i}^{(\mathrm{H})})$ in decoding $(128,64,22)$ eBCH code at different SNR, when $i=1$.}}
    			\label{Fig::VI::BCH128-minHD-Psuc}
    		\end{center}
    	\end{figure}

     We compare (\ref{equ::HDtech::CondDis::minHDpsuc}) with simulations in decoding the $(128,64,22)$ eBCH code at various SNRs in Fig. \ref{Fig::VI::BCH128-minHD-Psuc}. As can be seen, the Hamming distance after $i$-reprocessing can be an indicator of the decoding quality. Furthermore, the hard success probability of codeword $\widetilde{\mathbf{c}}_i$ tends to 1 if the Hamming distance $d_i^{(\mathrm{H})}$ goes to 0.

\subsection{Stopping Rules} \label{sec::HDdistech::SR}
    In (\ref{equ::HDtech::CondDis::HDpsuc}) and (\ref{equ::HDtech::CondDis::minHDpsuc}), we have shown that the Hamming distances can be used to determine the \textit{a posterior} probability that the MRB errors can be eliminated. This section develops the decoding SR based on (\ref{equ::HDtech::CondDis::HDpsuc}) and (\ref{equ::HDtech::CondDis::minHDpsuc}), attempting to reduce the decoding complexity of OSD.
    
    Let us assume that at the receiver, a sequence of the samples of $[\widetilde{A}]_1^n$ is given by $\widetilde{\bm\alpha} = [\widetilde\alpha]_1^n$, i.e., the receiver receives a signal sequence $\mathbf{r}$ with reliabilities $\widetilde{\bm\alpha}$. Thus, conditioning on $\widetilde{A}_u = \widetilde\alpha_u$, the error probability of the $u$-th ($1\leq u \leq n$) bit of $\widetilde{\mathbf{y}}$ can be obtained as
        \begin{equation} \label{equ::HDdistech::Pebit::Cond}
            \mathrm{Pe}(u|\widetilde{A}_u = \widetilde\alpha_u) = \frac{f_R(-\widetilde\alpha_u)}{f_R(-\widetilde\alpha_u)+f_R(\widetilde\alpha_u)},
        \end{equation}
        where $f_R(x)$ is given by Eq. (\ref{equ::OrderStat::pdfofR}). For simplicity, we denote $\mathrm{Pe}(u|\widetilde{A}_u = \widetilde\alpha_u)$ as $\mathrm{Pe}(u|\widetilde\alpha_u)$. Then, the joint error probability of $u$-th and $v$-th ($1\leq u<v \leq n$) bits can be derived as 
        \begin{align}  \label{equ::HDdistech::Pebit2::Cond}
            \mathrm{Pe}(u,v|\widetilde\alpha_u,\widetilde\alpha_v) &=\frac{f_R(-\widetilde\alpha_u)}{f_R(-\widetilde\alpha_u)+f_R(\widetilde\alpha_u)}\cdot\frac{f_R(-\widetilde\alpha_v)}{f_R(-\widetilde\alpha_v)+f_R(\widetilde\alpha_v)} \notag\\
            &= \mathrm{Pe}(u|\widetilde\alpha_u)\mathrm{Pe}(v|\widetilde\alpha_v). 
        \end{align}
    From (\ref{equ::HDdistech::Pebit2::Cond}), we can see that although the bit-wise error probabilities of ordered received symbols are dependent as shown in (\ref{equ::OrderStat::jointpdfRij}), the conditional error probabilities are independent and $\mathrm{Pe}(u,v|\widetilde\alpha_u,\widetilde\alpha_v) = \mathrm{Pe}(u|\widetilde\alpha_u)\mathrm{Pe}(v|\widetilde\alpha_v)$ holds. Next, we introduce the SR design based on the reliabilities $\widetilde{\bm\alpha}$, which is obtained from the channel as \textit{a priori} information.

    \subsubsection{Hard Individual Stopping Rule (HISR)}   \label{sec::HDdistech::SR::HISR}
    Given the ordered reliabilities of received symbols, i.e., $\widetilde{\bm\alpha} = [\widetilde\alpha]_1^n$, the conditional correct probability $\mathrm{Pe}(\mathbf{e}|\widetilde{\bm\alpha})$ of TEP $\mathbf{e}$ can be simply derived as 
		\begin{equation} \label{equ::HDtech::SR::Pe(e)::Cond}
		    \mathrm{Pe}(\mathbf{e}|\widetilde{\bm\alpha}) = \prod_{\substack{0 < u \leq k\\e_u \neq 0}}  \mathrm{Pe}(u|\widetilde\alpha_u) \prod_{\substack{0 < u \leq k\\e_u = 0}} (1- \mathrm{Pe}(u|\widetilde\alpha_u)).
		\end{equation}
		We can also estimate conditional $\mathrm{pmf}$ of $E_{a}^{b}$, denoted by $p_{E_{a}^{b}}(j|\widetilde{\bm\alpha})$ (i.e., the number of errors over $[\widetilde{y}]_a^b$), as
 		\begin{equation}  \label{equ::HDtech::SR::Eab::Cond}
 		\begin{split}
             p_{E_{a}^{b}}(j|\widetilde{\bm\alpha})& = \binom{b-a+1}{j}
		     \left(\frac{1}{b-a+1}\sum_{u=a}^{b}\mathrm{Pe}(u|\widetilde\alpha_u)\right)^j \\
		     &\cdot \left(1 - \frac{1}{b-a+1}\sum_{u=a}^{b}\mathrm{Pe}(u|\widetilde\alpha_u)\right)^{b-a+1-j} .    
 		\end{split}
		\end{equation}   
		Accordingly, when $[\widetilde{A}]_1^n = [\widetilde\alpha]_1^n$, the hard success probability $	\mathrm{P}_{\mathbf{e}}^{\mathrm{suc}}(d_{\mathbf{e}}^{(\mathrm{H})}|\widetilde{\bm\alpha})$ can be simplt obtained as
		\begin{equation}  \label{equ::HDtech::SR::HDPsuc::Cond}
			\mathrm{P}_{\mathbf{e}}^{\mathrm{suc}}(d_{\mathbf{e}}^{(\mathrm{H})}|\widetilde{\bm\alpha})= \mathrm{Pe}(\mathbf{e}|\widetilde{\bm\alpha})\frac{p_{E_{k+1}^{n}}(d_{\mathbf{e}}^{(\mathrm{H})}-w(\mathbf{e})|\widetilde{\bm\alpha})}{p_{D_{\mathbf{e}}^{(\mathrm{H})}}(d_{\mathbf{e}}^{(\mathrm{H})}-w(\mathbf{e})|\widetilde{\bm\alpha})},
	    \end{equation}
	    where $p_{D_{\mathbf{e}}^{(\mathrm{H})}}(j|\widetilde{\bm\alpha})$ is given by (\ref{equ::HDtech::CondDis::HDforTEPe}), but in which $\mathrm{Pe}(\mathbf{e})$ is replaced by $\mathrm{Pe}(\mathbf{e}|\widetilde{\bm\alpha})$, and $p_{E_{1}^{k}}(j)$ and $p_{E_{k+1}^{n}}(j)$ are replaced with $p_{E_{1}^{k}}(j|\widetilde{\bm\alpha})$ and $p_{E_{k+1}^{n}}(j|\widetilde{\bm\alpha})$, respectively. {\color{black}Despite the complicated form, in Section {\ref{sec::Discussion::Implementation}}, we will show that {(\ref{equ::HDtech::SR::Eab::Cond})} can be computed with $O(n)$ floating-pointing operations (FLOPs) when $\mathcal{C}(n,k)$ has the binomial-like weight spectrum.}
	    
        We now introduce the hard individual stopping rule (HISR). Given a predetermined threshold success probability $\mathrm{P}_t^{\mathrm{suc}} \in [0,1]$, if the Hamming distance $d_{\mathbf{e}}^{(\mathrm{H})}$ between $\widetilde{\mathbf{c}}_{\mathbf{e}}$ and $\widetilde{\mathbf y} $ satisfies the following condition 
        \begin{equation} \label{equ::HDtech::SR::HISR}
            \mathrm{P}_{\mathbf{e}}^{\mathrm{suc}}(d_{\mathbf{e}}^{(\mathrm{H})}|\bm{\widetilde\alpha}) \geq \mathrm{P}_t^{\mathrm{suc}},
        \end{equation}
        the codeword $\hat{\mathbf{c}}_{\mathbf{e}} = \pi_1^{-1}(\pi_2^{-1}(\widetilde{\mathbf{c}}_{\mathbf{e}}))$ is selected as the decoding output, and the decoding is terminated. Therefore, the probability that errors in MRB are eliminated is lower bounded by $\mathrm{P}_t^{\mathrm{suc}}$ because of (\ref{equ::HDtech::SR::HISR}).

        Next, we give the performance bound and complexity analysis for an order-$m$ OSD decoding that only applies the HISR technique, attempting to characterize the complexity improvements and error rate performance loss introduced by the HISR.
        For an arbitrary TEP $\mathbf{e}$, there exists a maximum $d_{\mathbf{e}}^{(\mathrm{H})}$, referred to as $d_{\max,\mathbf{e}}^{(\mathrm{H})}$, satisfying  $\mathrm{P}_{\mathbf{e}}^{\mathrm{suc}}(d_{\mathbf{e}}^{(\mathrm{H})}|\bm{\widetilde\alpha}) \geq \mathrm{P}_t^{\mathrm{suc}}$, i.e., $d_{\max,\mathbf{e}}^{(\mathrm{H})} = \max\{d_{\mathbf{e}}^{(\mathrm{H})} \, |\, \mathrm{P}_{\mathbf{e}}^{\mathrm{suc}}(d_{\mathbf{e}}^{(\mathrm{H})}|\bm{\widetilde\alpha}) \geq \mathrm{P}_t^{\mathrm{suc}}\}$. It can be seen that $d_{\max,\mathbf{e}}^{(\mathrm{H})}$ depends on the values of reliabilities $\widetilde{\bm{\alpha}}$. Thus, we define $d_{b,\mathbf{e}}^{(\mathrm{H})}$ as the mean of $d_{\max,\mathbf{e}}^{(\mathrm{H})}$ with respect to $\widetilde{\bm{\alpha}}$, i.e., $d_{b,\mathbf{e}}^{(\mathrm{H})} = \mathbb{E}[d_{\max,\mathbf{e}}^{(\mathrm{H})}]$. Because $\widetilde{\bm{\alpha}}$ is a random vector with dependent distributions, $d_{b,\mathbf{e}}^{(\mathrm{H})}$ can be hardly determined. Thus, we give an approximation of $d_{b,\mathbf{e}}^{(\mathrm{H})}$ using $\mathrm{P}_{\mathbf{e}}^{\mathrm{suc}}(d_{\mathbf{e}}^{(\mathrm{H})})$ to enable the subsequent analysis . Let $\mathrm{P}_{\mathbf{e}}^{\mathrm{suc},-1}(x)$ and $\mathrm{P}_{\mathbf{e}}^{\mathrm{suc},-1}(x|\bm{\widetilde\alpha})$ denote the inverse functions of (\ref{equ::HDtech::CondDis::HDpsuc}) and (\ref{equ::HDtech::SR::HDPsuc::Cond}), respectively. It can be seen that $\mathrm{P}_{\mathbf{e}}^{\mathrm{suc}}(x|\bm{\widetilde\alpha})$ is a decreasing function and accordingly $\mathrm{P}_{\mathbf{e}}^{\mathrm{suc},-1}(x|\bm{\widetilde\alpha})$ is a decreasing function. In addition, $\mathrm{P}_{\mathbf{e}}^{\mathrm{suc},-1}(x)$ is also a 
       decreasing function. For the sake of brevity, we omit the proof of the monotonicity of $\mathrm{P}_{\mathbf{e}}^{\mathrm{suc},-1}(x)$ and $\mathrm{P}_{\mathbf{e}}^{\mathrm{suc},-1}(x|\bm{\widetilde\alpha})$, which can also be observed in Fig. \ref{Fig::VI::BCH128-HD-Psuc}. Note that $\mathrm{P}_{\mathbf{e}}^{\mathrm{suc},-1}(x)$ and $\mathrm{P}_{\mathbf{e}}^{\mathrm{suc},-1}(x|\bm{\widetilde\alpha})$ are discrete functions, i.e., $x$ cannot be a continuous real number, and it is possible that $\mathrm{P}_t^{\mathrm{suc}}$ is not in the domains of $\mathrm{P}_{\mathbf{e}}^{\mathrm{suc},-1}(x)$ and $\mathrm{P}_{\mathbf{e}}^{\mathrm{suc},-1}(x|\bm{\widetilde\alpha})$. In this regard, let us define $\mathrm{P}_{t'}^{\mathrm{suc}}$ as $\mathrm{P}_{t'}^{\mathrm{suc}}=\min\{x|x\geq \mathrm{P}_{t}^{\mathrm{suc}}, x \text{ is in the domain of } \mathrm{P}_{\mathbf{e}}^{\mathrm{suc},-1}(x) \}$ and define $\mathrm{P}_{t'}^{\mathrm{suc}}(\widetilde{\bm{\alpha}})$ as $\mathrm{P}_{t'}^{\mathrm{suc}}(\widetilde{\bm{\alpha}})\!=\!\min\{x|x \!\geq\! \mathrm{P}_{t}^{\mathrm{suc}}, x \text{ is in the domain of } \mathrm{P}_{\!\mathbf{e}}^{\mathrm{suc},-1}(x|\widetilde{\bm{\alpha}}) \}$. Based on these definitions, we can notice that 
        \begin{equation} \label{equ::HDtech::SR::HISR::db::equiv}
        \begin{split}
            d_{\max,\mathbf{e}}^{(\mathrm{H})} &= \max\{d_{\mathbf{e}}^{(\mathrm{H})} \, |\, \mathrm{P}_{\mathbf{e}}^{\mathrm{suc}}(d_{\mathbf{e}}^{(\mathrm{H})}|\bm{\widetilde\alpha}) \geq \mathrm{P}_t^{\mathrm{suc}}\}\\
            &=  \mathrm{P}_{\mathbf{e}}^{\mathrm{suc},-1}(\mathrm{P}_{t'}^{\mathrm{suc}}(\widetilde{\bm{\alpha}})|\bm{\widetilde\alpha}) 
        \end{split}
        \end{equation}
        and the difference between $\mathrm{P}_{t'}^{\mathrm{suc}}(\widetilde{\bm{\alpha}})$ and $\mathrm{P}_{t'}^{\mathrm{suc}}$ is upper bounded by
        \begin{equation} \label{equ::HDtech::SR::HISR::db::bound}
        \begin{split}
            & |\mathrm{P}_{t'}^{\mathrm{suc}}- \mathrm{P}_{t'}^{\mathrm{suc}}(\widetilde{\bm{\alpha}})| \\
            &\leq \max\left\{\left|\Delta_{1}\mathrm{P}_{\mathbf{e}}^{\mathrm{suc}}(\mathrm{P}_{t'}^{\mathrm{suc}})\right|,\left|\Delta_{1}\mathrm{P}_{\mathbf{e}}^{\mathrm{suc}}(\mathrm{P}_{t'}^{\mathrm{suc}}(\widetilde{\bm{\alpha}})|\widetilde{\bm{\alpha}})\right|\right\},            
        \end{split}
        \end{equation}
        where $\Delta_{1}\mathrm{P}_{\mathbf{e}}^{\mathrm{suc}}(j) = \mathrm{P}_{\mathbf{e}}^{\mathrm{suc}}(j+1) - \mathrm{P}_{\mathbf{e}}^{\mathrm{suc}}(j)$. Therefore, for $\mathrm{P}_{t}^{\mathrm{suc}}$ close to 0 or 1 (recall Fig. \ref{Fig::VI::BCH128-HD-Psuc}), we simply take $\mathrm{P}_{t'}^{\mathrm{suc}} \approx \mathrm{P}_{t'}^{\mathrm{suc}}(\widetilde{\bm{\alpha}})$. Then, $d_{b,\mathbf{e}}^{(\mathrm{H})}$ can be approximated as
        \begin{equation} \label{equ::HDtech::SR::HISR::db}
            \begin{split}
            d_{b,\mathbf{e}}^{(\mathrm{H})} & = 
            \mathbb{E}\left[\mathrm{P}_{\mathbf{e}}^{\mathrm{suc},-1}(\mathrm{P}_{t'}^{\mathrm{suc}}(\widetilde{\bm{\alpha}})|\bm{\widetilde\alpha})\right] \\
            & \approx \mathbb{E}\left[\mathrm{P}_{\mathbf{e}}^{\mathrm{suc},-1}(\mathrm{P}_{t'}^{\mathrm{suc}}|\bm{\widetilde\alpha}) \right]\\
            & \overset{(a)}{=}  \mathrm{P}_{\mathbf{e}}^{\mathrm{suc},-1}(\mathrm{P}_{t'}^{\mathrm{suc}}) \\
            & \overset{(b)} {=}  \max\{d_{\mathbf{e}}^{(\mathrm{H})} \, |\, \mathrm{P}_{\mathbf{e}}^{\mathrm{suc}}(d_{\mathbf{e}}^{(\mathrm{H})}|\bm{\widetilde\alpha}) \geq \mathrm{P}_t^{\mathrm{suc}}\}.
            \end{split}
        \end{equation}
        Step (a) of (\ref{equ::HDtech::SR::HISR::db}) follows from that $\mathrm{P}_{\mathbf{e}}^{\mathrm{suc},-1}(x) = \mathbb{E}[\mathrm{P}_{\mathbf{e}}^{\mathrm{suc},-1}(x|\bm{\widetilde\alpha})]$, and step (b) applies the equivalence (\ref{equ::HDtech::SR::HISR::db::equiv}) over $\mathrm{P}_{\mathbf{e}}^{\mathrm{suc},-1}(x)$.

        let $\overline{\mathrm{P}}_{\mathbf{e}}^{\mathrm{suc}}$ denote the expectation of the hard success probability of $\widetilde{\mathbf{c}}_{\mathbf{e}}$ with respect to $D_{\mathbf{e}}^{(\mathrm{H})}\!\leq \! d_{b,\mathbf{e}}^{(\mathrm{H})}$, i.e., $\overline{\mathrm{P}}_{\mathbf{e}}^{\mathrm{suc}} = \mathrm{Pr}(\mathbf{e} = \widetilde{\mathbf{e}}_{\mathrm{B}}|D_{\mathbf{e}}^{(\mathrm{H})}\!\leq \! d_{b,\mathbf{e}}^{(\mathrm{H})})$. Thus, if $\widetilde{\mathbf{c}}_{\mathbf{e}}$ satisfies the HISR, $\overline{\mathrm{P}}_{\mathbf{e}}^{\mathrm{suc}}$ is derived as
        \begin{equation} \label{equ::HDtech::CondDis::HISR::avePsuc}
            \overline{\mathrm{P}}_{\mathbf{e}}^{\mathrm{suc}} =
            \left(\sum_{j=w(\mathbf{e})}^{d_{b,\mathbf{e}}^{(\mathrm{H})}}\mathrm{P}_{\mathbf{e}}^{\mathrm{suc}}(j) p_{D_{\mathbf{e}}^{(\mathrm{H})}}(j)\right)\left(\sum_{j = w(\mathbf{e})}^{d_{b,\mathbf{e}}^{(\mathrm{H})}} p_{D_{\mathbf{e}}^{(\mathrm{H})}}(j)\right)^{-1}.
        \end{equation}
        
        On the other hand, given a specific reprocessing sequence $\{\mathbf{e}_1,\mathbf{e}_2,\ldots,\mathbf{e}_{b_{0:m}^{k}}\}$ (i.e., the decoder processes TEPs sequentially from $\mathbf{e}_1$ to $\mathbf{e}_{b_{0:m}^{k}}$), for any $j$, $1 < j \leq b_{0:m}^k$, the probability that $\hat{\mathbf{c}}_{\mathbf{e}_j} = \pi_1^{-1}(\pi_2^{-1}(\widetilde{\mathbf{c}}_{\mathbf{e}_j}))$ is identified and output by the HISR is given by
        \begin{equation} \label{equ::HDtech::SR::HISR::PsucEffect}
            \mathrm{P}_{\mathbf{e}_j} = \left(\sum_{u = w(\mathbf{e}_j)}^{d_{b,\mathbf{e}_j}^{(\mathrm{H})}} p_{D_{\mathbf{e}_j}^{(\mathrm{H})}}(u)\right)\prod_{v=1}^{j-1}\left(1-\sum_{u = w(\mathbf{e}_v)}^{d_{b,\mathbf{e}_v}^{(\mathrm{H})}} p_{D_{\mathbf{e}_v}^{(\mathrm{H})}}(u) \right) .
        \end{equation}
        Particularly, $\mathrm{P}_{\mathbf{e}_1} = \sum_{u = w(\mathbf{e}_1)}^{d_{b,\mathbf{e}_1}^{(\mathrm{H})}} p_{D_{\mathbf{e}_1}^{(\mathrm{H})}}(u)$. 
        
        {\color{black}Generally, the overall decoding error probability of an original OSD can be upper bounded by \cite{Fossorier1995OSD}
        \begin{equation} \label{equ::OSD_errorRate}
            \epsilon_e  \leq \mathrm{P}_{\mathrm{list}} + \mathrm{P}_{\mathrm{ML}} ,
        \end{equation}      
        where $\mathrm{P}_{\mathrm{ML}}$ is the error rate of maximum-likelihood decoding (MLD), and $\mathrm{P}_{\mathrm{list}}$ is the probability that the error pattern $\widetilde{\mathbf{e}}_{\mathrm{B}}$ (recall the definition of $\widetilde{\mathbf{e}}_{\mathrm{B}}$ in the proof of Lemma \ref{lem::HDdis::0phase}) is excluded in the list of TEPs of OSD, i.e., the probability that OSD does not eliminate the errors in MRB, which can be derived as  $\mathrm{P}_{\mathrm{list}} = 1-\sum_{i=0}^{m} p_{E_1^k}(i)$. For an order-$m$ OSD employing the HISR with the threshold success probability $\mathrm{P}_t^{\mathrm{suc}}$, the error rate upper bounded as
        \begin{equation} \label{equ::HDtech::CondDis::HISR::errorRate}
            \epsilon_e^{\mathrm{HISR}}  \leq \mathrm{P}_{\mathrm{list}} + \mathrm{P}_{\mathrm{HISR}} + \mathrm{P}_{\mathrm{ML}} ,
        \end{equation}  
        where $\mathrm{P}_{\mathrm{HISR}}$ is the probability that the HISR outputs an incorrect codeword estimate, which introduces performance degradation in $\epsilon_e^{\mathrm{HISR}}$ compared to $\epsilon_e$. Considering the probabilities given by (\ref{equ::HDtech::CondDis::HISR::avePsuc}) and (\ref{equ::HDtech::SR::HISR::PsucEffect}), $\mathrm{P}_{\mathrm{HISR}}$ can be derived as
        \begin{equation} \label{equ::HDtech::CondDis::HISR::PHISR}
            \mathrm{P}_{\mathrm{HISR}} = \sum_{j=1}^{b_{0:m}^{k}}\mathrm{P}_{\mathbf{e}_j} \left( 1- \overline{\mathrm{P}}_{\mathbf{e}_j}^{\mathrm{suc}}\right) .
        \end{equation}  
        Then, if the second permutation $\pi_2$ is omitted, by substituting (\ref{equ::HDtech::CondDis::HISR::PHISR}) into (\ref{equ::HDtech::CondDis::HISR::errorRate}), we can finally obtain the error rate upper bound of an order-$m$ OSD applying the HISR, i.e.,
        \begin{equation} \label{equ::HDtech::SR::HISR::errorRate2}
        \begin{split}
               \epsilon_e^{\mathrm{HISR}}  &\leq 1 - \sum_{j=1}^{m} p_{E_1^k}(j) + \sum_{j=1}^{b_{0:m}^{k}}\mathrm{P}_{\mathbf{e}_j} \left( 1- \overline{\mathrm{P}}_{\mathbf{e}_j}^{\mathrm{suc}}\right) + \mathrm{P}_{\mathrm{ML}} \\
               & = 1 - (1- \theta_{\mathrm{HISR}})\sum_{j=0}^{m}p_{E_1^{k}}(j) + \mathrm{P}_{\mathrm{ML}}
        \end{split}
        \end{equation}
        where $\theta_{\mathrm{HISR}}$ is defined as the error performance loss factor of the HISR, which is given by
        \begin{equation} \label{equ::HDtech::SR::HISR::loss}
            \theta_{\mathrm{HISR}} = \frac{\sum_{j=1}^{b_{0:m}^{k}}\mathrm{P}_{\mathbf{e}_j} \left( 1- \overline{\mathrm{P}}_{\mathbf{e}_j}^{\mathrm{suc}}\right)}{\sum_{j=0}^{m}p_{E_1^{k}}(j)}
        \end{equation}
        It can be noticed that the performance loss rate $\theta_{\mathrm{HISR}}$ is controlled by $\mathrm{P}_t^{\mathrm{suc}}$ and the value of $\theta_{\mathrm{HISR}}$ is bounded by
        \begin{equation}
             0   \leq  \theta_{\mathrm{HISR}} \leq \frac{1-p_{E_1^{k}}(0)}{\sum_{j=0}^{m}p_{E_1^{k}}(j)}. 
        \end{equation}
        We elaborate on the impact of $\mathrm{P}_{t}^{\mathrm{suc}}$ over the error rate as follows
        \begin{itemize}
            \item When $\mathrm{P}_t^{\mathrm{suc}}$ goes to $1$, $\mathrm{P}_{\mathbf{e}_j}$ goes to 0 for any $j$, which implies that no TEP will satisfy the HISR. In this case, $\theta_{\mathrm{HISR}}$ goes to 0, and (\ref{equ::HDtech::SR::HISR::errorRate2}) tends to be the performance upper bound of the original OSD.
            
            \item When $\mathrm{P}_t^{\mathrm{suc}}$ goes to 0, $\mathrm{P}_{\mathbf{e}_1} = \sum_{u = w(\mathbf{e}_1)}^{d_{b,\mathbf{e}_1}^{(\mathrm{H})}} p_{D_{\mathbf{e}_1}^{(\mathrm{H})}}(u)$ goes to 1 as $d_{b,\mathbf{e}_1}^{(\mathrm{H})}$ tends to be as large as $n$, which implies that the decoder will only process the first TEP (i.e., 0-reprocessing). When $d_{b,\mathbf{e}_1}^{(\mathrm{H})}$ goes to $n$, $\overline{\mathrm{P}}_{\mathbf{e}_1}^{\mathrm{suc}}$ given in (\ref{equ::HDtech::CondDis::HISR::avePsuc}) tends to be $\overline{\mathrm{P}}_{\mathbf{e}_1}^{\mathrm{suc}} = \mathrm{Pe}(\mathbf{e}_1) = p_{E_1^{k}}(0)$. In this case, we can observe that $\theta_{\mathrm{HISR}} = \frac{1 - p_{E_1^{k}}(0)}{\sum_{j=0}^{m}p_{E_1^{k}}(j)}$ and $\epsilon_e^{\mathrm{HISR}} = 1 - p_{E_1^{k}}(0) + \sum_{m+1}^{k} p_{E_1^{k}}(0) + \mathrm{P}_{\mathrm{ML}} $, which upper bounds the error rate of the 0-reprocessing OSD.
        \end{itemize}
        }
        
\begin{figure}
    		\begin{center}
    			\includegraphics[scale=0.6] {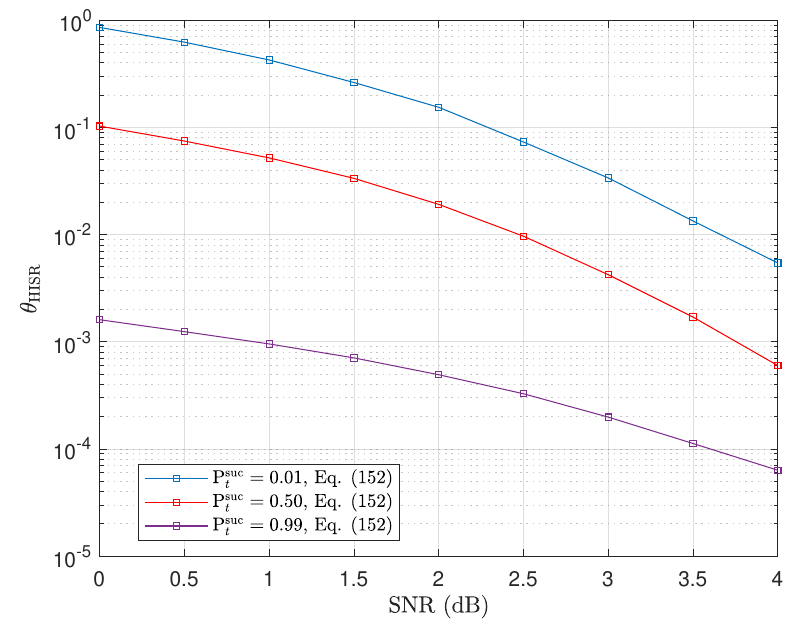}
    			\caption{The performance loss factor $\theta_{\mathrm{HISR}}$ of decoding $(64,30,14)$ eBCH code with an order-$1$ OSD applying the HISR.}
    			\label{Fig::VI::theta_HISR}
    		\end{center}
    	\end{figure}

        We illustrate the performance loss factor $\theta_{\mathrm{HISR}}$ with different values of $\mathrm{P}_t^{\mathrm{suc}}$ in the order-$1$ decoding of $(64,30,14)$ eBCH code in Fig. \ref{Fig::VI::theta_HISR}. It is worth mentioning that even for small $\mathrm{P}_t^{\mathrm{suc}}$ (e.g., 0.1 or 0.5), the loss $\theta_{\mathrm{HISR}}$ tends to be decreased significantly as SNR increases. For $\mathrm{P}_t^{\mathrm{suc}} = 0.99$, it can be seen that only less than 0.1\% of error correction probability is lost (recall that $1-\theta_{\mathrm{HISR}}$ is the coefficient of $\sum_{j=0}^{m}p_{E_1^{k}}(j)$ in (\ref{equ::HDtech::SR::HISR::errorRate2})).

        {\color{black}
        Regarding the decoding complexity, given a specific reprocessing sequence $\{\mathbf{e}_1,\mathbf{e}_2,\ldots,\mathbf{e}_{b_{0:m}^{k}}\}$ and considering the probability given by (\ref{equ::HDtech::SR::HISR::PsucEffect}), the average number of re-encoded TEPs, denoted by $N_a$, can be derived as
        \begin{equation} \label{equ::HDtech::SR::HISR::Na}
            \begin{split}
                N_a &=  \sum_{j=1}^{b_{0:m}^{k}}j\cdot \mathrm{P}_{\mathbf{e}_j} +b_{0:m}^{k} \left(1-\sum_{j=1}^{b_{0:m}^{k}}\mathrm{P}_{\mathbf{e}_j}\right).
            \end{split}
        \end{equation} 
        It can be seen that when $\mathrm{P}_t^{\mathrm{suc}}$ goes to $1$, $N_a$ goes to $b_{0:m}^{k}$, which is the number of TEPs required in the original OSD. In contrast, when $\mathrm{P}_t^{\mathrm{suc}}$ goes to 0, $N_a$ goes to 1 as $\mathrm{P}_{\mathbf{e}_1}$ goes to 1, which indicates that only one TEP is re-encoded. }
        
        Compared to the conventional approaches of maximum-likelihood decoding or OSD decoding, the HISR finds the decoding output by calculating the Hamming distance rather than comparing the WHD for every re-encoding products. Furthermore, the HISR can find the promising decoding result during the reprocessing and terminate the decoding without traversing all the TEP. This reduces the decoding complexity. Note that $\{\mathbf{e}_1,\mathbf{e}_2,\ldots,\mathbf{e}_{b_{0:m}^{k}}\}$ is non-exchangeable in (\ref{equ::HDtech::SR::HISR::errorRate2}) and (\ref{equ::HDtech::SR::HISR::Na}) as different reprocessing sequences may result in different decoding complexity and loss rate. According to (\ref{equ::HDtech::SR::HISR::errorRate2}) and (\ref{equ::HDtech::SR::HISR::Na}), the best sequence solution should be always prioritizing TEP $\mathbf{e}_j$, $1\leq j \leq b_{0:m}^k$, with higher $\sum_{u = w(\mathbf{e}_j)}^{d_{b,\mathbf{e}_j}^{(\mathrm{H})}} p_{d_{\mathbf{e}_j}^{(\mathrm{H})}}(u)$.
            
\begin{figure}[t]
	    	\vspace{-0.8em}
            \centering
            \subfigure[Frame error rate]
            {
                \includegraphics[scale = 0.65]{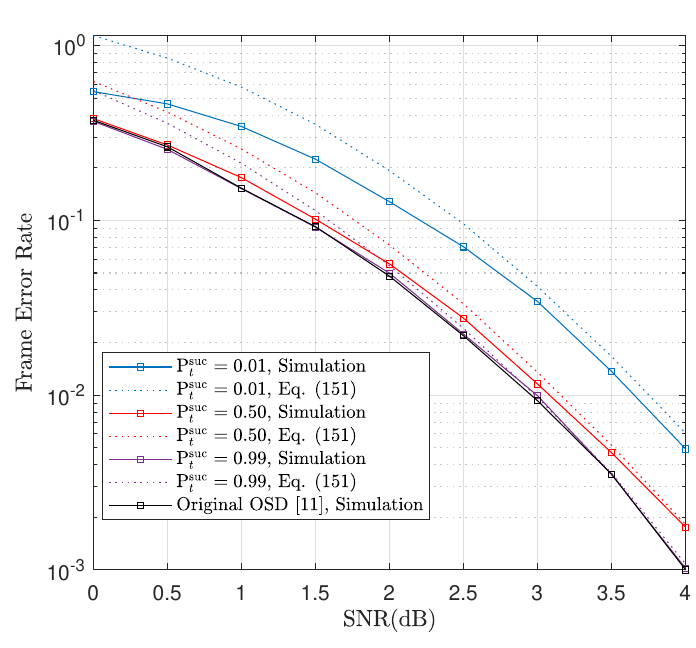}
                \label{Fig::VI::HISR_Pe}
            }
            \vspace{-1ex}
            \subfigure[Average number of TEPs]
            {
                \includegraphics[scale = 0.65]{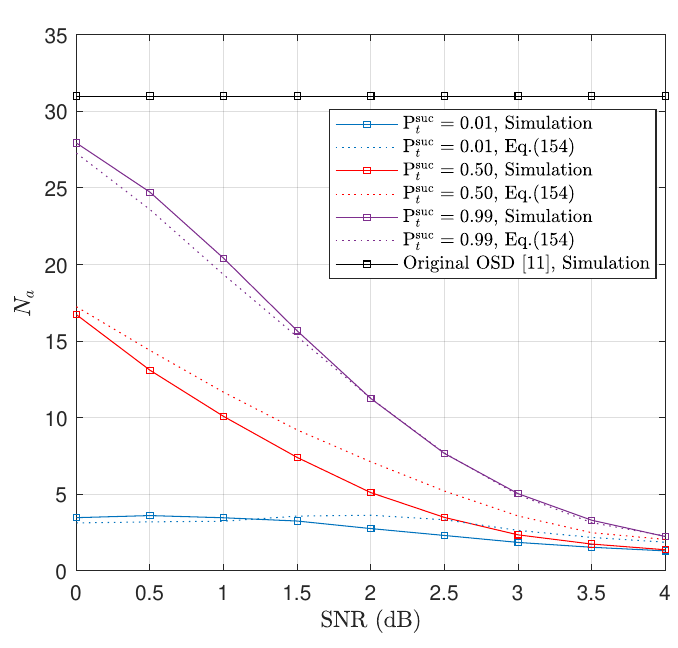}
                \label{Fig::VI::HISR_Na}
            }

            \caption{Decoding $(64,30,14)$ eBCH code with an order-$1$ OSD applying the HISR.}
            \label{Fig::HISR}
        \end{figure}

     We consider the implementation of an order-1 OSD algorithm applying the HISR. The decoding error performance and the average number of TEPs is compared with different threshold $\mathrm{P}_t^{\mathrm{suc}}$ settings in decoding $(64,30,14)$ eBCH code, as depicted in Fig. \ref{Fig::VI::HISR_Pe} and Fig. \ref{Fig::VI::HISR_Na}, respectively. As can be seen, HISR can be an effective stopping condition to reduce complexity, even if it is calculated based on the Hamming distance. In particular, with a high $\mathrm{P}_t^{\mathrm{suc}}$ (e.g., 0.99), the average number of TEPs $N_a$ is also significantly reduced at high SNRs. At the same time, the error performance is almost identical to the original OSD. {\color{black} It needs to be noted that the approximation in (\ref{equ::HDtech::SR::HISR::db}) introduces the discrepancies between (\ref{equ::HDtech::SR::HISR::Na}) and the simulations in Fig. \ref{Fig::VI::HISR_Na}. As explained in (\ref{equ::HDtech::SR::HISR::db::bound}), the approximation may lose tightness particularly for medium $\mathrm{P}_t^{\mathrm{suc}}$.}

    \subsubsection{Hard Group Stopping Rule} \label{sec::HDtech::SR::HGSR}
        Although the HISR can accurately evaluate the successful probabilities of TEPs, $\mathrm{P}_{\mathbf{e}}^{\mathrm{suc}}(d_{\mathbf{e}}^{(\mathrm{H})}|\widetilde{\bm\alpha})$ needs to be determined for each TEP individually and the reprocessing TEP sequence should also be carefully considered. We further propose a hard group stopping rule (HGSR) based on Theorem \ref{the::HDdis::iphase} and Corollary \ref{cor::HDtech::CondDis::minHDpsuc} as an alternative efficient implementation. Given the \textit{a prior} information $[\widetilde{A}]_1^n = [\widetilde{\alpha}]_1^n  = \widetilde{\bm\alpha}$, we can simplify (\ref{equ::HDtech::CondDis::minHDpsuc}) in Corollary \ref{cor::HDtech::CondDis::minHDpsuc} as
		\begin{equation} \label{equ::HDtech::SR::HGSR::Psuc}
		\begin{split}
			\mathrm{P}_{i}^{\mathrm{suc}}(d_{i}^{(\mathrm{H})}|\widetilde{\bm{\alpha}})&= 1- \left(1-\sum_{v=0}^{i}p_{E_1^{k}}(v|\widetilde{\bm\alpha}) \right)\\
			&\cdot\frac{\sum\limits_{v=0}^{n-k}  p_{E_{k+1}^{n}}(v|\widetilde{\bm\alpha}) p_{\widetilde W_{\mathbf{c}_{\mathrm{P}}}}(d_{i}^{(\mathrm{H})}\!\!-i,b_{0:i}^{k} | i^{(>)}\!,v,\widetilde{\bm{\alpha}}) 	}{p_{D_i^{(\mathrm{H})}}(d_{i}^{(\mathrm{H}}|\widetilde{\bm{\alpha}})},		    
		\end{split}
	    \end{equation}
	    where $p_{E_1^{k}}(v|\widetilde{\bm\alpha})$ and $p_{E_{k+1}^{n}}(v|\widetilde{\bm\alpha})$ are derived from $p_{E_a^{b}}(v|\widetilde{\bm{\alpha}})$ in (\ref{equ::HDtech::SR::Eab::Cond}). In (\ref{equ::HDtech::SR::HGSR::Psuc}), $p_{\widetilde W_{\mathbf{c}_{\mathrm{P}}}}(j-i,b_{0:i}^{k} | i^{(>)},v,\widetilde{\bm{\alpha}})$ and $p_{D_i^{(\mathrm{H})}}(d_{i}^{(\mathrm{H})}|\widetilde{\bm{\alpha}})$ are the conditional $\mathrm{pmf}$s obtained by replacing all $p_{E_a^{b}}(j)$ with $p_{E_a^{b}}(j|\widetilde{\bm{\alpha}})$ inside $p_{\widetilde W_{\mathbf{c}_{\mathrm{P}}}}(j-i,b_{0:i}^{k} | i^{(>)},v)$ and $p_{D_i^{(\mathrm{H})}}(d_{i}^{(\mathrm{H})})$, respectively, where $p_{D_{i}^{(\mathrm{H})}}(d)$ is given by (\ref{equ::HDdis::iphase}) and $p_{\widetilde W_{\mathbf{c}_{\mathrm{P}}}}(j-i,b_{0:i}^{k} | i^{(>)},v) $ is given by (\ref{equ::HDdis::iphase::Wcp}). {\color{black}Despite the complicated form of (\ref{equ::HDtech::SR::HGSR::Psuc}), in Section \ref{sec::Discussion::Implementation}, we further show that it can be implemented with $O(n^2)$ FLOPs if $C(n,k)$ has the binomial-like weight spectrum. }

	    Therefore, we can calculate the hard success probability $\mathrm{P}_{i}^{\mathrm{suc}}(d_{i}^{(\mathrm{H})}|\widetilde{\bm{\alpha}})$ according to (\ref{equ::HDtech::SR::HGSR::Psuc}) for the entire reprocessing stage, rather than calculating $\mathrm{P}_{\mathbf{e}}^{\mathrm{suc}}(d_{\mathbf{e}}^{(\mathrm{H})}|\widetilde{\bm{\alpha}})$ for each TEP $\mathbf{e}$ individually as in the HISR. All TEPs in the first $i$ phases of reprocessing can be regarded as a group and $\mathrm{P}_{i}^{\mathrm{suc}}(d_{i}^{(\mathrm{H})}|\widetilde{\bm{\alpha}})$ is calculated after each reprocessing. If $\mathrm{P}_{i}^{\mathrm{suc}}(d_{i}^{(\mathrm{H})}|\widetilde{\bm{\alpha}})$ is larger than a determined parameter, the decoder can be terminated. This approach is referred to as the HGSR.
        
        The HGSR is described as follows. Given a predetermined threshold success probability $\mathrm{P}_t^{\mathrm{suc}} \in [0,1]$, after the $i$-reprocessing ($0\leq i\leq m$) of an order-$m$ OSD, if the minimum Hamming distance $d_{i}^{(\mathrm{H})}$ satisfies the following condition
        \begin{equation} \label{equ::HDtech::SR::HGSR}
        	\mathrm{P}_{i}^{\mathrm{suc}}(d_{i}^{(\mathrm{H})}|\widetilde {\bm\alpha}) \geq \mathrm{P}_t^{\mathrm{suc}},
        \end{equation}
        the decoding is terminated and the codeword estimate found best so far, $\hat{\mathbf{c}}_{i} = \pi_1^{-1}(\pi_2^{-1}(\widetilde{\mathbf{c}_{i}}))$, is claimed as the decoding output. If $\hat{\mathbf{c}}_{i}$ is output, the probability that the errors in MRB are eliminated is lower bounded by $\mathrm{P}_t^{\mathrm{suc}}$ according to (\ref{equ::HDtech::SR::HGSR}). 
        
        Next, we consider an order-$m$ ($m \geq 1$) OSD decoding employing the HGSR with a given threshold success probability $\mathrm{P}_t^{\mathrm{suc}}$, and derive an upper bound on the error rate $\epsilon_{e}^{\mathrm{HGSR}}$ in a similar approach as described in Section \ref{sec::HDdistech::SR::HISR}. For the sake of brevity, we omit some details of the derivations in the analysis that follows in this section.
        
        For the $i$-reprocessing ($0\leq i\leq m$), there exists a maximum $d_{i}^{(\mathrm{H})}$, referred to as $d_{\max,i}^{(\mathrm{H})}$ , satisfying $\mathrm{P}_{i}^{\mathrm{suc}}(d_{i}^{(\mathrm{H})}|\widetilde {\bm\alpha}) \geq \mathrm{P}_t^{\mathrm{suc}}$, i.e., $d_{\max,i}^{(\mathrm{H})} = \max\{d_{i}^{(\mathrm{H})} \,|\,\mathrm{P}_{i}^{\mathrm{suc}}(d_{i}^{(\mathrm{H})}|\widetilde {\bm\alpha}) \geq \mathrm{P}_t^{\mathrm{suc}} \}$. We define $d_{b,i}$ as the mean of $d_{\max,i}^{(\mathrm{H})}$, which can be derived as
        \begin{equation}
            \begin{split}
                d_{b,i}^{(\mathrm{H})} \approx \max\{d_{i}^{(\mathrm{H})} \,|\,\mathrm{P}_{i}^{\mathrm{suc}}(d_{i}^{(\mathrm{H})}) \geq \mathrm{P}_t^{\mathrm{suc}} \},
            \end{split}
        \end{equation}
        where the approximation takes the same approach as (\ref{equ::HDtech::SR::HISR::db}). Then, the probability that $\widetilde{\mathbf{c}}_{i}$ ($1\leq i\leq m$) satisfies the HGSR can be derived as
        \begin{equation} \label{equ::HDtech::SR::HGSR::Pi}
            \mathrm{P}_i = \left( \sum_{u = 0}^{d_{b,i}^{(\mathrm{H})}}  p_{D_{i}^{(\mathrm{H})}}(u)\right)\prod_{v=0}^{i-1}\left(1-\sum_{u = 0}^{d_{b,v}^{(\mathrm{H})}} p_{D_{v}^{(\mathrm{H})}}(u) \right).
        \end{equation}
        Particularly, $\mathrm{P}_0 =  \sum_{u = 0}^{d_{b,0}^{(\mathrm{H})}}  p_{D_{0}^{(\mathrm{H})}}(u)$. Let $\overline{\mathrm{P}}_{i}^{\mathrm{suc}}$ denote the mean of the hard success probability of $i$-reprocessing conditioning on $D_{i}^{(\mathrm{H})} \leq d_{b,i}^{(\mathrm{H})}$, i.e., $\overline{\mathrm{P}}_{i}^{\mathrm{suc}} = \mathrm{Pr}(w(\widetilde{\mathbf{e}}_{\mathrm{B}})\leq i | D_{i}^{(\mathrm{H})} \leq d_{b,i}^{(\mathrm{H})})$, then $\overline{\mathrm{P}}_{i}^{\mathrm{suc}}$ can be derived as 
        \begin{equation}
            \overline{\mathrm{P}}_{i}^{\mathrm{suc}} = \frac{\sum_{u=0}^{d_{b,i}^{(\mathrm{H})}}\mathrm{P}_{i}^{\mathrm{suc}}(u) p_{D_{i}^{(\mathrm{H})}}(u)}{\sum_{u=0}^{d_{b,i}^{(\mathrm{H})}} p_{D_{i}^{(\mathrm{H})}}(u)}.
        \end{equation}
        
        {\color{black} Next, let us define $\mathrm{P}_{\mathrm{HGSR}}$ as the probability that the HGSR outputs an incorrect codeword estimate. Similar to obtaining (\ref{equ::HDtech::SR::HISR::errorRate2}), the error rate $\epsilon_{e}^{\mathrm{HGSR}}$ of an order-$m$ OSD applying the HGSR is upper bounded as
    	\begin{equation} \label{equ::HDtech::SR::HGSR::errorRate}
    	\begin{split}
    	     \epsilon_{e}^{\mathrm{HGSR}} &\leq  \mathrm{P}_{\mathrm{list}} + \mathrm{P}_{\mathrm{HGSR}} + \mathrm{P}_{\mathrm{ML}} \\
    	     & = 1 - \sum_{j=1}^{m} p_{E_1^k}(j) + \sum_{j=0}^{i}\mathrm{P}_{j} \left( 1- \overline{\mathrm{P}}_{j}^{\mathrm{suc}}\right) + \mathrm{P}_{\mathrm{ML}} \\
               & = 1 - (1- \theta_{\mathrm{HGSR}})\sum_{j=0}^{m}p_{E_1^{k}}(j) + \mathrm{P}_{\mathrm{ML}},
    	\end{split}
    	\end{equation}
    	 where $\theta_{\mathrm{HGSR}}$ is the error performance loss rate given by
        \begin{equation} \label{equ::HDtech::SR::HGSR_loss}
            \theta_{\mathrm{HGSR}} = \frac{\sum_{j=0}^{i}\mathrm{P}_{j} \left( 1- \overline{\mathrm{P}}_{j}^{\mathrm{suc}}\right)}{\sum_{j=0}^{m}p_{E_1^{k}}(j)}.
        \end{equation}
        }

\begin{figure}
    		\begin{center}
    			\includegraphics[scale=0.6] {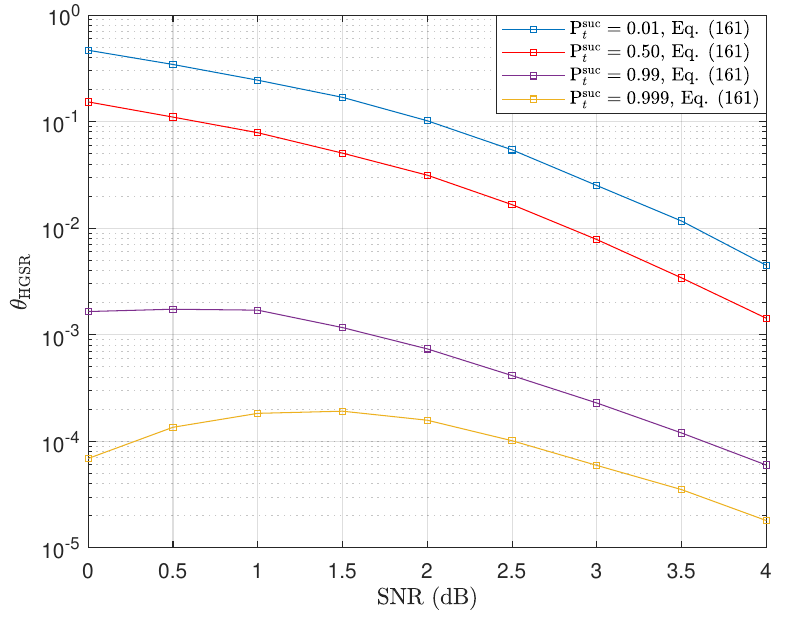}
    			\caption{The performance loss rate $\theta_{\mathrm{HGSR}}$ of decoding $(64,30,14)$ eBCH code with an order-$2$ OSD applying the HGSR.}
    			\label{Fig::VI::theta_HGSR}
    		\end{center}
    	\end{figure}

         Similar to the HISR, when $\mathrm{P}_t^{\mathrm{suc}}$ goes to $1$, (\ref{equ::HDtech::SR::HGSR::errorRate}) tends to be the performance upper bound of the original OSD, i.e., $ \epsilon_{e}^{\mathrm{HGSR}} \leq 1 - \sum_{j=0}^{m}p_{E_1^{k}}(j) + \mathrm{P}_{\mathrm{ML}}$. In contrast, when $\mathrm{P}_t^{\mathrm{suc}}$ goes to 0, (\ref{equ::HDtech::SR::HGSR::errorRate}) goes to $\epsilon_{e}^{\mathrm{HGSR}} = \frac{1 - p_{E_1^{k}}(0)}{\sum_{j=0}^{m}p_{E_1^{k}}(j)}$, indicating that the OSD only performs the $0$-reprocessing. We illustrate the performance loss $\theta_{\mathrm{HGSR}}$ with different values of $\mathrm{P}_t^{\mathrm{suc}}$ in decoding a $(64,30.14)$ eBCH code with an order-$2$ OSD applying the HGSR, as depicted in Fig. \ref{Fig::VI::theta_HGSR}.
         
        {\color{black}
        Considering the probability $\overline{\mathrm{P}}_{i}^{\mathrm{suc}}$ given by (\ref{equ::HDtech::SR::HGSR::Pi}), the average number of TEPs $N_a$ can be derived as
        \begin{equation} \label{equ::HDtech::SR::HGSR::Na}
            \begin{split}
                N_a &=  b_{0:m}^{k}\left(1 - \sum_{j=0}^{m}\mathrm{P}_i\right)+ \sum_{j=0}^{m}b_{0:j}^{k}\cdot\mathrm{P}_j.
            \end{split}
        \end{equation}
        }
        
\begin{figure}[t]
	    	\vspace{-0.8em}
            \centering
            \subfigure[Frame error rate]
            {
                \includegraphics[scale = 0.65]{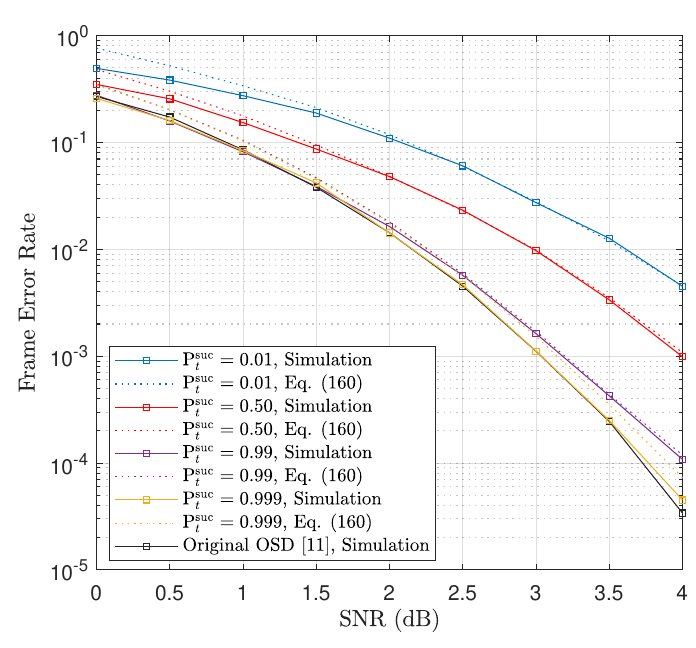}
                \label{Fig::VI::HGSR_Pe}
            }
            \vspace{-1ex}
            \subfigure[Average number of TEPs]
            {
                \includegraphics[scale = 0.65]{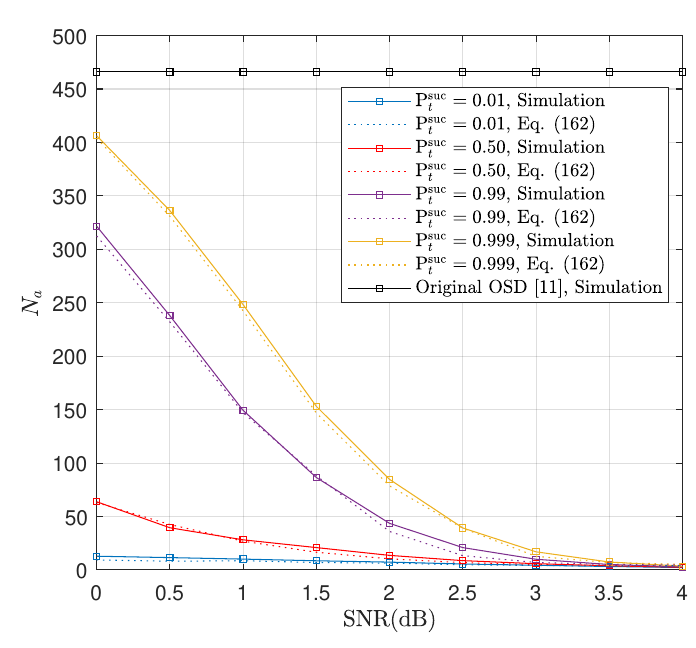}
                \label{Fig::VI::HGSR_Na}
            }

            \caption{Decoding $(64,30,14)$ eBCH code with an order-$2$ OSD applying the HGSR.}
            \label{Fig::HGSR}
        \end{figure}

         We consider the implementation of an order-$2$ OSD algorithm applying the HGSR. The decoding error performance and the average number of TEPs is compared in decoding $(64,30,14)$ eBCH code, as depicted in Fig. \ref{Fig::VI::HGSR_Pe} and Fig. \ref{Fig::VI::HGSR_Na}, respectively. From the simulation, it can be seen that HGSR is also effective in reducing complexity. Compared to the HISR, HGSR does not need to consider the sequence order of TEPs, and it only calculates the hard success probability after each round of reprocessing, thus is more suitable for high-order OSD implementations. %With different performance and complexity requirements, different selection of $\mathrm{P}_{t}^{\mathrm{suc}}$ guarantees the trade-off.}

\subsection{Discarding Rules}\label{sec::HDtech::DR}
    Although OSD looks for the best codeword by finding the minimum WHD, if a codeword estimate $\hat{\mathbf{c}}_{\mathbf{e}}$ could provide a better estimation, its Hamming distance $d_{\mathbf{e}}^{(\mathrm{H})}$ from $\mathbf y$ should be less than or around the minimum Hamming weight $d_{\mathrm{H}}$ of the code. According to \cite[Theorem 10.1]{lin2004ECC}, if and only if $d_{\mathbf{e}}^{(\mathrm{H})}\leq d_{\mathrm{H}}$, the correct codeword estimate is possible to be located in the region $\mathcal R \triangleq \{\hat{\mathbf{c}}_{\mathbf{e}'} \in \mathcal{C}(n,k) : d^{(\mathrm{H})}(\hat{\mathbf{c}}_{\mathbf{e}'},\hat{\mathbf{c}}_{\mathbf{e}})\leq d_{\mathrm{H}}\}$ \cite[Corollary 10.1.1]{lin2004ECC}. In other words, if $d_{\mathbf{e}}^{(\mathrm{H})}\leq d_{\mathrm{H}}$, the codeword estimate $\hat{\mathbf{c}}_{\mathbf{e}}$ is likely to be the correct estimate. In this section, we introduce a DR to discard unpromising TEP by evaluating the probability of producing a valid codeword estimate based on the Hamming distance, which is referred to as Hard Discarding Rule (HDR).
        
    In the decoding of one received signal sequence with the OSD algorithm, if the samples of ordered reliabilities sequence $[\widetilde{A}]_1^n$ are given by $\widetilde {\bm\alpha} = [\widetilde {\alpha}]_1^n$ and the minimum Hamming weight of $\mathcal{C}(n,k)$ is given by $d_{\mathrm{H}}$, for the re-encoding of an arbitrary TEP $\mathbf{e}$, the probability that $D_{\mathbf{e}}^{(\mathrm{H})}$ is less than or equal to $d_{\mathrm{H}}$ is given by
        \begin{equation} \label{equ::HDtech::DR::Ppro}
            \mathrm{P}_{\mathbf{e}}^{\mathrm{pro}}(d_{\mathrm{H}}|\widetilde {\bm\alpha}) = \sum_{j=0}^{d_{\mathrm{H}}} p_{D_{\mathbf{e}}^{(\mathrm{H})}}(j|\widetilde {\bm\alpha}),
        \end{equation}
        which is referred to as the \textit{hard promising probability}. In (\ref{equ::HDtech::DR::Ppro}), $p_{D_{\mathbf{e}}^{(\mathrm{H})}}(j|\widetilde{\bm\alpha})$ is given by (\ref{equ::HDtech::CondDis::HDforTEPe}), but in which $\mathrm{Pe}(\mathbf{e})$ is replaced by $\mathrm{Pe}(\mathbf{e}|\widetilde{\bm\alpha})$, and $p_{E_{1}^{k}}(j)$ and $p_{E_{k+1}^{n}}(j)$ are replaced with $p_{E_{1}^{k}}(j|\widetilde{\bm\alpha})$ and $p_{E_{k+1}^{n}}(j|\widetilde{\bm\alpha})$, respectively.
        
        The HDR is described as follows. Given a threshold of the promising probability $\mathrm{P}_{t}^{\mathrm{pro}} \in [0,1]$ and the minimum Hamming weight $d_{\mathrm{H}}$, if the hard promising probability of $\mathbf{e} $ satisfies the following condition 
        \begin{equation} \label{equ::HDtech::DR::HDR}
            \mathrm{P}_{\mathbf{e}}^{\mathrm{pro}}(d_{\mathrm{H}}|\widetilde {\bm\alpha}) \leq \mathrm{P}_{t}^{\mathrm{pro}},
        \end{equation}
        the TEP $\mathbf{e}$ is discarded without reprocessing.
        
        We further define $\mathrm{P}_{\mathbf{e}}^{\mathrm{pro}}(d_{\mathrm{H}})$ as
        \begin{equation}
            \mathrm{P}_{\mathbf{e}}^{\mathrm{pro}}(d_{\mathrm{H}}) =  \sum_{j=0}^{d_{\mathrm{H}}} p_{D_{\mathbf{e}}^{(\mathrm{H})}}(j),
        \end{equation}
        where $p_{D_{\mathbf{e}}^{(\mathrm{H})}}(j)$ is given by (\ref{equ::HDtech::CondDis::HDforTEPe}). It can be seen that $\mathrm{P}_{\mathbf{e}}^{\mathrm{pro}}(d_{\mathrm{H}})$ is the mean of $\mathrm{P}_{\mathbf{e}}^{\mathrm{pro}}(d_{\mathrm{H}}|\widetilde {\bm\alpha})$ with respect to $\widetilde {\bm\alpha}$, i.e., $\mathrm{P}_{\mathbf{e}}^{\mathrm{pro}}(d_{\mathrm{H}}) = \mathbb{E}[\mathrm{P}_{\mathbf{e}}^{\mathrm{pro}}(d_{\mathrm{H}}|\widetilde {\bm\alpha})]$ .
        
        For a linear block code $\mathcal{C}(n,k)$ with truncated binomial weight spectrum as described in (\ref{equ::HDdis::BinSpectrum}), it is unnecessary for the decoder to check the HDR for each TEP. For the hard promising probability defined by (\ref{equ::HDtech::DR::Ppro}), we have the following property.
        \begin{proposition} \label{pro::HDtech::DR::PproIncreasing}
            
            In the decoding of $\mathcal{C}(n,k)$ with truncated binomially distributed weight spectrum, for an arbitrary TEP $\mathbf{e}$ with the Hamming weight $w(\mathbf{e})$, $\mathrm{P}_{\mathbf{e}}^{\mathrm{pro}}(d_{\mathrm{H}}|\widetilde {\bm\alpha})$ is a monotonically increasing function of $\mathrm{Pe}(\mathbf{e}|\widetilde {\bm\alpha})$.
        \end{proposition}
        \begin{IEEEproof}
            The proof is provided in Appendix \ref{app::proof::HDtech::PproIncreasing}.
        \end{IEEEproof}
        Note that it is also easy to prove that the monotonicity given in Proposition \ref{pro::HDtech::DR::PproIncreasing} also holds for $\mathrm{P}_{\mathbf{e}}^{\mathrm{pro}}(d_{\mathrm{H}})$. We omit the proof for brevity. {\color{black}Section \ref{sec::Discussion::Implementation} will show that (\ref{equ::HDtech::DR::Ppro}) can be computed with complexity $O(n)$ FLOPs when $\mathcal{C}(n,k)$ has a binomial-like weight spectrum.}
        
        Next, we consider the decoding performance and complexity of HDR. In order to find the decoding performance of HDR, the TEP $\mathbf{e}$ which first satisfies the HDR check in the $i$-reprocessing needs to be determined. Assume that the decoder reprocesses TEPs with a specific sequence $\{\mathbf{e}_{i,1},\mathbf{e}_{i,2},\ldots,\mathbf{e}_{i,\binom{k}{i}}\}$. Given the threshold promising probability $ \mathrm{P}_{t}^{\mathrm{pro}}$, there exists a non-negative integer $\beta_i^{\mathrm{HDR}}$, such that 
        \begin{equation}  \label{equ::HDtech::DR::beta}
            \beta_i^{\mathrm{HDR}}=\sum_{j=1}^{\binom{k}{i}}\mathbf{1}_{[\mathrm{P}_{t}^{\mathrm{pro}},+\infty]}  \mathrm P_{\mathbf e_{i,j}}^{\mathrm{pro}}(d_{\mathrm{H}}|\bm{\widetilde\alpha})  
        \end{equation}
        where $\beta_i^{\mathrm{HDR}}$ in fact represents the number of TEPs re-encoded in the $i$-reprocessing conditioning on $[\widetilde{A}]_1^n = [\widetilde{\alpha}]_1^n$. Then, the mean of $\beta_i^{\mathrm{HDR}}$ can be represented as
    	\begin{equation}
    	    \mathbb{E}[\beta_i^{\mathrm{HDR}}] = \underbrace{\int_{0}^{\infty} \cdots \int_{0}^{\infty}}_{n} \beta_i^{\mathrm{HDR}} f_{[\widetilde{A}]_1^n}(\widetilde{\alpha}_1,\widetilde{\alpha}_2,\ldots,\widetilde{\alpha}_n) \prod_{u=1}^{n} d\widetilde{\alpha}_u.
    	\end{equation}
        where $ f_{[\widetilde{A}]_1^n}(x_{1},x_{2},\ldots,x_{n})$ is the joint distribution of random variables $[\widetilde{A}]_1^n$, which can be derived as \cite{balakrishnan2014order}
        \begin{equation}
            	 f_{[\widetilde{A}]_1^n}(x_{1},x_{2},\ldots,x_{n})= 
            	n! \prod_{u=1}^{n} f_{A}(x_u)\prod_{u=2}^{n} \mathbf{1}_{[0,x_{u-1}]}(x_u).
        \end{equation} 
        Similar to the approximation in (\ref{equ::HDtech::SR::HISR::db}), by considering $\mathbb{E}[\mathrm P_{\mathbf e_{i,j}}^{\mathrm{pro}}(d_{\mathrm{H}}|\bm{\widetilde\alpha})] = \mathrm P_{\mathbf e_{i,j}}^{\mathrm{pro}}(d_{\mathrm{H}})$ with respect to $\bm{\widetilde\alpha}$, $\mathbb{E}[\beta_i^{\mathrm{HDR}}]$ can be approximated by
        \begin{equation}
            \mathbb{E}[\beta_i^{\mathrm{HDR}}]\approx \sum_{j=1}^{\binom{k}{i}}\mathbf{1}_{[\mathrm{P}_{t}^{\mathrm{pro}},+\infty]}  \mathrm P_{\mathbf e_{i,j}}^{\mathrm{pro}}(d_{\mathrm{H}}) .
        \end{equation}
        Therefore, the average number of re-encoded TEP $N_a$ can be easily derived as
        \begin{equation}
            N_a =  \sum_{i=0}^{m} \mathbb{E}[\beta_i^{\mathrm{HDR}}].
        \end{equation}
       
        In the $i$-reprocessing with the HDR, the probability that the MRB errors $\widetilde{\mathbf{e}}_{\mathrm{B}}$ are eliminated can be lower bounded by (\ref{equ::HDtech::DR::Pfound}) on the top of the next page.

         \begin{table*} [t]
            \centering
            \begin{minipage}{\textwidth}
            \begin{equation}  \label{equ::HDtech::DR::Pfound}
                \begin{split}
                    \mathrm{P_{found}}(i) &=  \underbrace{\int_{0}^{\infty} \cdots \int_{0}^{\infty}}_{n}\left(p_{E_1^k}(i|\bm{\widetilde\alpha}) - \sum_{j=1}^{\binom{k}{i}}\left(\mathbf{1}_{[0,\mathrm{P}_{t}^{\mathrm{pro}}]}  \mathrm P_{\mathbf e_{i,j}}^{\mathrm{pro}}(d_{\mathrm{H}}|\bm{\widetilde\alpha}) \right)\mathrm{Pe}(\mathbf{e}_{i,j}|\bm{\widetilde\alpha})\right) f_{[\widetilde{A}]_1^n}(\widetilde{\alpha}_1,\widetilde{\alpha}_2,\ldots,\widetilde{\alpha}_n) \prod_{u=1}^{n} dx_u \\
                    &  = p_{E_1^k}(i) -\underbrace{\int_{0}^{\infty} \cdots \int_{0}^{\infty}}_{n}\left( \sum_{j=1}^{\binom{k}{i}}\left(\mathbf{1}_{[0,\mathrm{P}_{t}^{\mathrm{pro}}]}\mathrm P_{\mathbf e_{i,j}}^{\mathrm{pro}}(d_{\mathrm{H}}|\bm{\widetilde\alpha})\right) \mathrm{Pe}(\mathbf{e}_{i,j}|\bm{\widetilde\alpha})\right) f_{[\widetilde{A}]_1^n}(\widetilde{\alpha}_1,\widetilde{\alpha}_2,\ldots,\widetilde{\alpha}_n) \prod_{u=1}^{n} dx_u.
                \end{split}
            \end{equation}
            
            \medskip
            \hrule
            \end{minipage}
        \end{table*}        
        
        Therefore, the decoding error performance is upper bounded by
        \begin{equation}
            \begin{split}
            \epsilon_e^{\mathrm{HDR}} & \leq  \left(1- \sum_{i=0}^{m} \mathrm{P_{found}}(i)\right) + \mathrm{P}_{\mathrm{ML}}. \\
                 & \leq  \ 1 - \sum_{i=0}^m\left( p_{E_1^k}(i) - \eta_{\mathrm{HDR}}(i) \right) + \mathrm{P}_{\mathrm{ML}}  ,
            \end{split}
        \end{equation}
        where $\eta_{\mathrm{HDR}}(i)$ is the degradation factor of $i$-reprocessing given by
        \begin{equation}   \label{equ::HDtech::DR::HDReta}
        \begin{split}
            \eta_{\mathrm{HDR}}(i) &= \! \underbrace{\int_{0}^{\infty} \cdots }_{n}\left( \sum_{j=1}^{\binom{k}{i}}\left(\mathbf{1}_{[0,\mathrm{P}_{t}^{\mathrm{pro}}]}\mathrm P_{\mathbf e_{i,j}}^{\mathrm{pro}}(d_{\mathrm{H}}|\bm{\widetilde\alpha})\right)\mathrm{Pe}(\mathbf{e}_{i,j}|\bm{\widetilde\alpha})\right)\\
            & \cdot f_{[\widetilde{A}]_1^n}(\widetilde{\alpha}_1,\widetilde{\alpha}_2,\ldots,\widetilde{\alpha}_n) \prod_{u=1}^{n} dx_u.
        \end{split}
        \end{equation}
        If $\mathrm{P}_t^{\mathrm{pro}} = 1$, because $\mathbf{1}_{[0,\mathrm{P}_{t}^{\mathrm{pro}}]}\mathrm P_{\mathbf e_{i,j}}^{\mathrm{pro}}(d_{\mathrm{H}}|\bm{\widetilde\alpha}) = 1$ for $1\leq j \leq \binom{k}{i}$, it can be noticed that $\eta_{\mathrm{HDR}}(i) = \sum_{j=1}^{\binom{k}{i}}\mathrm{Pe}(\mathbf{e}_{i,j}) = p_{E_1^k}(i)$, which indicates the worst error rate performance and $\epsilon_e^{\mathrm{HDR}} \leq 1 + \mathrm{P}_{\mathrm{ML}} $. Furthermore, $\eta_{\mathrm{HDR}}(i)$ decreases as $\mathrm{P}_t^{\mathrm{pro}}$ decreases. This is because the smaller $\mathrm{P}_t^{\mathrm{pro}}$, the smaller $\sum_{j=1}^{\binom{k}{i}} \mathbf{1}_{[0,\mathrm{P}_{t}^{\mathrm{pro}}]}\mathrm P_{\mathbf e_{i,j}}^{\mathrm{pro}}(d_{\mathrm{H}}|\bm{\widetilde\alpha})$. In particular, if $\mathrm{P}_t^{\mathrm{pro}} = 0$, $\mathbf{1}_{[0,\mathrm{P}_{t}^{\mathrm{pro}}]}\mathrm P_{\mathbf e_{i,j}}^{\mathrm{pro}}(d_{\mathrm{H}}|\bm{\widetilde\alpha}) = 0$ for $1\leq j \leq \binom{k}{i}$ and $\eta_{\mathrm{HDR}}(i) = 0$, indicating the error rate performance is the same as the original OSD, i.e., $\epsilon_e^{\mathrm{HDR}} \leq 1 - \sum_{i=0}^m p_{E_1^k}(i) + \mathrm{P}_{\mathrm{ML}} $.
        
        {\color{black}
        If the weight spectrum of $\mathcal{C}(n,k)$ is binomial as described by (\ref{equ::HDdis::BinSpectrum}), the monotonicity described in Proposition \ref{pro::HDtech::DR::PproIncreasing} holds. Thus, in (\ref{equ::HDtech::DR::HDReta}), for each $\mathrm{Pe}(\mathbf{e}_{i,j}|\bm{\widetilde\alpha})$ satisfying $\mathrm P_{\mathbf e_{i,j}}^{\mathrm{pro}}(d_{\mathrm{H}}|\bm{\widetilde\alpha})\leq \mathrm{P}_{t}^{\mathrm{pro}}$, we can find the following inequity by referring to the definition of the HDR
        \begin{equation}
        \begin{split}
            \left\{\mathrm P_{\mathbf e_{i,j}}^{\mathrm{pro}}(d_{\mathrm{H}}|\bm{\widetilde\alpha}) \leq \mathrm{P}_{t}^{\mathrm{pro}} \right\} 
            \equiv  \left\{\mathrm{Pe}(\mathbf{e}_{i,j}|\bm{\widetilde\alpha}) \leq \mathrm{P}_{\mathbf{e}}^{\mathrm{pro},-1}(\mathrm{P}_{t}^{\mathrm{pro}}|\widetilde {\bm\alpha}) \right\}
        \end{split}
        \end{equation}
        where $\mathrm{P}_{\mathbf{e}}^{\mathrm{pro},-1}(\mathrm{P}_{t}^{\mathrm{pro}}|\widetilde {\bm\alpha})$ is the inverse function of $\mathrm P_{\mathbf e}^{\mathrm{pro}}(d_{\mathrm{H}}|\bm{\widetilde\alpha})$ with respect to $\mathrm{Pe}(\mathbf{e}|\bm{\widetilde\alpha})$. The equivalence naturally holds because of Proposition \ref{pro::HDtech::DR::PproIncreasing}. Thus, for  $\mathrm{P}_{\mathbf{e}}^{\mathrm{pro},-1}(\mathrm{P}_{t}^{\mathrm{pro}}|\widetilde {\bm\alpha}) \geq 0$, the degradation factor $\eta_{\mathrm{HDR}}(i)$ can be scaled by 
        \begin{equation}   \label{equ::HDtech::DR::HDReta::scaled}
        \begin{split}
            \eta_{\mathrm{HDR}}(i) &\leq \underbrace{\int_{0}^{\infty} \cdots \int_{0}^{\infty}}_{n} \beta_i^{\mathrm{HDR}} \cdot  \mathrm{P}_{\mathbf{e}}^{\mathrm{pro},-1}(\mathrm{P}_{t}^{\mathrm{pro}}|\widetilde {\bm\alpha}) \\
            & \quad\cdot f_{[\widetilde{A}]_1^n}(\widetilde{\alpha}_1,\widetilde{\alpha}_2,\ldots,\widetilde{\alpha}_n) \prod_{u=1}^{n} dx_u \\
            &\overset{(a)}{\leq} \binom{k}{i} \mathbb{E}[\mathrm{P}_{\mathbf{e}}^{\mathrm{pro},-1}(\mathrm{P}_{t}^{\mathrm{pro}}|\widetilde {\bm\alpha})] \\
            & = \binom{k}{i} \mathrm{P}_{\mathbf{e}}^{\mathrm{pro},-1}(\mathrm{P}_{t}^{\mathrm{pro}}),
        \end{split}
        \end{equation}
        where step (a) follows from $\beta_i^{\mathrm{HDR}}\leq \binom{k}{i}$ as shown by (\ref{equ::HDtech::DR::beta}). Particularly when $\mathrm{P}_{\mathbf{e}}^{\mathrm{pro},-1}(\mathrm{P}_{t}^{\mathrm{pro}}|\widetilde {\bm\alpha})< 0$, $\beta_i^{\mathrm{HDR}} = 0$ and $\eta_{\mathrm{HDR}}(i) = 0$. $ \mathrm{P}_{\mathbf{e}}^{\mathrm{pro},-1}(\mathrm{P}_{t}^{\mathrm{pro}})$ is the inverse function of $\mathrm P_{\mathbf e}^{\mathrm{pro}}(d_{\mathrm{H}})$ with respect to $\mathrm{Pe}(\mathbf{e})$, which is derived as
        \begin{equation} \label{equ::HDR_eta_app}
            \mathrm{P}_{\mathbf{e}}^{\mathrm{pro},-1}(\mathrm{P}_{t}^{\mathrm{pro}}) =  \frac{\mathrm{P}_t^{\mathrm{pro}} -  \sum_{j=i}^{d_{\mathrm{H}}}p_{W_{\mathbf{e},\mathbf{c}_{\mathrm{P}}}}(j-i|i)}{\sum_{j=i}^{d_{\mathrm{H}}}\left(p_{E_{k+1}^{n}}(j-i) - p_{W_{\mathbf{e},\mathbf{c}_{\mathrm{P}}}}(j-i|i)  \right)},
        \end{equation}
        where $p_{E_{k+1}^{n}}(j)$ is given by (\ref{equ::OrderStat::Eab}) and $p_{W_{\mathbf{e},\mathbf{c}_{\mathrm{P}}}}(j|i)$ is given by (\ref{equ::HDtech::CondDis::Wecp::w(e)=v}). }

\begin{figure}[t]
	    	\vspace{-0.8em}
            \centering
            \subfigure[Frame error rate]
            {
                \includegraphics[scale = 0.65]{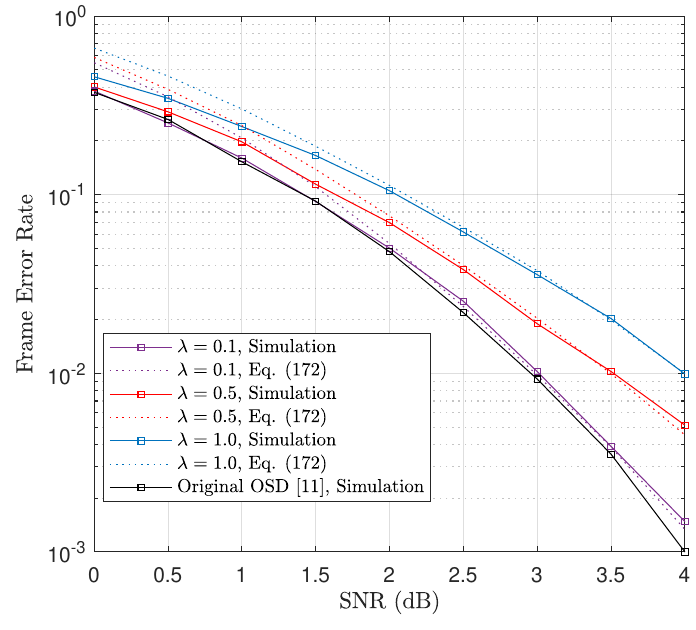}
                \label{Fig::VI::HDR_Pe}
            }
            \vspace{-1ex}
            \subfigure[Average number of TEPs]
            {
                \includegraphics[scale = 0.65]{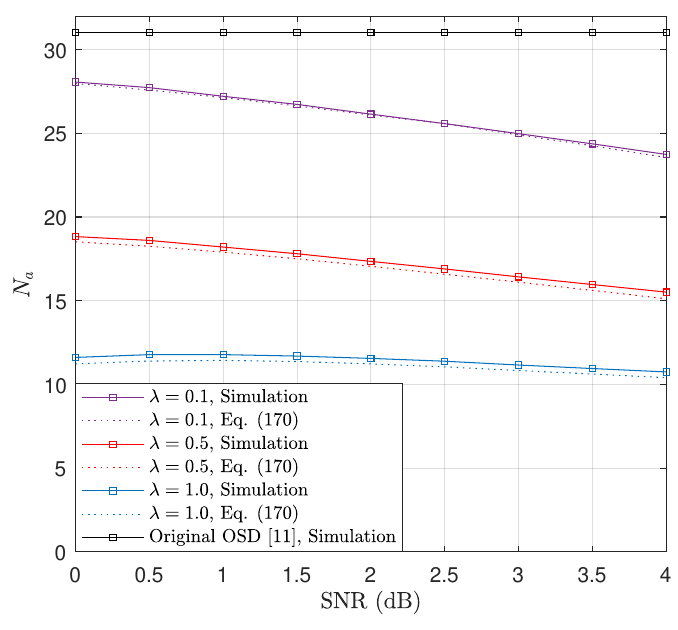}
                \label{Fig::VI::HDR_Na}
            }

            \caption{Decoding $(64,30,14)$ eBCH code with an order-$1$ OSD applying the HDR.}
            \label{Fig::HDR}
        \end{figure}

\begin{figure}
    		\begin{center}
    			\includegraphics[scale=0.6] {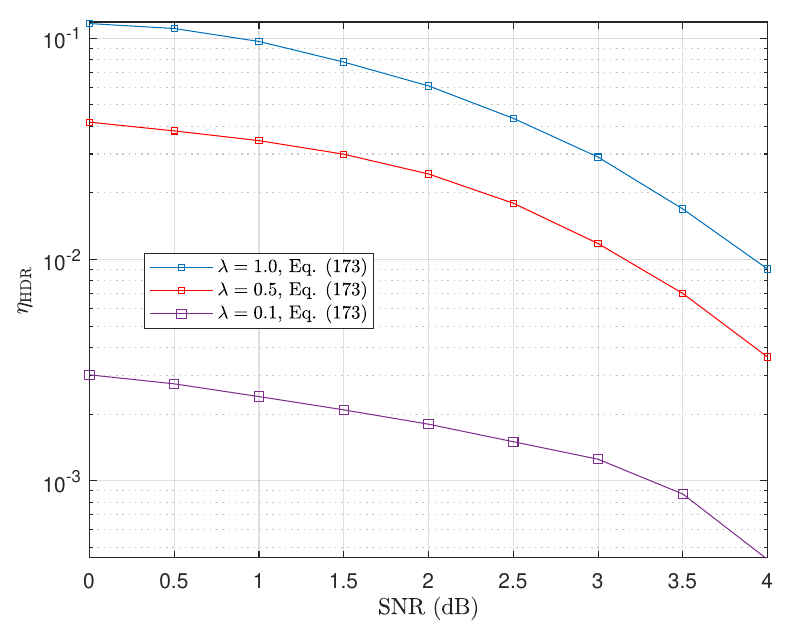}
    			\caption{The performance degradation factor $\eta_{\mathrm{HGSR}}$ of decoding $(64,30,14)$ eBCH code with an order-$1$ OSD applying HDR.}
    			\label{Fig::VI::eta_HDR}
    		\end{center}
    	\end{figure}

        We consider an order-1 OSD algorithm applying HDR in decoding a $(64,30,14)$ eBCH code. According to (\ref{equ::HDtech::DR::HDReta::scaled}) and (\ref{equ::HDR_eta_app}), the threshold promising probability is set to $\mathrm{P}_t^{\mathrm{pro}} =  \frac{\lambda}{\binom{k}{i}} p_{E_1^k}(i) +  \sum_{j=i}^{d_{\mathrm{H}}}p_{W_{\mathbf{e},\mathbf{c}_{\mathrm{P}}}}(j-i|i)$ in the $i$ reprocessing to adapt to the channel conditions, where $\lambda$ is a non-negative real parameter. The comparisons of error performance and average number of TEPs $N_a$ are depicted in Fig.\ref{Fig::VI::HDR_Pe} and Fig.\ref{Fig::VI::HDR_Na}, respectively. The performance degradation $\eta_{\mathrm{HDR}}$ with different $\lambda$ is also illustrated in Fig. \ref{Fig::VI::eta_HDR}. As can be seen, the trade-off between error performance and decoding complexity can be maintained by changing $\lambda$. The decoding complexity decreases and the frame error rate suffers more degradation when $\lambda$ increases, and vice versa. Compared with the HISR or HGSR, HDR has better error performance at low SNRs but worse error performance at high SNRs with the same level of $N_a$, which implies that one can combine HDR as a DR and HISR or HGSR as SRs to reduce the decoding complexity in both low and high SNR scenarios.

\section{Soft-decision Decoding Techniques Based on WHD Distribution} \label{Sec::SoftTech}

\subsection{Soft Success Probability of codeword estimate} \label{sec::SoftTech::Cond}
    
Based on the WHD distribution we derived in Section \ref{sec::WHD}, we can also propose different SRs and DRs for improving the decoding efficiency of OSD. Different from the hard-decision decoding techniques proposed in Section \ref{sec::HDdistech}, the soft-decision decoding techniques can make better use of the \textit{a priori} information.
        
We first investigate the distribution of WHD $D_{\mathbf{e}}^{(\mathrm{W})}$ between $\widetilde{\mathbf{c}}_{\mathbf{e}} = [\widetilde{\mathbf{y}}_{\mathrm{B}}\oplus \mathbf{e}]\widetilde{\mathbf{G}}$ and $\widetilde{\mathbf{y}}$. For a specific TEP $\mathbf{e} = [e]_1^k$, let $\mathbf{t}_{\mathbf{e}}^{\mathrm{B}} = [t^{\mathrm{B}}]_1^{w(\mathbf{e})}$ represent the positions indices of nonzero elements of $\mathbf{e}$. Also, following in the same definition of $\mathbf{t}_h^{\mathrm{P}}$ in Section \ref{sec::WHD}, let us consider an index vector $\mathbf{t}_{\mathbf{e}}^{h}$ defined as  $\mathbf{t}_{\mathbf{e}}^{h} = [\mathbf{t}_{\mathbf{e}}^{\mathrm{B}} \ \ \mathbf{t}_h^{\mathrm{P}}]$ with the length $w(\mathbf{e}) + h$, where $\mathbf{t}_h^{\mathrm{P}} = [t^{\mathrm{P}}]_1^h$. Based on the Lemma \ref{lem::WHD::iphase::eB=e} and Lemma \ref{lem::WHD::iphase::eB!=e}, we derive the distribution of $D_{\mathbf{e}}^{(\mathrm{W})}$ for a specific $\mathbf{e} = [e]_1^k$ in the following Corollary.
       {\color{black} 
       \begin{corollary} \label{cor::Stech::Condis::WHDforTEP}
    		Given a linear block code $\mathcal{C}(n,k)$ with its respective $p_{\mathbf{c}_{\mathrm{P}}}(u,q)$ and a specific TEP $\mathbf{e} = [e]_1^k$, the $\mathrm{pdf}$ of the weighted Hamming distance between $\widetilde{\mathbf{y}}$ and $\widetilde{\mathbf{c}}_{\mathbf{e}}$, denoted by $f_{D_{\mathbf{e}}^{(\mathrm{W})}}(x|\mathbf{e} \!=\! [e]_1^k)$, is given by
    		\begin{equation} \label{equ::Stech::Condis::WHDforTEP}
    		\begin{split}
     		    f_{D_{\mathbf{e}}^{(\mathrm{W})}}(x|\mathbf{e} \!=\! [e]_1^k) &= \sum_{h=0}^{n-k} \sum_{\substack{\mathbf{t}_h^{\mathrm{P}} \in \mathcal{T}_h^{\mathrm{P}}}} \mathrm{Pe}(\mathbf{t}_{\mathbf{e}}^{h})  f_{\widetilde { A}_{\mathbf{t}_{\mathbf{e}}^{h}}}\!\!(x) \\
     		    &+(1 - \mathrm{Pe}(\mathbf{e}))\sum_{h=0}^{n-k} \sum_{\substack{ \mathbf{t}_h^{\mathrm{P}} \in \mathcal{T}_h^{\mathrm{P}}}}  \mathrm{Pc}(\mathbf{t}_{\mathbf{e}}^{h})  f_{\widetilde { A}_{\mathbf{t}_{\mathbf{e}}^{h}}}\!\!(x),   
    		\end{split}
    		\end{equation}
    		where
            \begin{equation} \label{equ::Stech::Condis::WHDforTEP::eB=e}
            \begin{split}
                  \mathrm{Pe}(\mathbf{t}_{\mathbf{e}}^{h}) &=\underbrace{\int_{0}^{\infty}\! \cdots}_{n\!-\!h\!-\!w(\mathbf{e})}\underbrace{\int_{-\infty}^{0} \!\cdots}_{h\!+\!w(\mathbf{e})} \left(n! \prod_{v=1}^{n} f_{R}(x_v) \prod_{v=2}^{n} \mathbf{1}_{[0,|x_{v-1}|]}(|x_v|) \right) \\
                  &\cdot\prod_{\substack{1 \leq v \leq {n}\\v\in \mathbf{t}_{\mathbf{e}}^{h}}} dx_v \prod_{\substack{1 \leq v \leq {n}\\v\notin \mathbf{t}_{\mathbf{e}}^{h}}} dx_v 
            \end{split}
            \end{equation}
            and
            \begin{equation}  \label{equ::Stech::Condis::WHDforTEP::eB!=e}
                  \mathrm{Pc}(\mathbf{t}_{\mathbf{e}}^{h}) = \sum_{\mathbf{x}\in\{0,1\}^{n-k}}\mathrm{Pr}(\widetilde{\mathbf{c}}_{\mathbf{e},\mathrm{P}}' = \mathbf{z}_{\mathbf{t}_h^{\mathrm{P}}}\oplus\mathbf{x})\mathrm{Pr}(\widetilde{\mathbf{e}}_{\mathrm{P}} = \mathbf{x}).
            \end{equation} 
            where $\mathbf{x}$ is a length-$(n-k)$ binary vectors. The probability $\mathrm{Pr}(\widetilde{\mathbf{c}}_{\mathbf{e}}' = \mathbf{z}_{\mathbf{t}_h^{\mathrm{P}}}\oplus\mathbf{x})$ is given by
            \begin{equation}
                 \mathrm{Pr}(\widetilde{\mathbf{c}}_{\mathbf{e},\mathrm{P}}' \!\!=\! \mathbf{z}_{\mathbf{t}_h^{\mathrm{P}}}\oplus\mathbf{x}) \!=\!\! \sum_{q = 1}^{k}\!\! \sum_{\substack{\mathbf{x} \in \{0,1\}^{k}\\ w(\mathbf{e}\oplus \mathbf{x})=q}}\!\!\!\frac{\mathrm{Pr}(\widetilde{\mathbf{e}}_{\mathrm{B}}\!\! =\! \mathbf{x})}{\binom{n-k}{w(\mathbf{z}_{\mathbf{t}_h^{\mathrm{P}}}\oplus\mathbf{x})}} p_{\mathbf{c}_{\mathrm{P}}}\!(w(\mathbf{z}_{\mathbf{t}_h^{\mathrm{P}}}\oplus\mathbf{x}),q).
            \end{equation}
             where $\mathbf{x}$ is a length-$k$ binary vector satisfying $w(\mathbf{e}\oplus \mathbf{x})=q$, and $\mathrm{Pr}(\widetilde{\mathbf{e}}_{\mathrm{B}} = \mathbf{x})$ is given by (\ref{equ::WHD::iphase::eB!=e::Pc::eB=eksi}). The probability
            $\mathrm{Pe}(\mathbf{e})$ is given by (\ref{equ::HDtech::CondDis::Pe(e)}) and $f_{\widetilde { A}_{\mathbf{t}_{\mathbf{e}}^{h}}}(x)$ is the $\mathrm{pdf}$ of $\widetilde { A}_{\mathbf{t}_{\mathbf{e}}^{h}}= \sum_{v=1}^{w(\mathbf{e})}\widetilde{A}_{{t}_v^{\mathrm{B}}} +  \sum_{v=1}^{h} \widetilde{A}_{{t}_v^{\mathrm{P}}}$.
        \end{corollary}
        \begin{IEEEproof}
            The proof is provided in Appendix \ref{app::proof::Stech::Condis::WHDforTEP}
        \end{IEEEproof}
        
        Note that Corollary \ref{cor::Stech::Condis::WHDforTEP} is slightly different from a simple combination of Lemma \ref{lem::WHD::iphase::eB=e} and \ref{lem::WHD::iphase::eB!=e}, because Corollary \ref{cor::Stech::Condis::WHDforTEP} assumes that the TEP $\mathbf{e} =  [e]_1^k$ is known. However, Lemma \ref{lem::WHD::iphase::eB=e} and \ref{lem::WHD::iphase::eB!=e} assume that $\mathbf{e}$ is unknown to the decoder. Henceforth, we use $\{\widetilde{\mathbf{e}}_{\mathrm{B}} \!=\! \mathbf{e} \!=\! [e]_1^k\}$ to represent the condition that the MRB errors are eliminated by a TEP $\mathbf{e}$ and $\mathbf{e}$ is known as $\mathbf{e} \!=\! [e]_1^k$. By using a similar approach as in Section \ref{subsec::appWHD}, we approximate the distribution $f_{D_{\mathbf{e}}^{(\mathrm{W})}}(x|\mathbf{e} \!=\! [e]_1^k)$ of $D_{\mathbf{e}}^{(\mathrm{W})}$ as a mixture of Gaussian distributions, i.e.
        \begin{equation} \label{equ::Stech::Condis::WHDforTEP::app}
            \begin{split}
                &f_{D_{\mathbf{e}}^{(\mathrm{W})}}(x|\mathbf{e} \!=\! [e]_1^k) \\
                & =\! \mathrm{Pe}(\mathbf{e}) f_{\!D_{\mathbf{e}}^{(\mathrm{W})}}(x|\widetilde{\mathbf{e}}_{\mathrm{B}} \!=\! \mathbf{e} \!=\! [e]_1^k)  \\
                & + (1 - \mathrm{Pe}(\mathbf{e})) f_{\!D_{\mathbf{e}}^{(\mathrm{W})}}(x|\widetilde{\mathbf{e}}_{\mathrm{B}} \!\neq\! \mathbf{e} \!=\! [e]_1^k) \\
                & \approx \!\frac{\mathrm{Pe}(\mathbf{e})}{\sqrt{2\pi\sigma_{D_{\mathbf{e}}^{(\mathrm{w})}|\widetilde{\mathbf{e}}_{\mathrm{B}}= \mathbf{e} = [e]_1^k}^2}}\!\exp\!{\left(\!\!-\frac{(x\!-\!\mathbb{E}[D_{\mathbf{e}}^{(\mathrm{w})}|\widetilde{\mathbf{e}}_{\mathrm{B}}\!= \!\mathbf{e} \!=\! [e]_1^k])^2}{2\sigma_{D_{\mathbf{e}}^{(\mathrm{w})}|\widetilde{\mathbf{e}}_{\mathrm{B}}= \mathbf{e} = [e]_1^k}^2}\!\right)} \\
                & + \!\frac{1 - \mathrm{Pe}(\mathbf{e})}{\sqrt{2\pi\sigma_{D_{\mathbf{e}}^{(\mathrm{w})}|\widetilde{\mathbf{e}}_{\mathrm{B}}\neq \mathbf{e} = [e]_1^k}^2}}\!\exp{\!\left(\!\!-\frac{(x\!-\!\mathbb{E}[D_{\mathbf{e}}^{(\mathrm{w})}|\widetilde{\mathbf{e}}_{\mathrm{B}}\!\neq \!\mathbf{e} \!=\! [e]_1^k])^2}{2\sigma_{D_{\mathbf{e}}^{(\mathrm{w})}|\widetilde{\mathbf{e}}_{\mathrm{B}}\neq \mathbf{e} = [e]_1^k}^2}\!\right)}\!,
            \end{split}
        \end{equation}
        where $\mathbb{E}[D_{\mathbf{e}}^{(\mathrm{w})}|\widetilde{\mathbf{e}}_{\mathrm{B}}\!= \!\mathbf{e} \!=\! [e]_1^k]$ and $\sigma_{D_{\mathbf{e}}^{(\mathrm{w})}|\widetilde{\mathbf{e}}_{\mathrm{B}}= \mathbf{e} =[e]_1^k}^2$ are respectively given by
        \begin{equation} \label{equ::Stech::Condis::WHDforTEP::eB=e::mean}
            \begin{split}
                \mathbb{E}[D_{\mathbf{e}}^{(\mathrm{w})}|\widetilde{\mathbf{e}}_{\mathrm{B}}\!= \!\mathbf{e} \!=\! [e]_1^k] =\sum_{u=1}^{w(\mathbf{e})} \sqrt{\widetilde{\mathbf{E}}_{t_{u}^{\mathrm{B}},t_{u}^{\mathrm{B}}}} \! + \!\!  \sum_{u=k+1}^{n} \mathrm{Pe}(u)\sqrt{\widetilde{\mathbf{E}}_{u,u}} ,
            \end{split}
        \end{equation}
        and
        \begin{equation}  \label{equ::Stech::Condis::WHDforTEP::eB=e::var}
        	\begin{split}
        		\sigma_{D_{\mathbf{e}}^{(\mathrm{w})}|\widetilde{\mathbf{e}}_{\mathrm{B}}= \mathbf{e} =[e]_1^k}^2 & =  2 \sum_{u=1}^{w(\mathbf{e})} \sum_{v=k+1}^{{n}}\mathrm{Pe}(v) \left[\widetilde{\mathbf{E}}+ \widetilde{\mathbf\Sigma}\right]_{t_{u}^{\mathrm{B}},v} \\
        		&+ \sum_{u=k+1}^{{n}} \sum_{v=k+1}^{{n}} \mathrm{Pe}(u,v) \left[\widetilde{\mathbf{E}} + \widetilde{\mathbf\Sigma}\right]_{u,v} \\
        		&+\sum_{u=1}^{w(\mathbf{e})} \sum_{v=1}^{w(\mathbf{e})}  \left[\widetilde{\mathbf{E}} + \widetilde{\mathbf\Sigma}\right]_{t_{u}^{\mathrm{B}},t_{v}^{\mathrm{B}}} \\
        		&- \left(\mathbb{E}[D_{\mathbf{e}}^{(\mathrm{w})}|\widetilde{\mathbf{e}}_{\mathrm{B}}\!= \!\mathbf{e} \!=\! [e]_1^k]\right)^2 \ .
        	\end{split}
        \end{equation}
        Then $\mathbb{E}[D_{\mathbf{e}}^{(\mathrm{w})}|\widetilde{\mathbf{e}}_{\mathrm{B}}\!\neq \!\mathbf{e} \!=\! [e]_1^k]$ and $\sigma_{D_{\mathbf{e}}^{(\mathrm{w})}|\widetilde{\mathbf{e}}_{\mathrm{B}}\neq \mathbf{e} = [e]_1^k}^2$ are respectively given by
        \begin{align} \label{equ::Stech::Condis::WHDforTEP::eB!=e::mean}
                \mathbb{E}[D_{\mathbf{e}}^{(\mathrm{w})}|\widetilde{\mathbf{e}}_{\mathrm{B}}\!\neq \!\mathbf{e} \!=\! [e]_1^k] &=\sum_{u=1}^{w(\mathbf{e})} \sqrt{\widetilde{\mathbf{E}}_{t_{u}^{\mathrm{B}},t_{u}^{\mathrm{B}}}}  \\
                &+  \sum_{u=k+1}^{n} \mathrm{Pc}_{\mathbf{e}}(u|\mathbf{e} \!=\! [e]_1^k])\sqrt{\widetilde{\mathbf{E}}_{u,u}}, \notag
        \end{align}
        and 
        \begin{align}  \label{equ::Stech::Condis::WHDforTEP::eB!=e::var}
    		\sigma_{D_{\mathbf{e}}^{(\mathrm{w})}|\widetilde{\mathbf{e}}_{\mathrm{B}}\neq \mathbf{e} = [e]_1^k}^2 &= 2 \sum_{u=1}^{w(\mathbf{e})} \sum_{v=k+1}^{{n}}\mathrm{Pc}_{\mathbf{e}}(v|\mathbf{e} \!=\! [e]_1^k]) \left[\widetilde{\mathbf{E}}+ \widetilde{\mathbf\Sigma}\right]_{t_{u}^{\mathrm{B}},v} \notag \\
    		&+ \sum_{u=k+1}^{{n}} \sum_{v=k+1}^{{n}} \mathrm{Pc}_{\mathbf{e}}(u,v|\mathbf{e} \!=\! [e]_1^k]) \left[\widetilde{\mathbf{E}} + \widetilde{\mathbf\Sigma}\right]_{u,v} \notag \\
    		&+\sum_{u=1}^{w(\mathbf{e})} \sum_{v=1}^{w(\mathbf{e})}  \left[\widetilde{\mathbf{E}} + \widetilde{\mathbf\Sigma}\right]_{t_{u}^{\mathrm{B}},t_{v}^{\mathrm{B}}} \\
    		&-\left(\mathbb{E}[D_{\mathbf{e}}^{(\mathrm{w})}|\widetilde{\mathbf{e}}_{\mathrm{B}}\!\neq \!\mathbf{e} \!=\! [e]_1^k]\right)^2,\notag
        \end{align}
        where $\mathrm{Pc}_{\mathbf{e}}(u|\mathbf{e} \!=\! [e]_1^k])$ is the probability of $\widetilde{d}_{\mathbf{e},u}\neq 0$ given that $\mathbf{e} = [e]_1^k$, while $\mathrm{Pc}_{\mathbf{e}}(u,v|\mathbf{e} \!=\! [e]_1^k])$ is the joint conditional probability of $\widetilde{d}_{\mathbf{e},u}\neq 0$ and $\widetilde{d}_{\mathbf{e},v}\neq 0$. Similar to (\ref{equ::WHD::App::debit1::eB<=i}), $\mathrm{Pc}_{\mathbf{e}}(u|\mathbf{e} \!=\! [e]_1^k])$ can be derived as
        \begin{equation}
        \begin{split}
            &\mathrm{Pc}_{\mathbf{e}}(u|\mathbf{e} \!=\! [e]_1^k]) \\
            &= \sum_{q=1}^{k}\sum_{\substack{\mathbf{x} \in \{0,1\}^{k}\\ w(\mathbf{e}\oplus\mathbf{x})=q}}\mathrm{Pr}(\widetilde{\mathbf{e}}_{\mathrm{B}} = \mathbf{x}) p_{\mathbf{c}_{\mathrm{P}}}^{\mathrm{bit}}(u,q)(1-\mathrm{Pe}(u)) \\
            &+ (1 - p_{\mathbf{c}_{\mathrm{P}}}^{\mathrm{bit}}(u,q))\mathrm{Pe}(u) ,    
        \end{split}
        \end{equation}
        where $\mathrm{Pr}(\widetilde{\mathbf{e}}_{\mathrm{B}} = \mathbf{x})$ is given by (\ref{equ::WHD::iphase::eB!=e::Pc::eB=eksi}) and $p_{\mathbf{c}_{\mathrm{P}}}^{\mathrm{bit}}$ is given by (\ref{equ::WHD::App::PcPbit1}). The joint probability $\mathrm{Pc}_{\mathbf{e}}(u,v|\mathbf{e} \!=\! [e]_1^k])$ can be obtained similarly following the derivation of (\ref{equ::WHD::App::debit2::eB<=i}). We omit the details for the sake of brevity. }

        Based on Corollary \ref{cor::Stech::Condis::WHDforTEP}, for a specific TEP $\mathbf{e} = [e]_1^k $, the probability that the TEP $\mathbf{e}$ can eliminate the MRB errors $\widetilde{\mathbf{e}}_{\mathrm{B}}$ can be obtained if $D_{\mathbf{e}}^{(\mathrm{W})}$ is given by $d_{\mathbf{e}}^{(\mathrm{W})}$, i.e., $\mathrm{Pr}(\widetilde{\mathbf{e}}_{\mathrm{B}} = \mathbf{e}|D_{\mathbf{e}}^{(\mathrm{W})}\!\!=\!d_{\mathbf{e}}^{(\mathrm{W})})$. We refer to $ \widetilde{\mathrm{P}}_{\mathbf{e}}^{\mathrm{suc}}(d_{\mathbf{e}}^{(\mathrm{W})}) = \mathrm{Pr}(\widetilde{\mathbf{e}}_{\mathrm{B}} = \mathbf{e}|D_{\mathbf{e}}^{(\mathrm{W})}\!\!=\!d_{\mathbf{e}}^{(\mathrm{W})})$ as the \textit{soft success probability} of $\widetilde{\mathbf{c}}_{\mathbf{e}}$. After re-encoding $\widetilde{\mathbf{c}}_{\mathbf{e}} = [\widetilde{\mathbf{y}}_{\mathrm{B}}\oplus \mathbf{e}]\widetilde{\mathbf{G}}$, if the WHD between $\widetilde{\mathbf{c}}_{\mathbf{e}}$ and $\widetilde{\mathbf{y}}$ is given by $d_{\mathbf{e}}^{(\mathrm{W})}$, the soft success probability of $\widetilde{\mathbf{c}}_{\mathbf{e}}$ is given by
        \begin{equation} \label{equ::Stech::Condis::Psuce}
            \begin{split}
                \widetilde{\mathrm{P}}_{\mathbf{e}}^{\mathrm{suc}}(d_{\mathbf{e}}^{(\mathrm{W})}) &= \mathrm{Pe}(\mathbf{e})\frac{ f_{D_{\mathbf{e}}^{(\mathrm{W})}}(d_{\mathbf{e}}^{(\mathrm{W})}|\widetilde{\mathbf{e}}_{\mathrm{B}}\!=\!\mathbf{e} \!=\! [e]_1^k)}{f_{D_{\mathbf{e}}^{(\mathrm{W})}}(d_{\mathbf{e}}^{(\mathrm{W})}|\mathbf{e} \!=\! [e]_1^k)}, 
            \end{split}
        \end{equation}
        where $f_{D_{\mathbf{e}}^{(\mathrm{W})}}(x|\mathbf{e} \!=\! [e]_1^k)$ is given by (\ref{equ::Stech::Condis::WHDforTEP}). 
        The success probability $\widetilde{\mathrm{P}}_{\mathbf{e}}^{\mathrm{suc}}(d_{\mathbf{e}}^{(\mathrm{W})})$ can be approximately computed using the normal approximations introduced in (\ref{equ::Stech::Condis::WHDforTEP::app}).
        
\begin{figure}
    		\begin{center}
    			\includegraphics[scale=0.6] {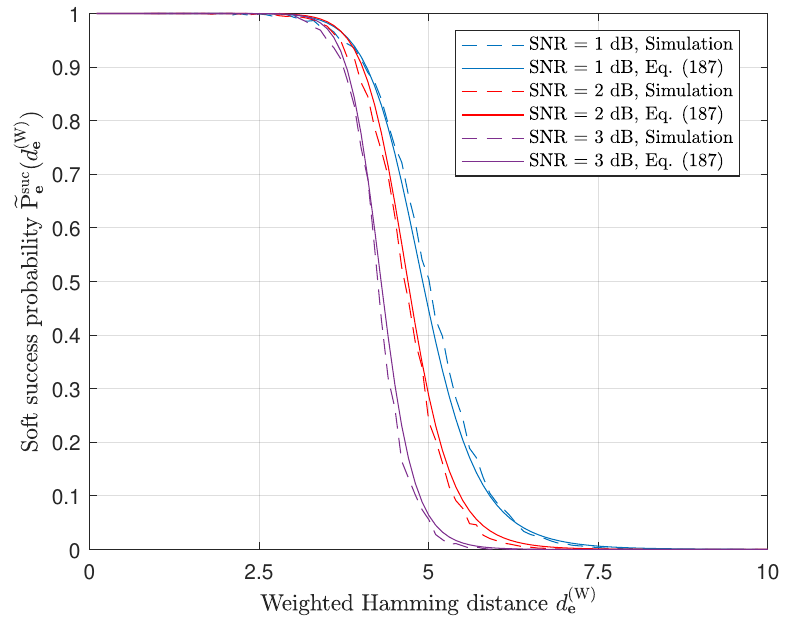}
    			\caption{{\color{black}$\widetilde{\mathrm{P}}_{\mathbf{e}}^{\mathrm{suc}}(d_{\mathbf{e}}^{(\mathrm{W})})$ in decoding $(64,30,14)$ eBCH code at different SNR, when $\mathbf{e} = [0,\ldots,0,1,0]$.}}
    			\label{Fig::VII::BCH64-WHD-Psuc}
    		\end{center}
    	\end{figure}

         We illustrate the result of $\widetilde{\mathrm{P}}_{\mathbf{e}}^{\mathrm{suc}}(d_{\mathbf{e}}^{(\mathrm{W})})$ as the function of $d_{\mathbf{e}}^{(\mathrm{W})}$ for TEP $\mathbf{e} = [0,\ldots,0,1,0]$ in decoding the $(64,30,14)$ eBCH code in Fig. \ref{Fig::VII::BCH64-WHD-Psuc}. As can be seen, when WHD $d_{\mathbf{e}}^{(\mathrm{W})}$ decreases, the success probability of $\widetilde{\mathbf{c}}_{\mathbf{e}}$ increases rapidly. At all SNRs, the success probability tends to be very close to 1 when the WHD $d_{\mathbf{e}}^{(\mathrm{W})}$ is less than 3. Therefore, the WHD of one codeword estimate can be a good indicator to identify promising decoding output.

         After the $i$-reprocessing ($0 \leq i \leq m $), if the recorded minimum WHD is given as $ d_{i}^{(\mathrm{W})}$, the conditional probability $\mathrm{Pr}(w(\widetilde{\mathbf{e}}_{\mathrm{B}})\leq i | D_{i}^{(\mathrm{W})} \!=\! d_{i}^{(\mathrm{W})})$ can also be calculated according to Theorem \ref{the::WHD::iphase}, which is referred to as the \textit{soft success probability} $\widetilde{\mathrm{P}}_{i}^{\mathrm{suc}}(d_{i}^{(\mathrm{W})})$ of codeword $\hat{\mathbf{c}}_{i}$, i.e.,
         \begin{equation}  \label{equ::Stech::Condis::Psuci}
         \begin{split}
              \widetilde{\mathrm{P}}_{i}^{\mathrm{suc}}(d_{i}^{(\mathrm{W})}) &= 1 - \left(1 - \sum_{v=0}^{i}p_{E_1^{k}}(v)\right)\\
              &\cdot\frac{f_{\widetilde{D}_i^{(\mathrm{W})}|\widetilde{\mathbf{e}}_{\mathrm{B}} \neq \mathbf{e}}\left(x,b_{0:i}^{k}|w(\widetilde{\mathbf{e}}) \geq i\right)}{ f_{D_i^{(\mathrm{W})}}(d_{i}^{(\mathrm{W})})},
         \end{split}
         \end{equation}
         where $f_{D_i^{(\mathrm{W})}}(x)$ is given by (\ref{equ::WHD::iphase}) and $ f_{\widetilde{D}_i^{(\mathrm{W})}|\widetilde{\mathbf{e}}_{\mathrm{B}} \neq \mathbf{e}}\left(x,b_{0:i}^{k}|w(\widetilde{\mathbf{e}}) \geq i\right)$ is given by (\ref{equ::WHD::iphase::DependOrder2}).
         
\begin{figure}
    		\begin{center}
    			\includegraphics[scale=0.6] {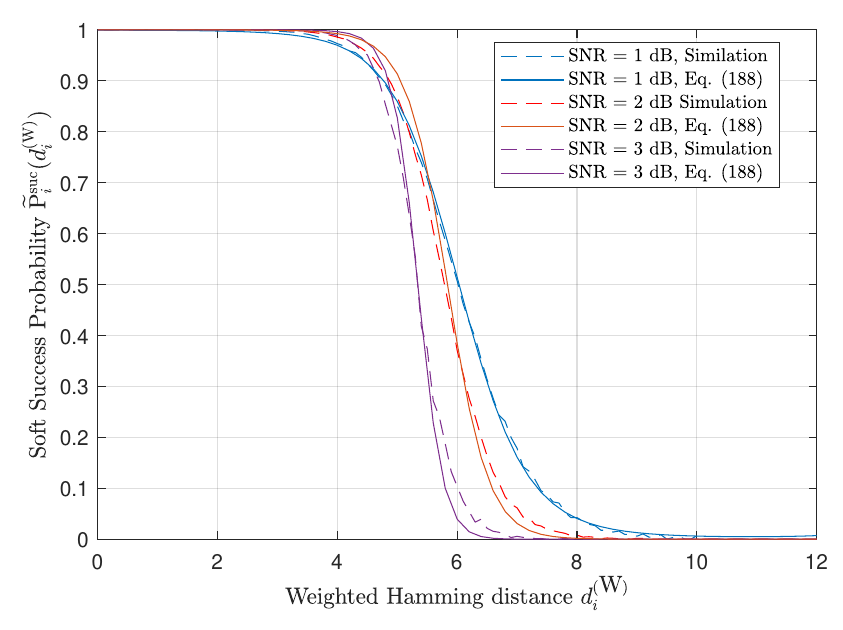}
    			\caption{{\color{black}$\widetilde{\mathrm{P}}_{i}^{\mathrm{suc}}(d_{i}^{(\mathrm{W})})$ in decoding $(64,30,14)$ eBCH code when $i=1$.}}
    			\label{Fig::VII::BCH64-minWHD-Psuc}
    		\end{center}
    	\end{figure}

        We illustrate the probability $ \widetilde{\mathrm{P}}_{i}^{\mathrm{suc}}(d_{i}^{(\mathrm{W})})$ as a function of $d_{i}^{(\mathrm{W})}$ in Fig. \ref{Fig::VII::BCH64-minWHD-Psuc}. It can be seen that the minimum WHD $d_{i}^{(\mathrm{W})}$ after the $i$-th reprocessing indicates the probability that the errors in MRB are eliminated by an OSD algorithm. It is worth noting that the discrepancies between the simulated curves and analytical curves are because of applying the approximation (\ref{equ::WHD::App::iphase::App}) in numerical computation of (\ref{equ::Stech::Condis::Psuci}).

    \subsection{Stopping Rules} \label{sec::Stech::SR}
        Next, we introduce the soft-decision SRs based on the success probabilities described in Section \ref{sec::SoftTech::Cond}. Soft-decision SRs give more accurate information of success probability than the hard-decision SRs introduced in Section \ref{sec::HDdis} because the soft information is utilized.

        \subsubsection{Soft Individual Stopping Rule}      
        
        Let us first re-consider the distribution of $D_{\mathbf{e}}^{(\mathrm{W})}$ if the reliability information $[\widetilde{A}]_1^n = [\widetilde{\alpha}]_1^n$ is given. Note that conditioning on $[\widetilde{A}]_1^n = [\widetilde{\alpha}]_1^n$, $D_{\mathbf{e}}^{(\mathrm{W})}$ is no longer a continuous random variable, but is a discrete random variable, and the sample space of $D_{\mathbf{e}}^{(\mathrm{W})}$ is all possible linear combinations of elements of $\widetilde{\bm{\alpha}} =  [\widetilde{\alpha}]_1^n$ with the coefficient 0 or 1. Given a 
       specific TEP $\mathbf{e} = [e]_1^k$, a sample of $D_{\mathbf{e}}^{(\mathrm{W})}$ can be represented as $d_{\mathbf{t}_{\mathbf{e}}^{h}}^{(\mathrm{W})} = [\mathbf{e} \   \mathbf{z}_{\mathbf{t}_h^{\mathrm{P}}}]{\widetilde{\bm\alpha}}^\mathrm{T}$ with $\mathbf{t}_h^{\mathrm{P}}\in \mathcal{T}_{h}^{\mathrm{P}}$, $1\leq h \leq n-k$. Based on Corollary \ref{cor::Stech::Condis::WHDforTEP}, we summarize the distribution of WHD $D_{\mathbf{e}}^{(\mathrm{W})}$ conditioning on $[\widetilde{A}]_1^n = [\widetilde{\alpha}]_1^n$ in the following Corollary.
       {\color{black}
        \begin{corollary} \label{cor::Stech::SR::WHDforTEP::alpha}
    		Given a linear block code $\mathcal{C}(n,k)$ and a specific TEP $\mathbf{e} = [e]_1^k$, if the ordered reliability is given by $\widetilde{\bm\alpha} = [\widetilde{\alpha}]_1^n$, the probability mass function of the Weighted Hamming distance between $\widetilde{\mathbf{y}}$ and $\widetilde{\mathbf{c}}_{\mathbf{e}}$ is given by
    		\begin{equation} \label{equ::Stech::SR::WHDforTEP::alpha}
    		\begin{split}
	      		    &p_{D_{\mathbf{e}}^{(\mathrm{W})}}(d_{\mathbf{t}_{\mathbf{e}}^{h}}^{(\mathrm{W})}|\widetilde{\bm\alpha}) \\
	      		    &= \mathrm{Pe}(\mathbf{e}|\widetilde{\bm\alpha}) \prod_{\substack{k < u \leq n\\u\in \mathbf{t}_h^{\mathrm{P}}}} \mathrm{Pe}(u|\widetilde{\alpha}_u)\prod_{\substack{k < u \leq n\\u\notin \mathbf{t}_h^{\mathrm{P}}}} (1- \mathrm{Pe}(u|\widetilde{\alpha}_u)) \\
	      		      &+ (1-\mathrm{Pe}(\mathbf{e}|\widetilde{\bm\alpha}))\prod_{\substack{k < u \leq n\\u\in \mathbf{t}_h^{\mathrm{P}}}} \mathrm{Pc}_{\mathbf{e}}(u|\widetilde{\alpha}_u) \prod_{\substack{k < u \leq n\\u\notin \mathbf{t}_h^{\mathrm{P}}}} (1- \mathrm{Pc}_{\mathbf{e}}(u|\widetilde{\alpha}_u)),  
    		\end{split}
    		\end{equation} 
    		where $\mathrm{Pe}(\mathbf{e}|\widetilde{\bm\alpha})$ is given by (\ref{equ::HDtech::SR::Pe(e)::Cond}), $\mathrm{Pe}(u|\widetilde{\alpha}_u)$ is given by (\ref{equ::HDdistech::Pebit::Cond}), and $\mathrm{Pc}_{\mathbf{e}}(u|\widetilde{\alpha}_u)$ is given by
    		\begin{equation}   \label{equ::Stech::SR::Pce::alpha}
    		\begin{split}
    		    \mathrm{Pc}_{\mathbf{e}}(u|\widetilde{\alpha}_u) &=\! \sum_{q=1}^{k}\!\sum_{\substack{\mathbf{x} \in \{0,1\}^{k}\\ w(\mathbf{e}\oplus\mathbf{x})=q}}\!\!\!\!\mathrm{Pr}(\widetilde{\mathbf{e}}_{\mathrm{B}} \!=\! \mathbf{x}|\widetilde{\bm{\alpha}}) p_{\mathbf{c}_{\mathrm{P}}}^{\mathrm{bit}}(u,q)(1\!-\!\mathrm{Pe}(u|\widetilde{\alpha}_u) \\
    		    &+ (1 - p_{\mathbf{c}_{\mathrm{P}}}^{\mathrm{bit}}(u,q))\mathrm{Pe}(u|\widetilde{\alpha}_u),    
    		\end{split}
    		\end{equation}
	        where $p_{\mathbf{c}_{\mathrm{P}}}^{\mathrm{bit}} (u,q)$ is given by (\ref{equ::WHD::App::PcPbit1}) and $\mathrm{Pr}(\widetilde{\mathbf{e}}_{\mathrm{B}} = \mathbf{x}|\widetilde{\bm{\alpha}})$ is derived as 
	        \begin{equation} \label{equ::Stech::SR::Pce::eb=x}
	            \mathrm{Pr}(\widetilde{\mathbf{e}}_{\mathrm{B}} = \mathbf{x}|\widetilde{\bm{\alpha}}) = \prod_{\substack{1 \leq u \leq k\\ x_u \neq 0}} \mathrm{Pe}(u|\widetilde{\alpha}_u) \prod_{\substack{1 \leq u \leq k\\ x_u = 0}} (1- \mathrm{Pe}(u|\widetilde{\alpha}_u)).
	        \end{equation}
        \end{corollary}
        \begin{IEEEproof}
            The proof is provided in Appendix \ref{app::proof::Stech::SR::WHDforTEP::alpha}.
        \end{IEEEproof}   }
    
        Corollary \ref{cor::Stech::SR::WHDforTEP::alpha} describes the $\mathrm{pmf}$ of $D_{\mathbf{e}}^{(\mathrm{W})}$ with respect to TEP $\mathbf{e}$ if channel reliabilities are known. It can be found that WHD $d_{\mathbf{t}_{\mathbf{e}}^{h}}^{(\mathrm{W})} = [\mathbf{e} \   \mathbf{z}_{\mathbf{t}_h^{\mathrm{P}}}]{\widetilde{\bm\alpha}}^\mathrm{T}$ is only determined by the TEP $\mathbf{e}$ and the positions $\mathbf{t}_h^{\mathrm{P}}$ that differ between $\widetilde {\mathbf{c}}_{\mathbf{e},\mathrm{P}}$ and $\widetilde{\mathbf{y}}_{\mathrm{P}}$. In other words, $D_{\mathbf{e}}^{(\mathrm{W})} = d_{\mathbf{t}_{\mathbf{e}}^{h}}^{(\mathrm{W})} $ when the difference pattern between $\widetilde {\mathbf{c}}_{\mathbf{e}}$ and $\widetilde{\mathbf{y}}$ is given by $[\mathbf{e} \   \mathbf{z}_{\mathbf{t}_h^{\mathrm{P}}}]$. Based on Corollary \ref{cor::Stech::SR::WHDforTEP::alpha}, we give the following Corollary about the soft success probability utilizing WHD.
        {\color{black}
        \begin{corollary} \label{cor::Stech::SR::Psuce::alpha}
    		Given a linear block code $\mathcal{C}(n,k)$ and the ordered reliability observation $\bm{\widetilde\alpha} = [\widetilde\alpha]_1^n$, for a specific TEP $\mathbf{e} = [e]_1^k$, if the difference pattern between $\widetilde {\mathbf{c}}_{\mathbf{e}}$ and $\widetilde {\mathbf y}$ is given by $\widetilde{\mathbf{d}}_{\mathbf{e}} = \widetilde {\mathbf{c}}_{\mathbf{e}}\oplus \widetilde {\mathbf y} = [\widetilde{d}_{\mathbf{e}}]_1^n$, the probability that the errors in MRB are eliminated by $\mathbf{e}$ is given by
    		\begin{equation} \label{equ::Stech::SR::Psuce::alpha}
    		\begin{split}
     			&\widetilde{\mathrm{P}}_{\mathbf{e}}^{\mathrm{suc}}(\widetilde{\mathbf{d}}_{\mathbf{e}}|\bm{\widetilde\alpha})\\
     			&= \!\Bigg(\!1\! +\! \frac{1\!-\!\mathrm{Pe}(\mathbf{e}|\bm{\widetilde\alpha})}{\mathrm{Pe}(\mathbf{e}|\bm{\widetilde\alpha})}
    			\!\!\prod\limits_{\substack{k < u \leq n\\ \widetilde{d}_{\mathbf{e},u} \neq 0 }}\!\!\frac{\mathrm{Pc}_{\mathbf{e}}(u|\widetilde\alpha_u)}{\mathrm{Pe}(u|\widetilde\alpha_u)}\!\prod\limits_{\substack{k < u \leq n\\ \widetilde{d}_{\mathbf{e},u} = 0 }}\!\frac{1\!-\!\mathrm{Pc}_{\mathbf{e}}(u|\widetilde\alpha_u)}{1\!-\!\mathrm{Pe}(u|\widetilde\alpha_u)} \Bigg)^{-1} 
    		\end{split}
    	    \end{equation}
        \end{corollary}
        \begin{IEEEproof}
             Following the same step as the proof of Corollary \ref{cor::HDtech::CondDis::HDpsuc} and using Corollary \ref{cor::Stech::SR::WHDforTEP::alpha}, (\ref{equ::Stech::SR::Psuce::alpha}) can be obtained.
        \end{IEEEproof}
        }
        
            We propose the soft individual stopping rule (SISR) to terminate the decoding in advance by utilizing the WHD. After each re-encoding, given a success probability threshold  $\mathrm{P}_{t}^{\mathrm{suc}} \in [0,1]$, if the difference pattern $ \widetilde{\mathbf{d}}_{\mathbf{e}} = \widetilde{\mathbf{c}}_{\mathbf{e}} \oplus \widetilde{\mathbf{y}}$ between the generated codeword $\widetilde{\mathbf{c}}_{\mathbf{e}}$ and $\widetilde{\mathbf{y}} $ satisfies the following condition
            \begin{equation} \label{equ::Stech::SR::SISR}
                \widetilde{\mathrm{P}}_{\mathbf{e}}^{\mathrm{suc}}(\widetilde{\mathbf{d}}_{\mathbf{e}}|\bm{\widetilde\alpha}) \geq \mathrm{P}_{t}^{\mathrm{suc}},
            \end{equation}
            the decoding is terminated and the codeword estimate $\hat{\mathbf{c}}_{\mathbf{e}} = \pi_1^{-1}(\pi_2^{-1}(\widetilde{\mathbf{c}}_{\mathbf{e}}))$ is selected as the decoding output, where $\widetilde{\mathrm{P}}_{\mathbf{e}}^{\mathrm{suc}}(\widetilde{\mathbf{d}}_{\mathbf{e}}|\bm{\widetilde\alpha})$ is given by (\ref{equ::Stech::SR::Psuce::alpha}). {\color{black}Section \ref{sec::Discussion::Implementation} will further show that (\ref{equ::Stech::SR::Psuce::alpha}) is computed with $O(n)$ FLOPs when $\mathcal{C}(n,k)$ has a binomial-like weight spectrum.}
            
             Compared with the HISR, SISR terminates the decoding based on the difference pattern, rather than the number of different positions (Hamming distance), making it more accurate for estimating the probability of decoding success.

            Next, using the similar approach in Section \ref{sec::HDdistech::SR::HISR}, we give an upper bound of the decoding error rate when applying the SISR. Let us consider an order-$m$ OSD applying the SISR with a threshold $\mathrm{P}_{t}^{\mathrm{suc}}$. Given a specific reprocessing sequence $\{\mathbf{e}_1,\mathbf{e}_2,\ldots,\mathbf{e}_{b_{0:m}^{k}}\}$ (i.e., the decoder processes TEPs sequentially from $\mathbf{e}_1$ to $\mathbf{e}_{b_{0:m}^{k}}$), for an arbitrary TEP $\mathbf{e}_{j}$ ($1\leq j \leq b_{0:m}^{k}$), there exists a maximum WHD $d_{\max,\mathbf{e}_j}^{(\mathrm{W})}$ with respect to $\mathbf{e}_j$ which satisfies $\widetilde{\mathrm{P}}_{\mathbf{e}}^{\mathrm{suc}}(\widetilde{\mathbf{d}}_{\mathbf{e}}|\bm{\widetilde\alpha}) \geq \mathrm{P}_{t}^{\mathrm{suc}}$, where $d_{\max,\mathbf{e}_j}^{(\mathrm{W})} = \widetilde{\mathbf{d}}_{\mathbf{e}}\widetilde{\bm\alpha}^{\mathrm{T}}$. Let us define $d_{b,\mathbf{e}_j}^{(\mathrm{W})}$ as the mean of $d_{\max,\mathbf{e}_j}^{(\mathrm{W})}$ with respect to $\widetilde{\bm\alpha}$, then similar to (\ref{equ::HDtech::SR::HISR::db}), $d_{b,\mathbf{e}_j}^{(\mathrm{W})}$ can be derived as
            \begin{equation}
                 d_{b,\mathbf{e}_j}^{(\mathrm{W})} = \mathrm{P}_{\mathbf{e}_j}^{\mathrm{suc},-1}(\mathrm{P}_t^{\mathrm{suc}}),
            \end{equation}
            where $\mathrm{P}_{\mathbf{e}_j}^{\mathrm{suc},-1}(x)$ is the inverse function of (\ref{equ::Stech::Condis::Psuce}). Then, similar to (\ref{equ::HDtech::SR::HISR::errorRate2}), we can obtain the error rate upper bound of an order-$m$ OSD applying the SISR as 
            \begin{equation} \label{equ::Stech::SR::SISR::errorRate}
            \begin{split}
                   \epsilon_e^{\mathrm{SISR}}   = 1 - (1- \theta_{\mathrm{SISR}})\sum_{j=0}^{m}p_{E_1^{k}}(j) + \mathrm{P}_{\mathrm{ML}},
            \end{split}
            \end{equation}
            where $\theta_{\mathrm{SISR}}$ is the error rate performance loss factor of the SISR, i.e.,
            \begin{equation} \label{equ::Stech::SR::SISR::loss}
                \theta_{\mathrm{SISR}} = \frac{\sum_{j=1}^{b_{0:m}^{k}}\widetilde{\mathrm{P}}_{\mathbf{e}_j} \left( 1- \overline{\widetilde{\mathrm{P}}}_{\mathbf{e}_j}^{\mathrm{suc}}\right)}{\sum_{j=0}^{m}p_{E_1^{k}}(j)}.
            \end{equation}
            In (\ref{equ::Stech::SR::SISR::loss}), $\overline{\widetilde{\mathrm{P}}}_{\mathbf{e}_j}^{\mathrm{suc}}$ is the mean of $\widetilde{\mathrm{P}}_{\mathbf{e}}^{\mathrm{suc}}(\widetilde{\mathbf{d}}_{\mathbf{e}}|\bm{\widetilde\alpha})$ with respect to $\bm{\widetilde\alpha}$ and conditioning on $D_{\mathbf{e}_j}^{(\mathrm{W})}\!\leq\! d_{b,\mathbf{e}_j}^{(\mathrm{W})}$, i.e., $\overline{\widetilde{\mathrm{P}}}_{\mathbf{e}}^{\mathrm{suc}} \!\!= \mathrm{Pr}(\mathbf{e} = \widetilde{\mathbf{e}}_{\mathrm{B}}|D_{\mathbf{e}}^{(\mathrm{W})}\!\leq \! d_{b,\mathbf{e}}^{(\mathrm{W})})$, which is given by
            \begin{equation}
                \overline{\mathrm{P}}_{\mathbf{e}_j}^{\mathrm{suc}} \!=\! \left(\!  \int_{0}^{d_{b,\mathbf{e}_j}^{(\mathrm{W})}}\! p_{D_{\mathbf{e}_j}^{(\mathrm{W})}}(x) \ dx \right) ^{-1}\!\! \int_{0}^{d_{b,\mathbf{e}_j}^{(\mathrm{W})}}\widetilde{\mathrm{P}}_{\mathbf{e}_j}^{\mathrm{suc}}(x) p_{D_{\mathbf{e}_j}^{(\mathrm{W})}}(x) \ dx,
            \end{equation}
            and $\widetilde{\mathrm{P}}_{\mathbf{e}_j} $ is the probability of that $\widetilde{\mathbf{c}}_{\mathbf{e}_j}$ ($1 \leq j \leq b_{0:m}^k$) satisfies the SISR, which is given by
            \begin{equation} \label{equ::Stech::SR::SISR::PsucEffect}
                \widetilde{\mathrm{P}}_{\mathbf{e}_j} = \prod_{v=1}^{j-1}\left(1 - \int_0^{d_{b,\mathbf{e}_v}^{(\mathrm{W})}}f_{D_{\mathbf{e}_v}^{(\mathrm{W})}}(x)dx \right)\int_0^{d_{b,\mathbf{e}_j}^{(\mathrm{W})}}f_{D_{\mathbf{e}_j}^{(\mathrm{W})}}(x)dx.
            \end{equation}
            Particularly, $\mathrm{P}_{\mathbf{e}_1} = \int_0^{d_{b,\mathbf{e}_1}^{(\mathrm{W})}}f_{D_{\mathbf{e}_1}^{(\mathrm{W})}}(x)dx$.
            
            Similar to (\ref{equ::HDtech::SR::HISR::Na}), given a specific reprocessing sequence $\{\mathbf{e}_1,\mathbf{e}_2,\ldots,\mathbf{e}_{b_{0:m}^{k}}\}$, the average number of re-encoded TEPs, denoted by $N_a$, is derived as
            \begin{equation} \label{equ::Stech::SR::SISR::Na}
                \begin{split}
                    N_a &=  b_{0:m}^{k} \left(1-\sum_{j=1}^{b_{0:m}^{k}}\widetilde{\mathrm{P}}_{\mathbf{e}_j}\right) + \sum_{j=1}^{b_{0:m}^{k}}j \widetilde{\mathrm{P}}_{\mathbf{e}_j}.
                \end{split}
            \end{equation}

\begin{figure}[t]
    	    	\vspace{-0.8em}
                \centering
                \subfigure[Frame error rate]
                {
                    \includegraphics[scale = 0.65]{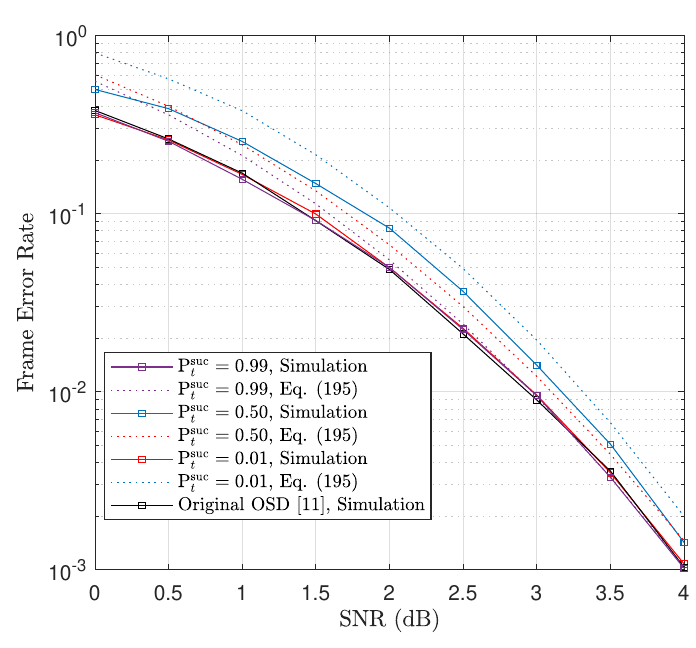}
                    \label{Fig::VII::SISR_Pe}
                }
                \vspace{-1ex}
                \subfigure[Average number of TEPs]
                {
                    \includegraphics[scale = 0.65]{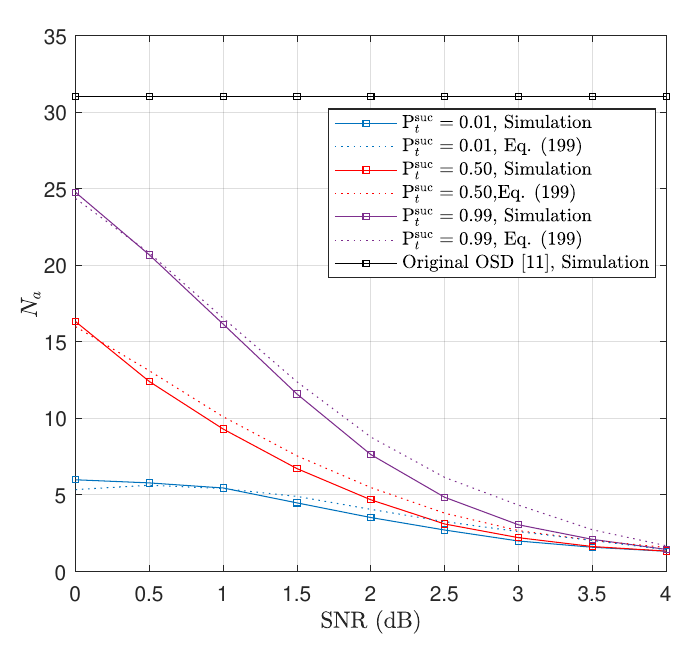}
                    \label{Fig::VII::SISR_Na}
                }
    
                \caption{Decoding $(64,30,14)$ eBCH code with an order-$1$ OSD applying the SISR.}
                \label{Fig::SISR}
            \end{figure}

            We compare the frame error rate and decoding complexity in terms of the number of TEPs $N_a$ in decoding the $(64,30,14)$ eBCH code with an order-1 OSD applying the SISR in Fig. \ref{Fig::VII::SISR_Pe} and Fig. \ref{Fig::VII::SISR_Na}, respectively. As can be seen in Fig. \ref{Fig::VII::SISR_Pe}, even for $\mathrm{P}_t^{\mathrm{suc}}=0.5$, the frame error performance exhibits no performance loss compared with the original OSD, while the number of re-encoded TEPs $N_a$ is significantly reduced. It is because for an arbitrary TEP $\mathbf{e}$, $\mathrm{P}_{\mathbf{e}}^{\mathrm{suc}} \geq \mathrm{P}_t^{\mathrm{suc}} \geq 0.5$ can ensure the codeword estimate $\widetilde{\mathbf{c}}_{\mathbf{e}}$ has higher \textit{a posterior} correct probability than other candidates. In other words, $\mathrm{P}_{\mathbf{e}}^{\mathrm{suc}} \geq \mathrm{P}_t^{\mathrm{suc}} \geq 0.5$ can be regarded as a sufficient condition of $\widetilde{\mathbf{c}}_{\mathbf{e}}$ being the best codeword estimate. It is also worthy of noting that for $\mathrm{P}_t^{\mathrm{suc}} = 0.01$, the loss of coding gain is still smaller than $0.2$ dB compared with the original OSD at error rate $10^{-3}$.
            
\begin{figure}
        		\begin{center}
        			\includegraphics[scale=0.6] {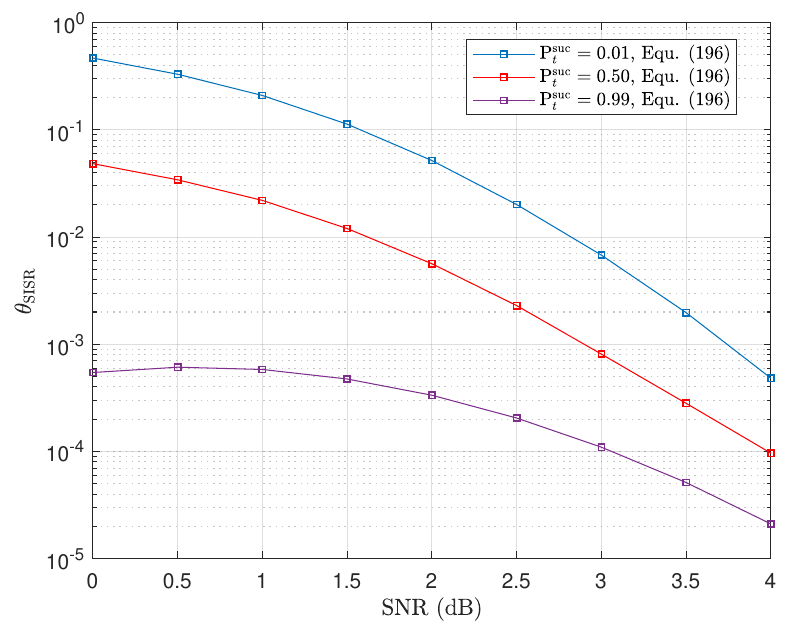}
        			\caption{The performance loss rate $\theta_{\mathrm{SISR}}$ of decoding $(64,30,14)$ eBCH code with an order-$1$ OSD applying the SISR.}
        			\label{Fig::VII::SISR_theta}
        		\end{center}
        	\end{figure}

            We illustrate the performance loss factor $\theta_{\mathrm{SISR}}$ in Fig. \ref{Fig::VII::SISR_theta}. Comparing $\theta_{\mathrm{SISR}}$ with $\theta_{\mathrm{HISR}}$ demonstrated in Fig \ref{Fig::VI::theta_HISR}, at the same channel SNR and $\mathrm{P}_t^{\mathrm{suc}}$, SISR has a lower performance loss and similar number of TEPs $N_a$. Further comparisons between SISR and HISR will be discussed in Section \ref{sec::Discussion}.
            
\subsubsection{Soft Group Stopping Rule}
        We first give an approximation of $i$-reprocessing success probability (\ref{equ::Stech::Condis::Psuci}) conditioning on $[\widetilde{A}]_1^n = [\widetilde{\alpha}]_1^n$. As introduced in Section \ref{subsec::appWHD}, the distribution of $i$-reprocessing WHD can be approximated to the ordered statistics of Gaussian distributions with positive correlation. However, given values of the ordered reliabilities $[\widetilde{A}]_1^n = [\widetilde{\alpha}]_1^n$, the WHDs between codeword estimates and the hard-decision vector are not correlated because the correlations introduced by $[\widetilde{A}]_1^n$ are removed. {\color{black}Then, based on Theorem \ref{the::WHD::iphase} and approximation (\ref{equ::WHD::App::iphase::App}), the $\mathrm{pdf}$ of $D_{i}^{(\mathrm{W})}$ after $i$-reprocessing ($0 \leq i \leq m$) can be approximated as 
        \begin{equation} \label{equ::Stech::SR::SGSR::iphase::alpha}
        	    \begin{split}
        	        &f_{D_{i}^{(\mathrm{W})}}(x|\widetilde{\bm\alpha}) \approx \sum_{v=0}^{i}p_{E_1^{k}}(v|\bm{\widetilde\alpha})\\
        	        &\cdot \left(f_{D_{\mathbf{e}}^{(\mathrm{W})}}^{\mathrm{app}}(x|\widetilde{\mathbf{e}}_{\mathrm{B}} \!=\! \mathbf{e},\widetilde{\bm\alpha}) \int_{x}^{\infty}f_{\widetilde{D}_{i}^{(\mathrm{W})}}^{\mathrm{app}}\left(u, b_{1:i}^{k}|w(\mathbf{e}_{\mathrm{B}})\!\leq \! i,\widetilde{\bm\alpha} \right) du \right. \\
        	        & + \left. f_{\widetilde{D}_{i}^{(\mathrm{W})}}^{\mathrm{app}}\left(x,b_{1:i}^{k}|w(\mathbf{e}_{\mathrm{B}})\!\leq \! i,\widetilde{\bm\alpha}\right) \int_{x}^{\infty}f_{D_{\mathbf{e}}^{(\mathrm{W})}}^{\mathrm{app}}(u|\widetilde{\mathbf{e}}_{\mathrm{B}} = \mathbf{e},\widetilde{\bm\alpha})du \right) \\
        	        &  +  \left(1 - \sum_{v=0}^{i}p_{E_1^{k}}(v|\bm{\widetilde\alpha})\right) f_{\widetilde{D}_{i}^{(\mathrm{W})}}^{\mathrm{app}}\left(x,b_{0:i}^{k} |w(\mathbf{e}_{\mathrm{B}})\!> \! i,\widetilde{\bm\alpha} \right) ,
        	    \end{split}
        	\end{equation}        
        	where $p_{E_1^{k}}(u|\bm{\widetilde\alpha})$ is given by (\ref{equ::HDtech::SR::Eab::Cond}), and $f_{\widetilde{D}_{i}^{(\mathrm{W})}}^{\mathrm{app}}\left(u, b_{1:i}^{k}|w(\mathbf{e}_{\mathrm{B}})\!\leq \! i,\widetilde{\bm\alpha} \right)$ is given by
        	\begin{equation} \label{equ::Stech::SR::SGSR::iphase::eB<i::app}
        	\begin{split}
        	    f_{\widetilde{D}_{i}^{(\mathrm{W})}}^{\mathrm{app}}\left(u, b|w(\mathbf{e}_{\mathrm{B}})\!\leq \! i,\widetilde{\bm\alpha} \right) &= \!b\!\left(\!1 \!-\! F_{D_{\mathbf{e}}^{(\mathrm{W})}}^{\mathrm{app}}(x|w(\mathbf{e}_{\mathrm{B}})\!\leq \! i,\widetilde{\bm\alpha} ) \!\right)^{b\!-\!1} \\
        	    &\cdot f_{D_{\mathbf{e}}^{(\mathrm{W})}}^{\mathrm{app}}(x|w(\mathbf{e}_{\mathrm{B}})\!\leq \! i,\widetilde{\bm\alpha}).    
        	\end{split}
        	\end{equation}
        	In (\ref{equ::Stech::SR::SGSR::iphase::eB<i::app}), $f_{D_{\mathbf{e}}^{(\mathrm{W})}}^{\mathrm{app}}(x|w(\mathbf{e}_{\mathrm{B}})\!\leq \! i,\widetilde{\bm\alpha})$ and $F_{D_{\mathbf{e}}^{(\mathrm{W})}}^{\mathrm{app}}(x|w(\mathbf{e}_{\mathrm{B}})\!\leq \! i,\widetilde{\bm\alpha})$ are respectively the $\mathrm{pdf}$ and cdf of the normal distribution $\mathcal{N}\left(\mathbb{E}[D_{\mathbf{e}}^{(\mathrm{W})}|\widetilde{\mathbf{e}}_{\mathrm{B}}\!\neq\!\mathbf{e},w(\widetilde{\mathbf{e}}_{\mathrm{B}}) \!\leq\! i,\widetilde{\bm\alpha}],\sigma_{D_{\mathbf{e}}^{(\mathrm{W})}|\widetilde{\mathbf{e}}_{\mathrm{B}}\neq\mathbf{e},w(\widetilde{\mathbf{e}}_{\mathrm{B}}) \!\leq\!i,\widetilde{\bm\alpha}}^2\right)$. In (\ref{equ::Stech::SR::SGSR::iphase::alpha}), $f_{D_{i}^{(\mathrm{W})}}^{\mathrm{app}}(x|w(\mathbf{e}_{\mathrm{B}})\!\leq \! i,\widetilde{\bm\alpha})$ is given by (\ref{equ::Stech::SR::SGSR::iphase::eB<i::app}) by replacing the condition $\{w(\mathbf{e}_{\mathrm{B}})\!\leq \! i\}$ with $\{w(\mathbf{e}_{\mathrm{B}})\! > \! i\}$ in each $\mathrm{pdf}$ and $\mathrm{cdf}$, and $f_{D_{\mathbf{e}}^{(\mathrm{W})}}^{\mathrm{app}}(x|\widetilde{\mathbf{e}}_{\mathrm{B}} \!=\! \mathbf{e},\widetilde{\bm\alpha})$ is the $\mathrm{pdf}$ of the normal distribution $\mathcal{N}\left(\mathbb{E}[D_{\mathbf{e}}^{(\mathrm{W})}|\widetilde{\mathbf{e}}_{\mathrm{B}} \!=\! \mathbf{e},\widetilde{\bm\alpha}],\sigma_{D_{\mathbf{e}}^{(\mathrm{W})}|\widetilde{\mathbf{e}}_{\mathrm{B}} \!=\! \mathbf{e},\widetilde{\bm\alpha}}^2\right)$. Therefore, to numerically compute (\ref{equ::Stech::SR::SGSR::iphase::eB<i::app}), the means and variances of $D_{\mathbf{e}}^{(\mathrm{W})}$ conditioning on $\{\widetilde{\mathbf{e}}_{\mathrm{B}}\!\neq\!\mathbf{e},w(\mathbf{e}_{\mathrm{B}})\!\leq \! i,\widetilde{\bm\alpha}\}$, $\{\widetilde{\mathbf{e}}_{\mathrm{B}}\!\neq\!\mathbf{e},w(\mathbf{e}_{\mathrm{B}})\!> \! i,\widetilde{\bm\alpha}\}$, and $\{\mathbf{e}_{\mathrm{B}} = \mathbf{e} ,\widetilde{\bm\alpha}\}$ need to be determined respectively. We take $\mathbb{E}[D_{\mathbf{e}}^{(\mathrm{W})}|\widetilde{\mathbf{e}}_{\mathrm{B}} \!=\! \mathbf{e}]$ and $\sigma_{D_{\mathbf{e}}^{(\mathrm{W})}|\widetilde{\mathbf{e}}_{\mathrm{B}} \!=\! \mathbf{e},\widetilde{\bm\alpha}}^2$ as examples; they can be respectively approximated as
        	\begin{equation}  \label{equ::Stech::SR::SGSR::iphase::eB=e::Mean}
        	\begin{split}
        	    \mathbb{E}[D_{\mathbf{e}}^{(\mathrm{W})}|\widetilde{\mathbf{e}}_{\mathrm{B}} \!=\! \mathbf{e},\widetilde{\bm\alpha}] &\approx\! \left(\!1\!-\!\frac{p_{E_1^k}(i|\widetilde{\bm\alpha})}{\sum_{v=0}^{i}p_{E_1^k}(v|\widetilde{\bm\alpha})}\!\right)\! \sum_{u=1}^{k} \mathrm{Pe}(u|\widetilde\alpha_u) \widetilde\alpha_u \\
        	    &+ \sum_{u=k+1}^{n} \mathrm{Pe}(u|\widetilde\alpha_u) \widetilde\alpha_u ,    
        	\end{split}
        	\end{equation}
        	and
   	        \begin{align}  \label{equ::Stech::SR::SGSR::iphase::eB=e::Var}
        	     & \sigma_{D_{\mathbf{e}}^{(\mathrm{W})}|\widetilde{\mathbf{e}}_{\mathrm{B}} = \mathbf{e},\widetilde{\bm\alpha}}^2 \notag\\
        	     & \approx \!\left(\!1 \!-\! \frac{p_{E_1^{k}}(i|\widetilde{\bm\alpha})\!+\!p_{E_1^{k}}(i\!-\!1|\widetilde{\bm\alpha})}{\sum_{\ell=0}^{i} p_{E_1^{k}}(\ell|\widetilde{\bm\alpha})}\!\right)\!\! \sum_{u\!=\!1}^{k}\!\sum_{v\!=\!1}^{k}\! \mathrm{Pe}(u,v|\widetilde\alpha_u,\widetilde\alpha_v)\widetilde\alpha_u\widetilde\alpha_v \notag\\
                & + 2\left(\!1 \!-\! \frac{p_{E_1^{k}}(i|\widetilde{\bm\alpha})}{\sum_{\ell=0}^{i} p_{E_1^{k}}(\ell|\widetilde{\bm\alpha})}\!\right)\!\sum_{u=1}^{k}\!\sum_{v=k\!+\!1}^{n} \!\mathrm{Pe}(u,v|\widetilde\alpha_u,\widetilde\alpha_v)\widetilde\alpha_u\widetilde\alpha_v\\
                 &+\sum_{u=k+1}^{n}\sum_{v=k+1}^{n} \mathrm{Pe}(u,v|\widetilde\alpha_u,\widetilde\alpha_v)\widetilde\alpha_u\widetilde\alpha_v\notag\\
                 &- \left(\mathbb{E}[D_{\mathbf{e}}^{(\mathrm{W})}|\widetilde{\mathbf{e}}_{\mathrm{B}} = \mathbf{e},\widetilde{\bm\alpha}]\right)^2,\notag
        	\end{align}
        	where $\mathrm{Pe}(u,v|\widetilde\alpha_u,\widetilde\alpha_v)$ is given by (\ref{equ::HDdistech::Pebit2::Cond}). Eq. (\ref{equ::Stech::SR::SGSR::iphase::eB=e::Mean}) and (\ref{equ::Stech::SR::SGSR::iphase::eB=e::Var}) follows from considering $[\widetilde{A}]_1^n = [\widetilde{\alpha}]_1^n$ in (\ref{equ::WHD::App::eB=e::Mean::App}) and (\ref{equ::WHD::App::eB=e::Var::App}), respectively. On the conditions $\{\widetilde{\mathbf{e}}_{\mathrm{B}}\!\neq\!\mathbf{e},w(\mathbf{e}_{\mathrm{B}})\!\leq \! i,\widetilde{\bm\alpha}\}$ and $\{\widetilde{\mathbf{e}}_{\mathrm{B}}\!\neq\!\mathbf{e},w(\mathbf{e}_{\mathrm{B}})\!> \! i,\widetilde{\bm\alpha}\}$, the means and variances of 
        	$D_{\mathbf{e}}^{(\mathrm{W})}$ can be obtained similarly based on (\ref{equ::WHD::App::eB!=e::eB<=i::Mean::App}) and (\ref{equ::WHD::App::eB!=e::eB<=i::Var::App}). We omit the detailed expressions for the sake of brevity.
            }
            
            From (\ref{equ::Stech::SR::SGSR::iphase::alpha}), we can obtain the soft success probability conditioning on $[\widetilde{A}]_1^n = [\widetilde{\alpha}]_1^n$. After the $i$-reprocessing, if the minimum WHD is calculated as $d_{i}^{(\mathrm{W})}$, the soft success probability of the codeword estimate $\widetilde{\mathbf{c}}_{i}$ corresponding to the minimum WHD can be calculated as
            \begin{equation} \label{equ::Stech::SR::SGSR::Psuci::alpha}
            \begin{split}
                \widetilde{\mathrm{P}}_{i}^{\mathrm{suc}}(d_{i}^{(\mathrm{W})}|\bm{\widetilde\alpha}) &= 1- \left(1 - \sum_{v=0}^{i}p_{E_1^{k}}(v|\bm{\widetilde\alpha})\right)\\
                &\cdot\frac{ f_{D_{i}^{(\mathrm{W})}}^{\mathrm{app}}\left(x,b_{0:i}^{k} |w(\mathbf{e}_{\mathrm{B}})\!> \! i,\widetilde{\bm\alpha} \right)}{f_{D_{i}^{(\mathrm{W})}| \widetilde{\bm\alpha}}(d_{i}^{(\mathrm{W})}|\widetilde{\bm\alpha})}.                
            \end{split}
            \end{equation}
            Note that $\widetilde{\mathrm{P}}_{i}^{\mathrm{suc}}(d_{i}^{(\mathrm{W})}|\bm{\widetilde\alpha})$ defined in (\ref{equ::Stech::SR::SGSR::Psuci::alpha}) is only an approximation of $\mathrm{Pr}(w(\widetilde{\mathbf{e}})\leq i|D_{i}^{(\mathrm{W})} \! = \! d_{i}^{(\mathrm{W})})$, by using the approximated $\mathrm{pdf}$ (\ref{equ::Stech::SR::SGSR::iphase::alpha}). {\color{black}In Section \ref{sec::Discussion::Implementation}, we will further show that (\ref{equ::Stech::SR::SGSR::Psuci::alpha}) can be computed with $O(n^2)$ FLOPs with simplifications.}

            Based on (\ref{equ::Stech::SR::SGSR::Psuci::alpha}), we can propose a soft group stopping rule (SGSR), which checks the success probability only after each reprocessing. With the help of the SGSR, a high-order OSD does not need to perform all reprocessing stages, but only adaptively performs several low-order reprocessings. The SGSR is described as follows. Given a predetermined threshold success probability $\mathrm{P}_t^{\mathrm{suc}} \in [0,1]$, after the $i$-reprocessing $(0\leq i \leq m)$ of an order-$m$ OSD, if the minimum WHD $d_{i}^{(\mathrm{W})} $ satisfies
            \begin{equation} \label{equ::Stech::SR::SGSR}
                \widetilde{\mathrm{P}}_{i}^{\mathrm{suc}}(d_{i}^{(\mathrm{W})}|\bm{\widetilde\alpha}) \geq \mathrm{P}_t^{\mathrm{suc}}
            \end{equation}
            the decoding is terminated and the codeword $\hat{\mathbf{c}}_{i} = \pi_1^{-1}(\pi_2^{-1}(\widetilde{\mathbf{c}}_{i}))$ is output as the decoding result, where $\widetilde{\mathrm{P}}_{i}^{\mathrm{suc}}(d_{i}^{(\mathrm{W})}|\bm{\widetilde\alpha})$ is given by (\ref{equ::Stech::SR::SGSR::Psuci::alpha}).

           {\color{black} 
            We next give an upper bound on the error rate of an order-$m$ OSD algorithm applying the SGSR. For the $i$-reprocessing ($0\leq i\leq m$), we define $d_{b,i}^{(\mathrm{W})}$ as the mean of $d_{\max,i}^{(\mathrm{W})} = \max\{d_{i}^{(\mathrm{W})} \,|\,\widetilde{\mathrm{P}}_{i}^{\mathrm{suc}}(d_{i}^{(\mathrm{W})}|\widetilde {\bm\alpha}) \geq \mathrm{P}_t^{\mathrm{suc}} \}$ with respect to $\widetilde{\bm{\alpha}}$. By considering that $\widetilde{\mathrm{P}}_{i}^{\mathrm{suc}}(x|\widetilde {\bm\alpha})$ is the variant of $\widetilde{\mathrm{P}}_{i}^{\mathrm{suc}}(x)$ given by (\ref{equ::Stech::Condis::Psuci}) conditioning on $[\widetilde{A}]_1^n = [\widetilde{\alpha}]_1^n$, $d_{b,i}^{(\mathrm{W})}$ can be derived as
            \begin{equation} \label{equ::Stech::SR::SGSR::boundWHD}
                d_{b,i}^{(\mathrm{W})} = \widetilde{\mathrm{P}}_i^{\mathrm{suc},-1}(\mathrm{P}_t^{\mathrm{suc}}),
            \end{equation}
            where $\widetilde{\mathrm{P}}_i^{\mathrm{suc},-1}(x)$ is the inverse function of $\widetilde{\mathrm{P}}_i^{\mathrm{suc}}(x)$. Then, following the approach of obtaining (\ref{equ::HDtech::SR::HGSR::errorRate}) in Section \ref{sec::HDtech::SR::HGSR}, the error rate of an order-$m$ OSD applying the SGSR, denoted by $\epsilon_{e}^{\mathrm{SGSR}}$, is upper bounded by
        	\begin{equation} \label{equ::Stech::SR::SGSR::errorRate}
        	\begin{split}
        	     \epsilon_{e}^{\mathrm{SGSR}} &\leq  1 - (1- \theta_{\mathrm{SGSR}})\sum_{j=0}^{m}p_{E_1^{k}}(j) + \mathrm{P}_{\mathrm{ML}}.
        	\end{split}
        	\end{equation}
        	 where $\theta_{\mathrm{SGSR}}$ is the error performance loss rate given by
            \begin{equation} \label{equ::Stech::SR::SGSR_loss}
                \theta_{\mathrm{SGSR}} = \frac{\sum_{j=0}^{i}\widetilde{\mathrm{P}}_{j} \left( 1- \overline{\widetilde{\mathrm{P}}}_{j}^{\mathrm{suc}}\right)}{\sum_{j=0}^{m}p_{E_1^{k}}(j)}.
            \end{equation}
            In (\ref{equ::Stech::SR::SGSR_loss}), $\widetilde{\mathrm{P}}_{j}$ and $\overline{\widetilde{\mathrm{P}}}_{j}^{\mathrm{suc}}$ are respectively given by
            \begin{equation}
                \widetilde{\mathrm{P}}_j = \prod_{v=1}^{j-1}\left(1-\int_{0}^{d_{b,v}^{(\mathrm{W})}} f_{D_{v}^{(\mathrm{W})}}(x)dx \right) \int_{0}^{d_{b,j}^{(\mathrm{W})}}  f_{D_{j}^{(\mathrm{W})}}(x) dx,
            \end{equation}
            and
            \begin{equation}
                \overline{\widetilde{\mathrm{P}}}_{j}^{\mathrm{suc}} = \int_{0}^{d_{b,j}^{(\mathrm{W})}}\!\!\widetilde{\mathrm{P}}_{j}^{\mathrm{suc}}(x) f_{D_{j}^{(\mathrm{W})}}(x)dx \left(\int_{0}^{d_{b,j}^{(\mathrm{W})}}\!\! f_{D_{j}^{(\mathrm{W})}}(x)dx\right)^{-1} \!\!.
            \end{equation}
            where $f_{D_{j}^{(\mathrm{W})}}(x)$ is the $\mathrm{pdf}$ of $D_{j}^{(\mathrm{W})}$ given by (\ref{equ::WHD::iphase}). In particular, $\mathrm{P}_0 =  \int_{0}^{d_{b,0}^{(\mathrm{W})}}  p_{D_{0}^{(\mathrm{W})}}(x) dx$. 
         
            Similar to (\ref{equ::HDtech::SR::HGSR::Na}), for an order-$m$ OSD applying the SGSR, the average number of TEPs, denoted by $N_a$, can be derived as
            \begin{equation} \label{equ::Stech::SR::SGSR::Na}
                \begin{split}
                    N_a &=  b_{0:m}^{k}\left(1 - \sum_{j=0}^{m}\widetilde{\mathrm{P}}_i\right)+ \sum_{j=0}^{m}b_{0:j}^{k}\cdot\widetilde{\mathrm{P}}_j.
                \end{split}
            \end{equation}
            }
            
\begin{figure}[t]
    	    	\vspace{-0.8em}
                \centering
                \subfigure[Frame error rate]
                {
                    \includegraphics[scale = 0.65]{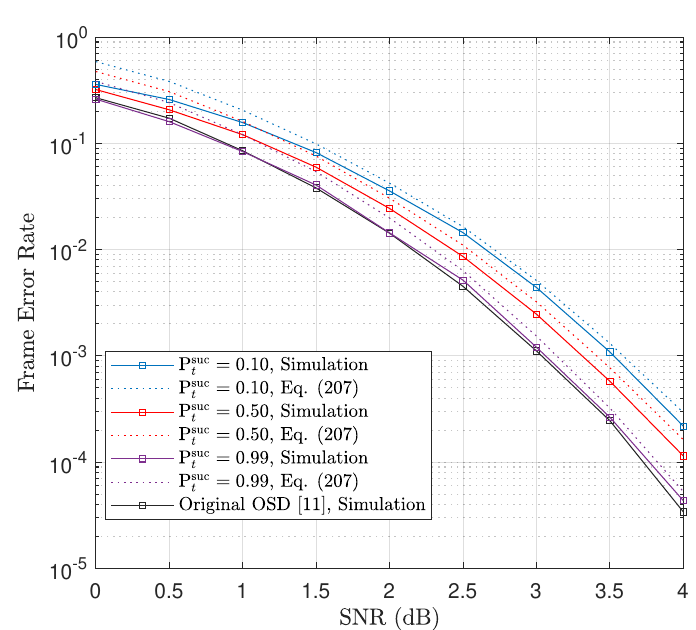}
                    \label{Fig::VII::SGSR_Pe}
                }
                \vspace{-1ex}
                \subfigure[Average number of TEPs]
                {
                    \includegraphics[scale = 0.65]{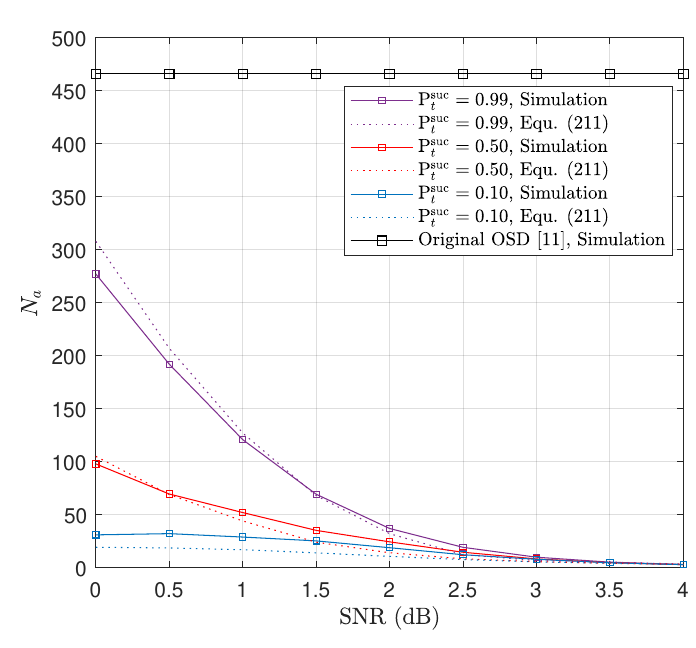}
                    \label{Fig::VII::SGSR_Na}
                }
    
                \caption{Decoding $(64,30,14)$ eBCH code with an order-$2$ OSD applying the SGSR.}
                \label{Fig::SGSR}
            \end{figure}

\begin{figure}
        		\begin{center}
        			\includegraphics[scale=0.6] {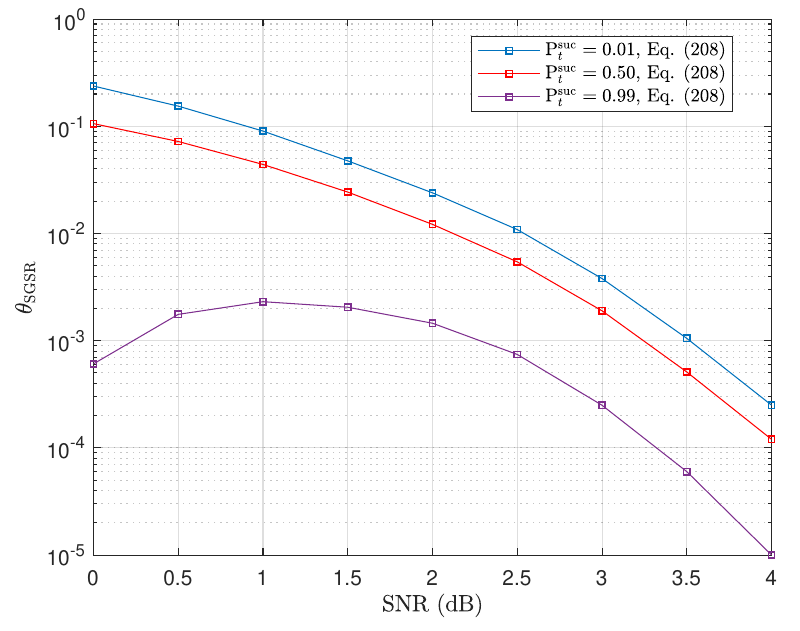}
        			\caption{The performance loss rate $\theta_{\mathrm{SGSR}}$ of decoding $(64,30,14)$ eBCH code with an order-$2$ OSD applying the HGSR.}
        			\label{Fig::VII::SGSR_theta}
        		\end{center}
        	\end{figure}

        We implemented an order-2 OSD algorithm applying the SGSR as the decoding stopping rule, where the decoder has the opportunity to be terminated early at the end of 0-reprocessing or 1-reprocessing. We illustrate the frame error rate $\epsilon_{e}^{\mathrm{SGSR}}$ and decoding complexity in terms of the average number of TEPs $N_a$ in decoding the $(64,30,12)$ eBCH code in Fig. \ref{Fig::VII::SGSR_Pe} and Fig. \ref{Fig::VII::SGSR_Na}, respectively. As can be seen in Fig. \ref{Fig::VII::SGSR_Pe}, the decoder has almost the same error rate performance as the original OSD when the threshold $\mathrm{P}^{\mathrm{suc}}_{t}$ is set to $ 0.99$, while $N_a$ is significantly reduced. In particular, $N_a$ is shown to be less than 10, when SNR reaches 3.5 dB and $\epsilon_{e}^{\mathrm{SGSR}}$ reaches $10^{-4}$. Compared with the HGSR, SGSR can help the decoder reach better error performance with a smaller $N_a$. We also illustrate the loss factor of SGSR $\theta_{\mathrm{SGSR}}$ in Fig. \ref{Fig::VII::SGSR_theta}. It can be seen that when $\mathrm{P}_t^{\mathrm{suc}} = 0.99$, the loss factor $\theta_{\mathrm{SGSR}}$ can reach $10^{-5}$ at SNR = 4 dB, indicating that SGSR has a negligible effect on the error performance according to (\ref{equ::Stech::SR::SGSR::errorRate}).

    \subsection{Discarding Rule}    
        In this Section, we introduce the soft discarding rule (SDR) based on the distribution of WHD. Compared to HDR, SDR is more accurate since it calculates the promising probability directly from the WHD. However, the computational complexity is accordingly higher.
                
        According to Corollary \ref{cor::Stech::SR::WHDforTEP::alpha}, if the ordered reliabilities of the received signal is given by $[\widetilde{A}]_1^n = [\widetilde{\alpha}]_1^n$ and the recorded minimum WHD is given by $d_{\min}^{(\mathrm{W})}$, for a specific TEP $\mathbf{e}$, the probability that $D_{\mathbf{e}}^{(\mathrm{W})}$ is less than $d_{\min}^{(\mathrm{W})}$ is given by
                \begin{equation} \label{equ::Stech::DR::Ppro}
                    \widetilde{\mathrm{P}}_{\mathbf{e}}^{\mathrm{pro}}(d_{\min}^{(\mathrm{W})}|\bm{\widetilde\alpha}) = \sum_{h=0}^{n-k} \sum_{\substack{\mathbf{t}_h^{\mathrm{P}}\in\mathcal{T}_h^{\mathrm{P}} \\ d_{\mathbf{t}_{\mathbf{e}}^{h}}^{(\mathrm{W})}<d_{\min}^{(\mathrm{W})}}} p_{D_{\mathbf{e}}^{(\mathrm{W})}}(d_{\mathbf{t}_{\mathbf{e}}^{h}}^{(\mathrm{W})}|\bm{\widetilde\alpha}),
                \end{equation}
                where $ p_{D_{\mathbf{e}}^{(\mathrm{W})}}(d_{\mathbf{t}_{\mathbf{e}}^{h}}^{(\mathrm{W})}|\bm{\widetilde\alpha})$ is given by (\ref{equ::Stech::SR::WHDforTEP::alpha}). The probability $\widetilde{\mathrm{P}}_{\mathbf{e}}^{\mathrm{pro}}(d_{\min}^{(\mathrm{W})}|\bm{\widetilde\alpha})$ is referred to as the \textit{soft promising probability} of TEP $\mathbf{e}$. {\color{black}In Section \ref{sec::Discussion::Implementation}, we will show that by introducing an approximation of $\widetilde{\mathrm{P}}_{\mathbf{e}}^{\mathrm{pro}}(d_{\min}^{(\mathrm{W})}|\bm{\widetilde\alpha})$, (\ref{equ::Stech::DR::Ppro}) can be evaluated with complexity of $O(n)$ FLOPs.}
                
        The SDR is described as follows. Given the threshold promising probability $\mathrm{P}_t^{\mathrm{pro}} \in [0,1]$ and the current recorded minimum WHD $d_{\min}^{(\mathrm{W})}$, if the soft promising probability of $\mathbf{e}$ calculated by (\ref{equ::Stech::DR::Ppro}) satisfies
                \begin{equation} \label{equ::Stech::DR::SDR}
                    \widetilde{\mathrm{P}}_{\mathbf{e}}^{\mathrm{pro}}(d_{\min}^{(\mathrm{W})}|\bm{\widetilde\alpha}) < \mathrm{P}_t^{\mathrm{pro}},
                \end{equation}
                the TEP $\mathbf{e}$ can be discarded without reprocessing.
        
        For a linear block code $\mathcal{C}(n,k)$ with truncated binomial weight spectrum,  the soft promising probability increases when $\mathrm{Pe}(\mathbf{e}| \bm{\widetilde\alpha})$ increases, which is summarized in the following proposition.
        \begin{proposition} \label{pro::Stech::DR::PproIncreasing}
            In the $i$-reprocessing ($0<i\leq m$) of the decoding of $\mathcal{C}(n,k)$ with truncated binomially distributed weight spectrum, $\widetilde{\mathrm{P}}_{\mathbf{e}}^{\mathrm{pro}}(d_{\min}^{(\mathrm{W})}|\bm{\widetilde\alpha})$ is an increasing function of $\mathrm{Pe}(\mathbf{e}|\bm{\widetilde\alpha})$.
        \end{proposition}
        \begin{IEEEproof}
            The proof is provided in Appendix \ref{app::proof::Stech::DR::PproIncreasing}.
        \end{IEEEproof}
        
         From Proposition \ref{pro::Stech::DR::PproIncreasing}, it can be seen that for an order-$m$ OSD decoder that processes the TEPs in the order $\{\mathbf{e}_1,\mathbf{e}_2,\cdots, \mathbf{e}_{b_{0:m}^{k}}\}$ satisfying $\mathrm{Pe}(\mathbf{e}_1|\bm{\widetilde\alpha})\geq \cdots \geq \mathrm{Pe}(\mathbf{e}_{b_{0:m}^{k}}|\bm{\widetilde\alpha})$, if one TEP fails in the SDR check, all following TEPs in the list can be also discarded.
        
         Next, we give simple upper bounds on the frame error rate $\epsilon_e^{\mathrm{SDR}}$ and the average number of TEPs $N_a$ of for an order-$m$ OSD employing the SDR. We assume that the decoder processes TEPs in a specific order $\{\mathbf{e}_{i,1},\mathbf{e}_{i,2},\ldots,\mathbf{e}_{i,\binom{k}{i}}\}$ in the $i$-reprocessing. Then, for the TEP $\mathbf{e}_{i,j}$, $1\leq j \leq \binom{k}{i}$, the mean of its soft promising probability with respect to $\widetilde{\bm{\alpha}}$, denoted by $\overline{\widetilde{\mathrm{P}}}_{\mathbf{e}_{i,j}}^{\mathrm{pro}}$, can be derived as
                \begin{equation}
                \begin{split}
                    \overline{\widetilde{\mathrm{P}}}_{\mathbf{e}_{i,j}}^{\mathrm{pro}} =& \mathbb{E}[ \widetilde{\mathrm{P}}_{\mathbf{e}_{i,j}}^{\mathrm{pro}}(d_{i,j}^{(\mathrm{W})}|\bm{\widetilde\alpha})]\\
                    =&\mathbb{E}[\mathrm{Pr}(D_{\mathbf{e}_{i,j}}^{(\mathrm{W})} \!<\! D_{i,j}^{(\mathrm{W})}|\bm{\widetilde\alpha})]\\
                    = & \int_0^y\int_0^{\infty} f_{D_{\mathbf{e}_{i,j}}^{(\mathrm{W})}}(x|\mathbf{e}_{i,j} = [e]_1^k) f_{D_{i,j}^{(\mathrm{W})}}(y) dy\, dx
                \end{split}
                \end{equation}
        where $D_{i,j}^{(\mathrm{W})}$ is the random variable of the minimum WHD before that $\mathbf{e}_j$ is processed, with $\mathrm{pdf}$ $f_{D_{i,j}^{(\mathrm{W})}}(y)$, and $f_{D_{\mathbf{e}_{i,j}}^{(\mathrm{W})}}(x|\mathbf{e}_{i,j}=[e]_1^k)$ is the $\mathrm{pdf}$ of $D_{\mathbf{e}_{i,j}}^{(\mathrm{W})}$ given by (\ref{equ::Stech::Condis::WHDforTEP}). However, $f_{D_{i,j}^{(\mathrm{W})}}(y)$ is difficult to be characterized because it varies with $i$ and $j$. Note that $\mathbf{e}_{i,j}$ is a TEP to be processed in the $i$-reprocessing, thus $D_{i-1}^{(\mathrm{W})} \geq D_{i,j}^{(\mathrm{W})} \geq D_{i}^{(\mathrm{W})}$ holds, where $D_{i-1}^{(\mathrm{W})}$ and $D_{i}^{(\mathrm{W})}$ are random variables representing the minimum WHDs after $(i-1)$-reprocessing and $i$-reprocessing, respectively. Thus, $\overline{\widetilde{\mathrm{P}}}_{\mathbf{e}_{i,j}}^{\mathrm{pro}}$ can be bounded by
        \begin{equation}
        \begin{split}
           \overline{\widetilde{\mathrm{P}}}_{\mathbf{e}_{i,j}}^{\mathrm{pro}} &\geq \mathrm{Pr}(D_{\mathbf{e}_{i,j}}^{(\mathrm{W})} < D_{i}^{(\mathrm{W})}) \\
           &= \int_0^y\int_0^{\infty} f_{D_{\mathbf{e}_{i,j}}^{(\mathrm{W})}}(x|\mathbf{e}_{i,j} = [e]_1^k) f_{D_{i}^{(\mathrm{W})}}(y) dy\, dx,            
        \end{split}
        \end{equation}
        and
        \begin{equation} \label{equ::Stech::DR::AvePproUpp}   
        \begin{split}
           \overline{\widetilde{\mathrm{P}}}_{\mathbf{e}_{i,j}}^{\mathrm{pro}} &\leq \mathrm{Pr}(D_{\mathbf{e}_{i,j}}^{(\mathrm{W})}\\
           &< D_{i-1}^{(\mathrm{W})}) = \int_0^y\int_0^{\infty} f_{D_{\mathbf{e}_{i,j}}^{(\mathrm{W})}}(x|\mathbf{e}_{i,j} = [e]_1^k) f_{D_{i-1}^{(\mathrm{W})}}(y) dy\, dx,    
        \end{split}
        \end{equation}     
        where $f_{D_{i}^{(\mathrm{W})}}(y)$ and $f_{D_{i-1}^{(\mathrm{W})}}(y)$ are given by (\ref{equ::WHD::iphase}). 
        
        Therefore, the average number of re-encoded TEPs can be upper bounded by 
        \begin{equation} \label{equ::Stech::DR::Na}
            N_a \leq \sum_{i=0}^{m} \beta_i^{\mathrm{upper}}, 
        \end{equation}
        where $ \beta_i^{\mathrm{upper}}$ is given by
        \begin{equation}
    	     \beta_i^{\mathrm{upper}} = \sum_{j=1}^{\binom{k}{i}}\mathbf{1}_{[\mathrm{P}_{t}^{\mathrm{pro}},+\infty]}  \mathrm{Pr}(D_{\mathbf{e}_{i,j}}^{(\mathrm{W})} < D_{i-1}^{(\mathrm{W})}).
        \end{equation}
        It can be seen that $\beta_i^{\mathrm{upper}}$ is in fact the upper bound of the number of re-encoded TEPs in the $i$-reprocessing with threshold $\mathrm{P}_{t}^{\mathrm{pro}}$. 
        
        Utilizing the inequality (\ref{equ::Stech::DR::AvePproUpp}), the decoding error performance of an order-$m$ OSD algorithm applying the SDR can be upper bounded by 
        \begin{equation} \label{equ::Stech::DR::ErrorRate}
            \epsilon_e^{\mathrm{SDR}}  \leq  \left(\ 1 - \sum_{i=0}^m\left( p_{E_1^k}(i) - \eta_{\mathrm{SDR}}(i) \right)\right) + \mathrm{P}_{\mathrm{ML}} ,
        \end{equation}
        where $\eta_{\mathrm{SDR}}(i)$ is the SDR degradation factor of $i$-reprocessing, i.e.,
        \begin{equation} \label{equ::Stech::DR::eta}
            \eta_{\mathrm{HDR}}(i) = \sum_{j=1}^{\binom{k}{i}}\left(\mathbf{1}_{[0,\mathrm{P}_{t}^{\mathrm{pro}}]}  \mathrm{Pr}(D_{\mathbf{e}_{i,j}}^{(\mathrm{W})} \!<\! D_{i}^{(\mathrm{W})})\right)\mathrm{Pe}(\mathbf{e}_{i,j}),
        \end{equation}
        for $0<i<m$. In particular, $\eta_{\mathrm{HDR}}(0)=0$ because $d_{\min}^{(\mathrm{W})}$ has not been recorded in the 0-reprocessing. From (\ref{equ::Stech::DR::eta}), it can be seen that if $\mathrm{P}_{t}^{\mathrm{pro}}=1$, $\eta_{\mathrm{HDR}}(i) = p_{E_1^k}(i)$ for $0<i<m$, then $\epsilon_e^{\mathrm{SDR}}  \leq   1 - p_{E_1^k}(0) + \mathrm{P}_{\mathrm{ML}}$ upper bounds the error rate of the $0$-reprocessing decoding. In contrast, when $\mathrm{P}_{t}^{\mathrm{pro}}=0$ and $\eta_{\mathrm{HDR}}(i) = 0$ for $0\leq i \leq m$, $\epsilon_e^{\mathrm{SDR}}  \leq   1 - \sum_{i=0}^{m}p_{E_1^k}(i) + \mathrm{P}_{\mathrm{ML}}$ is the error rate upper bound of the order-$m$ original OSD.

\begin{figure}[t]
	    	\vspace{-0.8em}
            \centering
            \subfigure[Frame error rate]
            {
                \includegraphics[scale = 0.65]{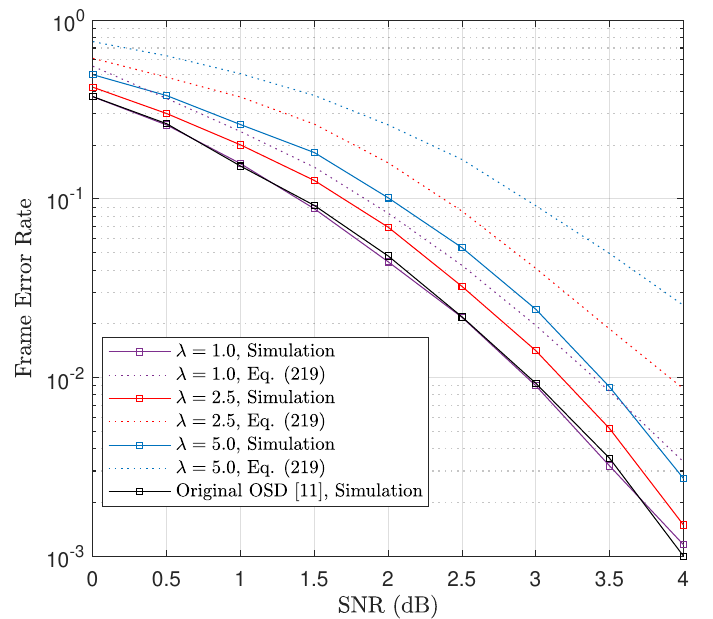}
                \label{Fig::VII::SDR_Pe}
            }
            \vspace{-1ex}
            \subfigure[Average number of TEPs]
            {
                \includegraphics[scale = 0.65]{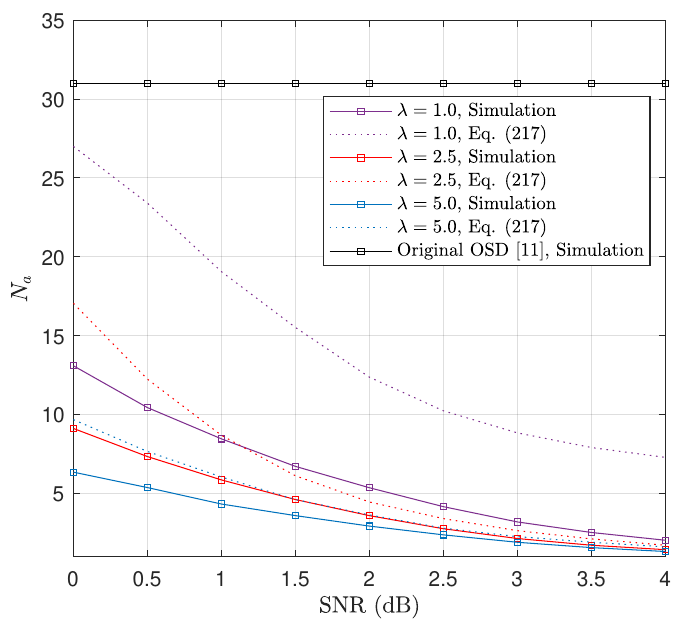}
                \label{Fig::VII::SDR_Na}
            }

            \caption{{\color{black}Decoding $(64,30,14)$ eBCH code with an order-$1$ OSD applying the SDR.}}
            \label{Fig::12864}
        \end{figure}

        Next, we demonstrate the performance of an order-1 OSD algorithm employing the SDR in terms of the decoding error probability and complexity. The threshold $\mathrm{P}_t^{\mathrm{pro}}$ is set as $\mathrm{P}_t^{\mathrm{pro}} = \lambda\frac{p_{E_1^k}(i)}{\binom{k}{i}}$, where $\lambda$ is a non-negative parameter. {\color{black}The frame error rate $\epsilon_e^{\mathrm{SDR}}$ and number of TEPs, $N_a$, with different parameter $\lambda$ in decoding the $(64,30,14)$ eBCH code are depicted in Fig. \ref{Fig::VII::SDR_Pe} and Fig. \ref{Fig::VII::SDR_Na}, respectively.} It can be seen that when $\lambda = 1$, the decoder with SDR has almost the same frame error rate performance as the original OSD, but the average number of TEPs $N_a$ is less than 5 at high SNRs, which is significantly decreased from 31 for the original OSD. Even for a higher $\lambda = 5$, the decoder can still maintain the error performance within only 0.5 dB gap to the original OSD at SNR as high as 4 dB, and the number TEP $N_a$ is reduced from 31 to less than 2. From the simulation, it can be concluded that the SDR can effectively decrease the complexity in terms of $N_a$ with a negligible loss of error performance, and the trade-off between $\epsilon_e^{\mathrm{SDR}}$ and $N_a$ can be adjusted by carefully tuning $\lambda$. However, it is hard to derive tight bounds for $\epsilon_e^{\mathrm{SDR}}$ and $N_a$ because of the difficulty in deriving $f_{D_{i,j}^{(\mathrm{W})}}(x)$. From Fig. \ref{Fig::VII::SDR_Pe} and Fig. \ref{Fig::VII::SDR_Na}, it can be seen that (\ref{equ::Stech::DR::Na}) and (\ref{equ::Stech::DR::ErrorRate}) only provide simple and loose upper bounds of $N_a$ and $\epsilon_e^{\mathrm{SDR}}$ , respectively, and they can be further tightened if $f_{D_{i,j}^{(\mathrm{W})}}(x)$ is derived accurately.

        {\color{black}
        \section{Implementation and Comparisons} \label{sec::Discussion}
        
        \subsection{Practical Implementation of the Proposed Decoding Techniques} \label{sec::Discussion::Implementation}
        
        Section \ref{sec::HDdistech} and Section \ref{Sec::SoftTech} proposed several decoding techniques to reduce the number of TEPs re-encoded in the OSD algorithm. However, it is worth to note that the overhead of the applied techniques also contributes to the overall decoding complexity. Thus, it is essential to analyze the overall complexity of the decoders when employing the proposed techniques. In this section, we show that the proposed techniques can be efficiently implemented when $\mathcal{C}(n,k)$ has a binomial-like weight spectrum. 
        
        \subsubsection{Implementation of the HISR and HGSR}
        If $\mathcal{C}(n,k)$ has the weight spectrum represented by the truncated binomial distribution, as described by (\ref{equ::HDdis::BinSpectrum}), we have obtained that $p_{\mathbf{c}_{\mathrm{P}}}(u,q)\approx \frac{1}{2^{n-k}}\binom{n-k}{u}$ in (\ref{equ::HDdis::0phase::pcpApp}) and $p_{\mathrm{W}_{\mathbf{c}_{\mathrm{P}}}}(j) \approx \frac{1}{2^{n-k}}\binom{n-k}{u}$ in (\ref{equ::HDdis::0phase::WcpApp}). Similarly, substituting $p_{\mathbf{c}_{\mathrm{P}}}(u,q)\approx \frac{1}{2^{n-k}}\binom{n-k}{j}$ into (\ref{equ::HDtech::CondDis::Wecp::w(e)=v}), we can also obtain that
        \begin{equation} \label{equ::discuss::hard::Wcep::App}
            p_{W_{\mathbf{e},\mathbf{c}_{\mathrm{P}}}}(j| w(\mathbf{e})=v) \approx \frac{1}{2^{n-k}}\binom{n-k}{j},
        \end{equation}
        which is independent of $w(\mathbf{e})=v$. Therefore, recall the HISR and the hard success probability $\mathrm{P}_{\mathbf{e}}^{\mathrm{suc}}(d_{\mathbf{e}}^{(\mathrm{H})}|\widetilde{\bm{\alpha}})$ given by (\ref{equ::HDtech::SR::HDPsuc::Cond}), $\mathrm{P}_{\mathbf{e}}^{\mathrm{suc}}(d_{\mathbf{e}}^{(\mathrm{H})}|\widetilde{\bm{\alpha}})$ can be further approximated by substituting (\ref{equ::discuss::hard::Wcep::App}) into (\ref{equ::HDtech::SR::HDPsuc::Cond}), i.e.,
        \begin{equation} \label{equ::discuss::hard::Psuce::App}
        \begin{split}
                &\mathrm{P}_{\mathbf{e}}^{\mathrm{suc}}(d_{\mathbf{e}}^{(\mathrm{H})}|\widetilde{\bm{\alpha}}) =  \mathrm{Pe}(\mathbf{e}|\widetilde{\bm\alpha})\frac{p_{E_{k+1}^{n}}(d_{\mathbf{e}}^{(\mathrm{H})}-w(\mathbf{e})|\widetilde{\bm\alpha})}{p_{D_{\mathbf{e}}^{(\mathrm{H})}}(d_{\mathbf{e}}^{(\mathrm{H})}-w(\mathbf{e})|\widetilde{\bm\alpha})}\\
                &\approx \!\!\left(\!1 \!+\! \left(\!\frac{1\!-\!\mathrm{Pe}(\mathbf{e}|\widetilde{\bm\alpha})}{\mathrm{Pe}(\mathbf{e}|\widetilde{\bm\alpha})}\!\right)\!\!\left(\!\frac{2^{k-n}}{p^{(d_{\mathbf{e}}^{(\mathrm{H})}\!-\!w(\mathbf{e}))}(1\!-\!p)^{(n\!-\!k\!-\!d_{\mathbf{e}}^{(\mathrm{H})}\!+\!w(\mathbf{e}))}}\!\right)\! \right)^{\!-1}\!\!\!\!,
        \end{split}
        \end{equation}
        where $p = \frac{1}{n-k}\sum_{u=k+1}^{n}\mathrm{Pe}(u|\widetilde{\alpha}_u)$ is the arithmetic mean of the bit-wise error probabilities of $\widetilde{\mathbf{y}}_{\mathrm{P}}$ conditioning on $[\widetilde{A}]_1^n = [\widetilde{\alpha}]_1^n$. Note that $p$ is independent of $\mathbf{e}$ and can be reused for the computations of the success probabilities of different TEPs. In addition, $\mathrm{Pe}(\mathbf{e}|\widetilde{\bm\alpha})$ given by (\ref{equ::HDtech::SR::Pe(e)::Cond}) can be computed with linear complexity in terms of the number of FLOPs. Therefore, it can be seen that by utilizing the approximation (\ref{equ::discuss::hard::Psuce::App}), the overhead of computing $\mathrm{P}_{\mathbf{e}}^{\mathrm{suc}}(d_{\mathbf{e}}^{(\mathrm{H})}|\widetilde{\bm{\alpha}})$ in checking the HISR is given by $O(n)$ FLOPs.

        In the HGSR, the hard success probability $\mathrm{P}_{i}^{\mathrm{suc}}(d_{i}^{(\mathrm{H})}|\widetilde{\bm{\alpha}})$ is calculated as (\ref{equ::HDtech::SR::HGSR::Psuc}). Eq. (\ref{equ::HDtech::SR::HGSR::Psuc}) can be simplified using the approximations of $p_{D_{i}^{(\mathrm{H})}}$ introduced in Section \ref{sec::HDdis::Numerical}. Specifically, when $\mathcal{C}(n,k)$ has a weight spectrum described as (\ref{equ::HDdis::BinSpectrum}), we have shown that the $\mathrm{pmf}$ of $D_{i}^{(\mathrm{H})}$ can be approximated by a continuous $\mathrm{pdf}$ $f_{D_{i}^{(\mathrm{H})}}(x)$ given by (\ref{equ::HDdis::iphase::App}). Then, $\mathrm{P}_{i}^{\mathrm{suc}}(d_{i}^{(\mathrm{H})}|\widetilde{\bm{\alpha}})$ in (\ref{equ::HDtech::SR::HGSR::Psuc}) can be approximated by $f_{D_{i}^{(\mathrm{H})}}(x)$, i.e., (\ref{equ::discuss::hard::Psuci::App}) on the top of the next page,
         \begin{table*} [t]
            \centering
            \begin{minipage}{\textwidth}
    		\begin{equation} \label{equ::discuss::hard::Psuci::App}
    		\begin{split}
    			\mathrm{P}_{i}^{\mathrm{suc}}(d_{i}^{(\mathrm{H})}|\widetilde{\bm{\alpha}}) &= 1- \left(1-\sum_{v=0}^{i}p_{E_1^{k}}(v|\widetilde{\bm\alpha}) \right)\frac{\sum\limits_{v=0}^{n-k}  p_{E_{k+1}^{n}}(v|\widetilde{\bm\alpha}) \cdot  p_{\widetilde W_{\mathbf{c}_{\mathrm{P}}}}(d_{i}^{(\mathrm{H})}-i,b_{0:i}^{k} | i^{(>)},v,\widetilde{\bm{\alpha}}) 	}{p_{D_i^{(\mathrm{H})}}(d_{i}^{(\mathrm{H}}|\widetilde{\bm{\alpha}})} \\
    			&\overset{(a)}{\approx} 1- \left(1-\sum_{v=0}^{i}p_{E_1^{k}}(v|\widetilde{\bm\alpha}) \right)\frac{f_{\widetilde{W}_{\mathbf{c}_{\mathrm{P}}}}(x,b_{0:i}^{k})}{f_{D_i^{(\mathrm{H})}}(d_{i}^{(\mathrm{H}}|\widetilde{\bm{\alpha}})},
    		\end{split}
    	    \end{equation}
            \medskip
            \hrule
            \end{minipage}
        \end{table*}   
	    where 
    	\begin{equation} \label{equ::discuss::hard::Psuci::App::Wcp}
    	\begin{split}
       		f_{\widetilde W_{\mathbf{c}_{\mathrm{P}}}}(x,b) &= b \  f_{W_{\mathbf{c}_{\mathrm{P}}}}(x) \left(1 - \int_{-\infty}^{x}f_{W_{\mathbf{c}_{\mathrm{P}}}}(v)dv \right)^{b-1} \\
       		&= b \  f_{W_{\mathbf{c}_{\mathrm{P}}}}(x) Q\left(\frac{2x-n+k}{\sqrt{n-k}}\right)^{b-1} .
    	\end{split}
    	\end{equation}
    	In (\ref{equ::discuss::hard::Psuci::App}), $f_{D_i^{(\mathrm{H})}}(d_{i}^{(\mathrm{H}}|\widetilde{\bm{\alpha}})$ is given by (\ref{equ::HDdis::iphase::App}) but with replacing $p_{E_1^{k}}(u)$ and $f_{E_{k+1}^{n}}(x)$ with $p_{E_1^{k}}(u|\widetilde{\bm\alpha})$ and $f_{E_{k+1}^{n}}(x|\widetilde{\bm\alpha})$, respectively, where $p_{E_1^{k}}(u|\widetilde{\bm\alpha})$ is given by (\ref{equ::HDtech::SR::Eab::Cond}) and $f_{E_{k+1}^{n}}(x|\widetilde{\bm\alpha})$ is the $\mathrm{pdf}$ of $\mathcal{N}((n-k)p,(n-k)p(1-p))$ for $p = \frac{1}{n-k}\sum_{u=k+1}^{n}\mathrm{Pe}(u|\widetilde{\alpha}_u)$. Step (a) of (\ref{equ::discuss::hard::Psuci::App}) follows from that $p_{\widetilde W_{\mathbf{c}_{\mathrm{P}}}}(j,b| i,v,\widetilde{\bm{\alpha}})$ is approximated to $f_{\widetilde W_{\mathbf{c}_{\mathrm{P}}}}(x,b)$, which is a $\mathrm{pdf}$ independent of $E_{k+1}^{n} \!=\! v$ and $[\widetilde{A}]_1^n \!=\! [\widetilde{\alpha}]_1^n$. In (\ref{equ::discuss::hard::Psuci::App::Wcp}), $f_{W_{\mathbf{c}_{\mathrm{P}}}}(x)$ is the $\mathrm{pdf}$ of $\mathcal{N}(\frac{1}{2}(n-k),\frac{1}{4}(n-k))$ given by (\ref{equ::HDdis::0phase::WcpNormalApp}). 

    	By using (\ref{equ::discuss::hard::Psuci::App}), the overhead of computing $\mathrm{P}_{i}^{\mathrm{suc}}(d_{i}^{(\mathrm{H})}|\widetilde{\bm{\alpha}})$ can be reduced. Precisely, the integral operation in computing $p_{\widetilde W_{\mathbf{c}_{\mathrm{P}}}}(j,b| i,v,\widetilde{\bm{\alpha}})$ inside $\mathrm{P}_{i}^{\mathrm{suc}}(d_{i}^{(\mathrm{H})}|\widetilde{\bm{\alpha}})$ is approximated to the $Q$-function as shown by (\ref{equ::discuss::hard::Psuci::App::Wcp}), which can be efficiently computed by its polynomial approximations, i.e., $Q(x) = e^{ax^2+bx+c}$ for $a = -0.385$, $b = -0.765$ and $c = -0.695$ \cite{QfunctionApp}. Thus, $f_{D_i^{(\mathrm{H})}}(d_{i}^{(\mathrm{H}}|\widetilde{\bm{\alpha}})$ dominates the overhead of computing (\ref{equ::discuss::hard::Psuci::App}), where the integral $\int_{x}^{\infty}f_{\widetilde W_{\mathbf{c}_{\mathrm{P}}}}(v,b) dv$ is involved (recall (\ref{equ::HDdis::iphase::App})). In the numerical integration of $\int_{x}^{\infty}f_{\widetilde W_{\mathbf{c}_{\mathrm{P}}}}(v,b) dv$, one can control the number of sub-intervals to limit complexity. For example, setting $n$ sub-intervals could provide acceptable accuracy and limit the overhead of (\ref{equ::discuss::hard::Psuci::App}) to $O(n^2)$ FLOPs.
    	
        \subsubsection{Implementation of the SISR and SGSR}

        When $\mathcal{C}(n,k)$ has the weight spectrum described by (\ref{equ::HDdis::BinSpectrum}), and $p_{\mathbf{c}_{\mathrm{P}}}(u,q)\approx \frac{1}{2^{n-k}}\binom{n-k}{u}$, the probability $p_{\mathbf{c}_{\mathrm{P}}}^{\mathrm{bit}}(\ell,q)$ given by (\ref{equ::WHD::App::PcPbit1}) can be approximated as
        \begin{equation}
        \begin{split}
              p_{\mathbf{c}_{\mathrm{P}}}^{\mathrm{bit}}(\ell,q) &= \sum_{u=0}^{n-k} \frac{u}{n-k} p_{\mathbf{c}_{\mathrm{P}}}(u,q) \\
              &\approx \sum_{u=0}^{n-k}\frac{u}{n-k}\cdot\frac{\binom{n-k}{u}}{2^{n-k}} = \frac{1}{2}.   
        \end{split}
        \end{equation}
        In other words, for an arbitrary parity bit of an arbitrary codeword from $\mathcal{C}(n,k)$, it approximately has the probability $\frac{1}{2}$ to be nonzero. Then, by taking $p_{\mathbf{c}_{\mathrm{P}}}^{\mathrm{bit}}(\ell,q) = \frac{1}{2}$ for any $k+1 \leq \ell \leq n$ and $1 \leq q \leq k$, the probability $\mathrm{Pc}_{\mathbf{e}}(u|\widetilde{\alpha}_u)$ given by (\ref{equ::Stech::SR::Pce::alpha}) can be approximated as 
        \begin{equation} \label{equ::discuss::soft::Pce::app}
        \begin{split}
             \mathrm{Pc}_{\mathbf{e}}(u|\widetilde{\alpha}_u) &\approx \sum_{q=1}^{k}\Big(\sum_{\xi = 1}^{\binom{k}{q}}\mathrm{Pr}(\widetilde{\mathbf{e}}_{\mathrm{B}} = \widetilde{\mathbf{e}}_{\mathrm{B}}^{\xi} | \widetilde{\bm{\alpha}} )\Big) \\
             &\cdot\Big(\frac{1}{2}(1-\mathrm{Pe}(u|\widetilde{\alpha}_u) + \frac{1}{2}\mathrm{Pe}(u|\widetilde{\alpha}_u) \Big) = \frac{1}{2}.
        \end{split}
        \end{equation}
        Then, substitute (\ref{equ::discuss::soft::Pce::app}) into (\ref{equ::Stech::SR::Psuce::alpha}) and the soft success probability $\widetilde{\mathrm{P}}_{\mathbf{e}}^{\mathrm{suc}}(\widetilde{\mathbf{d}}_{\mathbf{e}}|\bm{\widetilde\alpha})$ computed in the SISR can be approximated as
		\begin{equation}   \label{equ::discuss::soft::Psuce::App}
		\begin{split}
			&\widetilde{\mathrm{P}}_{\mathbf{e}}^{\mathrm{suc}}(\widetilde{\mathbf{d}}_{\mathbf{e}}|\bm{\widetilde\alpha}) \\
			&\approx \!\!\Bigg(\!\!1 \!+\! \frac{1\!-\!\mathrm{Pe}(\mathbf{e}|\bm{\widetilde\alpha})}{\mathrm{Pe}(\mathbf{e}|\bm{\widetilde\alpha})}
    			\!\!\prod\limits_{\substack{k < u \leq n\\ \widetilde{d}_{\mathbf{e},u} \neq 0 }}\frac{1}{2\mathrm{Pe}(u|\widetilde\alpha_u)}\!\prod\limits_{\substack{k < u \leq n\\ \widetilde{d}_{\mathbf{e},u} = 0 }}\frac{1}{2\!-\!2\mathrm{Pe}(u|\widetilde\alpha_u)} \!\Bigg)^{\!-1}\!\!\!\!,    
		\end{split}
	    \end{equation}
	    where $\mathrm{Pe}(\mathbf{e}|\widetilde{\bm\alpha})$ is given by (\ref{equ::HDtech::SR::Pe(e)::Cond}). As $\mathrm{Pe}(u|\widetilde\alpha_u)$ can be reused for computing $\widetilde{\mathrm{P}}_{\mathbf{e}}^{\mathrm{suc}}(\widetilde{\mathbf{d}}_{\mathbf{e}}|\bm{\widetilde\alpha})$ for different TEPs, it can be seen that (\ref{equ::discuss::soft::Psuce::App}) can be simply calculated with complexity $O(n)$ FLOPs.
	    
	    Similar to (\ref{equ::discuss::soft::Pce::app}), when $p_{\mathbf{c}_{\mathrm{P}}}(u,q)\approx \frac{1}{2^{n-k}}\binom{n-k}{u}$, we can also obtain that $p_{\mathbf{c}_{\mathrm{P}}}^{\mathrm{bit}}(\ell,h,q)\approx \frac{1}{4}$ for $k+1\leq \ell < h \leq n$. Then, recalling $\mathbb{E}[D_{\mathbf{e}}^{(\mathrm{W})}|\widetilde{\mathbf{e}}_{\mathrm{B}} \! \neq\! \mathbf{e},w(\widetilde{\mathbf{e}}_{\mathrm{B}})\!\leq\! i]$ and $\sigma^2_{D_{\mathbf{e}}^{(\mathrm{W})}|\widetilde{\mathbf{e}}_{\mathrm{B}} \neq \mathbf{e},w(\widetilde{\mathbf{e}}_{\mathrm{B}}) \leq i}$ respectively given by (\ref{equ::WHD::App::eB!=e::eB<=i::Mean::App}) and (\ref{equ::WHD::App::eB!=e::eB<=i::Var::App}), it can be observed that when $\widetilde{\mathbf{e}}_{\mathrm{B}} \!\neq\! \mathbf{e}$, the mean and variance of $D_{\mathbf{e}}^{(\mathrm{W})}$ tends to be unrelated to $w(\widetilde{\mathbf{e}}_{\mathrm{B}})$. Thus, we have $ \mathbb{E}[D_{\mathbf{e}}^{(\mathrm{W})}|\widetilde{\mathbf{e}}_{\mathrm{B}} \! \neq\! \mathbf{e}] \approx \mathbb{E}[D_{\mathbf{e}}^{(\mathrm{W})}|\widetilde{\mathbf{e}}_{\mathrm{B}} \! \neq\! \mathbf{e},w(\widetilde{\mathbf{e}}_{\mathrm{B}})\!\leq\! i] \approx \mathbb{E}[D_{\mathbf{e}}^{(\mathrm{W})}|\widetilde{\mathbf{e}}_{\mathrm{B}} \! \neq\! \mathbf{e},w(\widetilde{\mathbf{e}}_{\mathrm{B}})\!>\! i]$ and $\sigma^2_{D_{\mathbf{e}}^{(\mathrm{W})}|\widetilde{\mathbf{e}}_{\mathrm{B}} \!\neq\! \mathbf{e}} \approx \sigma^2_{D_{\mathbf{e}}^{(\mathrm{W})}|\widetilde{\mathbf{e}}_{\mathrm{B}} \neq \mathbf{e},w(\widetilde{\mathbf{e}}_{\mathrm{B}}) \leq i}\approx \sigma^2_{D_{\mathbf{e}}^{(\mathrm{W})}|\widetilde{\mathbf{e}}_{\mathrm{B}} \!\neq\! \mathbf{e},w(\widetilde{\mathbf{e}}_{\mathrm{B}})\! > \! i}$. Therefore, the $\mathrm{pdf}$ $f_{D_{i}^{(\mathrm{W})}}(x|\widetilde{\bm\alpha})$ given by (\ref{equ::Stech::SR::SGSR::iphase::alpha}) can be further approximated as 
       \begin{equation}  \label{equ::discuss::soft::WHD::iphase::app}
	    \begin{split}
	         & f_{D_{i}^{(\mathrm{W})}}(x|\widetilde{\bm\alpha}) \approx\sum_{v=0}^{i}p_{E_1^{k}}(v|\bm{\widetilde\alpha}) \\
	        &\cdot\left(f_{D_{\mathbf{e}}^{(\mathrm{W})}}^{\mathrm{app}}(x|\widetilde{\mathbf{e}}_{\mathrm{B}} \!=\! \mathbf{e},\widetilde{\bm\alpha}) \int_{x}^{\infty}f_{\widetilde{D}_{i}^{(\mathrm{W})}}^{\mathrm{app}}\left(u, b_{1:i}^{k}|\widetilde{\mathbf{e}}_{\mathrm{B}} \!\neq\! \mathbf{e},\widetilde{\bm\alpha} \right) du \right. \\
	        &+ \left. f_{\widetilde{D}_{i}^{(\mathrm{W})}}^{\mathrm{app}}\left(x,b_{1:i}^{k}|\widetilde{\mathbf{e}}_{\mathrm{B}} \!\neq\! \mathbf{e},\widetilde{\bm\alpha}\right) \int_{x}^{\infty}f_{D_{\mathbf{e}}^{(\mathrm{W})}}^{\mathrm{app}}(u|\widetilde{\mathbf{e}}_{\mathrm{B}} \!=\! \mathbf{e},\widetilde{\bm\alpha})du \right) \\
	        & +  \left(1 - \sum_{v=0}^{i}p_{E_1^{k}}(v|\bm{\widetilde\alpha})\right) f_{\widetilde{D}_{i}^{(\mathrm{W})}}^{\mathrm{app}}\left(x,b_{0:i}^{k} |\widetilde{\mathbf{e}}_{\mathrm{B}} \!\neq\! \mathbf{e},\widetilde{\bm\alpha} \right) ,
	    \end{split}
	\end{equation} 
	where
	\begin{equation} 
	\begin{split}
	    &f_{\widetilde{D}_{i}^{(\mathrm{W})}}^{\mathrm{app}}\left(x, b|\widetilde{\mathbf{e}}_{\mathrm{B}} \!\neq\! \mathbf{e},\widetilde{\bm\alpha} \right)\\
	    &= b\left(1 - F_{D_{\mathbf{e}}^{(\mathrm{W})}}^{\mathrm{app}}(x|\widetilde{\mathbf{e}}_{\mathrm{B}} \!\neq\! \mathbf{e},\widetilde{\bm\alpha} ) \right)^{b-1} f_{D_{\mathbf{e}}^{(\mathrm{W})}}^{\mathrm{app}}(x|\widetilde{\mathbf{e}}_{\mathrm{B}} \!\neq\! \mathbf{e},\widetilde{\bm\alpha}),
	\end{split}
	\end{equation}
	and $f_{D_{\mathbf{e}}^{(\mathrm{W})}}^{\mathrm{app}}(x|\widetilde{\mathbf{e}}_{\mathrm{B}} \!\neq\! \mathbf{e},\widetilde{\bm\alpha})$ and $F_{D_{\mathbf{e}}^{(\mathrm{W})}}^{\mathrm{app}}(x|\widetilde{\mathbf{e}}_{\mathrm{B}} \!\neq\! \mathbf{e},\widetilde{\bm\alpha})$ are respectively the $\mathrm{pdf}$ and cdf of  $\mathcal{N}\left(\mathbb{E}[D_{\mathbf{e}}^{(\mathrm{W})}|\widetilde{\mathbf{e}}_{\mathrm{B}}\!\neq\!\mathbf{e},\widetilde{\bm\alpha}],\sigma_{D_{\mathbf{e}}^{(\mathrm{W})}|\widetilde{\mathbf{e}}_{\mathrm{B}}\neq\mathbf{e},\\\widetilde{\bm\alpha}}^2\right)$. Based on (\ref{equ::WHD::App::eB!=e::eB<=i::Mean::App}) and (\ref{equ::WHD::App::eB!=e::eB<=i::Var::App}), $\mathbb{E}[D_{\mathbf{e}}^{(\mathrm{W})}|\widetilde{\mathbf{e}}_{\mathrm{B}}\!\neq\!\mathbf{e},\widetilde{\bm\alpha}]$ and $\sigma_{D_{\mathbf{e}}^{(\mathrm{W})}|\widetilde{\mathbf{e}}_{\mathrm{B}}\neq\mathbf{e},\\\widetilde{\bm\alpha}}^2$ are given by
	\begin{equation}   \label{equ::discuss::soft::iphase::eB!=e::mean}
        \mathbb{E}[D_{\mathbf{e}}^{(\mathrm{W})}|\widetilde{\mathbf{e}}_{\mathrm{B}} \!\neq \! \mathbf{e},\widetilde{\bm\alpha}] = \frac{b_{0:(i-1)}^{k-1}}{b_{0:i}^{k}}\sum_{u=1}^{k} \widetilde{\alpha}_u + \sum_{u=k+1}^{n} \frac{\widetilde{\alpha}_u}{2},
    \end{equation}
    and
   \begin{equation}   \label{equ::discuss::soft::iphase::eB!=e::var}
       \begin{split}
        \sigma_{D_{\mathbf{e}}^{(\mathrm{W})}|\widetilde{\mathbf{e}}_{\mathrm{B}} \neq \mathbf{e},\widetilde{\bm\alpha}}^2 &=   
         + \frac{b_{0:(i-1)}^{k-1}}{b_{0:i}^{k}}\sum_{u=1}^{k}\widetilde{\alpha}_u^2 + \frac{b_{0:(i-1)}^{k-1}}{b_{0:i}^{k}}\sum_{u=1}^{k}\sum_{v=k+1}^{n} \widetilde{\alpha}_u\widetilde{\alpha}_v \\
        & +  2\frac{b_{0:(i-2)}^{k - 2}}{b_{0:i}^{k}}\sum_{u=1}^{k-1}\sum_{v=u+1}^{k}\widetilde{\alpha}_u\widetilde{\alpha}_v \\
        &+ \sum_{u=k+1}^{n-1}\sum_{v=u}^{n}\frac{\widetilde{\alpha}_u\widetilde{\alpha}_v}{2}  \\\\
        &- \left(\mathbb{E}[D_{\mathbf{e}}^{(\mathrm{W})}|\widetilde{\mathbf{e}}_{\mathrm{B}} \!\neq \! \mathbf{e},\widetilde{\bm\alpha}]\right)^2.
        \end{split},
    \end{equation}
    Therefore, the approximation (\ref{equ::discuss::soft::WHD::iphase::app}) can be used in computing the soft success probability, i.e., $\widetilde{\mathrm{P}}_{i}^{\mathrm{suc}}(d_{i}^{(\mathrm{W})}|\bm{\widetilde\alpha})$ given in (\ref{equ::Stech::SR::SGSR::Psuci::alpha}), in the SGSR. As can be shown, $ \mathbb{E}[D_{\mathbf{e}}^{(\mathrm{W})}|\widetilde{\mathbf{e}}_{\mathrm{B}} \!\neq \! \mathbf{e},\widetilde{\bm\alpha}]$ in (\ref{equ::discuss::soft::iphase::eB!=e::mean}) is computed with complexity $O(n)$ and $\sigma_{D_{\mathbf{e}}^{(\mathrm{W})}|\widetilde{\mathbf{e}}_{\mathrm{B}} \neq \mathbf{e},\widetilde{\bm\alpha}}^2$ in (\ref{equ::discuss::soft::iphase::eB!=e::var}) is computed with complexity $O(n^2)$. In $(\ref{equ::discuss::soft::WHD::iphase::app})$, the terms $\int_{x}^{\infty}f_{D_{\mathbf{e}}^{(\mathrm{W})}}^{\mathrm{app}}(u|\widetilde{\mathbf{e}}_{\mathrm{B}} = \mathbf{e},\widetilde{\bm\alpha})du$ and $1 - F_{D_{\mathbf{e}}^{(\mathrm{W})}}^{\mathrm{app}}(x|\widetilde{\mathbf{e}}_{\mathrm{B}} \!\neq\! \mathbf{e},\widetilde{\bm\alpha}) $ can be both efficiently computed utilizing the polynomial approximation of the $Q$-function \cite{QfunctionApp}. Thus, the overhead of computing $\widetilde{\mathrm{P}}_{i}^{\mathrm{suc}}(d_{i}^{(\mathrm{W})}|\bm{\widetilde\alpha})$ will be dominated by the numerical integration $\int_{x}^{\infty}f_{\widetilde{D}_{i}^{(\mathrm{W})}}^{\mathrm{app}}\left(u, b_{1:i}^{k}|\widetilde{\mathbf{e}}_{\mathrm{B}} \!\neq\! \mathbf{e},\widetilde{\bm\alpha} \right) du$ in (\ref{equ::discuss::soft::WHD::iphase::app}). One can set the maximum number of sub-intervals to $n$ in the numerical integration, and therefore limit the overhead of computing (\ref{equ::discuss::hard::Psuci::App}) to $O(n^2)$ FLOPs.
    
    \subsubsection{Implementation of the HDR and SDR}

    Similar to (\ref{equ::discuss::hard::Psuci::App}) and (\ref{equ::discuss::hard::Psuci::App::Wcp}), after approximating $p_{E_{k+1}^{n}}(j|\widetilde{\bm\alpha})$ and $p_{W_{\mathbf{e},\mathbf{c}_{\mathrm{P}}}}(j| w(\mathbf{e})=v)$ to $f_{E_{k+1}^{n}}(x|\widetilde{\bm\alpha})$ and $f_{W_{\mathbf{c}_{\mathrm{P}}}}(x)$, respectively, the hard promising probability, i.e., $\mathrm{P}_{\mathbf{e}}^{\mathrm{pro}}(d_{\mathrm{H}}|\widetilde {\bm\alpha})$ given by (\ref{equ::HDtech::DR::Ppro}), can also be approximated as
    \begin{align}  \label{equ::discuss::hard::Ppro::app}
          \mathrm{P}_{\mathbf{e}}^{\mathrm{pro}}(d_{\mathrm{H}}|\widetilde {\bm\alpha}) &= \sum_{j=0}^{d_{\mathrm{H}}} p_{D_{\mathbf{e}}^{(\mathrm{H})}}(j|\widetilde {\bm\alpha}) \notag\\     
           & \overset{(a)}{\approx} \mathrm{Pe}(\mathbf{e}|\widetilde{\bm{\alpha}})\int_{-\infty}^{d_{\mathrm{H}}}f_{E_{k+1}^{n}}(x|\widetilde{\bm\alpha}) dx  \notag\\
           &+ (1 - \mathrm{Pe}(\mathbf{e}|\widetilde{\bm{\alpha}})) \int_{-\infty}^{d_{\mathrm{H}}} f_{W_{\mathbf{c}_{\mathrm{P}}}}(x) dx \\
           & = \mathrm{Pe}(\mathbf{e}|\widetilde{\bm{\alpha}})\left(1 \!-\! Q\left(\frac{d_{\mathrm{H}} - (n-k)p}{\sqrt{((n-k)p(1-p))}}\right)\!\right) \notag\\
           &+ (1 - \mathrm{Pe}(\mathbf{e}|\widetilde{\bm{\alpha}})) \left(1 - Q\left(\frac{2d_{\mathrm{H}}-n+k}{\sqrt{n-k}}\right)\right), \notag
    \end{align}
    where $p = \frac{1}{n-k}\sum_{u=k+1}^{n}\mathrm{Pe}(u|\widetilde{\alpha}_u)$. Thus, by using the polynomial approximations of $Q(x)$, i.e., $Q(x) = e^{ax^2+bx+c}$\cite{QfunctionApp}, $\mathrm{P}_{\mathbf{e}}^{\mathrm{pro}}(d_{\mathrm{H}}|\widetilde {\bm\alpha})$ is efficiently evaluated with complexity $O(n)$ FLOPs. Note that the approximation (\ref{equ::discuss::hard::Ppro::app}) can be further tightened by truncating the domain $\{x < 0 \}$ for $f_{E_{k+1}^{n}}(x|\widetilde{\bm\alpha})$ and $f_{W_{\mathbf{c}_{\mathrm{P}}}}(x)$ in step (a).
    
    In the SDR, the soft promising probability $\widetilde{\mathrm{P}}_{\mathbf{e}}^{\mathrm{pro}}(d_{\min}^{(\mathrm{W})}|\bm{\widetilde\alpha})$ is computed as (\ref{equ::Stech::DR::Ppro}). However, it can be noticed that (\ref{equ::Stech::DR::Ppro}) is involved with a large number of summations, which makes it hard to implement with acceptable overhead when the parity part length $n-k$ is large. Therefore, approximations have to be introduced for efficient implementation. For example, in (\ref{equ::Stech::DR::Ppro}), the $\mathrm{pmf}$ $p_{D_{\mathbf{e}}^{(\mathrm{W})}}(d_{\mathbf{t}_{\mathbf{e}}^{h}}^{(\mathrm{W})}|\bm{\widetilde\alpha})$ of $D_{\mathbf{e}}^{(\mathrm{W})}$ for a specific TEP $\mathbf{e}$ conditioning on $[\widetilde{A}]_1^n = [\widetilde{\alpha}]_1^n$ can be approximated by a continuous $\mathrm{pdf}$ using the similar approach to obtain (\ref{equ::Stech::Condis::WHDforTEP::app}). Specifically, we approximate the distribution of $D_{\mathbf{e}}^{(\mathrm{W})}$ by a $\mathrm{pdf}$ $f_{D_{\mathbf{e}}^{(\mathrm{W})}}(x|\mathbf{e} \!=\! [e]_1^k,\bm{\widetilde\alpha})$ given by (\ref{equ::discuss::soft::WHDofTEP}) on the top of the next page.

         \begin{table*} [t]
            \centering
            \begin{minipage}{\textwidth}
                \begin{equation} \label{equ::discuss::soft::WHDofTEP}
                \begin{split}
                    &f_{D_{\mathbf{e}}^{(\mathrm{W})}}(x|\mathbf{e} \!=\! [e]_1^k,\bm{\widetilde\alpha}) \\
                    &= \mathrm{Pe}(\mathbf{e}|\widetilde{\bm{\alpha}}) f_{D_{\mathbf{e}}^{(\mathrm{W})}}(x| \widetilde{\mathbf{e}}_{\mathrm{B}}\!=\!\mathbf{e} \!=\! [e]_1^k,\bm{\widetilde\alpha})  +(1-\mathrm{Pe}(\mathbf{e}|\widetilde{\bm{\alpha}}))f_{D_{\mathbf{e}}^{(\mathrm{W})}}(x| \widetilde{\mathbf{e}}_{\mathrm{B}}\!=\!\mathbf{e} \!=\! [e]_1^k,\bm{\widetilde\alpha}) \\
                     &\!=\! \frac{\mathrm{Pe}(\mathbf{e}|\widetilde{\bm{\alpha}})}{\sqrt{2\pi\sigma_{D_{\mathbf{e}}^{(\mathrm{W})}|\widetilde{\mathbf{e}}_{\mathrm{B}}=\mathbf{e} = [e]_1^k,\bm{\widetilde\alpha}}^2}}\exp\left(-\frac{(x- \mathbb{E}[D_{\mathbf{e}}^{(\mathrm{W})}|\widetilde{\mathbf{e}}_{\mathrm{B}}\!=\!\mathbf{e} \!=\! [e]_1^k,\bm{\widetilde\alpha}])^2}{2\sigma_{D_{\mathbf{e}}^{(\mathrm{W})}|\widetilde{\mathbf{e}}_{\mathrm{B}}=\mathbf{e} = [e]_1^k,\bm{\widetilde\alpha}}^2}\right) \\
                     & + \frac{1 - \mathrm{Pe}(\mathbf{e}|\widetilde{\bm{\alpha}})}{\sqrt{2\pi\sigma_{D_{\mathbf{e}}^{(\mathrm{W})}|\widetilde{\mathbf{e}}_{\mathrm{B}}\neq\mathbf{e} = [e]_1^k,\bm{\widetilde\alpha}}^2}}\exp\left(-\frac{(x- \mathbb{E}[D_{\mathbf{e}}^{(\mathrm{W})}|\widetilde{\mathbf{e}}_{\mathrm{B}}\!\neq\!\mathbf{e} \!=\! [e]_1^k,\bm{\widetilde\alpha}])^2}{2\sigma_{D_{\mathbf{e}}^{(\mathrm{W})}|\widetilde{\mathbf{e}}_{\mathrm{B}}\neq\mathbf{e} = [e]_1^k,\bm{\widetilde\alpha}}^2}\right),
                \end{split}
                \end{equation}
            \medskip
            \hrule
            \end{minipage}
        \end{table*} 
    Note that in (\ref{equ::discuss::soft::WHDofTEP}), the conditions $\{\widetilde{\mathbf{e}}_{\mathrm{B}}\!=\!\mathbf{e} \!=\! [e]_1^k,\bm{\widetilde\alpha}\}$ and $\{\widetilde{\mathbf{e}}_{\mathrm{B}}\!\neq\!\mathbf{e} \!=\! [e]_1^k,\bm{\widetilde\alpha}\}$ are different from the conditions $\{\widetilde{\mathbf{e}}_{\mathrm{B}}\!=\!\mathbf{e},\bm{\widetilde\alpha}\}$ and $\{\widetilde{\mathbf{e}}_{\mathrm{B}}\!\neq\!\mathbf{e},\bm{\widetilde\alpha}\}$ in (\ref{equ::Stech::SR::SGSR::iphase::alpha}) and (\ref{equ::discuss::soft::WHD::iphase::app}). Specifically, we assume that $\mathbf{e}$ is unknown to the decoder in (\ref{equ::Stech::SR::SGSR::iphase::alpha}) and (\ref{equ::discuss::soft::WHD::iphase::app}), while (\ref{equ::discuss::soft::WHDofTEP}) assumes that $\mathbf{e} = [e]_1^k$ is specified. Then, based on (\ref{equ::Stech::Condis::WHDforTEP::eB=e::mean}), (\ref{equ::Stech::Condis::WHDforTEP::eB=e::var}), \ref{equ::Stech::Condis::WHDforTEP::eB!=e::mean}, and (\ref{equ::Stech::Condis::WHDforTEP::eB!=e::var}) and considering $[\widetilde{A}]_1^n = [\widetilde{\alpha}]_1^n$, we can obtain that 
        \begin{equation}  
            \begin{split}
                \mathbb{E}[D_{\mathbf{e}}^{(\mathrm{w})}|\widetilde{\mathbf{e}}_{\mathrm{B}}\!= \!\mathbf{e} \!=\! [e]_1^k,\widetilde{\bm{\alpha}}] =\sum_{\substack{1\leq u\leq k \\ e_u \neq 0}} \widetilde{\alpha}_u +  \sum_{u=k+1}^{n} \mathrm{Pe}(u|\widetilde{\alpha}_u) \widetilde{\alpha}_u ,
            \end{split}
        \end{equation}
        \begin{equation}  
        	\begin{split}
        		\sigma_{D_{\mathbf{e}}^{(\mathrm{w})}|\widetilde{\mathbf{e}}_{\mathrm{B}}= \mathbf{e} = [e]_1^k,\widetilde{\bm{\alpha}}}^2 & = \sum_{u=k+1}^{{n}} \sum_{v=k+1}^{{n}} \mathrm{Pe}(u,v|\widetilde{\alpha}_u,\widetilde{\alpha}_v) \widetilde{\alpha}_u\widetilde{\alpha}_v  \\
        		&- \left(\sum_{u=k+1}^{n} \mathrm{Pe}(u|\widetilde{\alpha}_u) \widetilde{\alpha}_u \right)^2 \ ,
        	\end{split}
        \end{equation}
        and
        \begin{equation}
            \begin{split}
                \mathbb{E}[D_{\mathbf{e}}^{(\mathrm{w})}|\widetilde{\mathbf{e}}_{\mathrm{B}}\!\neq \!\mathbf{e} \!=\! [e]_1^k,\widetilde{\bm{\alpha}}] =\sum_{\substack{1\leq u\leq k \\ e_u \neq 0}} \widetilde{\alpha}_u +  \sum_{u=k+1}^{n} \frac{\widetilde{\alpha}_u}{2},
            \end{split}
        \end{equation}
        \begin{equation} 
        \begin{split}
    		\sigma_{D_{\mathbf{e}}^{(\mathrm{w})}|\widetilde{\mathbf{e}}_{\mathrm{B}}\neq \mathbf{e} = [e]_1^k,\widetilde{\bm{\alpha}}}^2 &=   \sum_{u=k+1}^{{n-1}} \sum_{v=u}^{{n}} \frac{ \widetilde{\alpha}_u\widetilde{\alpha}_v}{2}   - \left(\sum_{u=k+1}^{n} \frac{\widetilde{\alpha}_u}{2}\right)^2 \!\!\!,
        \end{split}
        \end{equation}
        where $\mathrm{Pe}(u|\widetilde{\alpha}_u)$ and $\mathrm{Pe}(u,v|\widetilde{\alpha}_u,\widetilde{\alpha}_v)$ are respectively given by (\ref{equ::HDdistech::Pebit::Cond}) and (\ref{equ::HDdistech::Pebit2::Cond}). Particularly, $\mathrm{Pe}(u,v|\widetilde{\alpha}_u,\widetilde{\alpha}_v) = \mathrm{Pe}(u|\widetilde{\alpha}_u)$ for $u=v$. 
        
        Using $f_{D_{\mathbf{e}}^{(\mathrm{W})}}(x|\mathbf{e} \!=\! [e]_1^k,\bm{\widetilde\alpha})$ given by (\ref{equ::discuss::soft::WHDofTEP}), we approximate the soft promising probability given by (\ref{equ::Stech::DR::Ppro}) as
        \begin{align}  \label{equ::discuss::soft::Ppro::app}
                &\widetilde{\mathrm{P}}_{\mathbf{e}}^{\mathrm{pro}}(d_{\min}^{(\mathrm{W})}|\bm{\widetilde\alpha}) \notag\\
                &= \sum_{h=0}^{n-k} \sum_{\substack{\mathbf{t}_h^{\mathrm{P}}\in\mathcal{T}_h^{\mathrm{P}} \\ d_{\mathbf{t}_{\mathbf{e}}^{h}}^{(\mathrm{W})}<d_{\min}^{(\mathrm{W})}}} p_{D_{\mathbf{e}}^{(\mathrm{W})}}(d_{\mathbf{t}_{\mathbf{e}}^{h}}^{(\mathrm{W})}|\bm{\widetilde\alpha}) \notag\\
                &\overset{(a)}{\approx} \int_{-\infty}^{d_{\min}^{(\mathrm{W})}} f_{D_{\mathbf{e}}^{(\mathrm{W})}}(x|\mathbf{e} \!=\! [e]_1^k,\bm{\widetilde\alpha}) dx \\
                & \overset{(b)}{=} \mathrm{Pe}(\mathbf{e}|\widetilde{\bm{\alpha}})\left(1 - Q\left(\frac{d_{\min}^{(\mathrm{W})} - \mathbb{E}[D_{\mathbf{e}}^{(\mathrm{w})}|\widetilde{\mathbf{e}}_{\mathrm{B}}\!= \!\mathbf{e} \!=\! [e]_1^k,\widetilde{\bm{\alpha}}]}{\sigma_{D_{\mathbf{e}}^{(\mathrm{w})}|\widetilde{\mathbf{e}}_{\mathrm{B}}= \mathbf{e} = [e]_1^k,\widetilde{\bm{\alpha}}}}\right)\right) \notag\\
                &+ \!(1 \!-\! \mathrm{Pe}(\mathbf{e}|\widetilde{\bm{\alpha}})) \!\left(\!1 \!-\! Q\!\left(\!\frac{d_{\min}^{(\mathrm{W})}\!-\!\mathbb{E}[D_{\mathbf{e}}^{(\mathrm{w})}|\widetilde{\mathbf{e}}_{\mathrm{B}}\!\neq \!\mathbf{e} \!=\! [e]_1^k,\widetilde{\bm{\alpha}}]}{\sigma_{D_{\mathbf{e}}^{(\mathrm{w})}|\widetilde{\mathbf{e}}_{\mathrm{B}}\neq \mathbf{e} = [e]_1^k,\widetilde{\bm{\alpha}}}}\!\right)\!\right)\!,\notag
        \end{align}
        where step (a) approximate the summation of the $\mathrm{pmf}$ of a discrete variable to the $\mathrm{cdf}$ of a continuous distribution. Note that although $D_{\mathbf{e}}^{(\mathrm{W})} \geq 0$, step (a) does not exclude the domain $\{x<0\}$ for $f_{D_{\mathbf{e}}^{(\mathrm{W})}}(x|\mathbf{e} \!=\! [e]_1^k,\bm{\widetilde\alpha})$ for the sake of simplicity, and (\ref{equ::discuss::soft::Ppro::app}) can be further tightened by truncating the domain $\{x<0\}$. Step (b) of (\ref{equ::discuss::soft::Ppro::app}) converts the $\mathrm{cdf}$s of normal distributions as $Q$-functions, which can be efficiently computed by the polynomial approximation \cite{QfunctionApp}. Therefore, considering that $\sigma_{D_{\mathbf{e}}^{(\mathrm{w})}|\widetilde{\mathbf{e}}_{\mathrm{B}}= \mathbf{e} = [e]_1^k,\widetilde{\bm{\alpha}}}^2 $ and $\sigma_{D_{\mathbf{e}}^{(\mathrm{w})}|\widetilde{\mathbf{e}}_{\mathrm{B}}= \mathbf{e} = [e]_1^k,\widetilde{\bm{\alpha}}}^2 $ are independent of TEP $\mathbf{e}$ and can be reused in computing (\ref{equ::discuss::soft::Ppro::app}), and $\mathbb{E}[D_{\mathbf{e}}^{(\mathrm{w})}|\widetilde{\mathbf{e}}_{\mathrm{B}}\!= \!\mathbf{e} \!=\! [e]_1^k,\widetilde{\bm{\alpha}}]$ and $\mathbb{E}[D_{\mathbf{e}}^{(\mathrm{w})}|\widetilde{\mathbf{e}}_{\mathrm{B}}\!\neq \!\mathbf{e} \!=\! [e]_1^k,\widetilde{\bm{\alpha}}]$ are simply computed with $O(n)$ FLOPs, the overhead of computing $\mathrm{P}_{\mathbf{e}}^{\mathrm{pro}}(d_{\mathrm{H}}|\widetilde {\bm\alpha})$ in the SDR can be as low as $O(n)$ FLOPs
        
        \subsection{Overall Complexity Analysis} \label{sec::Discussion::Complexity}
        Next, we evaluate the overall computational complexity of OSD algorithms applying the proposed decoding techniques when $\mathcal{C}(n,k)$ has the binomial weight spectrum as described in (\ref{equ::HDdis::BinSpectrum}). Let the $C_{\mathrm{total}}$ denote the computational complexity of an OSD algorithm applying one of stopping rules (including the HISR, HGSR, SISR, and SGSR) and one of discarding rules (including the HDR and SDR). $C_{\mathrm{total}}$ can be derived as
        \begin{equation}
        \begin{split}
            C_{\mathrm{total}} &=   O(n) + \underbrace{O(n\log n)}_{\text{sorting (FLOP)}} + \underbrace{O(n\min(n,n-k))}_{\text{Gaussian elimination (BOP)}} \\
            &+ N_{a}\underbrace{O(k+k(n-k))}_{\text{re-encoding (BOP)}} + C_{\mathrm{SR}} + C_{\mathrm{DR}},        
        \end{split}
        \end{equation}
        where $N_{a}$ is the average number of re-encoded TEPs, $C_{\mathrm{SR}}$ and $C_{\mathrm{DR}}$ are the complexity of checking stopping rules and discarding rules, respectively, and other terms are the complexity of various stages in the original OSD \cite{Fossorier1995OSD}. Stopping rules and discarding rules are used to reduce the number of TEPs, $N_a$, so that the total number of re-encodings, each with complexity of $O(k+k(n-k))$ binary operations (BOPs), can be decreased. Let $N_{a} = N_{\max} - N_{s}$, where $N_{\max}$ is the maximum TEP number (i.e., number of TEPs required of original OSD) and $N_{s}$ is the number of TEPs reduced by applying stopping rules and discarding rules. The simulations in Section \ref{sec::HDdistech} and Section \ref{Sec::SoftTech} have shown that the proposed techniques can significantly reduce the number of re-encoded TEPs, i.e., $N_{a} \ll N_{s} < N_{\max}$. Therefore, if $C_{\mathrm{SR}} + C_{\mathrm{DR}}$ is negligible compared to $N_{s}\cdot O(k+k(n-k))$, i.e., $C_{\mathrm{SR}} + C_{\mathrm{DR}} \ll N_{s}\cdot O(k+k(n-k))$ , the overall computational complexity can be effectively reduced compared to the original OSD, i.e.,
        \begin{equation}
        \begin{split}
            C_{\mathrm{OSD}} &=   O(n) + \underbrace{O(n\log n)}_{\text{sorting (FLOP)}} + \underbrace{O(n\min(n,n-k))}_{\text{Gaussian elimination (BOP)}}\\
            &+ N_{\max}\underbrace{O(k+k(n-k))}_{\text{re-encoding (BOP)}}.           
        \end{split}
        \end{equation}  
           
        \subsubsection{Complexity Introduced by Stopping Rules} In our paper, the stopping rules can be implemented by one of the HISR, HGSR, SISR and SGSR. Commonly, in these four different techniques, a success probability is calculated first, and then the success probability is compared with a threshold. Let us denote the complexity $ C_{\mathrm{SR}}$ of stopping rules as
           \begin{equation}
                C_{\mathrm{SR}} = N_{\mathrm{suc}} \cdot C_{\mathrm{suc}},
           \end{equation}
           where $C_{\mathrm{suc}}$ is the complexity of calculating a single success probability, and $N_{\mathrm{suc}}$ is the number of success probabilities that are calculated.
           
         In the HISR and SISR as described in (\ref{equ::HDtech::SR::HISR}) and (\ref{equ::Stech::SR::SISR}), the success probabilities are calculated for each generated codeword estimate, and is compared with a threshold parameter to determine whether the best codeword estimate has been found. Thus, the success probabilities in the HISR and SISR can only be computed for a codeword estimate $\widetilde{\mathbf{c}}_{\mathbf{e}}$, when $\widetilde{\mathbf{c}}_{\mathbf{e}}$ results in a lower WHD $d_{\mathbf{e}}^{(\mathrm{W})}$ compared to the recorded minimum WHD $d_{\min}^{(\mathrm{W})}$. This is because $\widetilde{\mathbf{c}}_{\mathbf{e}}$ cannot be the best output if $d_{\mathbf{e}}^{(\mathrm{W})} > d_{\min}^{(\mathrm{W})}$. Therefore, for the HISR and SISR, it can be concluded that $N_{\mathrm{suc}} < N_{a} \ll N_{s} < N_{\max}$. 
            
         Furthermore, in the HISR and SISR, the success probabilities can be calculated according to (\ref{equ::discuss::hard::Psuce::App}) and (\ref{equ::discuss::soft::Psuce::App}), respectively, each with complexity $O(n)$ FLOPs. Usually, it is a few times slower to run a FLOP than a BOP by a modern processor; nevertheless, modern processors have narrowed the gap between FLOPs and BOPs with float process units (FPU) \cite{langer1998comparison}. Thus, let us assume that $O(n)_{(\mathrm{FLOP})} \approx O(k+k(n-k))_{(\mathrm{BOP})}$, i.e., we roughly take that the FLOP is about $\frac{n}{4}$ times slower than the BOP for $k \approx \frac{n}{2}$, which is reasonable when $n$ is not too small. Then, it can be still observed that $C_{\mathrm{SR}} = N_{\mathrm{suc}} \cdot O(n) \ll N_{s}\cdot O(k+k(n-k)) $ as $N_{\mathrm{suc}}  \ll N_{s}$. Therefore, the HISR and SISR can be implemented in OSD to effectively reduce the overall decoding complexity.
           
        In the HGSR and SGSR, as described in (\ref{equ::HDtech::SR::HGSR}) and (\ref{equ::Stech::SR::SGSR}), the success probability is calculated at the end of each order of reprocessing, so that $N_{\mathrm{suc}} \leq m$, where $m$ is the maximum reprocessing order of OSD. Thus, only a small number of success probabilities need to be calculated in the HGSR and SGSR, because the decoder is asymptotically optimal when $m = \lfloor d_{\mathrm{H}}/4-1 \rfloor$ \cite{Fossorier1995OSD}. Then, it can be found that $N_{\mathrm{suc}} \ll N_{\mathrm{a}} \ll N_{s} < N_{\max}$. However, the success probabilities calculated in the HGSR and SGSR could be time-consuming. As shown by (\ref{equ::discuss::hard::Psuci::App::Wcp}) and (\ref{equ::discuss::soft::WHD::iphase::app}), the success probabilities in the HGSR and SGSR involve numerical integration and could be computed with $O(n^2)$ FLOPs when limiting the maximum number of sub-intervals to $n$. Recall $N_{\mathrm{suc}} \leq m$, then it can be seen that $C_{\mathrm{SR}}$ for the HGSR and SGSR will be negligible compared with $N_{s}\cdot O(k+k(n-k))$ when $\frac{m}{N_{s}} \ll \frac{O(k+(n-k)k)_{(\mathrm{BOP})}}{O(n^2)_{(\mathrm{FLOP})}}$. By assuming the FLOP is about $\frac{n}{4}$ times slower than the BOP for $k \approx \frac{n}{2}$, it can be approximately obtained that $ \frac{O(k+(n-k)k)_{(\mathrm{BOP})}}{O(n^2)_{(\mathrm{FLOP})}}\approx \frac{1}{n}$. Therefore, when $nm \ll N_s$, the HGSR and SGSR could effectively reduce the overall decoding complexity. For example, as shown in Fig. \ref{Fig::VII::SGSR_Na}, the SGSR reduces the number of TEPs from over 450 to less than 10 in decoding $(64,30,14)$ eBCH code with $m=2$. In this case, $N_s = 440 > nm = 128$ and the SGSR could indeed reduce the overall complexity.
           
        \subsubsection{Complexity Introduced by Discarding Rules} The discarding rules can be implemented by one of the HDR and SDR. As described in (\ref{equ::HDtech::DR::HDR}) and (\ref{equ::Stech::DR::SDR}), a promising probability is calculated in HDR and SDR before re-encoding a TEP, and the promising probability is compared with a threshold to determine whether the TEP can be discarded. Thus, let us denote the complexity $ C_{\mathrm{DR}}$ of discarding rules as
           \begin{equation}
                C_{\mathrm{DR}} = N_{\mathrm{pro}} \cdot C_{\mathrm{pro}},
           \end{equation}
           where $C_{\mathrm{pro}}$ is the complexity of calculating a single promising probability and $N_{\mathrm{pro}}$ is the number of promising probabilities that are being calculated. 
           
            According to Proposition 1 and Proposition 2, the promising probabilities in the HDR and SDR are monotonically increasing functions of the reliability of TEPs. Thus, if the decoder re-encodes TEPs in descending order of their reliabilities, the HDR and SDR do not need to calculate the promising probability for each TEPs, but can discard all following TEPs when one TEP fails in the promising probability check. In this case, we can see that $N_{\mathrm{pro}} = N_a \ll N_s$. Note that TEPs are ordered according to the received reliability (channel outputs), which can be efficiently implemented following the algorithm introduced in \cite{improvedTEP}. For long block codes, it is also possible to use the algorithm in \cite{improvedTEPwithMean} to further improve the efficiency.
            
           Furthermore, utilizing the monotonicity of the promising probabilities, the decoder can further reduce $N_{\mathrm{pro}}$. Precisely, the promising probabilities can be calculated every $\ell$ TEPs, where $\ell$ is a positive integer, so that $ N_{\mathrm{pro}}$ can be as low as $\frac{N_a}{\ell}$, but the average complexity will not be apparently increased for $\ell\ll N_{a}$. We refer to this implementation as ``$\ell$-step discarding rule''. For example, let us assume $\ell = 5$, and the decoder calculates the promising probability every 5 TEPs. Because the decoder will discard all following TEPs when a TEP fails in the discarding rule check, the $N_a$ will not be apparently affected by the ``5-step discarding rule'' implementation. However, $N_{\mathrm{pro}}$ is reduced to $N_{\mathrm{pro}} = \frac{N_a}{5} \ll N_s$, and $C_{\mathrm{DR}} = N_{\mathrm{pro}} \cdot C_{\mathrm{pro}}$ can be reduced by 5 times accordingly.
           
        In both HDR and SDR, we have shown in Section \ref{sec::Discussion::Implementation} that the promising probability can be calculated with $O(n)$ FLOPs. Therefore, the overhead satisfies $C_{\mathrm{DR}}  = \frac{N_a}{\ell}\cdot O(n)_{(\mathrm{FLOP})} \ll  N_{s}\cdot O(k+k(n-k))_{(\mathrm{BOP})} $ by assuming that the FLOP is about $\frac{n}{4}$ times slower than the BOP. Hence, the HDR and SDR can effectively reduce the overall decoding complexity . 
        
        \subsection{Comparisons with state of the art} \label{sec::Discussion::Cmp}
        
        \subsubsection{Comparison of Stopping Rules}
        
        In this section, we compare the stopping rules proposed in Section \ref{sec::HDdistech::SR} and Section \ref{sec::Stech::SR} with previous approaches introduced in \cite{lin2004ECC} and \cite{jin2006probabilisticConditions}. In \cite[Theorem 10.1]{lin2004ECC}, a decoding optimality condition was proposed to terminate the decoding early. Specifically, it has been proved that for a codeword estimate $\widetilde{\mathbf{c}}_{\mathbf{e}}$ in OSD, if the following condition
        \begin{equation}  \label{equ::Discussion::Cmp::ConOpt}
            d_{\mathbf{e}}^{(\mathrm{W})} \leq g(\widetilde{\mathbf{c}}_{\mathbf{e}} , d_{\mathrm{H}}),
        \end{equation}
        is satisfied, $\hat{\mathbf{c}}_{\mathbf{e}} = \pi_1^{-1} (\pi_2^{-1}(\widetilde{\mathbf{c}}_{\mathbf{e}}$)) is the maximum-likelihood estimate of the received sequence, where $d_{\mathrm{H}}$ is the minimum distance of $\mathcal{C}(n,k)$, and $g(\widetilde{\mathbf{c}}_{\mathbf{e}} , d_{\mathrm{H}})$ is given by \cite[Eq. (10.31)]{lin2004ECC}. It has been proved that (\ref{equ::Discussion::Cmp::ConOpt}) is a rigorous sufficient condition of the maximum-likelihood decoding \cite{lin2004ECC}. On the other hand, the trade-off between complexity and error rate cannot be tuned as no parameters are introduced. In the subsequent comparisons, we refer to (\ref{equ::Discussion::Cmp::ConOpt}) as the decoding optimality condition (DOC).
        
        In \cite{jin2006probabilisticConditions}, a probabilistic sufficient condition (PSC) on optimality for reliability based decoding was proposed. The PSC was also integrated with the decoder proposed in \cite{NewOSD-5GNR}. In the PSC, a syndrome-like index is calculated as 
        \begin{equation}
            p_{sc} = [\widetilde{\mathbf{y}}_{\mathrm{B}}\oplus \mathbf{e} \  \ \widetilde{\mathbf{y}}_{\mathrm{P}}]\widetilde{\mathbf{H}}^{\mathrm{T}},
        \end{equation}
        where $\widetilde{\mathbf{H}}$ is the ordered parity matrix corresponding to $\widetilde{\mathbf{G}}$. Then, $p_{sc}$ is compared with a parameter $\tau$, and the decoding is terminated if $w(p_{sc}) \leq \tau$. Authors of \cite{jin2006probabilisticConditions} have shown that the probability of the ``False alarm'' of PSC can be negligible when $\tau$ is carefully selected. Furthermore, $\tau$ provides the flexibility between the complexity and error rate.
        
        Next, we compare the complexity of decoders with different stopping rules. The DOC \cite{lin2004ECC} and PSC \cite{jin2006probabilisticConditions} are included as benchmarks and the HISR, HGSR, SISR, SGSR are compared. We consider the order-3 decoding of $(64,30,14)$ eBCH codes, which reaches the near-maximum-likelihood error performance \cite{Fossorier1995OSD}. All decoders are fine-tuned to reach the same error performance as the original OSD \cite{Fossorier1995OSD} which applies no stopping conditions, and the sequence of TEPs are arranged in descending order of the reliabilities. The average number of processed TEPs are compared in Fig. \ref{Fig::VIII::SR-Na}. As can be seen, the proposed stopping techniques can significantly reduce the number of required TEPs compared to the DOC \cite{lin2004ECC} and PSC \cite{jin2006probabilisticConditions}. Furthermore, the soft conditions (i.e., SISR and SGSR) outperform the hard conditions (i.e., HIHR and HGSR). 
        
        The average decoding times for decoding a single codeword are further compared using MATLAB implementation on a 3.0 GHz CPU, as depicted in Fig. \ref{Fig::VIII::SR-time}. It can be seen that the SISR and SGSR can reduce the decoding time to less than 10 ms. However, the HGSR is not competitive in decoding time as it has the worst performance at low SNRs, where its overhead undermines the advantages.  It is worth noting that the HGSR, DOC, and PSC require a longer time to decode a codeword than the original OSD at low SNRs.
        
        The numbers of TEPs and decoding times of applying different stopping rules are recorded in Table \ref{tab::SR}.
        
        \begin{figure}[t]
	    	\vspace{-0.8em}
            \centering
            \subfigure[Average Number of TEP]
            {
                \includegraphics[scale = 0.65]{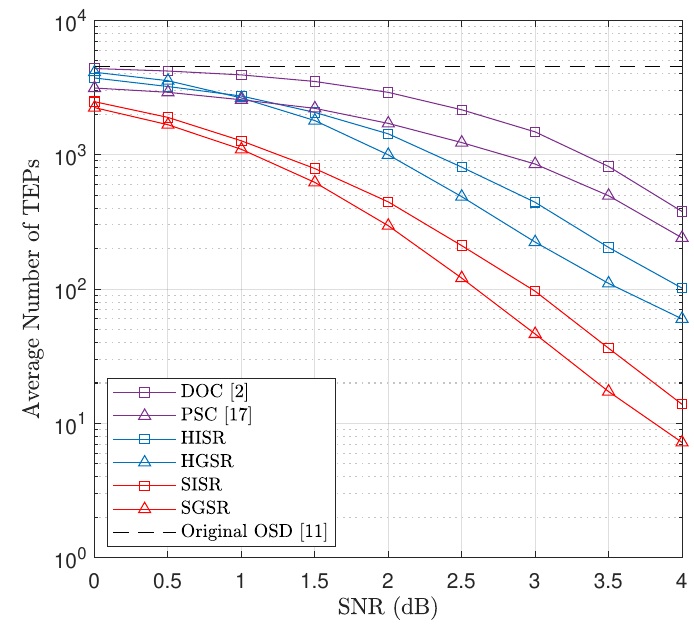}
                \label{Fig::VIII::SR-Na}
            }
            \vspace{-1ex}
            \subfigure[Average Decoding Time]
            {
                \includegraphics[scale = 0.65]{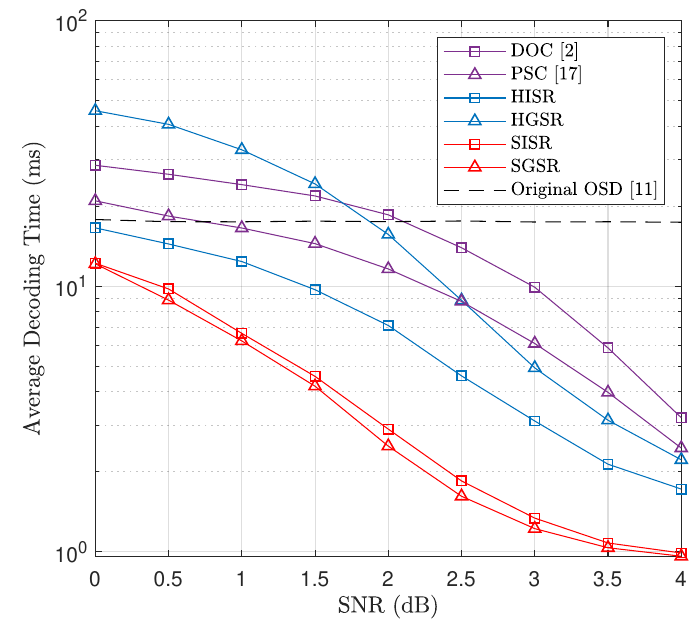}
                \label{Fig::VIII::SR-time}
            }

            \caption{{\color{black}Decoding $(64,30,14)$ eBCH code with order-$3$ OSD algorithms applying different stopping rules.}}
            \label{Fig::SR}
        \end{figure}

        \begin{table}[]
        \small	
    	\centering
    	\tabcolsep=0.11cm
    	\caption{{\color{black}Decoding $(64,30,14)$ eBCH code with order-$3$ OSD algorithms applying different stopping rules.}}
        \begin{tabular}{|c|c|c|c|c|c|c|} 	
        \hline
        \multicolumn{2}{|c|}{SNR (dB)}                                                                & 0     & 1     & 2     & 3    & 4    \\ \hline \hline
        \multirow{2}{*}{\begin{tabular}[c]{@{}c@{}}Original\\ OSD\cite{Fossorier1995OSD}\end{tabular}} & Ave. TEP       & \multicolumn{5}{c|}{4526}           \\ \cline{2-7} 
                                                                                & Time (ms) & \multicolumn{5}{c|}{17.45}          \\ \hline \hline
        \multirow{2}{*}{DOC\cite{lin2004ECC}}                                                    & Ave. TEP       & 4377  & 3924  & 2909  & 1477 & 377  \\ \cline{2-7} 
                                                                                & Time (ms) & 28.47 & 24.87 & 18.53 & 9.90 & 3.20 \\ \hline \hline
        \multirow{2}{*}{PSC\cite{jin2006probabilisticConditions}}                                                    & Ave. TEP       & 3134  & 2564  & 1709  & 851  & 240  \\ \cline{2-7} 
                                                                                & Time (ms) & 20.77 & 17.71 & 12.04 & 6.38 & 2.40 \\ \hline \hline
        \multirow{2}{*}{HISR}                                                   & Ave. TEP       & 3690  & 2712  & 1391  & 446  & 101  \\ \cline{2-7} 
                                                                                & Time (ms) & 16.65 & 12.62 & 7.09  & 3.21 & 1.72 \\ \hline \hline
        \multirow{2}{*}{HGSR}                                                   & Ave. TEP       & 4107  & 2644  & 997   & 233  & 60   \\ \cline{2-7} 
                                                                                & Time (ms) & 45.69 & 32.57 & 15.63 & 4.92 & 2.21 \\ \hline \hline
        \multirow{2}{*}{SISR}                                                   & Ave. TEP       & 2479  & 1267  & 445   & 96   & 13   \\ \cline{2-7} 
                                                                                & Time (ms) & 12.19 & 6.63  & 2.89  & 1.33 & 0.99 \\ \hline \hline
        \multirow{2}{*}{SGSR}                                                   & Ave. TEP       & 2240  & 1095  & 296   & 46   & 7    \\ \cline{2-7} 
                                                                                & Time (ms) & 12.12 & 6.21  & 2.49  & 1.22 & 0.96 \\ \hline
        \end{tabular}  \label{tab::SR}
        \end{table}
            
        \subsubsection{Comparison of Discarding Rules}
        
        We consider the discarding rules proposed in \cite{Wu2007OSDMRB} as the benchmark, which can discard the unpromising TEPs before performing the re-encoding, to reduce the decoding complexity. In \cite{Wu2007OSDMRB}, a decoding necessary condition (DNC) was proposed as follows. A lower bound of the reliabilities of the TEPs is first estimated based on the so-far recorded WHD $d_{\min}^{(\mathrm{W})}$, i.e.,
        \begin{equation}
            \ell^{*} = \frac{ d_{\min}^{(\mathrm{W})} \sum_{u = 1}^{k} \widetilde{\alpha}_u}{\sum_{u = 1}^{k} \widetilde{\alpha}_u + \lambda \sum_{u = k+1}^{n} \widetilde{\alpha}_u},
        \end{equation}
        where $\lambda$ is a parameter to be chosen. Then, for an arbitrary TEP $\mathbf{e}$, if the reliability of $\mathbf{e}$, i.e., $\ell(\mathbf{e}) = \sum_{\substack{1\leq u\leq k \\ e_{u}\neq 0 }}\widetilde{\alpha}_u$, satisfies $\ell(\mathbf{e}) \geq \ell^{*} $, $\mathbf{e}$ is discarded without re-encoding.
        
        Next, we compare the complexity of decoders with different discarding rules. The DNC \cite{Wu2007OSDMRB} is considered as the benchmark and the HDR and SDR are compared. We consider the order-3 decoding of $(64,30,14)$ eBCH codes. All parameters in the simulated decoder are carefully selected to ensure that they can reach the same error rate as the original OSD \cite{Fossorier1995OSD}, and the sequence of TEPs are ordered in descending order of the reliabilities. As discussed in Section \ref{sec::Discussion::Complexity}, we further adopt the ``5-step'' implementation for the HDR and SDR to reduce the overhead, i.e., checking the conditions every 5 TEPs.
        
        The average numbers of re-encoded TEPs are compared in Fig. \ref{Fig::VIII::NC-Na}. It can be seen that the proposed SDR can significantly reduce the number of re-encoded TEPs, and a notable improvement is shown compared to the DNC \cite{Wu2007OSDMRB}, especially at low SNRs. However, the HDR is the worst among its counterparts. This is because the soft information (i.e., channel reliabilities) are not well utilized to determine the likelihoods of TEPs in the HDR. In addition, the average decoding times of decoding a single codeword are compared in Fig. \ref{Fig::VIII::NC-time}. As shown, each simulated approach can significantly reduce the decoding time compared to the original OSD in both low and high SNR regimes. The main reason is that as shown in  Section \ref{sec::Discussion::Implementation}, the HDR and SDR can be efficiently implemented with $O(n)$ FLOPs, and the overhead is further reduced by $\ell$ times with the``$\ell$-step'' implementation. We can also conclude that the SDR and DNC have similar decoding time at high SNRs, close to 1 ms; nevertheless, the SDR outperforms at low SNRs. The numbers of TEPs and decoding times of different decoders are recorded in Table \ref{tab::NC}.
        
        \begin{figure}[t]
	    	\vspace{-0.8em}
            \centering
            \subfigure[Average Number of TEP]
            {
                \includegraphics[scale = 0.65]{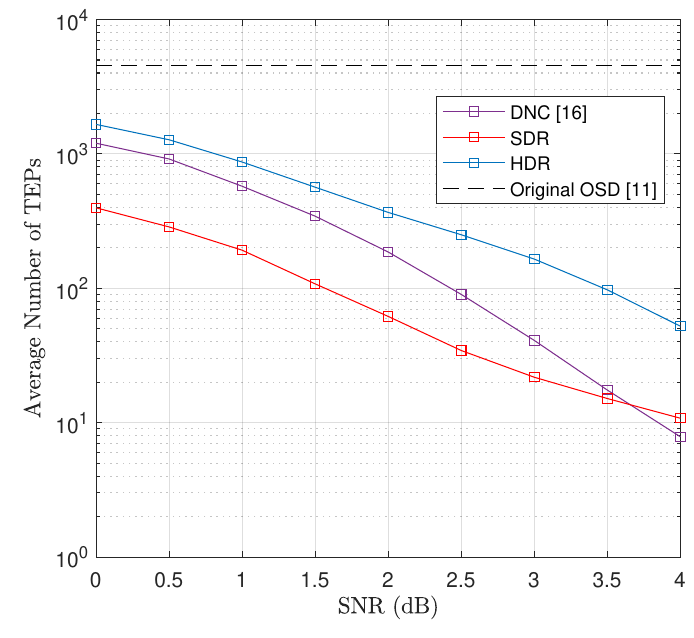}
                \label{Fig::VIII::NC-Na}
            }
            \vspace{-1ex}
            \subfigure[Average Decoding Time]
            {
                \includegraphics[scale = 0.65]{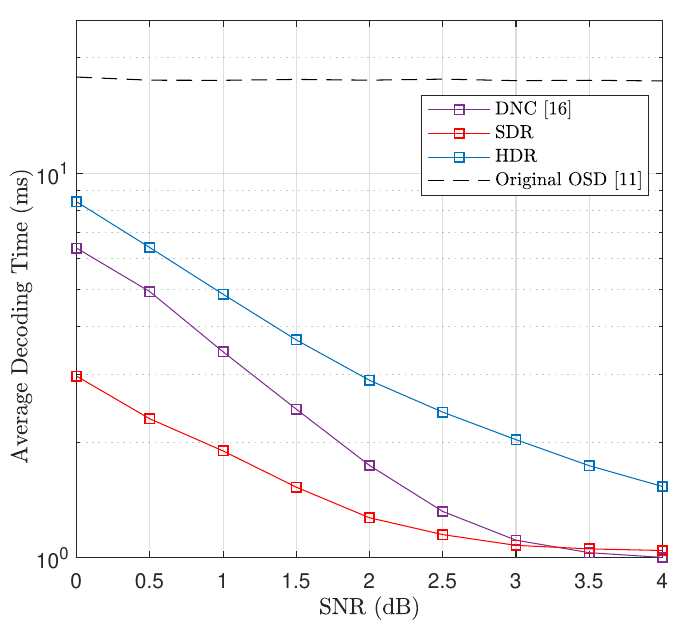}
                \label{Fig::VIII::NC-time}
            }

            \caption{{\color{black}Decoding $(64,30,14)$ eBCH code with order-$3$ OSD algorithms different discarding rules.}}
            \label{Fig::NC}
        \end{figure}
        
        \begin{table}[]
            \small	
        	\centering
	    	\tabcolsep=0.11cm
        	\caption{{\color{black}Decoding $(64,30,14)$ eBCH code with order-$3$ OSD algorithms applying different discarding rules.}}
            \begin{tabular}{|c|c|c|c|c|c|c|}
            \hline
            \multicolumn{2}{|c|}{SNR (dB)}                                                                & 0     & 1     & 2     & 3    & 4    \\ \hline \hline
            \multirow{2}{*}{\begin{tabular}[c]{@{}c@{}}Original\\ OSD\cite{Fossorier1995OSD}\end{tabular}} & Ave. TEP       & \multicolumn{5}{c|}{4526}           \\ \cline{2-7} 
                                                                                    & Time (ms) & \multicolumn{5}{c|}{17.45}          \\ \hline \hline
            \multirow{2}{*}{DNC\cite{Wu2007OSDMRB}}                                 & Ave. TEP       & 1200  & 574  & 186  & 40 & 8  \\ \cline{2-7} 
                                                                                    & Time (ms) & 6.39 & 3.43 & 1.73 & 1.11 & 1.00 \\ \hline \hline
            \multirow{2}{*}{SDR}                                                   & Ave. TEP       & 396  & 192  & 61   & 21   & 10   \\ \cline{2-7} 
                                                                                    & Time (ms) & 2.96 & 1.89  & 1.27  & 1.07 & 1.04 \\ \hline \hline
            \multirow{2}{*}{HDR}                                                   & Ave. TEP       & 1657  & 870  & 366   & 164   & 52    \\ \cline{2-7} 
                                                                                    & Time (ms) & 8.43 & 4.84  & 2.89  & 2.02 & 1.53 \\ \hline
            \end{tabular}  \label{tab::NC}
        \end{table}
         
         }
            
\section{Conclusion} \label{Sec::Conclusion}
In this paper, we revisited the ordered statistics decoding algorithm as a promising decoding approach for short linear block codes approaching maximum-likelihood performance. We investigated and characterized the statistical properties of the Hamming distance and weighted Hamming distance in the reprocessing stages of the ordered statistics decoding (OSD) algorithm. The derived statistical properties can give insights into the relationship between the decoding quality and the distance in the decoding process. According to the derived Hamming and weighted Hamming distance (WHD) distributions, we proposed two classes of decoding techniques, namely hard and soft techniques, to improve the decoding complexity of the OSD algorithm. These decoding techniques are analyzed and simulated. It is shown that they can significantly reduce the complexity in terms of the number of test error patterns (TEPs), with a negligible error performance loss in comparison with the original OSD. For example, from the numerical results of decoding $(64,30,14)$ eBCH code, the hard individual stopping rule (HISR) and hard group stopping rule (HGSR) with parameter $\mathrm{P}_t^{\mathrm{suc}} = 0.99$ can maintain the error performance of the original OSD, while reducing the TEP numbers from 31 to around 2 for the order-1 decoding and from 466 to around 4 for the order-2 decoding at high SNRs, respectively. The same improvement can also be observed by using the soft individual stopping rule (SISR) and soft group stopping rule (SGSR) with $\mathrm{P}_t^{\mathrm{suc}} = 0.5$. The hard discarding rule (HDR) with $\lambda =0.1$ can reduce the TEP numbers from 31 to 24 of the order-1 decoding of $(64,30,14)$ eBCH code with slight error performance loss, and the soft discarding rule (SDR) with $\lambda =0.1$ can reduce the TEP numbers from 21 to around 5 of the order-1 decoding of $(30,21,16)$ eBCH code with virtually the same error performance with the original OSD. Comparisons are further performed with approaches from the literature. As shown, the proposed techniques outperform the state of the art in terms of the number of TEPs and the run-time of decoding a single codeword.
     
These decoding techniques can be adopted to design reduced-complexity OSD algorithms in particular for short BCH codes in ultra-reliable and low-latency communications. For example, considering the hard techniques introduced in Section \ref{sec::HDdistech}, HISR and HGSR can serve as the stopping rule (SR) to terminate decoding early, and the HDR can serve as the TEP discarding rule (DR) to further improve the decoding efficiency. Applying the soft techniques introduced in Section \ref{Sec::SoftTech}, the soft-technique decoder can be designed, where the SISR and SGSR can serve as SRs and the SDR can serve as a DR. Compared to hard techniques, soft techniques exhibit better error performance however with a slightly increased overhead due to the calculation of WHD distribution. All techniques proposed in this paper can be easily combined with other OSD techniques and approaches to further reduce the decoding complexity.

% or


\appendices
\section{The approximation of $f_{\widetilde{A}_u}(x)$} \label{App::Aapp}

    For a real number $t>0$, we note the equivalence between events $\{\widetilde{ A}_u \geq t\}$ and $ \left\{\sum_{v=1}^{n} \mathbf{1}_{[0,t]}( A_v) \leq n-u \right\}$, where $\mathbf{1}_{\mathcal{X}}(x) = 1$ if $x \in \mathcal{X}$ and $\mathbf{1}_{\mathcal{X}}(x) = 0$, otherwise. We define a new random variable $Z_n$ as
    \begin{equation}
        Z_n = \sum_{v=1}^{n}\mathbf{1}_{[0,t]}( A_v),
    \end{equation} 
    which is a random variable with a binomial distribution $\mathcal B(n,F_{ A}(t))$. By using the Demoivre-Laplace theorem \cite{papoulis2002probability}, $Z_{n}$ can be approximated by a normal distribution ${\mathcal N}(\mathbb{E}[Z_{n}],\sigma^2_{Z_{n}})$ with mean
    \begin{equation} \label{equ::meanZn}
        \mathbb{E}[Z_n] = n F_{ A}(t),
    \end{equation} 
     and variance
    \begin{equation} \label{equ::varZn}
        \sigma^2_{Z_n} = n F_{ A}(t)(1-F_{ A}(t)).
    \end{equation}
    For a particular $t \leq 0$ and a large $n$ satisfying $ n^3 F_{A}^2(t)(1-F_{ A}(t)) \gg 1$, the above normal approximation ${\mathcal N}(\mathbb{E}[Z_{n}],\sigma^2_{Z_{n}})$ holds \cite[equation 3-27]{papoulis2002probability}. To find an approximation independent of $t$, we first define a random variable dependent on $t$ as
    \begin{equation}
        W(t)=\frac{t(n-Z_n)}{u}.
    \end{equation}
    Therefore, we can observe the following equivalence.
    \begin{equation}
        \{\widetilde{A}_u \geq t \}  \equiv \{Z_n \leq n-u \}  \equiv \{W(t) \geq t \}.
    \end{equation}
    Because $Z_n$ is a normal random variable with mean and varianve given by (\ref{equ::meanZn}) and (\ref{equ::varZn}), respectively, $W(t)$ is also a normal random variable with mean and variance respective given by
    \begin{equation}
        \mathbb{E}[W(t)] = \frac{t{n}(1-F_{A}(t))}{u},
    \end{equation}
    and 
    \begin{equation}
        \sigma_{W(t)}^2 = \frac{t^2{n}F_{ A}(t)(1-F_{ A}(t))}{u^2}.
    \end{equation}
    Finally, we can observe the following equivalence between $\widetilde{ A}_u$ and $W(t)$ as
    \begin{equation}  \label{equ::Normal APP with t}
        \begin{split}
            \{\widetilde{ A}_u \geq t \} \equiv & \{W(t) \geq t \} \\
            \equiv & \!\left\{ \!{\mathcal N} \!\left(\!\frac{t{n}(1\!-\!F_{ A}(t))}{u}, \frac{t^2 {n}F_{ A}(t)(1\!-\!F_{ A}(t))}{u^2} \!\right) \!\geq\! t \!\right\}\!.
        \end{split}
    \end{equation} 
    Despite the equivalence of (\ref{equ::Normal APP with t}), the mean and variance of $\widetilde{ A}_u$ itself should be independent of $t$. Assume that $\widetilde{ A}_u$ follows a normal distribution $\mathcal N(\mathbb{E}[\widetilde{ A}_u],\sigma_{\widetilde{ A}_u}^2)$, and we have the following equivalence
    \begin{equation}
    \begin{split}
        &\left\{\mathcal N(\mathbb{E}[\widetilde{ A}_u],\sigma_{\widetilde{ A}_u}^2) \geq t \right\} \\
        &\equiv \left\{ {\mathcal N} \left(\frac{t{n}(1-F_{ A}(t))}{u}, \frac{t^2 {n}F_{ A}(t)(1-F_{ A}(t))}{u^2} \right) \geq t \right\}.    
    \end{split}
    \end{equation} 
    In other words
    \begin{equation}
    \begin{split}
        &\mathrm{Pr}\left(\mathcal N(\mathbb{E}[\widetilde{ A}_u],\sigma_{\widetilde{ A}_u}^2) \geq t \right)\\
        &= \mathrm{Pr}\left( {\mathcal N} \left(\frac{t{n}(1-F_{ A}(t))}{u}, \frac{t^2 {n}F_{ A}(t)(1-F_{ A}(t))}{u^2} \right) \geq t \right).        
    \end{split}
    \end{equation} 
    Let $t = t_0 = \mathbb{E}[\widetilde{ A}_u]$, and it can be obtained that 
    \begin{equation}
    \begin{split}
        &\mathrm{Pr}\left(\mathcal N(t_0,\sigma_{\widetilde{ A}_u}^2) \geq t_0 \right) \\
        &= \mathrm{Pr}\!\left( \!{\mathcal N} \!\left(\!\frac{t_0{n}(1\!-\!F_{ A}(t_0))}{u}, \frac{t_0^2 {n}F_{ A}(t_0)(1\!-\!F_{ A}(t_0))}{u^2} \right) \!\geq \! t_0 \!\right) \\
        &= \frac{1}{2},        
    \end{split}
    \end{equation} 
    and
    \begin{equation}
        \frac{t_0{n}(1-F_{ A}(t_0))}{u} = t_0.
    \end{equation}
    Therefore, the mean of $\widetilde  A_u$ is derived as
    \begin{equation}
        \mathbb{E}[\widetilde{ A}_u] = t_0 = F_{ A}^{-1}\left(1-\frac{u}{n}\right).
    \end{equation}
    
    From (\ref{equ::Normal APP with t}), we can also observe that
    \begin{align}  \label{equ::Normal APP without t}
            \{\widetilde{ A}_u \geq t \} \equiv & \left\{{\mathcal N}(0,1) \geq \frac{u-{n}+{n}F_{ A}(t)}{\sqrt{ {n} F_{ A}(t)(1-F_{ A}(t))}}   \right\} \notag\\
            \equiv & \left\{ {\mathcal N}(0,1) \geq - \frac{(u-{n}(1-F_{ A}(t)))}{(t-t_0)\sqrt{{n}F_{ A}(t)(1\!-\!F_{ A}(t))}}t_0 \right. \notag\\
            +&  \left.\frac{(u-{n}(1-F_{ A}(t)))}{(t-t_0)\sqrt{{n}F_{ A}(t)(1\!-\!F_{ A}(t))}}t  \right\}.
    \end{align} 
    Thus, the variance is given by 
    \begin{equation}
        \begin{split}
            \sigma_{\widetilde{A}_u}^2 =& \lim\limits_{t \to t_0} \frac{(t-t_0)^2{n}F_{ A}(t)(1-F_{ A}(t))}{(u-{n}(1-F_{ A}(t)))^2} \\
            = & \pi N_0\frac{({n}-u)u}{{n}^3} \left(e^{-\frac{(t_0+1)^2}{N_0}} + e^{-\frac{(t_0-1)^2}{N_0}}  \right)^{-2}.
        \end{split}
    \end{equation}
    
    Therefore, the $u$-th ordered reliability can be approximated by a Normal distribution ${{\mathcal N}}(\mathbb{E}[\widetilde  A_u],\sigma_{\widetilde  A_u}^2)$, where
    \begin{equation} 
        \mathbb{E}[\widetilde  A_u] = t_0 = F_{ A}^{-1}\left(1-\frac{u}{{n}}\right)
    \end{equation}
    and
    \begin{equation} 
        \sigma_{\widetilde  A_u}^2 = \pi N_0\frac{({n}-u)u}{{n}^3} \left(e^{-\frac{(t_0+1)^2}{N_0}} + e^{-\frac{(t_0-1)^2}{N_0}}  \right)^{-2}.
    \end{equation}

\section{The approximation of $f_{\widetilde{A}_u,\widetilde{A}_v}(x,y)$}            \label{App::jointAapp}
    
    For $0<u<v$ and $0 \leq t\leq x \leq {n}$, we observe the equivalence between events $\{\widetilde{ A}_v \geq t |{\widetilde{ A}_u = x}\}$ and $\{\sum_{\ell = u}^{{n}} \mathbf{1}_{[t,x]}( A_{\ell}) \geq v-u\}$. Let the random variable $S_{n} = \sum_{\ell = u}^{n} \mathbf{1}_{[t,x]}( A_{\ell})$, and according to the central limit theorem, we have 
    \begin{equation} 
        \begin{split}
             & \left\{\widetilde{ A}_v \geq t |{\widetilde{ A}_u = x}\right\} \equiv  \left\{S_{n} \leq v-u\right\} \\
             &\equiv \!\left\{\!\mathcal N \!\left(\!\frac{t(u\!+\!(n\!-\!u)\gamma_x(t))}{v},\frac{t^2(n\!-\!u)\gamma_x(t)(1\!-\!\gamma_x(t))}{v^2}\!\right)\! \geq t \! \right\} \! ,
        \end{split}
    \end{equation}
    where
    \begin{equation}
        \gamma_{x}(t) = \frac{F_{ A}(x) -  F_{ A}(t)}{F_{ A}(x)}.
    \end{equation}
    
    Similarly as the approximation of $f_{\widetilde{A}_u}(x)$, the mean and variance of $\widetilde{ A}_v$ on the condition that $\widetilde{ A}_u = x $ can be obtained as 
    \begin{equation}
        \mathbb{E}[\widetilde{ A}_v | {\widetilde{ A}_u = x}] = t_1 = \gamma_{x}^{-1}(\frac{v-u}{{n}-u}).
    \end{equation}
    and
    \begin{equation}
        \begin{split}
           \sigma_{\widetilde{ A}_v | {\widetilde{ A}_u = x}}^2 =& \lim\limits_{t \to t_1} \frac{(t-t_1)^2 ({n}-u)\gamma_x(t)(1-\gamma_x(t))}{(v-u-({n}-u)\gamma_x(t))^2} \\
           =& \pi N_0 \frac{(n\!-\!v)(v\!-\!u)}{(n\!-\!u)^3}\left(\!\frac{e^{\frac{-(t_1\!-\!1)^2}{N_0}}\!\!+\!e^{\frac{-(t_1\!+\!1)^2}{N_0})}}{F_a(x)}\!\right) ^{-2}\!\!\!\!,
        \end{split}
    \end{equation}
    respectively. Therefore, for $ 0 < u < v \leq {n}$, the joint distribution of $\widetilde{ A}_u$ and $\widetilde{ A}_v$ can be approximated as
    \begin{equation}
    \begin{split}
        f_{\widetilde{ A}_u,\widetilde{ A}_v}&(x,y) \\
        &\approx \frac{1}{2 \pi \sigma_{\!\widetilde  A_u} \!\sigma_{\!\widetilde{ A}_v |{\widetilde{ A}_u=x}}}\!\exp\!\left(\!-\frac{(x\!-\!t_0)^2}{2\sigma_{\!\widetilde  A_u}^2}\!-\!\frac{(y\!-\!t_1)^2}{2\sigma_{\!\widetilde{ A}_v |{\widetilde{ A}_u=x}}^2}\!\right)\!. 
    \end{split}
    \end{equation}
    
\section{Proof of Theorem \ref{the::HDdis::iphase}} \label{app::proof::HDdis::iphase}
        Similar to Lemma \ref{lem::HDdis::0phase}, we first consider the composition of the Hamming distance in $i$-reprocessing ($0<i\leq m$). For the hard-decision results $
			\widetilde{\mathbf{y}}= [\widetilde{\mathbf{c}}_{\mathrm{B}}\oplus \widetilde{\mathbf{e}}_{\mathrm{B}} \ \ \widetilde{\mathbf{c}}_{\mathrm{P}}\oplus \widetilde{\mathbf{e}}_{\mathrm{P}}]$, it is obvious that error pattern $\widetilde{\mathbf{e}}_{\mathrm{B}}$ is in the TEP list from 0-reprocessing to $i$-reprocessing if and only if $w(\widetilde{\mathbf{e}}_{\mathrm{B}}) \leq i$. 
			
		When $w(\widetilde{\mathbf{e}}_{\mathrm{B}}) > i$, the first $i$ reprocessings cannot decode the received signal correctly, and the codeword estimate generated by each re-encoding is given by $\widetilde{\mathbf{c}}_{\mathbf{e}} = [\widetilde{\mathbf{c}}_{\mathrm{B}} \oplus  \widetilde{\mathbf{e}}_{\mathrm{B}} \oplus \mathbf{e} \ \ \widetilde{\mathbf{c}}_{\mathbf{e}, \mathrm{P}}]$. Then, we can obtain that the difference pattern $\widetilde{\mathbf{d}}_{\mathbf{e}} = \widetilde{\mathbf{c}}_{\mathbf{e}} \oplus \widetilde{\mathbf{y}}$ is given by 
		\begin{equation}
		    \widetilde{\mathbf{d}}_{\mathbf{e}} = [\mathbf{e}  \ \ \widetilde{\mathbf{c}}_{\mathrm{P}}\oplus \widetilde{\mathbf{c}}_{\mathbf{e}, \mathrm{P}} \oplus \widetilde{\mathbf{e}}_{\mathrm{P}}].
		\end{equation}
		Note that $\widetilde{\mathbf{d}}_{\mathbf{e}_{\mathrm{P}}}= \widetilde{\mathbf{c}}_{\mathrm{P}}\oplus \widetilde{\mathbf{c}}_{\mathbf{e}, \mathrm{P}} \oplus \widetilde{\mathbf{e}}_{\mathrm{P}} = [\widetilde{\mathbf{e}}_{\mathrm{B}}\oplus\mathbf{e}]\widetilde{\mathbf{P}}\oplus \widetilde{\mathbf{e}}_{\mathrm{P}}$. Thus, the Hamming distance between $\widetilde{\mathbf{c}}_{\mathbf{e}}$ and $\widetilde{\mathbf{y}}$, denoted by the random variable $D_{\mathbf{e}}^{(\mathrm{H})}$, can be represented as $D_{\mathbf{e}}^{(\mathrm{H})}  = w(\mathbf{e}) + W_{\mathbf{e},\mathbf{c}_{\mathrm{P}}}$,
        where $w(\mathbf{e})$ is the Hamming weight of $\mathbf{e}$, and $W_{\mathbf{e},\mathbf{c}_{\mathrm{P}}}$ is the random variable introduced in Lemma \ref{lem::HDdis::iphase::Wecp}. It has been shown that when $w(\widetilde{\mathbf{e}}_{\mathrm{B}}) = u$ and $w(\widetilde{\mathbf{e}}_{\mathrm{P}}) = v$, the $\mathrm{pmf}$ of $W_{\mathbf{e},\mathbf{c}_{\mathrm{P}}}$, i.e., $p_{W_{\mathbf{e},\mathbf{c}_{\mathrm{P}}}}(j| u,v)$, is given by (\ref{equ::HDdis::iphase::Wecp}).
        
        Then, after the $i$-reprocessing, the minimum Hamming distance conditioning on $w(\widetilde{\mathbf{e}}_{\mathrm{B}}) > i$ is derived as
        \begin{equation}
		    D_i^{(\mathrm{H})} =  \min_{\forall \mathbf{e} : w(\mathbf{e}) \leq i} \{w(\mathbf{e}) + W_{\mathbf{e},\mathbf{c}_{\mathrm{P}}} \} .
		\end{equation}
	    Let us consider a sequence of i.i.d random variables $ [D_{\mathbf{e}}^{(\mathrm{H})}]_1^{b_{0:i}^{k}}$ with length $b_{0:i}^{k}$, and the minimum Hamming distance $D_i^{(\mathrm{H})}$ can be represented as the minimal element of $[D_{\mathbf{e}}^{(\mathrm{H})}]_1^{b_{0:i}^{k}}$. When $i \ll k$, $w(\mathbf{e})$ can be regarded as a constant $i$ since $b_{0:i-1}^{k} \ll \binom{k}{i}$. Therefore, let $p_{\widetilde W_{\mathbf{c}_{\mathrm{P}}}}(j,b| u,v)$ denote the $\mathrm{pmf}$ of the minimal element of $b$ samples of $W_{\mathbf{e},\mathbf{c}_{\mathrm{P}}}$ conditioning on $\{w(\widetilde{\mathbf{e}}_{\mathrm{B}}) \!=\! u, w(\widetilde{\mathbf{e}}_{\mathrm{B}}) \!=\ v \}$. According to the discrete ordered statistics theory \cite[Eq. (2.4.1)]{David2004Order}, $p_{\widetilde W_{\mathbf{c}_{\mathrm{P}}}}(j,b| u,v)$ can be derived as 
        \begin{equation} \label{equ::proof::HDdis::iphase::Wcp}
            p_{\widetilde W_{\mathbf{c}_{\mathrm{P}}}}(j,b| u,v) = b\! \int_{F_{W_{\mathbf{e},\mathbf{c}_{\mathrm{P}}}}(j|u,v)-p_{W_{\mathbf{e},\mathbf{c}_{\mathrm{P}}}}(j|u,v)}^{F_{W_{\mathbf{e},\mathbf{c}_{\mathrm{P}}}}(j|u,v)} (1\!-\!\ell)^{b\!-\!1} d \ell ,
        \end{equation}
        Thus, the $\mathrm{pmf}$ of $D_i^{(\mathrm{H})}$ conditioning on $\{w(\widetilde{\mathbf{e}}_{\mathrm{B}} )\!>\! i\}$ can be obtaining by combining all values of $\widetilde{\mathbf{e}}_{\mathrm{P}} = v$ and considering $b = b_{0:i}^{k}$, i.e., 
        \begin{equation}
            p_{D_{i}^{(\mathrm{H})}}(j-i|w(\widetilde{\mathbf{e}}_{\mathrm{B}} )\!>\! i) = \sum_{v=0}^{n-k} p_{E_{k+1}^{n}}(v)p_{\widetilde W_{\mathbf{c}_{\mathrm{P}}}}(j,b_{0:i}^{k}| i^{(>)},v).
        \end{equation}
    
    	When $w(\widetilde{\mathbf{e}}_{\mathrm{B}}) \leq i$, the error pattern $\widetilde{\mathbf{e}}_{\mathrm{B}} $ can be eliminated by reprocessing with the TEP $\mathbf{e} = \widetilde{\mathbf{e}}_{\mathrm{B}} $, and the generated codeword estimate is given by
		\begin{equation}
		    \widetilde{\mathbf{c}}_{\mathbf{e}} = [\widetilde{\mathbf{c}}_{\mathrm{B}} \oplus  \widetilde{\mathbf{e}}_{\mathrm{B}} \oplus \widetilde{\mathbf{e}}_{\mathrm{B}}]\widetilde{\mathbf{G}} = [\widetilde{\mathbf{c}}_{\mathrm{B}}\ \ \widetilde{\mathbf{c}}_{\mathrm{P}}],
		\end{equation}
		thus, if the error pattern $\widetilde{\mathbf{e}}_{\mathrm{B}} $ is eliminated, the Hamming distance $D_{\mathbf{e}}^{(\mathrm{H})}$ between $\widetilde{\mathbf{c}}_{\mathbf{e}}$ and $\widetilde{\mathbf{y}}$ can be derived as
		\begin{equation}
		    D_{\mathbf{e}}^{(\mathrm{H})} =  \lVert \widetilde{\mathbf{c}}_{\mathrm{B}} \oplus  \widetilde{\mathbf{e}}_{\mathrm{B}} \oplus \widetilde{\mathbf{c}}_{\mathrm{B}}  \rVert +  \lVert \widetilde{\mathbf{c}}_{\mathrm{P}}\oplus \widetilde{\mathbf{e}}_{\mathrm{P}} \oplus \widetilde{\mathbf{c}}_{\mathrm{P}} \rVert = w(\widetilde{\mathbf{e}}_{\mathrm{B}}) + E_{k+1}^{n} .
		\end{equation}
        Thus, after the $i$-reprocessing, the minimum Hamming distance is given by the minimum element of $w(\mathbf{e}) + E_{k+1}^{n}$ and $[w(\mathbf{e}) + W_{\mathbf{e},\mathbf{c}_{\mathrm{P}}}]_1^{b_{1:i}^{k}}$, i.e.,
		\begin{equation} \label{equ::proof::HDdis::iphase::Di::eB<=i}
		    D_i^{(\mathrm{H})}  = \min\{w(\widetilde{\mathbf{e}}_{\mathrm{B}}) + E_{k+1}^{n}, \min_{\substack{\forall \mathbf{e} : w(\mathbf{e}) \leq i\\\mathbf{e} \neq \widetilde{\mathbf{e}}_{\mathrm{B}}}} \{w(\mathbf{e}) + W_{\mathbf{c}_{\mathrm{P}}} \} \}.
		\end{equation}
		Conditioning on $\{w(\widetilde{\mathbf{e}}_{\mathrm{B}}) \!=\! u, w(\widetilde{\mathbf{e}}_{\mathrm{B}}) \!=\! v \}$, the $\mathrm{pmf}$ of $\min_{\substack{\forall \mathbf{e} : w(\mathbf{e}) \leq i\\\mathbf{e} \neq \widetilde{\mathbf{e}}_{\mathrm{B}}}} \{w(\mathbf{e}) + W_{\mathbf{c}_{\mathrm{P}}} \}$ can be simply obtained by (\ref{equ::proof::HDdis::iphase::Wcp}), i.e., $p_{\widetilde W_{\mathbf{c}_{\mathrm{P}}}}(j,b_{1:i}^{k}| u,v)$. Furthermore, we can observe that $w(\widetilde{\mathbf{e}}_{\mathrm{B}}) \!+\! E_{k+1}^{n} = v\!+\!u$ when $\{w(\widetilde{\mathbf{e}}_{\mathrm{B}}) \!=\! u, w(\widetilde{\mathbf{e}}_{\mathrm{B}}) \!=\! v \}$. Therefore, the $\mathrm{pdf}$ of $D_i^{(\mathrm{H})}$ given by (\ref{equ::proof::HDdis::iphase::Di::eB<=i}) can be derived as $p_{EW}(j|u,v)$ given by (\ref{equ::HDdis::iphase::EW}). Then, only conditioning on $\{w(\widetilde{\mathbf{e}}_{\mathrm{B}}) \!=\! u\}$, the $\mathrm{pmf}$ of $D_i^{(\mathrm{H})}$, denoted by $f_{D_i^{(\mathrm{H})}}(j|w(\widetilde{\mathbf{e}}_{\mathrm{B}}) \!=\! u)$, can be derived as
		\begin{equation} \label{equ::proof::HDdis::iphase::WEW}
		    f_{D_i^{(\mathrm{H})}}(j|w(\widetilde{\mathbf{e}}_{\mathrm{B}}) \!=\! u) = \sum_{v=0}^{n-k}p_{E_{k+1}^{n}}(v)p_{EW}(j|u,v).
		\end{equation}
		
		Finally, the $\mathrm{pmf}$ of the $D_i^{(\mathrm{H})}$ can be obtained by the law of total probability as
		\begin{equation} \label{equ::proof::HDdis::iphase}
	    \begin{split}
 		    p_{D_i^{(\mathrm{H})}}(j) &= \sum_{u=0}^{i} p_{E_1^k}(u)  f_{D_i^{(\mathrm{H})}}(j|w(\widetilde{\mathbf{e}}_{\mathrm{B}}) \!=\! u)\\
 		    &+ \sum_{u=i+1}^{k} p_{E_1^k}(u)p_{D_{i}^{(\mathrm{H})}}(j-i|w(\widetilde{\mathbf{e}}_{\mathrm{B}} )\!>\! i) .   
	    \end{split}
		\end{equation}
		By substituting (\ref{equ::proof::HDdis::iphase::Wcp}) and (\ref{equ::proof::HDdis::iphase::WEW}) into (\ref{equ::proof::HDdis::iphase}), we finally obtain (\ref{equ::HDdis::iphase}) and Theorem \ref{the::HDdis::iphase} is proved.
		
    \section{Proof of Theorem \ref{the::WHD::0phase}} \label{app::proof::WHD::0phase}
    	Given an arbitrary position indices vector $\mathbf{t}_h^{\mathrm{P}} \in {\mathcal{T}}_h^{\mathrm{P}}$, $0\leq h \leq (n-k)$ and the corresponding random variable $\widetilde{A}_{\mathbf{t}_h^{\mathrm{P}}} = \sum_{i=u}^{h} \widetilde{A}_{t_u^{\mathrm{P}}}$ with $\mathrm{pdf}$ $ f_{\widetilde{A}_{\mathbf{t}_h^{\mathrm{P}}}}(x)$, the $\mathrm{pdf}$ of the WHD $D_0^{(\mathrm{W})}$ in 0-reprocessing can be obtained by considering the mixture of all cases of possible $\mathbf{t}_h^{\mathrm{P}}$ with length $0\leq h \leq (n-k)$, which can be written as
        \begin{equation} \label{equ::proof_Overall0D0}
            f_{D_0^{(\mathrm{W})}}(x)  = \sum_{h=0}^{n-k} \sum_{\mathbf{t}_h^{\mathrm{P}} \in \mathcal{T}_h^{\mathrm{P}}} \mathrm{Pr}(\widetilde{\mathbf{d}}_{0,\mathrm{P}} = \mathbf{z}_{\mathbf{t}_h^{\mathrm{P}}})  f_{\widetilde{A}_{\mathbf{t}_h^{\mathrm{P}}}}(x),
        \end{equation}
        where $\mathrm{Pr}(\widetilde{\mathbf{d}}_{0,\mathrm{P}} = \mathbf{z}_{\mathbf{t}_h^{\mathrm{P}}})$ is the probability that only positions $\mathbf{t}_h^{\mathrm{P}} = [t^{\mathrm{P}}]_1^h$ in the vector $\widetilde{\mathbf{d}}_{0} = \widetilde{\mathbf{y}} \oplus \widetilde{\mathbf{c}}_{0}$ are nonzero. Based on the arguments in the Lemma \ref{lem::HDdis::0phase}, we re-write (\ref{equ::proof_Overall0D0}) in the form of conditional probability as
        \begin{equation} \label{equ::proof::WHD::0phase::cond}
        \begin{split}
            &f_{D_0^{(\mathrm{W})}}(x) \\
            & =\! \mathrm{Pr}(w(\widetilde{\mathbf{e}}_{\mathrm{B}})\!=\!0)  \sum_{h=0}^{n-k}\!\sum_{\substack{\mathbf{t}_h^{\mathrm{P}} \in \mathcal{T}_h^{\mathrm{P}}}} \!\!\!  \mathrm{Pr}\!\!\left(\widetilde{\mathbf{d}}_{0,\mathrm{P}} \!=\! \mathbf{z}_{\mathbf{t}_h^{\mathrm{P}}} |  w(\widetilde{\mathbf{e}}_{\mathrm{B}})\!=\!0\!\right)\! f_{\!\widetilde{A}_{\mathbf{t}_h^{\mathrm{P}}}}\!(x) \\
            &+ \! \mathrm{Pr}(w(\widetilde{\mathbf{e}}_{\mathrm{B}})\!\neq \!0)  \sum_{h=0}^{n-k}\!\sum_{\mathbf{t}_h^{\mathrm{P}} \in \mathcal{T}_h^{\mathrm{P}}} \!\! \! \mathrm{Pr}\!\!\left(\widetilde{\mathbf{d}}_{0,\mathrm{P}} \!=\! \mathbf{z}_{\mathbf{t}_h^{\mathrm{P}}} | w(\widetilde{\mathbf{e}}_{\mathrm{B}})\! \neq\! 0\!\right) \! f_{\!\widetilde{A}_{\mathbf{t}_h^{\mathrm{P}}}}\!(x),
        \end{split}
        \end{equation}
        where $\{w(\widetilde{\mathbf{e}}_{\mathrm{B}})\!=\!0\}$ is equivalent to $\{E_1^{k} \!=\! 0\}$, and $\mathrm{Pr}(w(\widetilde{\mathbf{e}}_{\mathrm{B}})=0)$ and $\mathrm{Pr}(w(\widetilde{\mathbf{e}}_{\mathrm{B}})\neq 0)$ are given by $p_{E_1^k}(0)$ and $1 - p_{E_1^k}(0)$ ($p_{E_1^k}(0)$ is previously given by (\ref{equ::HDdis::0phase::weightPe})), respectively. 
        
        When $w(\widetilde{\mathbf{e}}_{\mathrm{B}})=0$, the difference parttern $\widetilde{\mathbf{d}}_{0} = \widetilde{\mathbf{y}} \oplus \widetilde{\mathbf{c}}_{0}$ can be fully described by the hard-decision errors (recall Lemma \ref{lem::HDdis::0phase}), i.e., $\widetilde{\mathbf{d}}_{0}  = [\mathbf 0_{\mathrm{B}}  \ \  \widetilde{\mathbf{e}}_{\mathrm{P}}]$, where $\mathbf 0_{\mathrm{B}}$ is the zero vector with length $k$. Therefore, $\mathrm{Pr}(w(\widetilde{\mathbf{e}}_{\mathrm{B}})\!=\!0) \mathrm{Pr}\left(\widetilde{\mathbf{d}}_{0,\mathrm{P}}\! =\! \mathbf{z}_{\mathbf{t}_h^{\mathrm{P}}} |  w(\widetilde{\mathbf{e}}_{\mathrm{B}})\!=\!0\right)$ can be represented as
        \begin{equation}
        \begin{split}
              &\mathrm{Pr}(w(\widetilde{\mathbf{e}}_{\mathrm{B}})\!=\!0) \mathrm{Pr}\left(\widetilde{\mathbf{d}}_{0,\mathrm{P}}\! =\! \mathbf{z}_{\mathbf{t}_h^{\mathrm{P}}} |  w(\widetilde{\mathbf{e}}_{\mathrm{B}})\!=\!0\right)\\
              &= \mathrm{Pr}(\widetilde{\mathbf{e}} \!=\! [\mathbf{0}_{\mathrm{B}} \ \ \mathbf{z}_{\mathbf{t}_h^{\mathrm{P}}}] ) = \mathrm{Pe}(\mathbf{t}_h^{\mathrm{P}}),          
        \end{split}
        \end{equation}
        which is the probability that only positions of $\mathbf{t}_h^{\mathrm{P}}$ are in error over $\widetilde{\mathbf{y}}$. Thus, $\mathrm{Pe}(\mathbf{t}_h^{\mathrm{P}})$ can be given by 
        \begin{equation} \label{equ::proof::WHD::0phase::Pet}
        \begin{split}
            \mathrm{Pe}(\mathbf{t}_h^{\mathrm{P}}) &= \underbrace{\int_{0}^{\infty} \cdots }_{n-h}\underbrace{\int_{-\infty}^{0} \cdots }_{h} f_{[\widetilde{R}]_1^n}(x_{1},x_{2},\ldots,x_{n})  \\
            &\cdot\prod_{\substack{1 < v \leq n\\v\in \mathbf{t}_h^{\mathrm{P}}}} dx_v\prod_{\substack{1 < v \leq n\\v\notin \mathbf{t}_h^{\mathrm{P}}}} dx_v   ,          
        \end{split}
        \end{equation} 
        where $f_{[\widetilde{R}]_1^n}(x_{1},x_{2},\ldots,x_{n})$ is the joint $\mathrm{pdf}$ of ordered received signals $[\widetilde{R}]_1^n$, which can be derived as \cite{balakrishnan2014order}
        \begin{equation} \label{equ::proof::WHD::0phase::jointPDF::R1n}
            	f_{[\widetilde{R}]_1^n}(x_{1},x_{2},\ldots,x_{n})= 
            	n! \prod_{v=1}^{n} f_{R}(x_v) \prod_{v=2}^{n} \mathbf{1}_{[0,|x_{v-1}|]}(|x_v|) .
        \end{equation} 
        
        When $w(\widetilde{\mathbf{e}}_{\mathrm{B}})\neq 0$, it can be seen from Lemma \ref{lem::HDdis::0phase} that $\widetilde{\mathbf{d}}_{0} = [\mathbf{0}_{\mathrm{B}} \ \ \widetilde{\mathbf{c}}_{0,\mathrm{P}}'\oplus \widetilde{\mathbf{e}}_{\mathrm{P}}]$ where $\widetilde{\mathbf{c}}_{0,\mathrm{P}}'$ is the parity part of $\widetilde{\mathbf{c}}_{0}' = \widetilde{\mathbf{e}}_{\mathrm{B}}\widetilde{\mathbf{G}}$. Assume that the codebook and $p_{\mathbf{c}_{\mathrm{P}}}(u,q)$ of $\mathcal{C}(n,k)$ is unknown. We can re-write $\mathrm{Pr}\left(\widetilde{\mathbf{d}}_{0,\mathrm{P}} \!=\! \mathbf{z}_{\mathbf{t}_h^{\mathrm{P}}} | w(\widetilde{\mathbf{e}}_{\mathrm{B}}) \!\neq\! 0\right) $ as
        \begin{equation} \label{equ::proof_PIn|!eB}
            \mathrm{Pr}\left(\widetilde{\mathbf{d}}_{0,\mathrm{P}} = \mathbf{z}_{\mathbf{t}_h^{\mathrm{P}}} | w(\widetilde{\mathbf{e}}_{\mathrm{B}}) \neq 0\right)  = \mathrm{Pr}\left(\widetilde{\mathbf{c}}_{0,\mathrm{P}}'\oplus \widetilde{\mathbf{e}}_{\mathrm{P}} \!=\! \mathbf{z}_{\mathbf{t}_h^{\mathrm{P}}}\right) ,
        \end{equation}
        where $\mathrm{Pr}\left(\widetilde{\mathbf{c}}_{0,\mathrm{P}}'\oplus \widetilde{\mathbf{e}}_{\mathrm{P}} \!=\! \mathbf{z}_{\mathbf{t}_h^{\mathrm{P}}}\right)$ is denoted by $\mathrm{Pc}(\mathbf{t}_h^{\mathrm{P}})$ and previously given by (\ref{equ::WHD::0phase::Pct}) in Lemma \ref{lem::WHD::0phase::Pc}.
        Substituting (\ref{equ::proof::WHD::0phase::Pet}) and (\ref{equ::proof_PIn|!eB}) into (\ref{equ::proof::WHD::0phase::cond}), we can finally obtain (\ref{equ::WHD::0phase}). This completes the proof of theorem \ref{the::WHD::0phase}.
    
    \section{Proof of Lemma \ref{lem::WHD::iphase::eB=e}} \label{app::proof::WHD::iphase::eB=e}
        If the error pattern in hard-decision $\widetilde{\mathbf{e}}_{\mathrm{B}}$ is eliminated by the TEP $\mathbf{e}$, i.e., $\widetilde{\mathbf{e}}_{\mathrm{B}} = \mathbf{e}$, the codeword generated by re-encoding can be given by
		\begin{equation}
    	    \widetilde{\mathbf{c}}_{\mathbf{e}} = [\widetilde{\mathbf{c}}_{\mathrm{B}} \oplus \mathbf{e} \oplus \widetilde{\mathbf{e}}_{\mathrm{B}}]\widetilde{\mathbf{G}} = [\widetilde{\mathbf{c}}_{\mathrm{B}}\ \ \widetilde{\mathbf{c}}_{\mathrm{P}}].
        \end{equation}
        Recall that $\widetilde {\mathbf y} = [\widetilde{\mathbf{c}}_{\mathrm{B}} \oplus \widetilde{\mathbf{e}}_{\mathrm{B}}  \ \ \widetilde{\mathbf{c}}_{\mathrm{P}} \oplus \widetilde{\mathbf{e}}_{\mathrm{P}}]$, and we can re-write the WHD between $\widetilde{\mathbf{c}}_{\mathbf{e}}$ and $\widetilde{\mathbf{y}}$, denoted by a random variable $D_{\mathbf{e}}^{(\mathrm{W})}$, as
        \begin{equation}
            D_{\mathbf{e}}^{(\mathrm{W})}= \sum_{\substack{1\leq u\leq k\\\widetilde{e}_{\mathrm{B},u} \neq 0}} \widetilde{A}_u + \sum_{\substack{1\leq u\leq n - k\\\widetilde{e}_{\mathrm{P},u} \neq 0}} \widetilde{A}_u .
        \end{equation}
        Since the error pattern $\widetilde{\mathbf{e}}_{\mathrm{B}}$ can be eliminated by the first $i$ reprocessings in the order-$m$ OSD, it can be obtained that $w(\widetilde{\mathbf{e}}_{\mathrm{B}}) \leq i$, i.e., the condition $\{E_1^k \leq i\}$ holds. The probability that positions in $\mathbf{t}_{\ell}^{h}$ are different between $\widetilde{\mathbf{c}}_{\mathbf{e}}$ and $\widetilde{\mathbf{y}}$, denoted by $\mathrm{P}(\mathbf{t}_{\ell}^{h})$, is given by
        \begin{equation}
            \begin{split}
                \mathrm{P}(\mathbf{t}_{\ell}^{h})& = \mathrm{Pr}(\widetilde{\mathbf{e}} = \mathbf{z}_{\mathbf{t}_{\ell}^h}|E_1^{k} \!\leq\! i) \\ 
                &= \frac{\mathrm{Pr}(\widetilde{\mathbf{e}} = \mathbf{z}_{\mathbf{t}_{\ell}^h} , E_1^{k} \!\leq \! i)}{\mathrm{Pr}(E_1^{k}\leq i)}.
            \end{split}
        \end{equation}
        Moreover, for $0\leq \ell \leq i$, when the event $\{\widetilde{\mathbf{e}} = \mathbf{z}_{\mathbf{t}_{\ell}^h}\}$ occurs, the event $\{E_1^{k}\leq i\}$ must occur. Therefore, we obtain that $\mathrm{Pr}(\widetilde{\mathbf{e}} = \mathbf{z}_{\mathbf{t}_{\ell}^h} , E_1^{k} \!\leq \! i) = \mathrm{Pr}(\widetilde{\mathbf{e}} = \mathbf{z}_{\mathbf{t}_{\ell}^h})$ and
        \begin{equation}
            \mathrm{P}(\mathbf{t}_{\ell}^{h}) = \frac{\mathrm{Pr}(\widetilde{\mathbf{e}} = \mathbf{z}_{\mathbf{t}_{\ell}^h})}{\mathrm{Pr}(E_1^{k}\leq i)},
        \end{equation}
        where $\mathrm{Pr}(E_1^{k}\leq i)$ is simply given by $\sum_{v=0}^i p_{E_1^{k}}(v)$ according to Lemma \ref{lem::OrderStat::Eab}. Let us denote $\mathrm{Pr}(\widetilde{\mathbf{e}} = \mathbf{z}_{\mathbf{t}_{\ell}^h})$ as $\mathrm{Pe}(\mathbf{t}_{\ell}^{h})$. Similar to (\ref{equ::proof::WHD::0phase::Pet}), $\mathrm{Pe}(\mathbf{t}_{\ell}^{h})$ is derived as (\ref{equ::WHD::iphase::eb=e::Pet}) by using the joint $\mathrm{pdf}$ $f_{[\widetilde{R}]_1^n}(x_{1},x_{2},\ldots,x_{n})$. Finally, by considering all possible $\mathbf{t}_{\ell}^{h}$, we can obtain (\ref{equ::WHD::iphase::eb=e}). This completes the proof of lemma \ref{lem::WHD::iphase::eB=e}.
    
    \section{Proof of Lemma \ref{lem::WHD::iphase::eB!=e}} \label{app::proof::WHD::iphase::eB!=e}
        If the error pattern in hard-decision $\widetilde{\mathbf{e}}_{\mathrm{B}}$ is not eliminated by the TEP $\mathbf{e}$, i.e., $\widetilde{\mathbf{e}}_{\mathrm{B}} \neq \mathbf{e}$, the codeword generated by re-encoding can be given by
		\begin{equation}
    	    \widetilde{\mathbf{c}}_{\mathbf{e}} = [\widetilde{\mathbf{c}}_{\mathrm{B}} \oplus \mathbf{e} \oplus \widetilde{\mathbf{e}}_{\mathrm{B}}]\widetilde{\mathbf{G}} = [\widetilde{\mathbf{c}}_{\mathrm{B}} \oplus \mathbf{e} \oplus \widetilde{\mathbf{e}}_{\mathrm{B}}\ \ \widetilde{\mathbf{c}}_{\mathbf{e}, \mathrm{P}}].
        \end{equation}
        Thus, the difference pattern $\widetilde{\mathbf{d}}_{\mathbf{e}} = \widetilde{\mathbf{c}}_{\mathbf{e}} \oplus \widetilde{\mathbf{y}}$ can be obtained as 
        \begin{equation}
            \widetilde{\mathbf{d}}_{\mathbf{e}} = [\mathbf{e} \ \  \widetilde{\mathbf{c}}_{\mathrm{P}}\!\oplus\!\widetilde{\mathbf{c}}_{\mathbf{e}, \mathrm{P}}\!\oplus\! \widetilde{\mathbf{e}}_{\mathrm{P}}].
        \end{equation}
        Following the proof of Theorem \ref{the::HDdis::iphase}, we know that $\widetilde{\mathbf{c}}_{\mathrm{P}}\!\oplus\!\widetilde{\mathbf{c}}_{\mathbf{e}, \mathrm{P}}$ is in fact the parity part of the codeword $\widetilde{\mathbf{c}}_{\mathbf{e}}' = [\mathbf{e} \oplus \widetilde{\mathbf{e}}_{\mathrm{B}}]\widetilde{\mathbf{G}}$, i.e., $\widetilde{\mathbf{c}}_{\mathrm{P}}\!\oplus\!\widetilde{\mathbf{c}}_{\mathbf{e}, \mathrm{P}} = \widetilde{\mathbf{c}}_{\mathbf{e},\mathrm{P}}'$. Consider the position index vector $\mathbf{t}_{\ell}^{h}$. Then the probability $\mathrm{Pr}(\widetilde{\mathbf{d}}_{\mathbf{e}} = \mathbf{z}_{\mathbf{t}_{\ell}^{h}})$ can be represented as
        \begin{equation}  \label{equ::proof::WHD::iphase::Pc}
        \begin{split}
            \mathrm{Pr}(\widetilde{\mathbf{d}}_{\mathbf{e}} \!=\! \mathbf{z}_{\mathbf{t}_{\ell}^{h}}) = \mathrm{Pr}(\mathbf{e} \!=\! \mathbf{z}_{\mathbf{t}_{\ell}^{\mathrm{B}}}) \mathrm{Pr}(\widetilde{\mathbf{c}}_{\mathbf{e},\mathrm{P}}'\!\oplus\! \widetilde{\mathbf{e}}_{\mathrm{P}}\!=\! \mathbf{z}_{\mathbf{t}_{h}^{\mathrm{P}}} | \mathbf{e} \!=\!\mathbf{z}_{\mathbf{t}_{\ell}^{\mathrm{B}}}) .    
        \end{split}
        \end{equation} 
        By considering a random TEP $\mathbf{e}$ in the first $i$ reprocessings, it can be easily obtained that $\mathrm{Pr}(\mathbf{e} = \mathbf{z}_{\mathbf{t}_{\ell}^{\mathrm{B}}}) = \frac{1}{b_{0:i}^{k}}$. Furthermore, we consider $2^{n-k}$ pairs vectors, $\mathbf{x}$ and $\mathbf{x}\oplus \mathbf{z}_{\mathbf{t}_{h}^{\mathrm{P}}}$, with respect to an arbitrary length-$n-k$ binary vector $\mathbf{x}$. Then, $\mathrm{Pr}(\widetilde{\mathbf{c}}_{\mathbf{e},\mathrm{P}}'\oplus\widetilde{\mathbf{e}}_{\mathrm{P}} = \mathbf{z}_{\mathbf{t}_{h}^{\mathrm{P}}}|\mathbf{e} =\mathbf{z}_{\mathbf{t}_{\ell}^{\mathrm{B}}}) $ can be represented as 
        \begin{equation} 
        \begin{split}
           &\mathrm{Pr}(\widetilde{\mathbf{c}}_{\mathbf{e},\mathrm{P}}'\oplus\widetilde{\mathbf{e}}_{\mathrm{P}} = \mathbf{z}_{\mathbf{t}_{h}^{\mathrm{P}}}|\mathbf{e} =\mathbf{z}_{\mathbf{t}_{\ell}^{\mathrm{B}}}) \\
           &= \sum_{\mathbf{x}\in \{0,1\}^{n-k}}\mathrm{Pr}(\widetilde{\mathbf{c}}_{\mathbf{e},\mathrm{P}}' =  \mathbf{z}_{\mathbf{t}_h^{\mathrm{P}}}\oplus\mathbf{x}|\mathbf{e} =\mathbf{z}_{\mathbf{t}_{\ell}^{\mathrm{B}}})\mathrm{Pr}(\widetilde{\mathbf{e}}_{\mathrm{P}} = \mathbf{x}) .    
        \end{split}
        \end{equation}
        For $\mathrm{Pr}(\widetilde{\mathbf{c}}_{\mathbf{e},\mathrm{P}}' =  \mathbf{z}_{\mathbf{t}_h^{\mathrm{P}}}\oplus\mathbf{x}|\mathbf{e} =\mathbf{z}_{\mathbf{t}_{\ell}^{\mathrm{B}}})$, we can rewrite it as 
        \begin{equation} \label{equ::proof::WHD::iphase::eB!=e::CepExpand}
        \begin{split}
            \mathrm{Pr}(\widetilde{\mathbf{c}}_{\mathbf{e},\mathrm{P}}' = & \mathbf{z}_{\mathbf{t}_h^{\mathrm{P}}}\oplus\mathbf{x}|\mathbf{e} =\mathbf{z}_{\mathbf{t}_{\ell}^{\mathrm{B}}}) \\
            &= \sum_{q = 1}^{k} \mathrm{Pr}(w(\mathbf{e} \oplus \widetilde{\mathbf{e}}_{\mathrm{B}}) = q | \mathbf{e} = \mathbf{z}_{\mathbf{t}_{\ell}^{\mathrm{B}}})\\
            &\quad\cdot\mathrm{Pr}(\widetilde{\mathbf{c}}_{\mathbf{e},\mathrm{P}}' =  \mathbf{z}_{\mathbf{t}_h^{\mathrm{P}}}\oplus\mathbf{x} | w(\mathbf{e} \oplus \widetilde{\mathbf{e}}_{\mathrm{B}}) = q)\\
            &=  \sum_{q = 1}^{k} \mathrm{Pr}(w(\mathbf{e} \oplus \widetilde{\mathbf{e}}_{\mathrm{B}}) = q|\mathbf{e} =\mathbf{z}_{\mathbf{t}_{\ell}^{\mathrm{B}}}) \\
            &\quad \cdot \mathrm{Pr}(\widetilde{\mathbf{c}}_{\mathbf{e},\mathrm{P}}' = \mathbf{z}_{\mathbf{t}_h^{\mathrm{P}}}\oplus\mathbf{x}|w(\widetilde{\mathbf{c}}_{\mathbf{e},\mathrm{P}}') =  \ell) \\
            & \quad\cdot\mathrm{Pr}(w(\widetilde{\mathbf{c}}_{\mathbf{e},\mathrm{P}}') =  \ell | w(\mathbf{e} \oplus \widetilde{\mathbf{e}}_{\mathrm{B}}) = q).
        \end{split}
        \end{equation}
        In (\ref{equ::proof::WHD::iphase::eB!=e::CepExpand}), $\mathrm{Pr}(w(\widetilde{\mathbf{c}}_{\mathbf{e},\mathrm{P}}') =  \ell | w(\mathbf{e} \oplus \widetilde{\mathbf{e}}_{\mathrm{B}}) = q)$ is directly given by $p_{\mathbf{c}_{\mathrm{P}}}(\ell,q)$. It is important to note that $q \neq 0$ to ensure $\mathbf{e}\neq \mathbf{e}_{\mathrm{B}}$. Then, considering the columns of $\widetilde{\mathbf{G}}$ is randomly permuted according to the received sequence, it can be seen that $\mathrm{Pr}(\widetilde{\mathbf{c}}_{\mathbf{e},\mathrm{P}}' = \mathbf{z}_{\mathbf{t}_h^{\mathrm{P}}}\oplus\mathbf{x}|w(\widetilde{\mathbf{c}}_{\mathbf{e},\mathrm{P}}') =  \ell) = \frac{1}{\binom{n-k}{l}}$ for $\ell = w(\mathbf{z}_{\mathbf{t}_h^{\mathrm{P}}}\oplus\mathbf{x})$. It is worthy noting that (\ref{equ::proof::WHD::iphase::eB!=e::CepExpand}) does not have a summation over $\ell$ because $\ell = w(\mathbf{z}_{\mathbf{t}_h^{\mathrm{P}}}\oplus\mathbf{x})$ is determined by $\mathbf{x}$ and $\mathbf{z}_{\mathbf{t}_h^{\mathrm{P}}}$. Furthermore, $\mathrm{Pr}(w(\mathbf{e} \oplus \widetilde{\mathbf{e}}_{\mathrm{B}}) = q|\mathbf{e} =\mathbf{z}_{\mathbf{t}_{\ell}^{\mathrm{B}}})$ can be derived as
        \begin{equation}
            \mathrm{Pr}(w(\mathbf{e} \oplus \widetilde{\mathbf{e}}_{\mathrm{B}}) = q|\mathbf{e} =\mathbf{z}_{\mathbf{t}_{\ell}^{\mathrm{B}}}) = \sum_{\substack{\mathbf{x} \in \{0,1\}^{k}\\ w(\mathbf{z}_{t_{\ell}^{\mathrm{B}}}\oplus\mathbf{x})=q}}\mathrm{Pr}(\widetilde{\mathbf{e}}_{\mathrm{B}} = \mathbf{x}),
        \end{equation}
        where $\mathrm{Pr}(\widetilde{\mathbf{e}}_{\mathrm{B}} = \mathbf{x})$ is determined as (\ref{equ::WHD::iphase::eB!=e::Pc::eB=eksi}) by using the joint $\mathrm{pdf}$ of $[\widetilde{R}]_1^n$ given by (\ref{equ::proof::WHD::0phase::jointPDF::R1n}).
        
        When $\widetilde{\mathbf{d}}_{\mathbf{e}} = \mathbf{z}_{\mathbf{t}_{\ell}^{h}}$, the $\mathrm{pdf}$ of $D_{\mathbf{e}}^{(\mathrm{W})}$ is directly given by $f_{\widetilde{A}_{\mathbf{t}_{\ell}^{h}}}$. Let us take $ \mathrm{Pr}(\widetilde{\mathbf{d}}_{\mathbf{e}} = \mathbf{z}_{\mathbf{t}_{\ell}^{h}}) = \mathrm{Pc}(\mathbf{t}_{\ell}^{h})$. Thus,
        considering all possible $\mathbf{t}_{\ell}^{h}$ and using the law of total probability, we can finally obtain (\ref{equ::WHD::iphase::eB!=e}), which completes the proof of Lemma \ref{lem::WHD::iphase::eB!=e}.
        
    \section{Proof of Theorem \ref{the::WHD::iphase}} \label{app::proof::WHD::iphase}
        When $w(\widetilde{\mathbf{e}}_{\mathrm{B}}) > i $, i.e., $E_1^{k} > i$, the first $i$ reprocessings can not decode the received signal correctly. According to Lemma \ref{lem::WHD::iphase::eB!=e}, the minimum WHD on the condition that $E_i^{k} > i$ is given by
		\begin{equation}
	         D_i^{(\mathrm{W})}=  \min_{\forall \mathbf{e} : w(\mathbf{e}) \leq i} \{D_{\mathbf{e}}|\widetilde{\mathbf{e}}_{\mathrm{B}} \neq \mathbf{e}\} .
    	\end{equation}
        It is proved in Lemma \ref{Lem::positiveCov} that the covariance $\mathrm{cov}(\widetilde{A}_i,\widetilde{A}_j)$, $1\leq i < j \leq n$, is non-negative. From (\ref{equ::WHD::iphase::eB!=e::eB<=i::Cov}), we know that the covariance $\mathrm{cov}\left(D_{\mathbf{e}}^{(\mathrm{W})},D_{\mathbf{e}'}^{(\mathrm{W})}\right)$ is a linear combination of $\mathrm{cov}(\widetilde{A}_i,\widetilde{A}_j)$ with positive coefficients. Thus for any TEPs $\mathbf{e}$ and $\mathbf{e}'$ satisfying $\mathbf{e} \neq \widetilde{\mathbf{e}}_{\mathrm{B}}$ and $\mathbf{e}' \neq \widetilde{\mathbf{e}}_{\mathrm{B}}$, respectively , $\mathrm{cov}\left(D_{\mathbf{e}}^{(\mathrm{W})},D_{\mathbf{e}'}^{(\mathrm{W})}\right)$ and $\rho$ are also non-negative. Furthermore, we regard $D_{\mathbf{e}}^{(\mathrm{W})}$ as a normally distributed variable when $n$ is large because it is a large-number summation of random variables $[\widetilde{A}]_1^n$. Let $f_{\widetilde{D}_i^{(\mathrm{W})}}(x,b|w(\widetilde{\mathbf{e}}_{\mathrm{B}})\!>\!i)$ denote the $\mathrm{pdf}$ of the minimum element of a sequence of $b$ samples $d_{\mathbf{e}}^{(\mathrm{W})}$ of $D_{\mathbf{e}}^{(\mathrm{W})}$ , then $f_{\widetilde{D}_i^{(\mathrm{W})}}(x,b|w(\widetilde{\mathbf{e}}_{\mathrm{B}})\!>\!i)$ can be derived as (\ref{equ::WHD::iphase::DependOrder2}) by considering the ordered statistics of normal variables with positive correlation coefficient $\rho \in [0,1)$ \cite[Corollary 6.1.1]{tong2012multivariate}. Also, since in the first $i$ reprocessings, the overall number of checked TEP is $b_{0:i}^{k}$, we take $b = b_{0:i}^{k}$ in (\ref{equ::WHD::iphase::DependOrder2}). 

        When $w(\widetilde{\mathbf{e}}_{\mathrm{B}}) \leq i $, i.e., $E_1^{k} \leq i$, the first $i$ reprocessings can eliminate the errors in MRB positions by one TEP ${\mathbf{e}}$ which equals $\widetilde{\mathbf{e}}_{\mathrm{B}}$, while there are still $b_{1:i}^{k}$ TEPs that can not eliminate the error $\widetilde{\mathbf{e}}_{\mathrm{B}}$. Therefore, the munimum WHD on the condition $E_i^{k} \leq i$ is given by
    	\begin{equation}
    	     D_i^{(\mathrm{W})} =  \min_{\substack{\forall \mathbf{e} : w(\mathbf{e}) \leq i \\ \mathbf{e}\neq \widetilde{\mathbf{e}}_{\mathrm{B}}}} \{D_{\widetilde{\mathbf{e}}_{\mathrm{B}}}^{(\mathrm{W})},~D_{\mathbf{e}}^{(\mathrm{W})}\} .
    	\end{equation}
    
        Considering the ordered statistics over all possible $D_{\mathbf{e}}^{(\mathrm{W})}$, we obtain the $\mathrm{pdf}$ of $D_{i}^{(\mathrm{W})}$ conditioning on $\{w(\widetilde{\mathbf{e}}_{\mathrm{B}}) \leq i \}$ as
        \begin{equation}
        \begin{split}
           &f_{D_{\mathbf{e}}^{(\mathrm{W})}}(x|w(\widetilde{\mathbf{e}}_{\mathrm{B}}) \leq i)  \\
           &\quad= f_{D_{\mathbf{e}}^{(\mathrm{W})}}(x|\widetilde{\mathbf{e}}_{\mathrm{B}}\!= \!\mathbf{e})\!\!\int_{x}^{\infty}\!\!\!f_{\widetilde{D}_i^{(\mathrm{W})}}\left(u, b_{1:i}^{k}|w(\widetilde{\mathbf{e}}_{\mathrm{B}}) \!\leq\! i\right) du \\
           &\quad+ f_{\widetilde{D}_i^{(\mathrm{W})}}\left(u, b_{1:i}^{k}|w(\widetilde{\mathbf{e}}_{\mathrm{B}}) \!\leq\! i\right) \!\!\int_{x}^{\infty}\!\!\!f_{D_{\mathbf{e}}^{(\mathrm{W})}}(u|\widetilde{\mathbf{e}}_{\mathrm{B}}\! =\! \mathbf{e})du, 
        \end{split}
        \end{equation}
        where $f_{\widetilde{D}_i^{(\mathrm{W})}}\left(u, b_{1:i}^{k}|w(\widetilde{\mathbf{e}}_{\mathrm{B}}) \!\leq\! i\right)$ is derived as (\ref{equ::WHD::iphase::DependOrder2}) by using the arguments in \cite[Corollary 6.1.1]{tong2012multivariate}.
        Finally, we can obtain (\ref{equ::WHD::iphase}) by using the law of total probability, i.e.
        \begin{equation}
        \begin{split}
              f_{D_{\mathbf{e}}^{(\mathrm{W})}}(x) &= \mathrm{Pr}(w(\widetilde{\mathbf{e}}_{\mathrm{B}}) \leq i) f_{D_{\mathbf{e}}^{(\mathrm{W})}}(x|w(\widetilde{\mathbf{e}}_{\mathrm{B}}) \leq i) \\
              &+ \mathrm{Pr}(w(\widetilde{\mathbf{e}}_{\mathrm{B}}) > i) f_{D_{\mathbf{e}}^{(\mathrm{W})}}(x|w(\widetilde{\mathbf{e}}_{\mathrm{B}}) > i),  
        \end{split}
        \end{equation}
        where $\mathrm{Pr}(w(\widetilde{\mathbf{e}}_{\mathrm{B}}) \leq i) = \sum_{v=0}^{i}p_{E_1^{k}}(v)$ and $\mathrm{Pr}(w(\widetilde{\mathbf{e}}_{\mathrm{B}}) > i) = 1- \sum_{v=0}^{i}p_{E_1^{k}}(v)$ are obtained from Lemma \ref{lem::OrderStat::Eab}. This completes the proof of Theorem \ref{the::WHD::iphase}.
        
    \section{Proof of Proposition \ref{pro::HDtech::DR::PproIncreasing}} \label{app::proof::HDtech::PproIncreasing}
        Let us consider the derivative of $\mathrm{P}_{\mathbf{e}}^{\mathrm{pro}}(d_{\mathrm{H}}|\widetilde {\bm\alpha})$ with respect to $\mathrm{Pe}(\mathbf{e}|\widetilde {\bm\alpha})$, which can be derived as
        \begin{align} \label{equ::proof::HDtech::PproIncreasing}
            \frac{\partial \, \mathrm{P}_{\mathbf{e}}^{\mathrm{pro}}(d_{\mathrm{H}}|\widetilde {\bm\alpha})}{\partial \, \mathrm{Pe}(\mathbf{e}|\widetilde{\bm\alpha})} = & \sum_{j=w(\mathrm{e})}^{d_{\mathrm{H}}}p_{E_{k+1}^{n}}(j-w(\mathbf{e})|\widetilde{\bm\alpha}) \notag\\
            - &\sum_{j=w(\mathrm{e})}^{d_{\mathrm{H}}}p_{W_{\mathbf{c}_{\mathrm{P}}}}(j-w(\mathbf{e}))\\
             \overset{(a)}{=}& \sum_{j=0}^{d_{\mathrm{H}}-w(\mathrm{e})}\binom{n-k}{j}
	          \left(\mathbb{E}[\mathrm {Pe}] \right)^j \left(1 - \mathbb{E}[\mathrm {Pe}] \right)^{n\!-\!k\!-\!j}\notag\\
            -& \sum_{j=0}^{d_{\mathrm{H}}-w(\mathrm{e})}\binom{n-k}{j}\frac{1}{2^{n-k}},\notag
        \end{align}
        where 
        \begin{equation}
                    \mathbb{E}[\mathrm{Pe}] = \frac{1}{n -k}\sum_{j=k+1}^{n}\mathrm{Pe}(j|\widetilde\alpha_j) .   
        \end{equation}
        Step (a) of (\ref{equ::proof::HDtech::PproIncreasing}) follows from that $p_{E_{k+1}^{n}}(j-w(\mathbf{e})|\widetilde{\bm\alpha})$ is given by (\ref{equ::HDtech::SR::Eab::Cond}) and $p_{W_{\mathbf{c}_{\mathrm{P}}}}(j-w(\mathbf{e})) = p_{d}(j) =  \binom{n-k}{j}\frac{1}{2^{n-k}}$ under binomial code spectrum assumption.
        Using the regularized incomplete beta function $I_x(a,b)$, (\ref{equ::proof::HDtech::PproIncreasing}) can be represented as
        \begin{equation}
            \begin{split}
                \frac{\partial\, \mathrm{P}_{\mathbf{e}}^{\mathrm{pro}}(d_{\mathrm{H}}|\widetilde{\bm\alpha})}{\partial\, \mathrm{Pe}(\mathbf{e}|\widetilde{\bm\alpha})} &= I_{1- \mathbb{E}[\mathrm{Pe}]}(n\!-\!k\!-\!d_{\mathrm{H}}\!+\!w(\mathrm{e}),d_{\mathrm{H}}\!-\!w(\mathrm{e})\!+\!1) \\
                &\quad - \frac{1}{\beta}I_{\frac{1}{2}}(n\!-\!k\!-\!d_{\mathrm{H}}\!+\!w(\mathrm{e}),d_{\mathrm{H}}\!-\!w(\mathrm{e})\!+\!1) \\
                &\geq I_{1\!-\! \mathbb{E}[\mathrm{Pe}]}(n\!-\!k\!-\!d_{\mathrm{H}}\!+\!w(\mathrm{e}),d_{\mathrm{H}}\!-\!w(\mathrm{e})\!+\!1) \\
                &\quad- I_{\frac{1}{2}}(n\!-\!k\!-\!d_{\mathrm{H}}\!+\!w(\mathrm{e}),d_{\mathrm{H}}\!-\!w(\mathrm{e})\!+\!1) \\
                & = (n-k-d_{\mathrm{H}}+w(\mathrm{e}))\binom{n-k}{d_{\mathrm{H}}-w(\mathrm{e})} \\
                &\quad\cdot\int_{\frac{1}{2}}^{1\!-\! \mathbb{E}[\mathrm{Pe}]}t^{n\!-\!k\!-\!d_{\mathrm{H}}\!+\!w(\mathrm{e})\!-\!1} (1-t)^{d_{\mathrm{H}}\!-\!w(\mathrm{e})}d t .
            \end{split}
        \end{equation}
        Furthermore, it has been proved that $\mathrm{Pe}(j|\widetilde\alpha_j) < 1/2$ for $1\leq j \leq n$ \cite{Fossorier1995OSD}, so that we can obtain that $1-\mathbb{E}[\mathrm{Pe}] > 1/2$. Therefore, we can conclude that
        \begin{equation}
            \frac{\partial\, \mathrm{P}_{\mathbf{e}}^{\mathrm{pro}}(d_{\mathrm{H}}|\bm{\widetilde\alpha})}{\partial\, \mathrm{Pe}(\mathbf{e}|\bm{\widetilde\alpha})} > 0,
        \end{equation}
        and this completes the proof of Proposition \ref{pro::HDtech::DR::PproIncreasing}.
        
    \section{Proof of Corollary \ref{cor::Stech::Condis::WHDforTEP}} \label{app::proof::Stech::Condis::WHDforTEP}
            Given an arbitrary position indices vector $\mathbf{t}_h^{\mathrm{P}} \in \mathcal{T}_h^{\mathrm{P}}$, $0\leq h \leq (n-k)$ and the corresponding random variable $\widetilde { A}_{\mathbf{t}_{\mathbf{e}}^{h}}$ with $\mathrm{pdf}$ $f_{\widetilde { A}_{\mathbf{t}_{\mathbf{e}}^{h}}}(x)$, the $\mathrm{pdf}$ of the WHD $D_{\mathbf{e}}^{(\mathrm{W})}$ can be obtained by considering the mixture of all possible $\mathbf \mathbf{t}_h^{\mathrm{P}}$, $0\leq h \leq (n-k)$, i.e.,
            \begin{equation} \label{equ::proof_OveralleDe}
                f_{D_{\mathbf{e}}^{(\mathrm{W})}}(x|\mathbf{e} = [e]_1^k)  = \sum_{h=0}^{n-k} \sum_{\mathbf \mathbf{t}_h^{\mathrm{P}} \in \mathcal{T}_h^{\mathrm{P}}} \mathrm{Pr}(\widetilde{\mathbf{y}}_{\mathrm{P}} \!\oplus\! \widetilde{\mathbf{c}}_{0,\mathrm{P}} = \mathbf{z}_{ \mathbf{t}_h^{\mathrm{P}}}) f_{\widetilde{\mathbf{A}}_{\mathbf{t}_{\mathbf{e}}^{h}}}(x).
            \end{equation}
            We re-write (\ref{equ::proof_OveralleDe}) in the form of conditional probability, i.e.,
            \begin{equation} \label{equ::proof::Stech::Condis::WHDforTEP}
            \begin{split}
                &f_{D_{\mathbf{e}}^{(\mathrm{W})}}(x|\mathbf{e} = [e]_1^k) \\
                & = \mathrm{Pe}(\mathbf{e}) \sum_{h=0}^{n-k} \sum_{\mathbf \mathbf{t}_h^{\mathrm{P}} \in \mathcal{T}_h^{\mathrm{P}}} \!\! \mathrm{Pr}(\widetilde{\mathbf{y}}_{\mathrm{P}} \!\oplus\! \widetilde{\mathbf{c}}_{\mathbf{e},\mathrm{P}} \!=\! \mathbf{z}_{ \mathbf{t}_h^{\mathrm{P}}}| \widetilde{\mathbf{e}}_{\mathrm{B}} \!=\! \mathbf{e}) f_{\widetilde{A}_{\mathbf{t}_{\mathbf{e}}^{h}}}(x) \\
                &+(1 \!-\! \mathrm{Pe}(\mathbf{e})) \sum_{h=0}^{n-k} \sum_{\mathbf \mathbf{t}_h^{\mathrm{P}} \in \mathcal{T}_h^{\mathrm{P}}}  \!\!\mathrm{Pr}(\widetilde{\mathbf{y}}_{\mathrm{P}} \!\oplus\! \widetilde{\mathbf{c}}_{\mathbf{e},\mathrm{P}} \!=\! \mathbf{z}_{ \mathbf{t}_h^{\mathrm{P}}}|\widetilde{\mathbf{e}}_{\mathrm{B}} \!\neq\! \mathbf{e})  f_{\!\widetilde{A}_{\mathbf{t}_{\mathbf{e}}^{h}}}(x),
            \end{split}
            \end{equation}
            where $\mathrm{Pe}(\mathbf{e}) = \mathrm{Pr}(\widetilde{\mathbf{e}}_{\mathrm{B}}=\mathbf{e})$ is given by (\ref{equ::HDtech::CondDis::Pe(e)}). In (\ref{equ::proof::Stech::Condis::WHDforTEP}), we use $\mathrm{Pe}(\mathbf{t}_{\mathbf{e}}^{h})$ to denote $\mathrm{Pe}(\mathbf{e})  \mathrm{Pr}(\widetilde{\mathbf{y}}_{\mathrm{P}} \oplus \widetilde{\mathbf{c}}_{0,\mathrm{P}} = \mathbf{z}_{ \mathbf{t}_h^{\mathrm{P}}}| \widetilde{\mathbf{e}}_{\mathrm{B}} = \mathbf{e})$, i.e.,
            \begin{equation}
            \begin{split}
                   \mathrm{Pe}(\mathbf{t}_{\mathbf{e}}^{h}) &=  \mathrm{Pe}(\mathbf{e})  \mathrm{Pr}(\widetilde{\mathbf{y}}_{\mathrm{P}} \oplus \widetilde{\mathbf{c}}_{0,\mathrm{P}} = \mathbf{z}_{ \mathbf{t}_h^{\mathrm{P}}}| \widetilde{\mathbf{e}}_{\mathrm{B}} = \mathbf{e})\\ &\overset{(a)}{=} \mathrm{Pr}(\widetilde{\mathbf{e}} = \mathbf{z}_{ \mathbf{t}_{\mathbf{e}}^h}) ,               
            \end{split}
            \end{equation}
            where step (a) follows from that when $\widetilde{\mathbf{e}}_{\mathrm{B}} = \mathbf{e}$, the difference pattern between $\widetilde{\mathbf{y}}_{\mathrm{P}}$ and $\widetilde{\mathbf{c}}_{\mathbf{e},\mathrm{P}}$ is given by $\widetilde{\mathbf{e}}_{\mathrm{P}}$, as proved in Lemma \ref{lem::WHD::iphase::eB=e}. Thus, $\mathrm{Pe}(\mathbf{t}_{\mathbf{e}}^{h})$ is the probability that only positions $\mathbf{t}_{\mathbf{e}}^{h}$ are in error in $\widetilde{\mathbf{y}}$. Thus, $\mathrm{Pe}(\mathbf{t}_{\mathbf{e}}^{h})$ can be obtained as (\ref{equ::Stech::Condis::WHDforTEP::eB=e}) by considering the joint distribution of $[\widetilde{R}]_1^n$, i.e., $f_{[\widetilde{R}]_1^n}(x_1,\ldots,x_n)$ given by \ref{equ::proof::WHD::0phase::jointPDF::R1n}. For the second term of (\ref{equ::proof::Stech::Condis::WHDforTEP}), the conditional probability $\mathrm{Pr}(\widetilde{\mathbf{y}}_{\mathrm{P}} \oplus \widetilde{\mathbf{c}}_{0,\mathrm{P}} = \mathbf{z}_{ \mathbf{t}_h^{\mathrm{P}}}|\widetilde{\mathbf{e}}_{\mathrm{B}} \neq \mathbf{e})$ can be derived as (\ref{equ::Stech::Condis::WHDforTEP::eB!=e}) following the approach to obtain (\ref{equ::proof::WHD::iphase::Pc}) in Lemma  (\ref{lem::WHD::iphase::eB!=e}). This completes the proof of Corollary \ref{cor::Stech::Condis::WHDforTEP}.
            
        \section{Proof of Corollary \ref{cor::Stech::SR::WHDforTEP::alpha}} \label{app::proof::Stech::SR::WHDforTEP::alpha}
            The probability $\mathrm{Pr}(D_{\mathbf{e}}^{(\mathrm{W})}\!=\!d_{\mathbf{t}_{\mathbf{e}}^{h}}^{(\mathrm{W})}|\widetilde{\bm\alpha}) $ can be represented as
            \begin{equation} \label{equ::proof::Stech::Condis::WHDforTEP::alpha}
            \begin{split}
                  &\mathrm{Pr}(D_{\mathbf{e}}^{(\mathrm{W})}\!=\!d_{\mathbf{t}_{\mathbf{e}}^{h}}^{(\mathrm{W})}|\widetilde{\bm\alpha}) \\
                  &\quad=  \mathrm{Pe}(\mathbf{e} | \widetilde{\bm\alpha})  \mathrm{Pr}(\widetilde{\mathbf{y}}_{\mathrm{P}} \!\oplus\! \widetilde{\mathbf{c}}_{\mathbf{e},\mathrm{P}} = \mathbf{z}_{\mathbf{t}_h^{\mathrm{P}}}| \widetilde{\mathbf{e}}_{\mathrm{B}} \!=\! \mathbf{e},\widetilde{\bm\alpha})\\
                  &\quad+  (1- \mathrm{Pe}(\mathbf{e}| \widetilde{\bm\alpha}))  \mathrm{Pr}(\widetilde{\mathbf{y}}_{\mathrm{P}} \!\oplus\! \widetilde{\mathbf{c}}_{0,\mathrm{P}} \!=\! \mathbf{z}_{\mathbf{t}_h^{\mathrm{P}}}| \widetilde{\mathbf{e}}_{\mathrm{B}} \!\neq\! \mathbf{e}, \widetilde{\bm\alpha}).  
            \end{split}
            \end{equation}
            By considering that the bit-wise error probabilities conditioning on $[{\widetilde{A}}]_1^n = [\widetilde{\alpha}]_1^n$ are independent, $\mathrm{Pe}(\mathbf{e} | \widetilde{\bm\alpha}) $ is simply given by (\ref{equ::HDtech::SR::Pe(e)::Cond}). Furthermore, as proved in the proof of Lemma \ref{lem::WHD::iphase::eB=e}, when $\widetilde{\mathbf{e}}_{\mathrm{B}} = \mathbf{e}$, it can be obtained that $\widetilde{\mathbf{y}}_{\mathrm{P}} \!\oplus\! \widetilde{\mathbf{c}}_{\mathbf{e},\mathrm{P}} = \widetilde{\mathbf{e}}_{\mathrm{P}}$. Thus, it can be seen that $ \mathrm{Pr}(\widetilde{\mathbf{y}}_{\mathrm{P}} \!\oplus\! \widetilde{\mathbf{c}}_{0,\mathrm{P}} \!=\! \mathbf{z}_{\mathbf{t}_h^{\mathrm{P}}}| \widetilde{\mathbf{e}}_{\mathrm{B}} \!=\! \mathbf{e},\widetilde{\bm\alpha})$ is the probability that only positions indexed by $\mathbf{t}_h^{\mathrm{P}}$ are in error in $\widetilde{\mathbf{y}}_{\mathrm{P}}$ conditioning on $[{\widetilde{A}}]_1^n = [\widetilde{\alpha}]_1^n$, which can be derived as 
            \begin{equation} \label{equ::proof::Stech::Condis::WHDforTEP::alpha::Pet}
                 \mathrm{Pr}(\widetilde{\mathbf{y}}_{\mathrm{P}} \oplus \widetilde{\mathbf{c}}_{0,\mathrm{P}}\! =\! \mathbf{z}_{\mathbf{t}_h^{\mathrm{P}}}| \widetilde{\mathbf{e}}_{\mathrm{B}} \!=\! \mathbf{e}) \!= \!\!\!\!\prod_{\substack{k < u \leq n\\u\in \mathbf{t}_h^{\mathrm{P}}}}\!\!\!\! \mathrm{Pe}(u|\widetilde\alpha_u) \!\!\!\!\prod_{\substack{k < u \leq n\\u\notin \mathbf{t}_h^{\mathrm{P}}}} \!\!\!\!(1\!-\! \mathrm{Pe}(u|\widetilde\alpha_u)).
            \end{equation}
            On the other hand, according to Lemma \ref{lem::WHD::iphase::eB!=e}, $\mathrm{Pr}(\widetilde{\mathbf{y}}_{\mathrm{P}} \!\oplus\! \widetilde{\mathbf{c}}_{0,\mathrm{P}} \!=\! \mathbf{z}_{\mathbf{t}_h^{\mathrm{P}}}| \widetilde{\mathbf{e}}_{\mathrm{B}} \!\neq\! \mathbf{e}, \widetilde{\bm\alpha})$ can be represented as 
            \begin{equation}   \label{equ::proof::Stech::Condis::WHDforTEP::alpha::Pct}
            \begin{split}
                &\mathrm{Pr}(\widetilde{\mathbf{y}}_{\mathrm{P}} \!\oplus\! \widetilde{\mathbf{c}}_{0,\mathrm{P}} \!=\! \mathbf{z}_{\mathbf{t}_h^{\mathrm{P}}}| \widetilde{\mathbf{e}}_{\mathrm{B}} \!\neq\! \mathbf{e}, \widetilde{\bm\alpha}) \\
                &\qquad = \mathrm{Pr}(\widetilde{\mathbf{c}}_{\mathbf{e},\mathrm{P}}'\oplus\widetilde{\mathbf{e}}_{\mathrm{P}}   = \mathbf{z}_{\mathbf{t}_{h}^{\mathrm{P}}}| \widetilde{\mathbf{e}}_{\mathrm{B}} \!\neq\! \mathbf{e}, \widetilde{\bm\alpha}) .   
            \end{split}
            \end{equation}
            Note that when $[{\widetilde{A}}]_1^n = [\widetilde{\alpha}]_1^n$, for the $u$-th bit of $\widetilde{\mathbf{e}}$, $k<u\leq n$, we can obtain $\mathrm{Pr}(\widetilde{e}_u \neq 0|\widetilde{\alpha}_u) = \mathrm{Pe}(u|\widetilde{\alpha}_u)$. For the $u$-th bit of $\widetilde{\mathbf{c}}_{\mathbf{e}}'$, $k<u\leq n$, the probability $\mathrm{Pr}(\widetilde{c}_{\mathbf{e},u}'\neq 0|\widetilde{\alpha}_u)$ can be represented as 
            \begin{equation}
            \begin{split}
                    &\mathrm{Pr}(\widetilde{c}_{\mathbf{e},u}'\neq 0|\widetilde{\alpha}_u) \\
                    &\qquad = \mathrm{Pr}(\widetilde{c}_{\mathbf{e},u}'\neq 0|w(\widetilde{\mathbf{e}}_{\mathrm{B}}\oplus \mathbf{e})\!=\! q)\mathrm{Pr}(w(\widetilde{\mathbf{e}}_{\mathrm{B}}\oplus \mathbf{e})\! = \! q|\widetilde{\bm{\alpha}}),    
            \end{split}
            \end{equation}
            where $\mathrm{Pr}(\widetilde{c}_{\mathbf{e},u}'\neq 0|w(\widetilde{\mathbf{e}}_{\mathrm{B}}\oplus \mathbf{e})= q)$ is previously given by $p_{\mathbf{c}_{\mathrm{P}}}^{\mathrm{bit}}(u,q)$ in (\ref{equ::WHD::App::PcPbit1}). $\mathrm{Pr}(w(\widetilde{\mathbf{e}}_{\mathrm{B}}\oplus \mathbf{e})\! = \! q|\widetilde{\bm{\alpha}})$ can be derived by considering all length-$k$ vectors $\mathbf{x}$ satisfying $w(\mathbf{x}\oplus \mathbf{e}) = q$, i.e., $\mathrm{Pr}(w(\widetilde{\mathbf{e}}_{\mathrm{B}}\oplus \mathbf{e})\! = \! q|\widetilde{\bm{\alpha}}) = \sum_{\substack{\mathbf{x} \in \{0,1\}^{k}\\ w(\mathbf{e}\oplus\mathbf{x})=q}}\mathrm{Pr}(\widetilde{\mathbf{e}}_{\mathrm{B}} = \mathbf{x}|\widetilde{\bm{\alpha}})$, where $\mathrm{Pr}(\widetilde{\mathbf{e}}_{\mathrm{B}} = \mathbf{x}|\widetilde{\bm{\alpha}})$ can be easily derived as (\ref{equ::Stech::SR::Pce::eb=x}) by using the reliabilities $[\widetilde{\alpha}]_1^n$. Thus, for the $u$-th bit, $k<u\leq n$, of $\widetilde{\mathbf{c}}_{\mathbf{e},\mathrm{P}}'\oplus\widetilde{\mathbf{e}}_{\mathrm{P}}$, i.e., $\widetilde{c}_{\mathbf{e},u}'\oplus \widetilde{e}_u$, we have
            \begin{equation}
            \begin{split}
                 \mathrm{Pr}(\widetilde{c}_{\mathbf{e},u}'\oplus \widetilde{e}_u \!\neq\! 0 |\widetilde{\alpha}_u)& = \mathrm{Pe}(u|\widetilde{\alpha}_u) (1-\mathrm{Pr}(\widetilde{c}_{\mathbf{e},u}'\neq 0|\widetilde{\alpha}_u)) \\
                 &+ (1 -  \mathrm{Pe}(u|\widetilde{\alpha}_u))\mathrm{Pr}(\widetilde{c}_{\mathbf{e},u}'\neq 0|\widetilde{\alpha}_u).   
            \end{split}
            \end{equation}
            For the simplicity, we take $\mathrm{Pc}_{\mathbf{e}}(u|\widetilde{\alpha}_u) = \mathrm{Pr}(\widetilde{c}_{\mathbf{e},u}'\oplus \widetilde{e}_u \!\neq\! 0 |\widetilde{\alpha}_u) $. Then, $\mathrm{Pr}(\widetilde{\mathbf{y}}_{\mathrm{P}} \!\oplus\! \widetilde{\mathbf{c}}_{0,\mathrm{P}} \!=\! \mathbf{z}_{\mathbf{t}_h^{\mathrm{P}}}| \widetilde{\mathbf{e}}_{\mathrm{B}} \!\neq\! \mathbf{e}, \widetilde{\bm\alpha})$ given by (\ref{equ::proof::Stech::Condis::WHDforTEP::alpha::Pct}) is derived as
            \begin{equation} \label{equ::proof::Stech::Condis::WHDforTEP::alpha::Pct2}
            \begin{split}
                &\mathrm{Pr}(\widetilde{\mathbf{y}}_{\mathrm{P}} \!\oplus\! \widetilde{\mathbf{c}}_{0,\mathrm{P}} \!=\! \mathbf{z}_{\mathbf{t}_h^{\mathrm{P}}}| \widetilde{\mathbf{e}}_{\mathrm{B}} \!\neq\! \mathbf{e}, \widetilde{\bm\alpha}) \\
                &\qquad = \prod_{\substack{k < u \leq n\\u\in \mathbf{t}_h^{\mathrm{P}}}} \mathrm{Pc}_{\mathbf{e}}(u|\widetilde{\alpha}_u) \cdot\prod_{\substack{k < u \leq n\\u\notin \mathbf{t}_h^{\mathrm{P}}}} (1- \mathrm{Pc}_{\mathbf{e}}(u|\widetilde{\alpha}_u)).    
            \end{split}
            \end{equation}
            Substituting (\ref{equ::proof::Stech::Condis::WHDforTEP::alpha::Pet}) and (\ref{equ::proof::Stech::Condis::WHDforTEP::alpha::Pct2}) into (\ref{equ::proof::Stech::Condis::WHDforTEP::alpha}), we can finally obtain (\ref{equ::Stech::SR::WHDforTEP::alpha}). This completes the proof of Corollary \ref{cor::Stech::SR::WHDforTEP::alpha}.
            
        \section{Proof of Proposition \ref{pro::Stech::DR::PproIncreasing}} \label{app::proof::Stech::DR::PproIncreasing}
        
                Assume that there exist two arbitrary TEPs $\mathbf{e}_1$ and $\mathbf{e}_2$ to be processed in the $i$-reprocessing, satisfying $\mathrm{Pe}(\mathbf{e}_1|\bm{\widetilde\alpha}) > \mathrm{Pe}(\mathbf{e}_2|\bm{\widetilde\alpha})$. Let us define $\Delta \triangleq \widetilde{\mathrm{P}}_{\mathbf{e}_1}^{\mathrm{pro}}(d_{\min}^{(\mathrm{W})}|\bm{\widetilde\alpha}) - \widetilde{\mathrm{P}}_{\mathbf{e}_2}^{\mathrm{pro}}(d_{\min}^{(\mathrm{W})}|\bm{\widetilde\alpha})$, which can be obtained that
                \begin{equation}
                \begin{split}
                    \Delta & = \sum_{h=0}^{n-k}\!\!\!\!\sum_{\substack{\mathbf{t}_h^{\mathrm{P}}\in\mathcal{T}_h^{\mathrm{P}} \\d_{_{\mathbf{t}_{\mathbf{e}_1}^h}}^{(\mathrm{W})}\!< d_{\min}^{(\mathrm{W})}}} \!\!\!\!p_{D_{\mathbf{e}_1}^{(\mathrm{W})}}(d_{\mathbf{t}_{\mathbf{e}_1}^h}^{(\mathrm{W})}|\bm{\widetilde\alpha}) - \!\!\sum_{h=0}^{n-k}\!\!\!\!\sum_{\substack{\mathbf{t}_h^{\mathrm{P}}\in\mathcal{T}_h^{\mathrm{P}} \\ d_{_{\mathbf{t}_{\mathbf{e}_2}^h}}^{(\mathrm{W})}\!<d_{\min}^{(\mathrm{W})}}} \!\!\!\! p_{D_{\mathbf{e}_2}^{(\mathrm{W})}}(d_{\mathbf{t}_{\mathbf{e}_2}^h}^{(\mathrm{W})}|\bm{\widetilde\alpha})\\
                    & \overset{(a)}{\geq}  \sum_{h=0}^{n-k}\sum_{\substack{\mathbf{t}_h^{\mathrm{P}}\in\mathcal{T}_h^{\mathrm{P}} \\ d_{_{\mathbf{t}_{\mathbf{e}_2}^h}}^{(\mathrm{W})}\!< d_{\min}^{(\mathrm{W})}}} \left(p_{D_{\mathbf{e}_1}^{(\mathrm{W})}}(d_{\mathbf{t}_{\mathbf{e}_1}^h}^{(\mathrm{W})}|\bm{\widetilde\alpha}) - p_{D_{\mathbf{e}_2}^{(\mathrm{W})}}(d_{\mathbf{t}_{\mathbf{e}_2}^h}^{(\mathrm{W})}|\bm{\widetilde\alpha})\right)\\
                    & \overset{(b)}{=}  \sum_{h=0}^{n-k}\sum_{\substack{\mathbf{t}_h^{\mathrm{P}}\in\mathcal{T}_h^{\mathrm{P}} \\ d_{_{\mathbf{t}_{\mathbf{e}_2}^h}}^{(\mathrm{W})}\!< d_{\min}^{(\mathrm{W})}}}\left(\mathrm{Pe}(\mathbf{e}_1|\bm{\widetilde\alpha}) - \mathrm{Pe}(\mathbf{e}_2|\bm{\widetilde\alpha}) \right) \\
                    &\cdot \left( \prod_{\substack{k < u \leq n\\u\in \mathbf{t}_h^{\mathrm{P}}}} \mathrm{Pe}(u|\widetilde\alpha_u) \prod_{\substack{k < u \leq n\\u\notin \mathbf{t}_h^{\mathrm{P}}}} (1- \mathrm{Pe}(u|\widetilde\alpha_u)) - 2^{k-n}\right)\\
                    & \!\overset{(c)}{>}   \left(\mathrm{Pe}(\mathbf{e}_1|\bm{\widetilde\alpha}) \!-\! \mathrm{Pe}(\mathbf{e}_2|\bm{\widetilde\alpha}) \right)\!\left( \prod_{u = k+1}^{n} \!\!(1\!-\! \mathrm{Pe}(u|\widetilde\alpha_u)) \!-\! 2^{k\!-\!n}\!\right)\!,
                \end{split}
            \end{equation}
            where step (a) follows from that for a specific vector $\mathbf{t}_h^{\mathrm{P}}\in\mathcal{T}_h^{\mathrm{P}}$, inequality $d_{\mathbf{t}_{\mathbf{e}_1}^h}^{(\mathrm{W})} \geq d_{\mathbf{t}_{\mathbf{e}_2}^h}^{(\mathrm{W})}$ holds, step (b) follows from that in (\ref{equ::Stech::Condis::Psuce}), $\mathrm{Pc}_{\mathbf{e}}(u|\widetilde{\alpha}_{u}) = \frac{1}{2}$ when the weight spectrum of $\mathcal{C}(n,k)$ is binomial (see Eq. (\ref{equ::discuss::soft::Pce::app})), and step (c) takes $h=0$ and $\mathbf{t}_h^{\mathrm{P}} = \varnothing$.
            
            Furthermore, because $\mathrm{Pe}(u|\widetilde\alpha_u) < \frac{1}{2}$ holds for $1 \leq u \leq n$ \cite{Fossorier1995OSD}, the inequality
            \begin{equation}
                 \prod_{u = k+1}^{n} (1- \mathrm{Pe}(u|\widetilde\alpha_u)) - 2^{k-n} \geq 0
            \end{equation}
            holds. Therefore, it can be concluded that $\Delta > 0$, which completes the proof of Proposition \ref{pro::Stech::DR::PproIncreasing}.


\section*{Acknowledgment}
The authors would like to thank the reviewers for their efforts in reviewing this paper, which are of importance for improving the paper's quality.

\ifCLASSOPTIONcaptionsoff
  \newpage
\fi



%\bibliographystyle{IEEEtran}

%\bibliography{Refnormal/MyCollection_OSD,Refnormal/MyCollection_LP,Refnormal/MyCollection_Survey,Refnormal/MyCollection_Classic,Refnormal/MyCollection_ML,Refnormal/MyCollection_Math}
%\bibliographystyle{IEEEtran}
\bibliographystyle{IEEEtran}
\bibliography{RefAbrv/IEEEabrv,RefAbrv/OSDAbrv,RefAbrv/LPAbrv,RefAbrv/SurveyAbrv,RefAbrv/ClassicAbrv,RefAbrv/MLAbrv,RefAbrv/MathAbrv}

% If you have an EPS/$\mathrm{pdf}$ photo (graphicx package needed) extra braces are
\begin{IEEEbiographynophoto}{Chentao Yue}
(Student Member, IEEE) received his bachelor's degree in information engineering from Southeast University, China, in 2017. He is currently pursuing a Ph.D. degree at the Centre for IoT and Telecommunications, University of Sydney. His major research interests are error control coding, information theory, and wireless communications.
\end{IEEEbiographynophoto}

\begin{IEEEbiographynophoto}{Mahyar Shirvanimoghaddam}
(Senior Member, IEEE) is a Lecturer at Centre for IoT and Telecommunications, The University of Sydney. Prior to this role, he was with The School of Electrical Engineering and Computing, The University of Newcastle as a Research Fellow in Error Control Coding, where he currently holds a conjoint position. He received his Ph.D. in Electrical Engineering from The University of Sydney in 2015 with The University of Sydney Postgraduate Award and Norman I Prize. He received M.Sc. and B.Sc. both in Electrical Engineering with 1st Class Honor in 2010 and 2008, respectively from Sharif University of Technology and University of Tehran. Dr Shirvanimoghaddam was selected as one of the Top 50 Young Scientists in the World by the World Economic Forum in 2018 for his contribution to the 4th Industrial Revolution. His research interests include Coding and Information Theory, Rateless coding, Communication strategies for the Internet of Things, and Information-theoretic approaches to Machine Learning. He received the Best Paper Awards from the 2017 IEEE International Symposium on Personal, Indoor and Mobile Radio Communications (PIMRC). He serves as a Guest Editor for the Journal of Entropy and Transactions on Emerging Telecommunications Technologies. He is a Fellow of the Higher Education Academy.
\end{IEEEbiographynophoto}

\begin{IEEEbiographynophoto}{Branka Vucetic} (Life Fellow, IEEE) is currently an ARC Laureate Fellow and the Director of the Centre of Excellence for IoT and Telecommunications, The University of Sydney, Sydney, NSW, USA. Her current research work is in wireless networks and the Internet of Things. In the area of wireless networks, she works on ultrareliable low-latency communications (URLLC) and system design for millimeter-wave frequency bands. In the area of the Internet of Things, she works on providing wireless connectivity for mission-critical applications. Dr. Vucetic is a Fellow of the Australian Academy of Technological Sciences and Engineering and the Australian Academy of Science.
\end{IEEEbiographynophoto}

\begin{IEEEbiographynophoto}{Yonghui Li} (Fellow, IEEE) received the Ph.D. degree from the Beijing University of Aeronautics and Astronautics in November 2002. From 1999 to 2003, he was affiliated with Linkair Communication Inc., where he held a position of the Project Manager with responsibility for the design of physical layer solutions for the LAS-CDMA system. Since 2003, he has been with the Centre of Excellence in Telecommunications, The University of Sydney, Australia. He is currently a Professor with the School of Electrical and Information Engineering, The University of Sydney. His current research interests include wireless communications, with a particular focus on MIMO, millimeter wave communications, machine to machine communications, coding techniques, and cooperative communications. He holds a number of patents granted and pending in these fields. He was a recipient of the Australian Queen Elizabeth II Fellowship in 2008 and the Australian Future Fellowship in 2012. He received the Best Paper Awards from IEEE International Conference on Communications (ICC) 2014, IEEE PIMRC 2017, and IEEE Wireless Days Conferences (WD) 2014. He also served as a Guest Editor for several special issues of IEEE journals, such as IEEE Journal on Selected Areas in Communications Special Issue on Millimeter Wave Communications. He is also an Editor of IEEE Transactions on Communications and IEEE Transactions on Vehicular Technology.
\end{IEEEbiographynophoto}


%\newpage



%\vfill


\end{document}